%% file: dmclub_v3.tex
\documentclass[a4paper,11pt]{article}
\pdfoutput=1 

\usepackage{jheppub}                
\usepackage[all]{hypcap}            
\usepackage{slashed}                
\usepackage[usenames,table]{xcolor} 
\usepackage{feyndiag}               
\usepackage{parskip}                
\usepackage{multirow}               
\usepackage{array}                  
\usepackage[normalem]{ulem}         
\usepackage[section]{placeins}      
\usepackage{enumitem}               
\usepackage{moresize}               
\usepackage[utf8]{inputenc}         
\usepackage{fancyhdr}               

\newcommand{\subref}[2]{\hyperref[#1]{\ref*{#1}#2}}
\definecolor{Gray}{gray}{0.6}

\graphicspath{{./figures/}}
\title{The Coannihilation Codex}
\author{Michael~J.~Baker,}
\author{Joachim~Brod,}
\author{Sonia~El~Hedri,}
\author{Anna~Kaminska,}
\author{Joachim~Kopp,}
\author{Jia~Liu,}
\author{Andrea~Thamm,}
\author{Maikel~de~Vries,}
\author{Xiao-Ping~Wang,}
\author{Felix~Yu,}
\author[*]{Jos\'e~Zurita\note[*]{Since October 1, 2015 at Karlsruhe Institute of Technology, 76128 Karlsruhe, Germany.}}
\affiliation{PRISMA Cluster of Excellence \& Mainz Institute for Theoretical Physics, Johannes Gutenberg University, 55099 Mainz, Germany}
\emailAdd{micbaker@uni-mainz.de}
\emailAdd{joachim.brod@uni-mainz.de}
\emailAdd{elhed001@uni-mainz.de}
\emailAdd{akaminsk@uni-mainz.de}
\emailAdd{jkopp@uni-mainz.de}
\emailAdd{liuj@uni-mainz.de}
\emailAdd{athamm@uni-mainz.de}
\emailAdd{mdevrie@uni-mainz.de}
\emailAdd{xiaowang@uni-mainz.de}
\emailAdd{yu001@uni-mainz.de}
\emailAdd{jose.zurita@uni-mainz.de}
\preprint{MITP/15-078}
\arxivnumber{1510.03434}

\abstract{\input{sections/abstract_v11}}

\begin{document}

\maketitle
\clearpage
\pagestyle{fancy}


\input{sections/introduction_v16}
\input{sections/classification_v36}
\input{sections/phenomenology_v25}
\input{sections/casestudy_v35}

\input{sections/conclusions_v10}

\input{sections/acknowledgements_v2}

\appendix
\input{sections/appendix_flavor_v9}
\input{sections/appendix_muons_v11}

\bibliographystyle{JHEP}
\bibliography{dmclub}

\end{document}

%% file: sections/abstract_v11.tex
We present a general classification of simplified models that lead to dark matter (DM) coannihilation processes of the form DM + X $\rightarrow$ SM$_1$ + SM$_2$, where X is a coannihilation partner for the DM particle and SM$_1$, SM$_2$ are Standard Model fields.  Our classification also encompasses regular DM pair annihilation scenarios if DM and X are identical.  Each coannhilation scenario motivates the introduction of a mediating particle M that can either belong to the Standard Model or be a new field, whereby the resulting interactions between the dark sector and the Standard Model are realized as tree-level and dimension-four couplings. We construct a basis of coannihilation models, classified by the $SU(3)_C\times SU(2)_L\times U(1)_Y$ quantum numbers of DM, X and M. Our main assumptions are that dark matter is an electrically neutral color singlet and that all new particles are either scalars, Dirac or Majorana fermions, or vectors. We illustrate how new scenarios arising from electroweak symmetry breaking effects can be connected to our electroweak symmetric simplified models.  We offer a comprehensive discussion of the phenomenological features of our models, encompassing the physics of thermal freeze-out, direct and indirect detection constraints, and in particular searches at the Large Hadron Collider (LHC). Many novel signatures that are not covered in current LHC searches are emphasized, and new and improved LHC analyses tackling these signatures are proposed. We discuss how the coannihilation simplified models can be used to connect results from all classes of experiments in a straightforward and transparent way.  This point is illustrated with a detailed discussion of the phenomenology of a particular simplified model featuring leptoquark-mediated dark matter coannihilation.

%% file: sections/introduction_v16.tex
\section{Introduction}
\label{sec:introduction}
Dark matter is a fundamental, outstanding puzzle, and identifying its particle nature will grant us unprecedented access to new sectors and interactions beyond the Standard Model (SM). The breadth of the problem mirrors our extremely limited direct knowledge of the properties of dark matter. The many probes of possible dark matter interactions with the Standard Model, from direct detection experiments~\cite{Cushman:2013zza} to indirect searches~\cite{Bertone:2004pz} to colliders~\cite{Askew:2014kqa}, hold incredible promise for painting a comprehensive picture of its particle nature. As yet, though, the null results remind us that many theoretically motivated dark matter candidates have not been Nature's choice.

A plethora of cosmological and astrophysical probes, including the anisotropies of the cosmic microwave background, the dynamics of galaxy clusters and rotation curves of galaxies, among others, have firmly established that dark matter is cold, non-baryonic and electrically neutral, with a relic density $\Omega h^2 = 0.1198 \pm 0.0026$~\cite{Agashe:2014kda,Ade:2015xua}. The lack of undisputed signals in direct detection experiments looking for scattering of dark matter particles on atomic nuclei~\cite{Cushman:2013zza} as well as constraints on dark matter self-interactions~\cite{Kaplinghat:2015aga} strongly support the hypothesis that dark matter is colorless and uncharged under electromagnetism.

In light of the broad set of experimental probes available, we are compelled to pursue a comprehensive characterization of dark sector physics in order to synthesize the numerous constraints on its possible interactions.  Dark matter direct detection experiments have established strong constraints on elastic scattering cross sections of dark matter particles on nuclei~\cite{Cushman:2013zza}.  Indirect detection experiments, searching for gamma rays, cosmic ray electrons, positrons, antiprotons, and neutrinos, constrain the possible annihilation rates for dark matter in the Universe~\cite{Bertone:2004pz}.  Collider probes for events with large missing transverse energy have also tested dark matter production rates in many distinct final states~\cite{Askew:2014kqa}.

These separate results can only be sensibly combined in the context of concrete theoretical frameworks.  Studies in the context of ultraviolet (UV) complete models like the Minimal Supersymmetric Standard Model (MSSM) allow for the most comprehensive combination of experimental data sets~\cite{deVries:2015hva,Henrot-Versille:2013yma, Bechtle:2012zk, Fowlie:2012im,Cahill-Rowley:2013yla}, but it is usually difficult to generalize their results to other models. Effective field theories (EFTs), on the other hand, offer highly model-independent results, but their applicability is more restricted, especially in high energy processes like dark matter (DM) production at the Large Hadron Collider (LHC). There, the particles mediating dark sector--SM interactions might be produced on-shell, so that a description in terms of contact operators is likely to fail.  As center of mass energies at colliders grow, the contact operator invites ever higher levels of scrutiny.  A good compromise between model independence and accuracy is provided by simplified models~\cite{Alves:2011wf}, in which the sector connecting the Standard Model and the DM particle is explicitly modeled, albeit in a highly simplified way.

The goal of this paper is to provide a systematic classification of dark matter simplified models, applicable to cosmology as well as direct, indirect and collider searches for DM.  Our special focus is modeling the dark matter annihilation mechanism, including the possibility of coannihilation~\cite{Griest:1990kh}. We assume the DM abundance in the Universe is determined by thermal freeze-out, but we allow for the next-to-lightest dark sector particle X to be close in mass to the DM particle, so that each DM--DM, X--X, and DM--X (co)annihilation process can be important in determining the DM relic density.  The inclusion of coannihilation leads to an increase in the complexity of the simplified models and opens up many new possibilities for the dark sector field content and its phenomenology. We emphasize, however, that the generality of our approach guarantees that at least one of these simplified models is realized in Nature given our assumptions. The full enumeration of all of these simplified models is the first main result of this work.

Having established our framework for setting up simplified models of dark matter coannihilating to the Standard Model particles, we also discuss the general phenomenology of such models, exploring the connections between collider probes and direct and indirect detection strategies. Part of our focus is on the many new channels for production of dark sector particles at the LHC, either on-shell or off-shell. As many search channels have overlapping regions of sensitivity, our work provides a comprehensive framework for interpreting a future positive signal in one channel in connection with results from other channels.  Making such connections will be essential in verifying any experimental hints and for ultimately painting a comprehensive picture of the dark sector.

The literature offers an extensive portfolio of studies classifying dark sector--SM interactions, although the possibility of coannihilation has not been considered in most of these works (see, however,~\cite{Bell:2013wua} and~\cite{Izaguirre:2015zva}).  Of particular importance are works in the context of EFTs.  For instance, the physics relevant to DM--nucleon scattering in direct detection experiments can be completely captured by a non-relativistic EFT~\cite{Fan:2010gt, Fitzpatrick:2012ix, Fitzpatrick:2012ib, Hill:2013hoa, Hill:2014yka, Catena:2014uqa,Hisano:2015bma}.  When relating direct detection results to other probes of DM interactions, it is more convenient to work with manifestly Lorentz-covariant effective operators, and this has been standard practice in the field for decades~\cite{Jungman:1995df, Kopp:2009qt, DelNobile:2013sia, Buckley:2013jwa}.  In mapping UV-complete models onto the low-energy EFT, it is desirable to include renormalization group effects~\cite{Freytsis:2010ne, Haisch:2013uaa, Crivellin:2014qxa, Crivellin:2014gpa, DEramo:2014aba}.

At relativistic energies, the space of possible effective operators opens up considerably, and significant effort has gone into classifying certain subsets of them.  For instance, refs.~\cite{Goodman:2010qn, Cheung:2010ua, Cheung:2011nt, Rajaraman:2012db, Rajaraman:2012fu, Gustafsson:2013gca, Alves:2014yha} study operators relevant to specific indirect DM searches, while ref.~\cite{Cheung:2012gi} offers a global fit including also direct searches, collider constraints and cosmological limits. For instance, classifications based on additional assumptions on the underlying model have been presented for fermionic DM in~\cite{Krauss:2013wfa}, for scalar DM in~\cite{DelNobile:2011uf}, for DM coupling to gauge bosons or Higgs bosons in~\cite{Cotta:2012nj, Chen:2013gya, Fedderke:2013pbc}, for self-conjugate DM annihilating through $s$-channel interactions in~\cite{DeSimone:2013gj}, for asymmetric DM models in~\cite{MarchRussell:2012hi}, and for DM with spin 3/2 in~\cite{Ding:2012sm, Ding:2013nvx}. Particularly comprehensive classifications of models are also presented in~\cite{Cheung:2012gi, Duch:2014xda}.

In the context of collider searches, EFTs have been widely used to study the mono-jet~\cite{Beltran:2010ww, Goodman:2010yf, Bai:2010hh, Goodman:2010ku, Fox:2011pm,Khachatryan:2014rra,Aad:2015zva} and mono-photon~\cite{Fox:2011fx, Bartels:2012ui, Bartels:2012ex, Dreiner:2012xm, Chae:2012bq, Khachatryan:2014rwa,Aad:2014tda} final states.  Of course, at LHC energies the contact operator approximation may break down if the particle mediating dark sector--SM interactions has a mass around or below the typical partonic center of mass energies of the LHC~\cite{Fox:2011pm, Busoni:2013lha, Buchmueller:2013dya, Busoni:2014haa, Busoni:2014sya, Racco:2015dxa, Profumo:2013hqa}.  Therefore, in the recent literature simplified models are gaining in importance~\cite{Alves:2011wf, An:2013xka, DiFranzo:2013vra, deSimone:2014pda, Abdallah:2014dma, Buckley:2014fba, Harris:2014hga, Garny:2015wea, Abdallah:2015ter}. The usefulness of EFTs and simplified models is illustrated by studies of higher order QCD corrections to DM production at the LHC~\cite{Haisch:2012kf, Fox:2012ru, Haisch:2013ata, Backovic:2015soa}.  These studies show that higher order effects can be very important, but computing them for every single UV-complete model is clearly impractical.

The structure of this paper is as follows.  In section~\ref{sec:classification}, we motivate our minimal assumptions about the particle nature of dark matter and present our framework for dark matter simplified models.  Our framework is central to understanding the full breadth of possible experimental signatures for thermal dark matter particles.  In section~\ref{sec:phenomenology}, we take our set of dark matter simplified models and explore the general phenomenology of direct, indirect, and collider probes.  We adopt a leptoquark-mediated DM coannihilation case study for detailed analysis in section~\ref{sec:leptoquark}, emphasizing searches motivated by probing coannihilation that have not yet been considered in the dark matter context as well as novel LHC signatures, such as single leptoquark resonances in combination with missing transverse energy.  In appendix~\ref{app:flavor}, we give an extended discussion of indirect flavor probes for the leptoquark mediator case study.  In appendix~\ref{app:muons}, we present collider prospects for our case study with second generation leptoquark couplings.  We conclude in section~\ref{sec:conclusion}.

%% file: sections/classification_v36.tex
\section{Classification of dark matter simplified models}
\label{sec:classification}

\subsection{Building the framework}
\label{subsec:building}
Our coannihilating dark matter framework is built on the following assumptions:
\begin{itemize}
\item Dark matter is a colorless and electrically neutral particle.\footnote{Models where dark matter has tiny fractional electric charge are still viable, see refs.~\cite{Gabrielli:2015hua,Profumo:2015oya} and references therein.}
\item Dark matter is a thermal relic.
\item Dark matter (co)annihilation proceeds via a two-to-two process.
\item Interaction vertices are realized via tree-level, dimension-four Lagrangian terms.
\item New particles have spin 0, 1/2, or 1, where spin 1 particles are massive vectors. 
\item All gauge bosons follow the minimal coupling provision~\cite{Peskin:1995ev}.
\end{itemize}
With these assumptions, the dark matter field transforms under the Standard Model $SU(3)_C \times SU(2)_L \times U(1)_Y$ gauge groups as $(1, N, \beta)$, with the restriction that $\beta = 2k + 1 - N$, $k \in \{0, 1, \ldots, N-1 \}$, to ensure one $SU(2)_L$ component is electrically neutral after electroweak symmetry breaking (EWSB).\footnote{We define $Q \equiv T_3 + \tfrac{1}{2} Y$, where $Q$ is the electric charge, $T_3$ the third component of weak isospin, and $Y$ the hypercharge of the multiplet.}  We then iterate over all possible SM$_1$ and SM$_2$ combinations to determine the possible quantum numbers for the coannihilation partner X.  Our conventions for the SM field gauge charges are shown in table~\ref{tab:SMcharges}.  Note that we do not include right-handed neutrinos in our Standard Model fields.  If right-handed neutrinos exist in a given model, we treat them as new physics fields.  Moreover, we conduct our classification in the unbroken phase of electroweak symmetry, as this readily allows an informative and detailed understanding of the underlying dynamics of each simplified model.  In particular, by working in the unbroken phase, we can identify and isolate the necessary interactions for a dark matter coannihilation diagram to exist and treat the ramifications from EWSB separately.

\begin{figure}
	\centering
	\input{diagrams/coannihilationchannels}
	\caption{Coannihilation channels: \textbf{(a)} indicates the general process for DM X $\to$ SM$_1$ SM$_2$, where the specific processes are shown in \textbf{(b)} for $s$-channel, \textbf{(c)} for $t$-channel and \textbf{(d)} for four-point interactions. The arrows on the external and internal lines denote the flow of charge under the Standard Model gauge groups. In these diagrams the line style does not indicate a particular Lorentz nature of the fields involved and we allow for all Lorentz invariant assignments.}
	\label{fig:coannihilationchannels}
\end{figure}
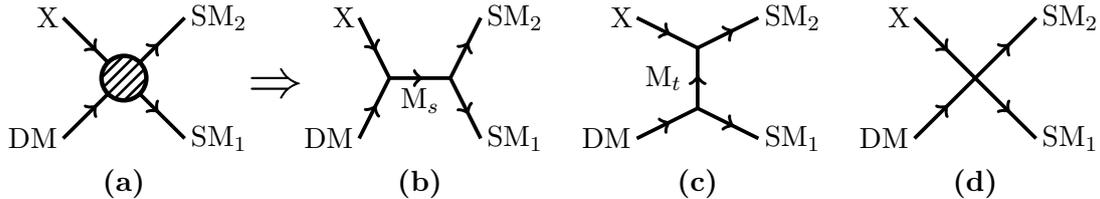

Having determined the entire set of possible Standard Model gauge representation assignments for X, DM, and their requisite SM$_1$ and SM$_2$ coannihilation products, we then explicitly resolve coannihilation diagrams with an $s$-channel mediator M$_s$ or a $t$-channel mediator M$_t$, as shown in figure~\ref{fig:coannihilationchannels}.  Note that arrows on the external and internal lines in figure~\ref{fig:coannihilationchannels} correspond to the flow of charge under the Standard Model gauge groups, which will be relevant for the quantum number assignments of fields in the simplified models to be presented in section~\ref{subsec:tables}.  We also note that the usual DM pair annihilation diagrams are included in our approach when X $\equiv$ DM.

\begin{table}
	\centering
	\footnotesize
	\input{tables/sm_charges_v2}
	\caption{Standard Model fields with $SU(3)_C \times SU(2)_L \times U(1)_Y$ charges specified. The electric charge is defined as $Q = T_3 + \tfrac{1}{2} Y$.}
	\label{tab:SMcharges}
\end{table}

This procedure adds at most three new fields (DM, X, M$_s$) or (DM, X, M$_t$) to the Standard Model and also comprises an exhaustive construction for the most well-motivated field content and corresponding coannihilation diagrams to test via dark matter probes.  Our final step is to specify the spin assignment of each new field: DM, X, and M$_s$ or M$_t$.  As previously mentioned, we will only consider spin assignments of 0, 1/2, or 1, and we will assume spin 1 particles are massive vectors.  Clearly, this presumes an understanding of the UV completion of our simplified models involving a Higgs mechanism or a strongly coupled sector, which generally requires more field content than is strictly necessary from the (co)annihilation diagram construction. Separately, although we define our simplified models in the unbroken phase of electroweak symmetry, we can readily translate our results to account for EWSB.  This will be addressed in detail in section~\ref{subsec:HiggsBreaking}.

We acknowledge that assuming (co)annihilating thermal relic dark matter precludes numerous interesting possibilities, such as asymmetric dark matter or more complicated dark matter scattering topologies~\cite{Petraki:2013wwa, Hochberg:2014dra}. Even these types of models might require coannihilation, however, if the relic density overshoots the measured value. For instance, in asymmetric DM scenarios, coannihilation could be pivotal for the efficient annihilation of the symmetric component of DM.  In such cases, our classification of simplified coannihilation models is still useful, even though the connection to the DM relic density is diluted and requires UV physics to become manifest. If our classification assumptions are satisfied in Nature, however, our framework explores new ground in studying coannihilation models, illuminating new connections between astrophysical probes and collider probes of dark matter. In particular, we emphasize that, given our assumptions, the dark matter field content of Nature is guaranteed to belong to at least one of the simplified models we consider.

While our prescription thus far is sufficient to detail the procedure for generating simplified models for dark matter coannihilation, it does not address many model-building subtleties or the issue of phenomenological viability. To this end, we augment our procedure with several detailed comments demonstrating that our framework gives an exhaustive set of viable and interesting simplified models.

\subsubsection{Benefits and drawbacks of simplified models}
\label{subsubsec:simplified}

The modern era of dark matter searches at colliders was inaugurated by a series of papers discussing effective operators for DM interactions with SM fields~\cite{Goodman:2010yf,Bai:2010hh,Goodman:2010ku,Fox:2011pm,Fox:2011fx}. While we can perform the same style of analysis by defining four-particle operators based on diagram figure~\subref{fig:coannihilationchannels}{(a)}, we eschew this approach in favor of simplified models~\cite{Alves:2011wf}.  Our reasons are manifold:
\begin{itemize}
	\item The numerous possible charge and spin assignments for $s$-channel and $t$-channel mediators, combined with the dark matter particle and its coannihilation partner, provide a complete characterization of dark sector scattering signatures, given our assumptions;
	\item Simplified models offer a consistent description of physics both for large and small mediator masses. In the former limit, EFTs would provide a more economical description and would moreover allow us to easily include loop-induced DM--SM couplings. We sacrifice these benefits in favor of describing both heavy and light dark sectors in a single framework.
	\item The explicit addition of the mediator leads to a much richer phenomenology than the inclusion of only X and DM and allows for a direct probe of the mediation mechanism;
	\item If new particles are found at the LHC or another experiment, our approach will allow for the interpretation of such particles in the (co)annihilating dark matter simplified model context.
\end{itemize}

We remark that in contrast to the spirit of ref.~\cite{Alves:2011wf}, our simplified model constructions do not originate as a minimalistic version of a UV-complete model, but rather as the minimal field content that allows a coannihilation diagram to exist. A complete dark matter model could have additional particles and interactions beyond the ones specified by our framework.  However, we stress that our simplified models will encapsulate in an economical manner the main phenomenological features associated with two-to-two coannihilation of dark sector particles to SM particles.

\subsubsection{Degeneracies in field content, dark matter decay, and dark sector parity}
\label{subsubsec:decay}

Our prescription for constructing explicit DM, X, and M$_s$ or M$_t$ matter content allows for degeneracies within the set of new fields as well as role reversal.  The fact that Standard Model quantum number assignments can be duplicated implies that our framework allows for DM pair annihilation whenever X $\equiv$ DM.  While models featuring DM pair annihilation are certainly more minimal and sufficient to account for all observations to date~\cite{Cirelli:2005uq, An:2013xka, Papucci:2014iwa}, we will focus on situations where X and DM are distinct fields, which give rise to interesting and novel collider signatures.

Permutations of our $s$-channel and $t$-channel models can also exhibit role reversal, where, for example, the X$_s$ coannihilation partner in one $s$-channel construction takes the role of the $t$-channel mediator M$_t$ and vice-versa. Thus, when we consider the possible Lagrangian terms involving the new fields, we can have X$_s$--DM--M$_s$, M$_s$--SM--SM, DM--M$_t$--SM, and X$_t$--M$_t$--SM interactions.  If we identify X$_s \equiv$ M$_t$ and M$_s \equiv$ X$_t$, then the simultaneous presence of all four interactions would lead to DM decay via DM $\to$ M$_t$ + SM, M$_t \to$ X$_t$ + SM, X$_t \equiv$ M$_s \to$ SM + SM.  While this decay width may be sufficiently suppressed to be phenomenologically viable on a case-by-case basis, we will instead appeal to a technically natural parity that distinguishes $s$-channel and $t$-channel coannihilation and protects against DM decay.  Namely, the $s$-channel interactions group together X$_s$ and DM as dark sector particles that can carry an odd $\mathbb{Z}_2$ charge, while M$_s$ and all SM fields are $\mathbb{Z}_2$ even. On the other hand, $t$-channel interactions would assign X$_t$, DM, and M$_t$ all to be $\mathbb{Z}_2$ odd.  Thus, in cases where role-reversal can interchange the mediators and coannihilation partners, the different topologies for the coannihilation mechanism can motivate specific two-dimensional slices of the four-dimensional coupling parameter space.  These two-dimensional slices exhibit an extra $\mathbb{Z}_2$ parity which prevents DM decay, as long as the DM is the lightest $\mathbb{Z}_2$ odd particle, and the $\mathbb{Z}_2$ parity ensures that $s$-channel and $t$-channel simplified models can be considered independently.  In the remainder of this paper, we will therefore assume the existence of the $\mathbb{Z}_2$ parity.

\subsubsection{New gauge and flavor symmetries}
\label{subsubsec:additional}

Additional gauge symmetries or horizontal symmetries generates further complications for our classification.  Needless to say, extra gauge groups, including gauge unification, and new flavor symmetries have each played a pivotal role for model building in the last decades and thus we comment on both.

We first consider additional gauge symmetries.  Our procedure groups models by their Standard Model gauge quantum numbers.  It is possible, however, to consider extra gauge groups, such as a dark photon model, or embedding the Standard Model gauge group into a grand unified theory.  Although we allow for extra vectors, we are agnostic about the corresponding charge assignments for the particles in our models.  Any UV embedding of the Standard Model, such as the embedding of SM multiplets into multiplets of $SU(5)$, generically introduces new fields in larger representations and also dictates concrete relations between couplings and field content.  These extra restrictions can be considered as a motivation for focusing on particular regions of parameter space, but we lose no generality by neglecting them.

We next discuss complications from both dark sector and Standard Model flavor symmetries.  We note that the dark sector particles or the mediators might indeed require some protection from the abundance of Standard Model flavor violation bounds in both the quark and lepton sectors.  To be safe from flavor constraints, we assume the simplest possible flavor structure, namely flavor universality, where appropriate.  If a different formulation is required, these symmetries can be reintroduced as additional ingredients at the Lagrangian level on a case by case basis.  For example, in models where the full set of Lagrangian couplings could induce dangerous proton decay operators, the additional imposition of a global $U(1)_{B-L}$ symmetry would be very beneficial to the phenomenological viability of such models. A complete discussion of all possible flavor aspects of each simplified model is beyond the scope of this work.  We will discuss relevant flavor aspects in the context of our concrete leptoquark case study in section~\ref{sec:leptoquark} and appendix~\ref{app:flavor}.

\subsubsection{Minimal coupling and gauge bosons}
\label{subsubsec:egg}
Simplified models with SM gauge bosons have interesting additional restrictions in our construction.  We remind the reader that we work in the unbroken phase of electroweak symmetry (we will extensively address the complications arising from EWSB in section~\ref{subsec:HiggsBreaking}).  Allowing at most dimension-four interactions and assuming minimal coupling, gauge bosons couple to fermions and scalars only via the covariant derivatives in Lagrangian kinetic terms, and interactions between spin-1 vectors in the unbroken phase result from field strength tensors.

For $s$-channel models, these working assumptions imply two selection rules that restrict the possibilities for the $s$-channel mediator:
\begin{enumerate}[leftmargin=4em]
	\item[(S1)] If either SM$_1$ (SM$_2$) is a SM gauge boson, then the mediator is a SM field and coincides with SM$_2$ (SM$_1$).  In this case, the simplified model only introduces at most two new fields, X and DM, and one new interaction, DM--X--SM.
	\item[(S2)] The mediator cannot be the SM gluon since DM is uncolored.
\end{enumerate}
These selection rules restrict the set of possible simplified models for dark matter coannihilation.  In particular, our procedure for defining the possible Standard Model gauge quantum numbers for X and M$_s$ or M$_t$ based on the Standard Model representations for DM, SM$_1$, and SM$_2$ can produce diagrams where Standard Model singlet contractions of interacting fields can be found but the minimal coupling provision is violated.  As an example, we can contract a color octet, triplet, and sextet to obtain a color singlet.  If the color octet field were the SM gluon, however, such an interaction could not occur via a kinetic term, violating the minimal coupling assumption.  As a result, such hypothetical models are removed.  Some of these models, especially those with final state electroweak gauge bosons, are recovered by considering the effects of EWSB, which we discuss in section~\ref{subsec:HiggsBreaking}.

We have analogous selection rules for $t$-channel models:
\begin{enumerate}[leftmargin=4em]
	\item[(T1)] If either SM$_1$ or SM$_2$ is a SM gauge boson, then the mediator M$_t$ is the same field as DM or X, respectively.  In this case, the only new vertex is again DM--X--SM.  These models coincide with the $s$-channel models defined by selection rule (S1).
	\item[(T2)] The SM$_1$ particle cannot be a gluon because DM is uncolored.
\end{enumerate}

We see that the selection rules (S1) and (T1) create a simpler subclass of coannihilation diagrams where the new physics content is characterized by two fields, DM and X, and one tree-level interaction between DM, X, and a SM particle, which we will label SM$_3$.  These more minimal constructions can be completed into both $s$-channel and $t$-channel coannihilation diagrams by using Standard Model gauge vertices, which result from the Standard Model gauge charges of DM and X.  We will call such models {\it hybrid models}, recognizing that both $s$-channel and $t$-channel coannihilation diagrams can be realized with only a subset of the fields and interactions generated by our framework.  Yet we retain the full flexibility of our complete classification by also extending these hybrid models to $s$-channel and $t$-channel constructions by adding a third new physics field and a second interaction vertex.  These hybrid models will be discussed in section~\ref{subsubsec:hybrid}.  A further consequence of having external gauge bosons is the fact that their couplings are restricted to be $g_s$, $g$ or $g'$ (multiplied by charge) for gluons, $W^i$, and $B$, respectively, which makes explicit the fact that hybrid models only require one new physics vertex.

\subsection{Catalog of simplified models}
\label{subsec:tables}

We now present the first main result of our work, namely the classification of all possible simplified models that arise if dark matter is a thermal relic, coannihilates with a new field X and satisfies the assumptions listed in the beginning of this section.  As discussed in section~\ref{subsec:building}, the $s$-channel and $t$-channel simplified models construct coannihilation diagrams with up to three new fields, DM, X, and M and two new couplings.  As alluded to in section~\ref{subsubsec:egg}, however, selection rules (S1) and (T1) lead to simpler coannihilation diagram constructions, which we call hybrid models, with only two new fields, X and DM, and one new coupling, DM--X--SM$_3$.  We will present hybrid models in section~\ref{subsubsec:hybrid}, $s$-channel models in section~\ref{subsubsec:schannel}, $t$-channel models in section~\ref{subsubsec:tchannel}, and briefly discuss four-point coannihilation interaction models in section~\ref{subsubsec:fourpoint}.  We note that the $\mathbb{Z}_2$ parity assignments for DM, X, and M$_s$ or M$_t$, as discussed in section~\ref{subsubsec:decay}, allow us to factorize the discussion of model content in this manner.

For a given set of quantum numbers for X, our tables will present all the allowed possibilities for the mediator M$_s$ or M$_t$ and the Standard Model fields SM$_1$, SM$_2$, as a function of the DM quantum numbers, $(1, N, \beta)$ and the hypercharge $\alpha$ of X.  Furthermore, we list the spins of the particles in a compact way, and we account for additional interactions beyond the coannihilation diagrams that are allowed by the Standard Model gauge charges of each model.

\subsubsection{Hybrid simplified models of DM coannihilation}
\label{subsubsec:hybrid}
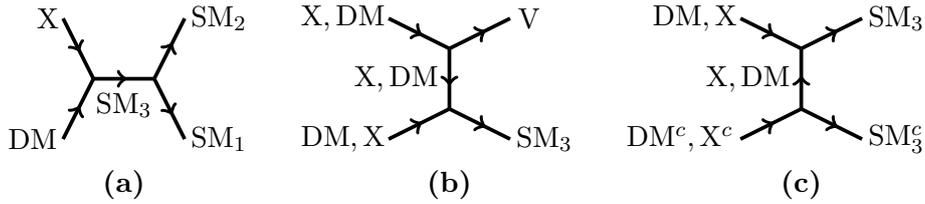
\begin{figure}[!ht]
	\centering
	\input{diagrams/coannihilationchannels_hybrid}
        \caption{Hybrid coannihilation models where there are both $s$-channel and $t$-channel diagrams. These are depicted for the $s$-channel in diagram \textbf{(a)} and for the $t$-channel in diagram \textbf{(b)}.  Diagram \textbf{(c)} accounts for the possibility of DM and X pair annihilation.  In the diagrams the label V represents a SM gauge boson ($g$, $W_i$ or $B$). As in figure~\ref{fig:coannihilationchannels}, the arrows indicate charge flow under the Standard Model gauge group.}
	\label{fig:coannihilationchannels:hybrid}
\end{figure}

In some models DM and X can directly couple to a SM particle, which we label SM$_3$. In this case coannihilation can proceed without the involvement of a mediator and the simplified model content is reduced to X and DM fields only. In the case of $s$-channel coannihilation this corresponds to the Standard Model mediators covered by selection rule (S1). The SM$_3$ particle will mediate coannihilation via its interactions with SM$_1$ and SM$_2$. For the $t$-channel models we require the existence of the SM$_3$ field as well as a coupling of either DM or X with one SM gauge boson. This coupling is always available as long as DM or X are not pure gauge singlets. These models correspond to the models removed from the list of pure $t$-channel models by selection rule (T1). The possible coannihilation processes are illustrated in figure~\ref{fig:coannihilationchannels:hybrid}.

In these models, the DM--X--SM$_3$ interaction vertex generates both $s$-channel and $t$-channel coannihilation diagrams and are therefore named hybrid models. In table~\ref{tab:classification:hybrid}, we classify the field content for DM and X for the hybrid models and indicate which SM field is involved. Each model has a unique identification tag of the form ``H'' plus model ID number, which is shown in the first column.  The second column lists possible Standard Model representations for the coannihilation partner X.  In correspondence with the conventions established in figure~\subref{fig:coannihilationchannels}{(b)}, the DM, X, and SM$_3$ hypercharges are $\beta$, $\alpha$, and $\alpha + \beta$, respectively.  Since DM is not colored, X carries the same color charge as SM$_3$. The spin assignments for the new particles are also determined by SM$_3$.  A bosonic SM$_3$ can couple to either two fermionic or two bosonic dark sector particles, while for a fermionic SM$_3$, one of DM or X has to be a fermion. For all of these hybrid models, it is still possible to add a new mediator field with the same quantum numbers as SM$_3$, which yields an $s$-channel or $t$-channel model and richer phenomenology than the hybrid model alone.  The $s$-channel and $t$-channel models connected to the hybrid models in this way are listed as ``Extensions'' in the last column of the table. In addition to coannihilation, the DM--X--SM$_3$ interaction also generates pair annihilation channels for DM and X via the diagram presented in figure~\subref{fig:coannihilationchannels:hybrid}{(c)}.

\begin{table}[!tb]
	\centering
	\tiny
	\input{tables/classification_hybrid_v3}
        \caption{Hybrid simplified models where coannihilation is mediated via both $s$-channel and $t$-channel processes, see figure~\ref{fig:coannihilationchannels:hybrid}. The representation of the dark matter particle DM is $(1, N, \beta)$. ``Extensions'' indicates which $s$-channel and $t$-channel models could be created from the given hybrid model by adding a mediator.}
	\label{tab:classification:hybrid}
\end{table}

\subsubsection{\texorpdfstring{$s$}{s}-channel simplified models of DM coannihilation}
\label{subsubsec:schannel}
We present our classification of $s$-channel coannihilating simplified models in tables~\ref{tab:classification:s-channel:1},~\ref{tab:classification:s-channel:3}, and~\ref{tab:classification:s-channel:8ex}.  In every table, the DM field has Standard Model quantum numbers $(1, N, \beta)$.  In most of the cases, the conjugate of a single model will give a new set of fields but with the same phenomenology.  Hence, for the sake of brevity, we do not include such a model in our classification. We organize these tables according to the color charges of X and M$_s$, which coincide since DM is uncolored.  Table~\ref{tab:classification:s-channel:1} lists the possibilities with X as a color singlet, table~\ref{tab:classification:s-channel:3} corresponds to X being a color triplet, and table~\ref{tab:classification:s-channel:8ex} shows X as a color octet or sextet.  No other color charge possibilities for X exist, given our assumption about two-to-two scattering to SM particles.

\begin{table}[!tb]
	\centering
	\tiny
	\input{tables/classification_schannel_1_v5}
	\caption{List of possible models which give rise to coannihilation diagrams in the $s$-channel, for DM representation $(1,N,\beta)$ and X a color singlet. In most of the cases conjugating a single row will lead to a different model with similar phenomenology.  Gray shaded entries denote models with fermionic mediators, while unshaded entries represent models with bosonic mediators.}
	\label{tab:classification:s-channel:1}
\end{table}

Each $s$-channel model has a unique identification tag starting with ``SU'' (for $s$-channel uncolored), ``ST'' ($s$-channel triplet), ``SO'' ($s$-channel octet), and ``SE'' ($s$-channel exotic, which correspond to color representations not in the Standard Model), which is shown in the first column.  The second column lists possible Standard Model representations for the coannihilation partner X. The possible values of $\alpha + \beta$ are discretized as a result of the final coannihilation products SM$_1$ and SM$_2$ and are listed in the third column.  The corresponding $s$-channel mediators are given in the fourth column, where a superscript $N \geq 2$ denotes the requirement that the DM cannot be an $SU(2)_L$ singlet.

\begin{table}[!tb]
	\centering
	\tiny
	\input{tables/classification_schannel_3_v4}
	\caption{Same as table~\ref{tab:classification:s-channel:1}, where X is a fundamental of $SU(3)_C$. }
	\label{tab:classification:s-channel:3}
\end{table}

In the $s$-channel, the spin structure of the diagram can be characterized by the spin of the mediator, namely bosonic (B) or fermionic (F), which is shown in the ``Spin'' column.  A final state with one SM fermion and one SM boson can only be reached via a fermionic mediator, due to Lorentz invariance.  All other cases require a bosonic M$_s$.  We note that bosonic mediators allow DM and X either to be both bosons or fermions. This distinction can have important consequences for the low-energy DM scattering and annihilation cross sections, which are probed by direct and indirect searches, respectively.  The magnitude of these cross sections depends strongly on the Lorentz structure of the dark sector--SM interaction. At colliders, however, the distinction between bosons and fermions is much less important and will not alter the general character of the relevant signatures (see section~\ref{sec:phenomenology}).  It will, however, alter the precise details of kinematic distributions, production cross sections, and branching ratios.  Throughout this paper, we will not distinguish between vector bosons and scalars unless only one of these spin states is allowed.  Models with scalars and vectors will share the same classification of collider signatures.  

The possible SM coannihilation products are listed in the ``$(\text{SM}_1 \, \text{SM}_2)$'' column.  These pairings play a critical role in dictating the collider signatures of DM, X, and M$_s$, as we will discuss in detail in section~\ref{sec:phenomenology}.  The last two columns in tables~\ref{tab:classification:s-channel:1},~\ref{tab:classification:s-channel:3}, and~\ref{tab:classification:s-channel:8ex} list additional vertices allowed by the gauge quantum numbers of DM, X, and M$_s$.  If a direct coupling of the form DM--X--SM is possible, the model can be associated to one of the hybrid models identified in~\ref{subsubsec:hybrid}.  The corresponding model is listed in the DM--X--SM$_3$ column of the tables. Separately, we examine whether two X fields (and/or their conjugates) can directly couple to the mediator.  Depending on the charges, this can be realized for X--X, $\overline{\text{X}}$--X or $\overline{\text{X}}$--$\overline{\text{X}}$.  We do not differentiate these possibilities explicitly, but if such a coupling is allowed, we indicate the possibility with a check mark ($\checkmark$) and give the allowed values of $\alpha$.

\begin{table}[!tb]
	\centering
	\tiny
	\input{tables/classification_schannel_ex_v4}
	\caption{Same as table~\ref{tab:classification:s-channel:1}, where X is an adjoint or an exotic of $SU(3)_C$.}
	\label{tab:classification:s-channel:8ex}
\end{table}

\subsubsection{\texorpdfstring{$t$}{t}-channel simplified models of DM coannihilation}
\label{subsubsec:tchannel}
We show the possible $t$-channel coannihilation simplified models in tables~\ref{tab:classification:t-channel:1},~\ref{tab:classification:t-channel:3}, and~\ref{tab:classification:t-channel:8ex}.  These tables are organized in the same way as the corresponding $s$-channel tables, where $t$-channel uncolored mediators are denoted ``TU'' and shown in table~\ref{tab:classification:t-channel:1}, color triplet mediators are tagged ``TT'' and are shown in table~\ref{tab:classification:t-channel:3}, and color octet mediators and exotic colored mediators are named ``TO'' and ``TE'', respectively, and are combined in table~\ref{tab:classification:t-channel:8ex}.  These model identifiers are listed in the first column of each table.  The second column lists the Standard Model gauge quantum numbers for X, which always has hypercharge $\alpha$.  As with $s$-channel models, the DM representation is $(1, N, \beta)$, and the possible hypercharge sums $\alpha + \beta$ are shown in the third column.  Each corresponding possibility for the $t$-channel mediator is shown in the fourth column.

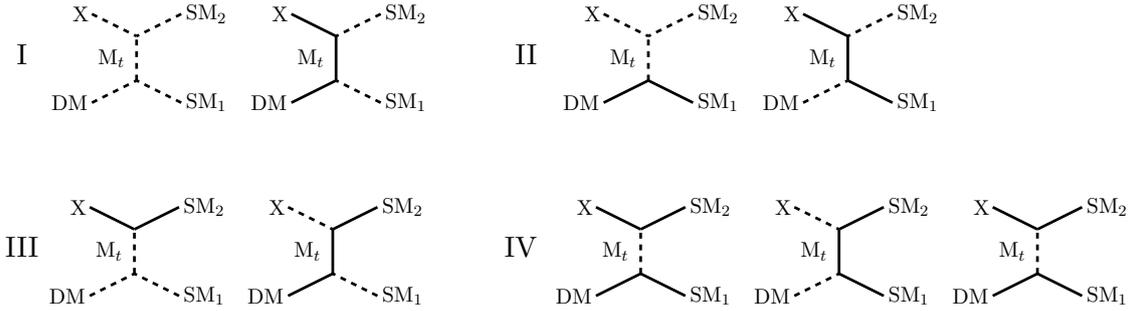
\begin{figure}[!tb]
	\input{diagrams/spinassignment_tchannel_v2}
	\caption{Different dark sector spin assignments for the $t$-channel processes. Dashed lines represent bosons (scalars and vectors), whereas solid lines represent fermions (either Dirac or Majorana).}
	\label{fig:spinassignment:t-channel}
\end{figure}

For $t$-channel models, there are four different possible classes for spin assignments of DM, X, and M$_t$.  We depict these classes in figure~\ref{fig:spinassignment:t-channel} (ordered according to the spins of the SM particles) and indicate the class for each simplified model in the ``Spin'' column of the tables.  As with the $s$-channel models, we do not differentiate between scalars and vectors, which are both represented as dashed lines in figure~\ref{fig:spinassignment:t-channel}.  The ``$(\text{SM}_1 \ \text{SM}_2)$'' column shows the possible SM pairings that complete each coannihilation diagram.

In $t$-channel models, DM, X, and M$_t$ are all odd under the assumed $\mathbb{Z}_2$ dark sector parity.  Hence, the only possible new dark sector--SM coupling beyond those used in the coannihilation diagram is the direct interaction DM--X--SM.  As for $s$-channel models, if this coupling is allowed, the model can be associated to a hybrid model, which is shown in the last column of tables~\ref{tab:classification:t-channel:1},~\ref{tab:classification:t-channel:3}, and~\ref{tab:classification:t-channel:8ex}.

\begin{table}[!tb]
	\centering
	\tiny
	\input{tables/classification_tchannel_1_v2}
	\caption{List of possible $t$-channel coannihilation simplified models, with DM as $(1,N,\beta)$ and X is uncolored.  We do not explicitly write the charge conjugate model for the sake of brevity.  Different cell shadings are used to differentiate the four spin structures shown in figure~\ref{fig:spinassignment:t-channel}.}
	\label{tab:classification:t-channel:1}
\end{table}

\begin{table}[!tb]
	\centering
	\tiny
	\input{tables/classification_tchannel_3_v2}
	\caption{Same as table~\ref{tab:classification:t-channel:1}, where X is a triplet of $SU(3)_C$.}
	\label{tab:classification:t-channel:3}
\end{table}

\begin{table}[!tb]
	\centering
	\tiny
	\input{tables/classification_tchannel_ex_v2}
	\caption{Same as table~\ref{tab:classification:t-channel:1}, where X is an octet or an exotic representation of $SU(3)_C$.}
	\label{tab:classification:t-channel:8ex}
\end{table}

\subsubsection{Four-point interactions}
\label{subsubsec:fourpoint}
Dark matter coannihilation can also proceed via a new four-point interaction.  For this discussion, we only focus on new physics interactions: even though Standard Model gauge interactions of new fields can also mediate coannihilation processes, such vertices are implicitly included in the Lagrangians for the simplified models presented above.  From figure~\ref{fig:coannihilationchannels}, we note that for each DM X $\to$ SM$_1$ SM$_2$ process in the $s$-channel or $t$-channel, the quantum numbers of the involved particles allow also for a four-point interaction, which in principle can contribute to or even dominate the coannihilation cross section.  Since we restrict ourselves to tree-level interactions, however, only a subset of our models will give rise to dimension-four vertices.  Such interactions arise from the kinetic terms for gauge bosons and scalars and from scalar quartic interactions, which ensures then that these vertices must consist of an even number of scalars and an even number of vectors. If the coannihilation partner X is colored, then its kinetic interactions do not involve DM.  Moreover, if X is a colored scalar, then there are no renormalizeable four-point couplings of the form DM--X--SM$_1$--SM$_2$, since there are no colored scalars in the Standard Model.  Hence, the only simplified coannihilation models where four-point coannihilation can play a role are those with an uncolored coannihilation partner X. The models leading to four-point coannihilation vertices will belong to two main categories.

\paragraph{Four-point interactions with one or two SM gauge bosons} If only one SM gauge boson is involved, the model has to involve new couplings between a dark sector gauge boson and a $W$ or $B$ boson. Such couplings imply that the electroweak gauge group arises from the breaking of a larger gauge group and will involve, in particular, extended Higgs scalar multiplets. If two SM gauge bosons are involved, DM and X could also both be scalars. In this case, due to selection rule (S1), the dark matter and its coannihilation partner would have to belong to the same multiplet. In both types of models, the existence of the four-point interactions will lead to vertices of the form DM--X--$W,B$ and/or DM--X--$H$, corresponding to hybrid models \hyperref[tab:classification:hybrid]{H1} and \hyperref[tab:classification:hybrid]{H3}.

\paragraph{Four-point interactions with two Higgs fields} Such interactions arise in all ``SU'' and ``TU'' models where SM$_1$ and SM$_2$ are both $H$ or $H^\dagger$.  Models where SM$_1 = H$, SM$_2 = H^\dagger$ are \hyperref[tab:classification:s-channel:1]{SU1}, \hyperref[tab:classification:s-channel:1]{SU3}, \hyperref[tab:classification:s-channel:1]{SU14}, \hyperref[tab:classification:t-channel:1]{TU1}, and \hyperref[tab:classification:t-channel:1]{TU26}, while models where SM$_1 = $ SM$_2 = H^\dagger$ are \hyperref[tab:classification:s-channel:1]{SU5}, \hyperref[tab:classification:s-channel:1]{SU7}, \hyperref[tab:classification:s-channel:1]{SU16}, \hyperref[tab:classification:t-channel:1]{TU9}, and \hyperref[tab:classification:t-channel:1]{TU31}.  We can realize a variant of such models by discarding the $s$-channel or $t$-channel mediator.  These four-point models still have a (co)annihilation diagram but with just DM and X.

\subsection{The broken phase of electroweak symmetry and Higgs-induced mixing}
\label{subsec:HiggsBreaking}
The models shown in tables~\ref{tab:classification:hybrid}--\ref{tab:classification:t-channel:8ex} are constructed in the unbroken phase of electroweak symmetry.  We now address the complications involved in taking our framework and incorporating EWSB.

In the Standard Model, electroweak symmetry is broken by the vacuum expectation value (vev) of the Higgs field. In the broken phase of electroweak symmetry, the Higgs vev thus acts as a spurion for each Higgs insertion needed in order to realize an electroweak symmetric interaction.  Even in multi-Higgs doublet models, it is always possible to choose a basis in which only one Higgs doublet carries a vev. Even in such models, it is therefore sufficient to treat $H$ and $H^\dagger$, for $H \sim (1, 2, 1)$ as the SM Higgs field, as the basic building blocks for counting $SU(2)_L \times U(1)_Y$ breaking. While more complicated EWSB patterns are possible and extensively researched, we will adopt the simple prescription that the SM Higgs field and its conjugate are sufficient to characterize the impact of EWSB.

In this section we will focus entirely on the new features in our classification introduced by EWSB. In particular, we will not comment on EWSB features already present in the Standard Model, such as $h W^+ W^-$, $h ZZ$, $hhh$ interactions and Dirac mass terms for SM fermions.

First, in section~\ref{subsubsec:modelsafterEWSB}, we will address how our models in tables~\ref{tab:classification:hybrid}--\ref{tab:classification:t-channel:8ex} transmute in the broken phase of electroweak symmetry.  Second, in sections~\ref{subsubsec:modelsrequireEWSB}--\ref{subsubsec:modelsrequireEWSBsummary}, we will discuss models that exist only thanks to EWSB and have no direct counterpart in the tables.

\subsubsection{The simplified models after electroweak symmetry breaking}
\label{subsubsec:modelsafterEWSB}

Electroweak symmetry breaking effects on our simplified models are fourfold.  First, after EWSB, the DM, X, and M$_s$ or M$_t$ fields generally devolve into nearly mass-degenerate multiplets of particles that differ by units of electromagnetic (EM) charge.  These EM charges $Q$ are prescribed by our convention $Q = T_3 + \frac{1}{2} Y$, where $T_3$ is the third component of weak isospin and $Y$ is the hypercharge of the multiplet.  One-loop electroweak corrections generally split the masses of the particles by a few hundred MeV, typically leaving the neutral particle, if present, as the lightest component~\cite{Thomas:1998wy, Cirelli:2005uq, Hill:2011be, Hill:2014yka}.  Since the resulting mass splittings are relatively small, we can treat all members of the DM, X and M multiplets as mass-degenerate when computing production rates at the LHC. A characteristic class of signatures arising from EWSB at the LHC is the decay of the heavier members of the new multiplets to the lighter ones, for instance $\text{DM}^{\pm n} \to \text{DM}^{\pm (n-1)} + W^*$, where $n$ denotes the electric charge.  Thanks to the small mass splitting between $\text{DM}^{\pm n}$ and $\text{DM}^{\pm (n-1)}$, the off-shell $W$ boson will manifest itself predominantly as a charged pion~\cite{Cirelli:2005uq}. The typical decay length of $\text{DM}^{\pm n}$ is of the order of centimeters, thus disappearing charged tracks will serve as an additional smoking gun signature of an extended DM multiplet.

Second, Higgs insertions also induce mixing between particles.  We note that, following the discussion of section~\ref{subsubsec:decay}, DM and X are $\mathbb{Z}_2$ odd in $s$-channel models, while M$_s$ and all SM fields are $\mathbb{Z}_2$ even.  Hence, in $s$-channel models, the Higgs vev can induce mixing between DM and X, or M$_s$ and SM fields.  In $t$-channel models, DM, X, and M$_t$ are all $\mathbb{Z}_2$ odd, and thus the only novel mixing effects from the Higgs vev occur in the dark sector.  In addition to the $\mathbb{Z}_2$ parity, Higgs-induced mixing is controlled by the spin assignments of the constituent fields.  For fermions, such mixing will appear in the simplified model Lagrangian as a Yukawa coupling, while mixing between scalars will appear as cubic or quartic scalar interactions. If the Higgs and SM fermions are charged under a new gauge symmetry, the new interactions might induce mixing between a SM and a new gauge boson, e.g.~$Z$--$Z'$ mixing.

As an example, model \hyperref[tab:classification:s-channel:1]{SU2} in table~\ref{tab:classification:s-channel:1} has DM $\sim (1, N, \beta)$, X $\sim (1, N, -\beta)$, and a fermionic M$_s \sim (1, 1, 0)$.  Here, we can identify M$_s$ as a right-handed neutrino, and the M$_s$--SM$_1$--SM$_2$ coupling in the coannihilation diagram in the unbroken electroweak phase is simply the well-known neutrino Yukawa coupling $y_{\nu} \tilde{H} \overline{L_L} \text{ M}_s$, where $\tilde{H} \equiv i \sigma^2 H^*$, which leads to a Dirac neutrino mass term.  After EWSB, the mediator mixes with the SM neutrinos.  If the mediator is purely Dirac, the active SM neutrinos supplant the $s$-channel mediator. On the other hand, if the mediator is Majorana, it can have a mass at the electroweak scale and possibly lead to electroweak gauge boson--lepton resonances in collider searches.  In this model, the fermionic nature of the mediator M$_s$ implies that one of X and DM is a fermion and the other is a boson, and hence the components of these electroweak multiplets cannot mix.  As a result, in the broken phase, model \hyperref[tab:classification:s-channel:1]{SU2} with a Majorana mass for M$_s$ exhibits coannihilation of DM and X via a heavy Majorana neutrino into $W^\pm \ell^\mp$, $Z \nu$, and $h \nu$ final states.

A third consequence of EWSB is related to four-point vertices involving the Higgs field. Such vertices are in general allowed in our simplified model Lagrangians.  These four-point interactions involve new physics couplings that are not dictated by the coannihilation diagram. As such, these couplings are very model-dependent and should be considered on a case-by-case basis whenever they arise.

Finally, there is the possibility that electroweak symmetry is broken not only by the SM Higgs field, but also by one of the new fields acquiring a vev. Note that the only field for which this is possible is an uncolored $s$-channel mediator M$_s$. All other new fields are charged under the dark sector $\mathbb{Z}_2$ parity, the breaking of which would lead to unacceptable DM decay.  If M$_s$ acquires a vev, this will on the one hand lead to mixing between the DM and X fields. A simple rediagonalization of the DM--X mass matrix can be used to absorb that effect. On the other hand, an M$_s$ vev leads to new mass terms as well. The associated phenomenology is identical to that of models with extended Higgs sectors, for instance two Higgs doublet models.

\subsubsection{Simplified models that require EWSB}
\label{subsubsec:modelsrequireEWSB}

\begin{figure}[!tb]
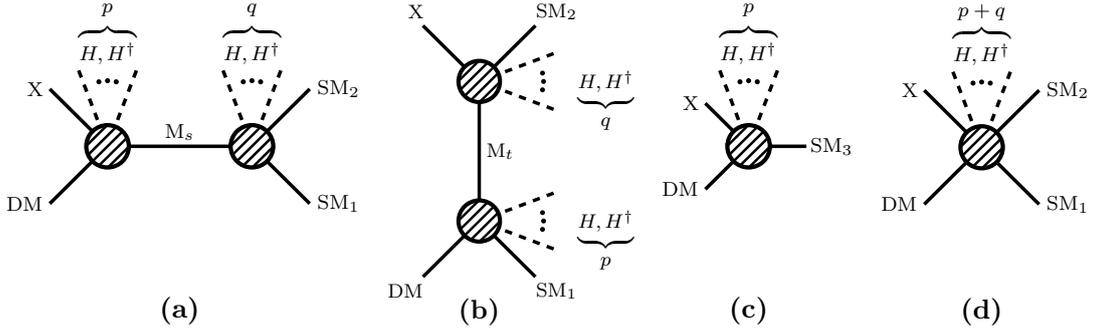

	\centering
	\include{diagrams/externalvevs_v3}
        \caption{Simplified models that admit two-to-two coannihilation only after EWSB.  We characterize the {\bf (a)} $s$-channel, {\bf (b)} $t$-channel, {\bf (c)} hybrid, and {\bf (d)} four-point models of this type by effective vertices with a total of $p+q$ Higgs insertions.}
	\label{fig:externalvevs}
\end{figure}

We now tackle simplified models for coannihilation that explicitly require EWSB.  In this section we demonstrate that for these models, all of the phenomenology arising from vertices in their coannihilation diagrams is identical to that coming from models in tables~\ref{tab:classification:hybrid}--\ref{tab:classification:t-channel:8ex}, up to group theory factors and mixing angles.  The only major exceptions are mixing between SM fields and new physics fields, discussed in section \ref{subsubsec:modelsrequireEWSBs-channel}, and certain $t$-channel models with coannihilation to gauge bosons. Simplified models that require EWSB are built from sets of fields that are distinct from those simplified models we specified in tables~\ref{tab:classification:hybrid}--\ref{tab:classification:t-channel:8ex}. They are still characterized by at most three new fields, DM, X and M$_s$ or M$_t$, but now these fields do not admit a two-to-two coannihilation diagram at tree level in the unbroken phase of electroweak symmetry.  However, such a diagram can be realized in the broken phase of electroweak symmetry when an electroweak symmetric diagram with Higgs insertions (see figure~\ref{fig:externalvevs}) is reduced by replacing the Higgs insertions by vevs. In this section, we will treat EWSB by $H$ and $H^\dagger$ as equivalent in order to simplify the discussion.

Allowing for an arbitrary number $p+q$ of Higgs insertions, as shown in figure~\ref{fig:externalvevs}, we will now repeat our classification procedure from section~\ref{subsec:building}.  Recall that in section~\ref{subsec:building} we found fields which satisfied
\begin{align}
  \begin{aligned}
    \pmb{n}_{\mathrm{X}} \otimes \pmb{n}_{\mathrm{DM}} \otimes \pmb{n}_{\mathrm{SM}_1} \otimes \pmb{n}_{\mathrm{SM}_2} &\supseteq \mathbf{1}\,, \\[0.2cm]
    Y_\mathrm{X}+Y_\mathrm{DM} - Y_{\mathrm{SM}_1} - Y_{\mathrm{SM}_2} &= 0\,,
  \end{aligned}
  \label{eq:unbrokensu2conditionX}
\end{align}
where $\pmb{n}$ denotes the $SU(2)_L$ representation and $Y$ the hypercharge.  Our goal, as before, is to characterize all possible coannihilation diagrams by the two new fields DM and X in hybrid or four-point models, or the three new fields DM, X and M$_s$ or M$_t$ in $s$-channel or $t$-channel models.  We show the corresponding coannihilation topologies in the electroweak symmetric phase in figure~\ref{fig:externalvevs}.  We can characterize coannihilation models that require EWSB by considering the effective vertices shown in this figure.  For $p+q$ Higgs insertions, each of which contributes a $\mathbf{2}$ of $SU(2)_L$, $\pmb{n}_{\mathrm{X}}$ and $\pmb{n}_{\mathrm{DM}}$ must now satisfy
\begin{align}
\label{eq:unbrokensu2conditionXt}
	\pmb{n}_\mathrm{X} \otimes \mathbf{2}^{p+q} \otimes \pmb{n}_{\mathrm{DM}} \otimes \pmb{n}_{\mathrm{SM}_1} \otimes \pmb{n}_{\mathrm{SM}_2} &\supseteq \mathbf{1} \,
\end{align}
and hypercharge must be conserved.

Our goal in the following sections is to show how effective models requiring EWSB, defined in terms of DM, X and M$_s$ or M$_t$, are related to the ones constructed in tables~\ref{tab:classification:hybrid}--\ref{tab:classification:t-channel:8ex}. In particular, we will show that in a broad category of cases the effective models illustrated in figure~\ref{fig:externalvevs} have UV completions whose fundamental interactions are among the ones we have already enumerated.  This is because the effects of the Higgs insertions can in many cases be encapsulated by field relabelings of DM, SM$_1$, and SM$_2$, allowing us to fold the EWSB effect into the representation of X.  The close connection between models that do and do not require EWSB implies that much of the phenomenology of these two classes of models is shared.  Of course, the UV completions of the diagrams in figure~\ref{fig:externalvevs} are not unique, and for any given valid set of DM, X and M$_s$ or M$_t$ fields, UV completions can be constructed that are not captured by our tables and have phenomenological consequences beyond those offered by our simplified models. We will also characterize the general structure of this category of UV completions.

\subsubsection{\texorpdfstring{$s$}{s}-channel models that require EWSB}
\label{subsubsec:modelsrequireEWSBs-channel}

\begin{figure}[!tb]
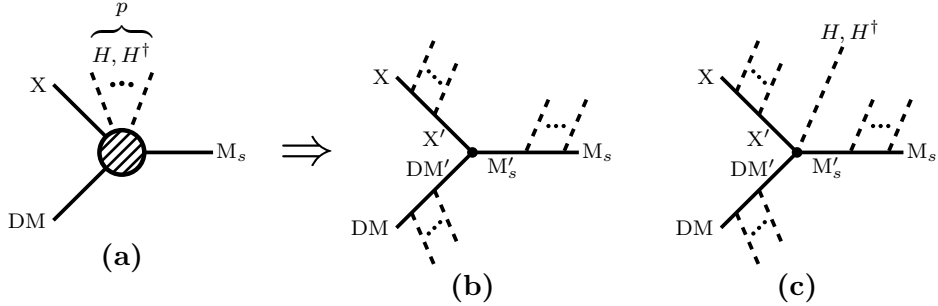

	\centering
	\include{diagrams/higgs_s_leftvertex_v2}
	\caption{Resolving the $s$-channel left effective vertex in figure~\subref{fig:externalvevs}{(a)} as tree-level vertices with Higgs insertions denoting EWSB.}
	\label{fig:Higgs_s_leftvertex}
\end{figure}

\begin{figure}[!tb]
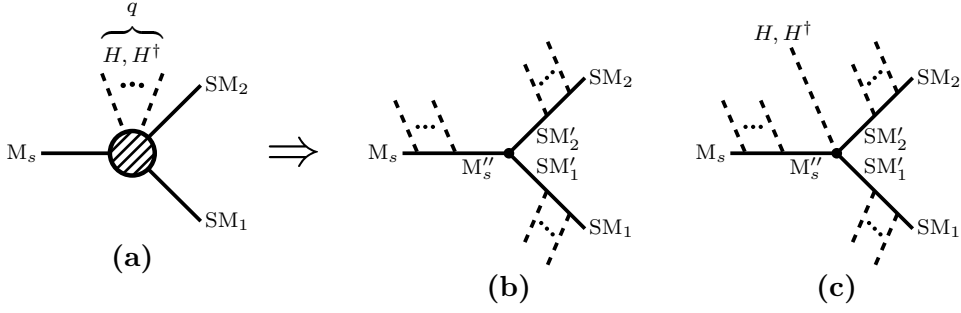

	\centering
	\include{diagrams/higgs_s_rightvertex_v2}
	\caption{Resolving the $s$-channel right effective vertex in figure~\subref{fig:externalvevs}{(a)} as tree-level vertices with Higgs insertions denoting EWSB.}
	\label{fig:Higgs_s_rightvertex}
\end{figure}

We begin by analyzing effective models with an $s$-channel coannihilation diagram, see figure~\subref{fig:externalvevs}{(a)}.  To construct the most general UV completion of such an effective model, we resolve the two effective vertices.  Considering first the dark sector vertex that couples DM, X and M$_s$, we note that we can attach the Higgs insertions to the DM leg, the X leg, the M$_s$ leg, or directly to the vertex, as depicted in figure~\ref{fig:Higgs_s_leftvertex}.  We can also resolve the right blob of figure~\subref{fig:externalvevs}{(a)} in a similar fashion, with Higgs insertions on the M$_s$ leg, SM$_1$ leg, SM$_2$ leg, or directly on the vertex, as shown in figure~\ref{fig:Higgs_s_rightvertex}. Note that the Higgs insertions on the external legs can be either three-point or four-point interactions, depending on the Lorentz nature of the external legs.  We implicitly assume the couplings for these Higgs insertions to be perturbative. We will denote the innermost vertices in figure~\subref{fig:Higgs_s_leftvertex}{(b)--(c)}, coupling DM$'$--X$'$--M$_s'$, and the innermost vertices in figure~\subref{fig:Higgs_s_rightvertex}{(b)--(c)}, coupling M$_s''$--SM$_1'$--SM$_2'$, the \emph{microscopic} vertices, as opposed to the effective \emph{macroscopic} vertices coupling DM--X--M$_s$ and M$_s$--SM$_1$--SM$_2$.

Let us now demonstrate that, among the many possible UV completions, there is at least one that is captured by tables \ref{tab:classification:hybrid}--\ref{tab:classification:t-channel:8ex}.  To this end, recall that almost all pairwise combinations of SM$_1$ and SM$_2$ were included in these tables.  Only scenarios where SM$_1$ or SM$_2$ are gauge bosons were forbidden.  When electroweak symmetry is broken, though, the possibility that SM$_1$ or SM$_2$ (or both) are electroweak gauge bosons cannot be discarded anymore.  Let us first assume that SM$_1$, SM$_2$ are not gauge bosons.  Then, given SM$_1$ and SM$_2$ in figure~\ref{fig:Higgs_s_rightvertex}, we can look up a corresponding mediator M$_s''$ such that the direct coupling M$_s''$--SM$_1$--SM$_2$ is possible in the electroweak symmetric phase.  Then, the $SU(2)_L$ gauge singlet condition on the macroscopic vertex,
\begin{align}
  \pmb{n}_{\mathrm{M}_s} \otimes \mathbf{2}^{q} \otimes
  \pmb{n}_{\mathrm{SM}_1} \otimes \pmb{n}_{\mathrm{SM}_2}  &\supseteq \mathbf{1} \ ,
\end{align}
can be satisfied by a UV completion where all $q$ Higgs insertions show up on the M$_s$--M$_s''$ leg, and the Standard Model gauge representations of M$_s$ and M$_s''$ are related by $q$ Higgs insertions.  The blob in figure~\subref{fig:Higgs_s_rightvertex}{(a)} is thus resolved in figure~\subref{fig:Higgs_s_rightvertex}{(b)} with $\text{SM}_1' = \text{SM}_1$ and $\text{SM}_2' = \text{SM}_2$.

If SM$_1$ and SM$_2$ are electroweak gauge bosons, an effective 3-point vertex coupling M$_s$--SM$_1$--SM$_2$ can be constructed if M$_s''$ is identified with the Higgs boson, and an extra Higgs insertion is added on the vertex, as in figure~\subref{fig:Higgs_s_rightvertex}{(c)}. Again, $q$ Higgs insertions on the mediator leg can be introduced to connect $\text{M}_s'' = H,\,H^\dag$ to M$_s$. Note, however, that the microscopic vertex constructed in this way involves only Standard Model fields. Without loss of generality, we can therefore cut the diagram on the $\text{M}_s'' = H$ line to obtain a new coannihilation diagram with (SM$_1$, SM$_2$) = ($H$, $H$), ($H$, $H^\dagger$), or ($H^\dagger$, $H^\dagger$). Since models with these final states are included in our tables, we can find a corresponding model there and proceed as described in the previous paragraph. Since all of our models are built on top of the Standard Model, all phenomenology associated with the gauge boson--Higgs coupling that we have formally cut away from the coannihilation diagram is of course still included. 

If only one of the SM fields, say SM$_1$, is an electroweak gauge boson, then SM$_1'$ must be this gauge boson since, by our minimal coupling assumptions, Higgs insertions cannot change its identity. Then, if SM$_2$ is a Standard Model fermion $f$, we identify SM$_2'$ and M$_s''$ with this fermion, and Higgs insertions then couple M$_s''$ to M$_s$. We can again cut the diagram on the M$_s''$ line to obtain a new diagram where SM$_1 = f$ and SM$_2$ is the Higgs field or its conjugate. 

Finally, if SM$_2$ is the Higgs field itself, we can identify M$_s''$ with the Higgs field as well and cut the diagram on the M$_s''$ line as before. Alternatively, we can identify M$_s''$ with the gauge boson SM$_1$ and use a microscopic 4-point vertex coupling M$_s''$--SM$_1$--SM$_2$--$H$. Thanks to minimal coupling, additional Higgs insertions on the mediator leg will not change the identity of the mediator, hence such models are viable only if $\text{M}_s = \text{M}_s''$. Thus, M$_s$ has to be a Standard Model gauge boson, and such a scenario would not be classified as an $s$-channel model, but as a hybrid model.

In resolving the left blob in figure~\subref{fig:externalvevs}{(a)}, we note that one possible UV completion is obtained by identifying M$_s'$ in figure~\subref{fig:Higgs_s_leftvertex}{(b)} with M$_s''$.  We have already identified a simplified model from tables \ref{tab:classification:s-channel:1}--\ref{tab:classification:s-channel:8ex} with M$_s''$ as its mediator in the previous paragraph when resolving the M$_s$--SM$_1$--SM$_2$ vertex. We therefore choose the dark sector vertex of this model with M$_s''$ as its mediator to be the microscopic DM$'$--X$'$--M$_s'$ vertex in figure~\subref{fig:Higgs_s_leftvertex}{(b)}.  This requires $q$ Higgs insertions on the $\text{M}_s' = \text{M}_s''$ leg to turn M$_s'$ into M$_s$. After EWSB, these Higgs insertions lead to mixing between M$_s$ and M$_s''$. We also need Higgs insertions on the DM$'$ leg to turn DM$'$ into DM, but these can readily be absorbed by simply relabeling DM$'$ as DM.  Again, the Higgs insertions will mix the symmetry eigenstates DM and DM$'$, leading to a physical (mass eigenstate) dark matter particle that is an admixture of DM, DM$'$ and any intermediate fields on the DM leg. The interactions of the physical dark matter particle are related to the interactions of the DM and DM$'$ fields by appropriate insertions of mixing angles and of group theory factors arising from the contractions of the $SU(2)_L$ representations of DM, DM$'$ and the Higgs insertions.  Finally, Higgs insertions will be needed on the $\text{X}'$ leg to turn $\text{X}'$ into $\text{X}$.  Again, these Higgs insertions mix $\text{X}$ and $\text{X}'$, implying that the interactions of the physical coannihilation partner are related to those of $\text{X}'$ by a mixing angle.  

The key point is that we have constructed a possible UV completion for the simplified model defined by DM, X, M$_s$, SM$_1$ and SM$_2$ that is based on a simplified model from tables \ref{tab:classification:s-channel:1}--\ref{tab:classification:s-channel:8ex} with particle content DM$'$, X$'$, M$_s' \equiv$ M$_s''$, SM$_1$ and SM$_2$.  The coannihilation cross section of the former model differs from that of the latter model only by the insertion of simple multiplicative factors, namely mixing angles and group theory factors.  Collider observables in the two models will differ also by mixing angles and group theory factors. Moreover, the different $SU(2)_L \times U(1)_Y$ representations of DM, X and M$_s$ imply that the multiplicities of new particles and the rates for production through gauge interactions will be different.  We emphasize, however, that for this UV completion the tree-level Lagrangian for the coannihilation microscopic interactions will be shared between these models.

Of course, the UV completion constructed in this way is not unique. First, the provision that there are no Higgs insertions on the SM$_1$ and SM$_2$ legs can be relaxed.  If SM$_1'$ and SM$_2'$ are Standard Model fields, we can simply relabel SM$_1'$ into SM$_1$ and SM$_2'$ into SM$_2$ and truncate the diagram at this level.  In the special case that SM$_1'$ (SM$_2'$) is the Higgs field, the diagram leads to mixing between M$_s''$ and SM$_2'$ (SM$_1'$).  We are then left with a hybrid model, the discussion of which we defer to section~\ref{subsubsec:modelsrequireEWSBhybrid}.  On the other hand, SM$_1'$ or SM$_2'$ could be new physics fields, for example fourth generation quarks or leptons mixing with Standard Model fermions through a Higgs insertion. Such a UV completion has no direct correspondence to any of the models in tables~\ref{tab:classification:hybrid}--\ref{tab:classification:t-channel:8ex}, and its phenomenology will typically be richer than that of the models in the tables. In particular, such models involve mixing between Standard Model particles and new fields, with possibly measurable consequences in precision experiments.  Moreover, the new particles SM$_1'$, SM$_2'$ themselves may be produced at the LHC and lead to a rich spectrum of novel final states, in particular because the Higgs boson is among their decay products.  Finally, models in which SM$_1'$, SM$_2'$ are heavy particles outside the Standard Model admit new coannihilation modes to 3-body and 4-body final states, for instance in the process DM X $\to$ SM$_1'$ ($\to H$ + SM$_1$) + SM$_2'$ ($\to H$ + SM$_2$). Here, the intermediate SM$_1'$ and SM$_2'$ particles can be either on-shell or off-shell, depending on their mass.

A further generalization of the UV completion of our effective simplified model arises if the microscopic M$_s''$--SM$_1'$--SM$_2'$ vertex is a four-point interaction (see figure~\subref{fig:Higgs_s_rightvertex}{(c)}).  This is only possible if M$_s''$, SM$_1'$ and SM$_2'$ are bosons since otherwise, an M$_s''$--SM$_1'$--SM$_2'$--$H$ vertex would be higher-dimensional and is therefore not allowed in a UV-complete model.  We are left with four-point interactions of bosonic fields.  If either SM$_1$ or SM$_2$ are gauge bosons, then by our minimal coupling assumption, M$_S''$ must be the SM Higgs, its conjugate, or the same gauge boson, since the microscopic vertex is written in the ultraviolet and all couplings to gauge bosons come from kinetic terms.  The model is then a hybrid model, the discussion of which we defer to section~\ref{subsubsec:modelsrequireEWSBhybrid}. The last possibility is that SM$_1'$ and SM$_2'$ are Higgs fields, and the microscopic vertex is simply a scalar quartic interaction.  This type of vertex is encapsulated by cubic scalar couplings from tables \ref{tab:classification:s-channel:1}--\ref{tab:classification:s-channel:8ex} if we shift the $SU(2)_L \times U(1)_Y$ quantum numbers of the mediator listed in the tables by a Higgs insertion.  We see that, as long as SM$_1'$ and SM$_2'$ are SM fields, the possible realizations for a microscopic M$_s''$--SM$_1$--SM$_2$--$H$ vertex are already described by the vertices from tables~\ref{tab:classification:hybrid}--\ref{tab:classification:s-channel:8ex}.

We are now left with categorizing all possible UV completions of the effective vertex DM--X--M$_s$. As before, Higgs insertions on the three legs of this vertex lead to mixing between DM, X, M$_s$ on the one side and DM$'$, X$'$, M$_s'$ on the other.  If the quantum numbers of M$_s'$ are such that this field allows for couplings to SM particles, the microscopic three-point vertex DM$'$--X$'$--M$_s'$ in figure~\subref{fig:Higgs_s_leftvertex}{(b)} appears in tables~\ref{tab:classification:hybrid}--\ref{tab:classification:s-channel:8ex}.  If the quantum numbers of M$_s'$ do not admit couplings to pairwise combinations of Standard Model particles (for instance, if M$_s'$ is in a very large representation of $SU(2)_L$), this particular UV completion is outside our classification.  If the microscopic vertex is four-point, as in figure~\subref{fig:Higgs_s_leftvertex}{(c)}, the corresponding macroscopic vertex is not directly listed in our tables, but it is encapsulated by a scalar three-point vertex in the tables involving DM$'$, X$'$, and a mediator field with the quantum numbers of M$_s'$, shifted by a Higgs insertion. As before, the necessary requirement is that the shifted quantum numbers admit a coupling to SM particles.

In the broken phase, we can straightforwardly identify the two-to-two coannihilation process DM X $\to$ SM$_1$ SM$_2$.  We will always implicitly assume that this two-to-two process in the broken phase dominates over any possible DM coannihilation to intermediate mediator UV states and external physical Higgs particles.  In this way, the particles participating in the coannihilation process in the broken phase of electroweak symmetry are matched onto fields in the coannihilation vertices in figures~\ref{fig:Higgs_s_leftvertex} and~\ref{fig:Higgs_s_rightvertex}, where all external Higgs insertions in these figures are taken to their vevs.

\subsubsection{\texorpdfstring{$t$}{t}-channel models that require EWSB}
\label{subsubsec:modelsrequireEWSBt-channel}

\begin{figure}[!tb]
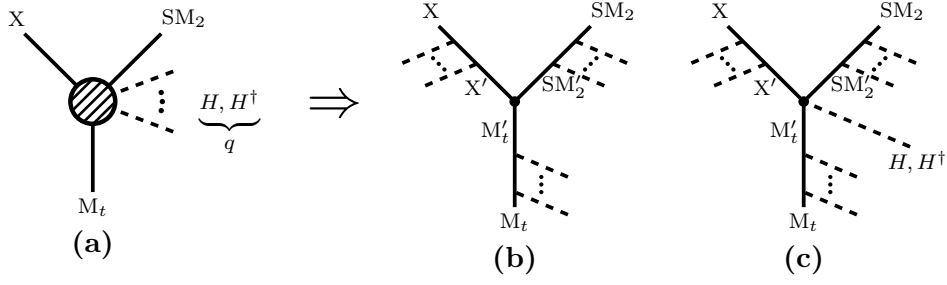

	\centering
	\include{diagrams/higgs_t_uppervertex}
	\caption{Resolving the upper effective vertex in figure~\subref{fig:externalvevs}{(b)} as tree-level vertices with Higgs insertions.}
	\label{fig:Higgs_t_uppervertex}
\end{figure}

\begin{figure}[!tb]
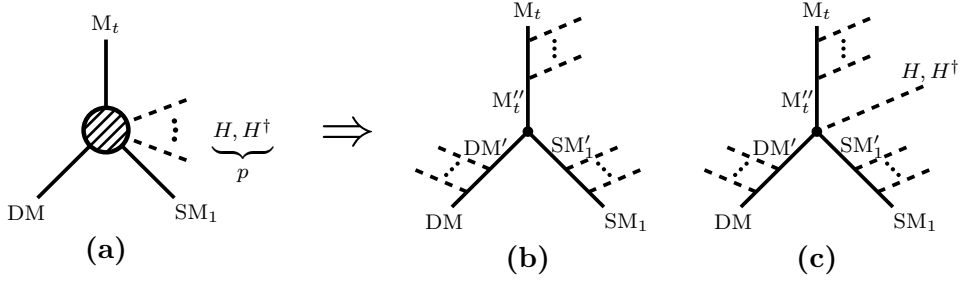

	\centering
	\include{diagrams/higgs_t_lowervertex_v2}
	\caption{Resolving the lower effective vertex in figure~\subref{fig:externalvevs}{(b)} as tree-level vertices with Higgs insertions.}
	\label{fig:Higgs_t_lowervertex}
\end{figure}

A similar procedure as for $s$-channel models requiring EWSB applies also for characterizing $t$-channel coannihilation models that require EWSB.  Again, we define the simplified model by the new fields DM, X, and M$_t$, as shown in figure~\subref{fig:externalvevs}{(b)}.  Since we assume tree-level vertices, the blobs resolve into the microscopic interactions shown in figures~\ref{fig:Higgs_t_uppervertex} and~\ref{fig:Higgs_t_lowervertex}.

We demonstrate again that there is at least one UV completion of any given $t$-channel simplified model requiring EWSB that is captured by tables~\ref{tab:classification:t-channel:1}--\ref{tab:classification:t-channel:8ex}. The only exception is the case where SM$_1$ or SM$_2$ is a gauge boson, which we discuss next. In the case that neither SM$_1$ or SM$_2$ is a gauge boson, we identify the dark matter multiplet of that model with the field DM$'$ in figure~\ref{fig:Higgs_t_lowervertex}, its coannihilation partner with X$'$ in figure~\ref{fig:Higgs_t_uppervertex}, and its $t$-channel mediator with both M$_t'$ and M$_t''$.  Suitable Higgs insertions on the mediator legs connect M$_t'$ and M$_t''$ to M$_t$ for matching onto the effective diagram from figure~\subref{fig:externalvevs}{(b)}.  Similarly, Higgs insertions are included on the external dark sector legs to connect DM$'$ to DM and X to X$'$.  After EWSB, the Higgs insertions lead to mixing of M$_t$ with M$_t'$ and M$_t''$, of DM with DM$'$, and of X with X$'$.  After rediagonalizing the mass matrices of the dark sector fields for the effective model requiring EWSB, defined by DM, X, M$_t$, SM$_1$, SM$_2$, the coannihilation amplitude is then given by the amplitude for the underlying model without EWSB, defined by DM$'$, X$'$, $\text{M}_t' = \text{M}_t''$, SM$_1$, SM$_2$, multiplied by appropriate mixing angles and group theory factors.

Let us now consider the case that SM$_1$ is a gauge boson. Then, also SM$_1'$ must be that same gauge boson, so without loss of generality we can assume that there are no Higgs insertions on the SM$_1$ leg. Now, if M$_t$ and DM are fermions, then  DM$'$ and M$_t'$ must also be fermions, and because of the minimal coupling prescription, we must have $\text{DM}' = \text{M}_t'$. Such a vertex is \emph{not} included in our tables, and we conclude that in this special case, no direct connection can be made between the model requiring EWSB and our tables. The same is true if M$_t$ and DM are scalars. Note that neither M$_t$ or DM can be a gauge boson since this would imply that M$_t'$ or DM$'$ must be identical to SM$_1$, which is forbidden by the $\mathbb{Z}_2$ parity. Completely analogous arguments can be made for the case that SM$_2$ is a gauge boson.

Again, the particular UV completion of the effective diagram in figure~\subref{fig:externalvevs}{(b)} is not unique. Therefore, let us now classify all possible UV completions, as we did for $s$-channel simplified models.  By arguments identical to those given in section~\ref{subsubsec:modelsrequireEWSBs-channel}, allowing for Higgs insertions on the SM$_1$ and SM$_2$ legs does not introduce new effects if SM$_1'$ and SM$_2'$ are Standard Model fields. In the special case that SM$_1'$ (SM$_2'$) is the Higgs, mixing between DM$'$ and M$_t''$ (X$'$ and M$_t'$) is induced. Models of this type still admit the regular $t$-channel coannihilation topology, with SM$_1$ and SM$_2$ in the final state, but after EWSB they also admit the topologies of hybrid models (see figure~\ref{fig:coannihilationchannels:hybrid} and the discussion in section~\ref{subsubsec:modelsrequireEWSBhybrid}).  In less minimal scenarios where SM$_1'$ or SM$_2'$ is not part of the Standard Model, we cannot make a direct connection to tables~\ref{tab:classification:t-channel:1}--\ref{tab:classification:t-channel:8ex}. We note, that in this case the new SM$_1'$ and/or SM$_2'$ fields must be even under the dark sector $\mathbb{Z}_2$ parity.

As a second generalization of the UV completion of figure~\subref{fig:externalvevs}{(b)}, consider the case that the microscopic vertex coupling DM$'$, M$_t''$ and SM$_1$ is a four-point interaction in the unbroken phase of electroweak symmetry, with an extra Higgs insertion on the vertex (see figure~\subref{fig:Higgs_t_lowervertex}{(c)}). In this case, DM$'$, M$_t''$ and SM$_1$ must be scalars because our assumption of minimal coupling and the accidental $\mathbb{Z}_2$ parity prevent SM$_1$ from being a gauge boson. In other words, SM$_1$ must be the Higgs field as well. After EWSB, the vertex thus leads to both DM X $\to H$ SM$_2$ coannihilation, as well as to hybrid model signatures.  The former process has a correspondence to at least one of the models in tables~\ref{tab:classification:t-channel:1}--\ref{tab:classification:t-channel:8ex}, if the quantum numbers of DM are shifted by one Higgs insertion.  The latter process will be discussed in section~\ref{subsubsec:modelsrequireEWSBhybrid} below.  If the upper effective vertex in figure~\subref{fig:externalvevs}{(b)} becomes a four-point interaction at the microscopic level (see figure~\subref{fig:Higgs_t_uppervertex}{(c)}), we can make exactly analogous arguments.

\subsubsection{Hybrid models that require EWSB}
\label{subsubsec:modelsrequireEWSBhybrid}

\begin{figure}[!tb]
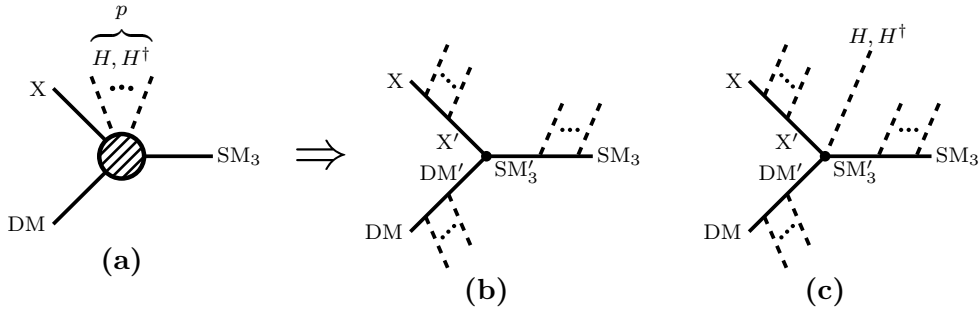

	\centering
	\include{diagrams/higgs_hybrid}
	\caption{Resolving the hybrid vertex in figure~\subref{fig:externalvevs}{(c)} as a tree-level vertex with Higgs insertions.}
	\label{fig:Higgs_hybrid}
\end{figure}

Our last category of simplified models is the hybrid class, where the minimal coannihilation diagram only introduces one new coupling between DM, X, and a SM field, as shown in figure~\ref{fig:Higgs_hybrid}.  To construct a particular UV completion of the effective vertex in figure~\subref{fig:Higgs_hybrid}{(a)}, we choose from table~\ref{tab:classification:hybrid} the model with the same SM$_3$. We then identify the DM--X--SM$_3$ vertex of that model with the microscopic vertex in figure~\subref{fig:Higgs_hybrid}{(b)}.  We add appropriate Higgs insertions on the DM$'$ and X$'$ legs to connect DM$'$ to DM and X$'$ to X. After EWSB, the phenomenological effect of these Higgs insertions can be captured by multiplying the interaction amplitudes of the selected model from table~\ref{tab:classification:hybrid} by appropriate mixing angles and group theory factors.

More general UV completions are readily constructed in analogy to sections~\ref{subsubsec:modelsrequireEWSBs-channel} and \ref{subsubsec:modelsrequireEWSBt-channel}: Higgs insertions on the SM$_3$ line in figure~\ref{fig:Higgs_hybrid} can be absorbed by a simple relabeling of SM$_3'$ into SM$_3$ if SM$_3'$ is a SM field. If it is not, the model falls outside our classification scheme.  If a Higgs insertion is added to the microscopic vertex (figure~\subref{fig:Higgs_hybrid}{(c)}), the minimal coupling prescription for gauge bosons and the accidental dark sector $\mathbb{Z}_2$ symmetry dictate that also SM$_3$ must be a Higgs doublet.  This class of models has been discussed in section~\ref{subsubsec:fourpoint}. Compared to the scenarios described there, the only new features arising from EWSB are possible Higgs insertions of the DM and X legs. As for $s$-channel and $t$-channel models, these merely lead to mixing within the dark sector and are accounted for by mixing angles and group theory factors, which are simple multiplicative factors in the interaction amplitudes.

\subsubsection{Examples of simplified models that require EWSB}
\label{subsubsec:modelsrequireEWSBexamples}

We will now demonstrate by means of two concrete examples how a coannihilation model that requires EWSB can be reduced to a model from tables~\ref{tab:classification:hybrid}--\ref{tab:classification:t-channel:8ex}.

\begin{figure}[!tb]
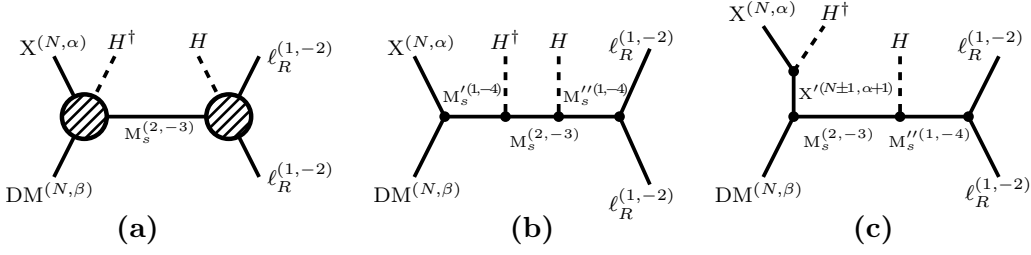

	\centering
	\include{diagrams/higgs_exampleB}
	\caption{Illustrative example for a coannihilation simplified model that requires EWSB.  Note the hypercharges are defined to flow into the vertex for X and DM and out of the vertex for $\ell_R$ and the Higgs insertions.  Each field is labelled with its $SU(2)_L \times U(1)_Y$ representation.}
	\label{fig:Higgs_example_s}
\end{figure}

For our first example, we take the products of coannihilation to be $(\text{SM}_1, \, \text{SM}_2) = (\ell_R, \ell_R)$, where we recall that the quantum numbers of $\ell_{R}$ in our conventions are $(1,1,-2)$.  For the DM particle and its coannihilation partner, we take the representations $\text{DM} \sim (1, N, \beta)$ and $\text{X} \sim (1, N, \alpha)$, with $\alpha + \beta = -4$. We take an $s$-channel mediator with quantum numbers $\text{M}_s \sim (1, 2, -3)$. This model is not present in tables \ref{tab:classification:s-channel:1}--\ref{tab:classification:s-channel:8ex} and requires EWSB to realize a coannihilation diagram. This is illustrated in figure~\subref{fig:Higgs_example_s}{(a)}. Each of the two vertices requires one Higgs insertion to ensure conservation of gauge quantum numbers. If we had chosen the hypercharge of M$_s$ differently, coupling it to SM$_1$ and SM$_2$ via Higgs insertions might or might not be possible. For instance, had we chosen $\text{M}_s \sim (1, 2, -4)$ instead, we would need equal numbers of $H$ and $H^\dagger$ insertions to ensure hypercharge conservation. Even numbers of $SU(2)_L$ doublets can never be contracted into a doublet, so even allowing for EWSB, a model with this combination of M$_s$, SM$_1$ and SM$_2$ cannot be constructed. Coming back to our original choice $\text{M}_s \sim (1, 2, -3)$, we now follow the procedure outlined in section~\ref{subsubsec:modelsrequireEWSBs-channel} to construct one UV completion that is captured by our tables. This UV completion is shown in figure~\subref{fig:Higgs_example_s}{(b)}. We note that, without EWSB, the only $s$-channel simplified model with $(\text{SM}_1, \text{SM}_2) = (\ell_R, \ell_R)$ in the tables is model \hyperref[tab:classification:s-channel:1]{SU9}. This model has the same DM and X representations as we take here, but a different $s$-channel mediator with quantum numbers $(1, 1, -4)$.  We identify this mediator with $\text{M}_s''$ in figure~\subref{fig:Higgs_example_s}{(b)} and include an insertion of a Higgs field $H \sim (1, 2, 1)$ to couple M$_s''$ to M$_s$. We observe that the dark sector vertex in figure~\subref{fig:Higgs_example_s}{(a)} also requires one Higgs insertion to satisfy $SU(2)_L \times U(1)_Y$ conservation.  We can choose to put this Higgs insertion on the mediator leg again and identify $\text{M}_s' \equiv \text{M}_s'' \sim (1, 1, -4)$. Thus, given the quantum numbers of DM, X, M$_s$, SM$_1$ and SM$_2$ specified above, we have constructed a coannihilation model in the broken phase of electroweak symmetry.  From our discussion, it is clear that, up to mixing angles from the mediator sector, all phenomenology depending on the coannihilation diagram vertices is identical to that of model \hyperref[tab:classification:s-channel:1]{SU9}. 

Other UV completions of the effective model shown in figure~\subref{fig:Higgs_example_s}{(a)} are obtained by including Higgs insertions on the DM, X, SM$_1$ or SM$_2$ lines, or directly on the vertices. For instance, in figure~\subref{fig:Higgs_example_s}{(c)} we show a UV completion with Higgs insertions on the X and M$_s$ legs.  The right hand vertex is still the same as in model \hyperref[tab:classification:s-channel:1]{SU9}, while the left hand vertex is present in model \hyperref[tab:classification:s-channel:1]{SU12}.  The phenomenology associated with the coannihilation vertices in this case would be related to that of both  models \hyperref[tab:classification:s-channel:1]{SU9} and  model \hyperref[tab:classification:s-channel:1]{SU12}.

\begin{figure}[!tb]
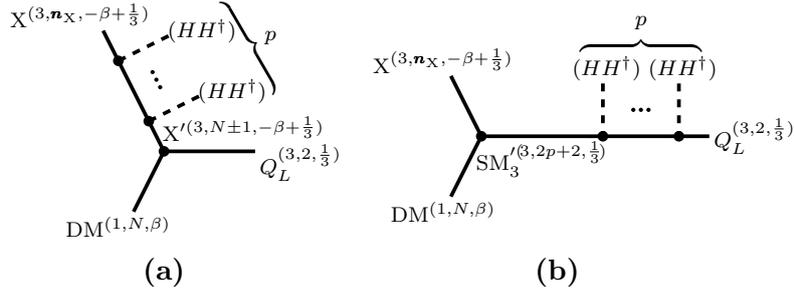

	\centering
	\include{diagrams/higgs_example_hybrid_v2}
	\caption{Illustrative examples for a coannihilation simplified model that requires EWSB, {\bf (a)}, and for one which falls outside of our classification, {\bf (b)}.  The $SU(2)_L$ representation of X, $\pmb{n}_{\mathrm{X}}$, can be any of $\{|N-2p-2|+1,|N-2p-2|+3, \ldots, N+2p+2-1\}$.}
	\label{fig:Higgs_example_h}
\end{figure}

As a second example, we consider a hybrid model with DM in representation $(1,N,\beta)$ and SM$_3 = Q_L$.  With $p$ insertions of ($H H^\dagger$) on the effective vertex, see figure~\subref{fig:Higgs_hybrid}{(a)}, the allowed $SU(2)_L$ representations for X are $\pmb{n}_{\mathrm{X}}\in\{|N-2p-2|+1,|N-2p-2|+3, \ldots, N+2p+2-1\}$.  This construction exemplifies how large $SU(2)_L$ representations can be connected to our classification.  This could be motivated by the fact that a large $SU(2)_L$ multiplet leads to multiply charged particles and to large modifications of loop-induced Higgs decay rates to $\gamma\gamma$ and $Z\gamma$. Moreover, it can lead to enhanced DM DM $\to \gamma \gamma$ annihilation, yielding a smoking gun gamma ray signature.  Furthermore, the members of a large $SU(2)_L$ multiplet have large production rates at the LHC in electroweak processes.  Their cascade decays of the form $\text{X}^{\pm n} \to \text{X}^{\pm (n-1)} + W^{(*)}$ are in principle observable.

Following the discussion of section~\ref{subsubsec:modelsrequireEWSBhybrid} we can connect this model to a hybrid model from table~\ref{tab:classification:hybrid} by making all Higgs insertions on the X leg, as shown in figure~\subref{fig:Higgs_example_h}{(a)}.  The microscopic interaction vertex then contains X$'\sim(3,N\pm1,-\beta+\frac{1}{3})$, so this class of models can be reduced to model \hyperref[tab:classification:hybrid]{H7}.  All phenomenology associated with the effective DM--X--SM$_3$ vertex is thus given by model \hyperref[tab:classification:hybrid]{H7}, up to group theory factors and mixing angles.  Moreover, the model can be extended to an $s$- or $t$-channel model in the same way as model \hyperref[tab:classification:hybrid]{H7}, see the last column of table~\ref{tab:classification:hybrid}.  In this extension either SM$_3$ is identified with the $s$-channel mediator or DM or X$'$ is identified with the $t$-channel mediator.

Alternatively, all of the Higgs insertions could be placed on the SM$_3$ leg, figure~\subref{fig:Higgs_example_h}{(b)}.  For $\pmb{n}_{\mathrm{X}}\neq N\pm1$ this requires the introduction of a new field SM$'_3\sim(3,2p+2,\frac{1}{3})$, which is outside the Standard Model.  The microscopic vertex of this UV completion does not correspond to any of the models in table~\ref{tab:classification:hybrid}, and is thus outside of our classification.

\subsubsection{Summary of simplified models that require EWSB}
\label{subsubsec:modelsrequireEWSBsummary}

In this section, we have shown that, for simplified models of coannihilation that explicitly require EWSB, all of the phenomenology associated with vertices in their coannihilation diagrams is identical to that coming from models in tables~\ref{tab:classification:hybrid}--\ref{tab:classification:t-channel:8ex}, up to group theory factors and mixing angles.  The only exception arises in UV models that cannot be reduced to a two-to-two coannihilation diagram to SM pairs, as discussed in sections~\ref{subsubsec:modelsrequireEWSBs-channel} and~\ref{subsubsec:modelsrequireEWSBt-channel}. For these UV completions, more model dependence from the ultraviolet model enters the coannihilation diagram.  In particular, new physics fields, denoted by SM$_1'$ and SM$_2'$, and a new Lagrangian coupling M$_s''$--SM$_1'$--SM$_2'$, would need to be specified in order to classify the resulting coannihilation process.

%% file: tables/sm_charges_v2.tex
\begin{tabular}{!{\vrule width 1pt} c | c !{\vrule width 1pt} c | c !{\vrule width 1pt}}
	\noalign{\hrule height 1pt}
	field & charges & field & charges \\
	\noalign{\hrule height 1pt}
	$Q_L$ & $(3,2,\tfrac{1}{3})$ & $L_L$ & $(1,2,-1)$ \\
	$u_R$ & $(3,1,\tfrac{4}{3})$ & $\ell_R$ & $(1,1,-2)$ \\
	$d_R$ & $(3,1,-\tfrac{2}{3})$ &  & \\
	\hline
	$\overline{Q_L}$ & $(\bar{3},2,-\tfrac{1}{3})$ & $\overline{L_L}$ & $(1,2,1)$ \\
	$\overline{u_R}$ & $(\bar{3},1,-\tfrac{4}{3})$ & $\overline{\ell_R}$ & $(1,1,2)$ \\
	$\overline{d_R}$ & $(\bar{3},1,\tfrac{2}{3})$ &  & \\
	\hline
	$H$ & $(1,2,1)$ & $H^\dagger$ & $(1,2,-1)$ \\
	\hline
	$g$ & $(8,1,0)$ & $W_i$ & $(1,3,0)$ \\
	$B$ & $(1,1,0)$ & & \\
	\noalign{\hrule height 1pt}
\end{tabular}

%% file: diagrams/coannihilationchannels.tex
\begin{tikzpicture}[line width=1.4pt, scale=1]
	\draw[fermionbar] (0.8,0.8)--(0,0);
	\draw[fermionbar] (0.8,-0.8)--(0,0);
	\draw[fermion] (-0.8,0.8)--(0,0);
	\draw[fermion] (-0.8,-0.8)--(0,0);
	\draw[fill=black] (0,0) circle (3.0mm);
	\draw[fill=white] (0,0) circle (2.9mm);
	\begin{scope}
		\clip (0,0) circle (3.0mm);
		\foreach \x in {-.9,-.75,...,.3}
		\draw[line width=1 pt] (\x,-.3) -- (\x+.6,.3);
	\end{scope}
	\node at (-1.0,0.8) {X};
	\node at (-1.2,-0.8) {DM};
	\node at (1.25,0.8) {SM$_2$};
	\node at (1.25,-0.8) {SM$_1$};
	\node at (0,-1.4) {\textbf{(a)}};
\end{tikzpicture} \hspace{-5mm}
\raisebox{13.4mm}{\begin{tikzpicture}[line width=1.4pt, scale=1]
	\node at (1.8,0.0) {\huge $\Rightarrow$};
\end{tikzpicture}} \hspace{-5mm}
\begin{tikzpicture}[line width=1.4pt, scale=1]
	\draw[fermionbar] (0.8,0.8)--(0.4,0);
	\draw[fermionbar] (0.8,-0.8)--(0.4,0);
	\draw[fermion] (-0.8,0.8)--(-0.4,0);
	\draw[fermion] (-0.8,-0.8)--(-0.4,0);
	\draw[fermion] (-0.4,0)--(0.4,0);
	\node at (-1.0,0.8) {X};
	\node at (-1.2,-0.8) {DM};
	\node at (1.25,0.8) {SM$_2$};
	\node at (1.25,-0.8) {SM$_1$};
	\node at (0,-0.3) {M$_s$};
	\node at (0,-1.4) {\textbf{(b)}};
\end{tikzpicture} \hspace{0mm}
\begin{tikzpicture}[line width=1.4pt, scale=1]
	\draw[fermionbar] (0.8,0.8)--(0,0.4);
	\draw[fermionbar] (0.8,-0.8)--(0,-0.4);
	\draw[fermion] (-0.8,0.8)--(0,0.4);
	\draw[fermion] (-0.8,-0.8)--(0,-0.4);
	\draw[fermion] (0,-0.4)--(0,0.4);
	\node at (-1.0,0.8) {X};
	\node at (-1.2,-0.8) {DM};
	\node at (1.25,0.8) {SM$_2$};
	\node at (1.25,-0.8) {SM$_1$};
	\node at (-0.45,0) {M$_t$};
	\node at (0,-1.4) {\textbf{(c)}};
\end{tikzpicture} \hspace{0mm}
\begin{tikzpicture}[line width=1.4pt, scale=1]
	\draw[fermionbar] (0.8,0.8)--(0,0);
	\draw[fermionbar] (0.8,-0.8)--(0,0);
	\draw[fermion] (-0.8,0.8)--(0,0);
	\draw[fermion] (-0.8,-0.8)--(0,0);
	\node at (-1.0,0.8) {X};
	\node at (-1.2,-0.8) {DM};
	\node at (1.25,0.8) {SM$_2$};
	\node at (1.25,-0.8) {SM$_1$};
	\node at (0,-1.4) {\textbf{(d)}};
\end{tikzpicture}

%% file: diagrams/coannihilationchannels_hybrid.tex
\begin{tikzpicture}[line width=1.4pt, scale=1]
	\draw[fermionbar] (0.8,0.8)--(0.4,0);
	\draw[fermionbar] (0.8,-0.8)--(0.4,0);
	\draw[fermion] (-0.8,0.8)--(-0.4,0);
	\draw[fermion] (-0.8,-0.8)--(-0.4,0);
	\draw[fermion] (-0.4,0)--(0.4,0);
	\node at (-1.0,0.8) {X};
	\node at (-1.2,-0.8) {DM};
	\node at (1.25,0.8) {SM$_2$};
	\node at (1.25,-0.8) {SM$_1$};
	\node at (0,-0.35) {SM$_3$};
	\node at (0,-1.4) {\textbf{(a)}};
\end{tikzpicture} \hspace{2mm}
\begin{tikzpicture}[line width=1.4pt, scale=1]
	\draw[fermionbar] (0.8,0.8)--(0,0.4);
	\draw[fermionbar] (0.8,-0.8)--(0,-0.4);
	\draw[fermion] (-0.8,0.8)--(0,0.4);
	\draw[fermion] (-0.8,-0.8)--(0,-0.4);
	\draw[fermionbar] (0,-0.4)--(0,0.4);
	\node at (-1.4,0.8) {X,\,DM};
	\node at (-1.4,-0.8) {DM,\,X};
	\node at (1.05,0.8) {V};
	\node at (1.25,-0.8) {SM$_3$};
	\node at (-0.7,0) {X,\,DM};
	\node at (0,-1.4) {\textbf{(b)}};
\end{tikzpicture} \hspace{2mm}
\begin{tikzpicture}[line width=1.4pt, scale=1]
	\draw[fermionbar] (0.8,0.8)--(0,0.4);
	\draw[fermionbar] (0.8,-0.8)--(0,-0.4);
	\draw[fermion] (-0.8,0.8)--(0,0.4);
	\draw[fermion] (-0.8,-0.8)--(0,-0.4);
	\draw[fermion] (0,-0.4)--(0,0.4);
	\node at (-1.4,0.8) {DM,\,X};
	\node at (-1.6,-0.8) {DM$^c$,\,X$^c$};
	\node at (1.25,0.8) {SM$_3$};
	\node at (1.25,-0.8) {SM$_3^c$};
	\node at (-0.7,0) {X,\,DM};
	\node at (0,-1.4) {\textbf{(c)}};
\end{tikzpicture}

%% file: tables/classification_hybrid_v3.tex
\begin{tabular}{ !{\vrule width 1pt} c | c | c | c | c !{\vrule width 1pt}} 
	\noalign{\hrule height 1pt}
	
	ID & X & $\alpha+\beta$ & SM$_3$ & Extensions \\
	
	\noalign{\hrule height 1pt}
	
	H1 & \multirow{2}{*}{$(1,N,\alpha)$} & $0$ & $B$, $W_i^{N\geq2}$ & \hyperref[tab:classification:s-channel:1]{SU1}, \hyperref[tab:classification:s-channel:1]{SU3}, \hyperref[tab:classification:t-channel:1]{TU1}, \hyperref[tab:classification:t-channel:1]{TU4}--\hyperref[tab:classification:t-channel:1]{TU8} \\
	
	\cline{1-1} \cline{3-5}
	
	H2 & & $-2$ & $\ell_R$ & \hyperref[tab:classification:s-channel:1]{SU6}, \hyperref[tab:classification:s-channel:1]{SU8}, \hyperref[tab:classification:t-channel:1]{TU10}, \hyperref[tab:classification:t-channel:1]{TU11} \\
	
	\noalign{\hrule height 1pt}
	
	H3 & \multirow{2}{*}{$(1,N\pm1,\alpha)$} & \multirow{2}{*}{$-1$} & $H^\dagger$ & \hyperref[tab:classification:s-channel:1]{SU10}, \hyperref[tab:classification:t-channel:1]{TU18}--\hyperref[tab:classification:t-channel:1]{TU23} \\
	
	\cline{1-1} \cline{4-5}
	
	H4 & & & $L_L$ & \hyperref[tab:classification:s-channel:1]{SU11}, \hyperref[tab:classification:t-channel:1]{TU16}, \hyperref[tab:classification:t-channel:1]{TU17} \\
	
	\noalign{\hrule height 1pt}
	
	H5 & \multirow{2}{*}{$(3,N,\alpha)$} & $\tfrac{4}{3}$ & $u_R$ & \hyperref[tab:classification:s-channel:3]{ST3}, \hyperref[tab:classification:s-channel:3]{ST5}, \hyperref[tab:classification:t-channel:3]{TT3}, \hyperref[tab:classification:t-channel:3]{TT4} \\
	
	\cline{1-1} \cline{3-5}
	
	H6 & & $-\tfrac{2}{3}$ & $d_R$ & \hyperref[tab:classification:s-channel:3]{ST7}, \hyperref[tab:classification:s-channel:3]{ST9}, \hyperref[tab:classification:t-channel:3]{TT10}, \hyperref[tab:classification:t-channel:3]{TT11} \\
	
	\noalign{\hrule height 1pt}
	
	H7 & $(3,N\pm1,\alpha)$ & $\tfrac{1}{3}$ & $Q_L$ & \hyperref[tab:classification:s-channel:3]{ST14}, \hyperref[tab:classification:t-channel:3]{TT28}--\hyperref[tab:classification:t-channel:3]{TT31} \\
	
	\noalign{\hrule height 1pt}
\end{tabular}

%% file: tables/classification_schannel_1_v5.tex
\begin{tabular}{!{\vrule width 1pt} c | c | c | c | c !{\vrule width 1pt} c | c  | c !{\vrule width 1pt}}
	\noalign{\hrule height 1pt}
	
	ID & X & $\alpha+\beta$ & M$_s$ & Spin & (SM$_1$ SM$_2$) & X--DM--SM$_3$ & M$_s$--X--X \\
	
	\noalign{\hrule height 1pt}
	
	\multirow{2}{*}{SU1} & \multirow{10}{*}{$(1,N,\alpha)$} & \multirow{5}{*}{$0$} & \multirow{3}{*}{$(1,1,0)$} & \multirow{2}{*}{B} & $(u_R \, \overline{u_R}), (d_R \, \overline{d_R}), (\ell_R \, \overline{\ell_R})$ &  \multirow{2}{*}{\hyperref[tab:classification:hybrid]{H1}}  & \multirow{2}{*}{$\checkmark$}  \\
	& & & & & $(Q_L \, \overline{Q_L}), (L_L \, \overline{L_L}), (H \, H^\dagger)$ & & \\
	\cline{1-1} \cline{5-8}
	\cellcolor[gray]{0.8}SU2 & & & & \cellcolor[gray]{0.8}F & $(L_L \, H)$ & &  \\
	\cline{1-1} \cline{4-8}
	SU3 & & & \multirow{2}{*}{$(1,3,0)^{N\geq2}$} & B & $(Q_L \, \overline{Q_L}), (L_L \, \overline{L_L}), (H \, H^\dagger)$ & \hyperref[tab:classification:hybrid]{H1} & $\checkmark$ \\
	\cline{1-1} \cline{5-8}
	\cellcolor[gray]{0.8}SU4 & & & & \cellcolor[gray]{0.8}F & $(L_L \, H)$ & & \\
	\cline{1-1} \cline{3-8}
	SU5 & & \multirow{4}{*}{$-2$} & \multirow{2}{*}{$(1,1,-2)$} & B & $(d_R \, \overline{u_R}), (H^\dagger \, H^\dagger), (L_L \, L_L)$ & & $\checkmark$ \\
	\cline{1-1} \cline{5-8}
	\cellcolor[gray]{0.8}SU6 & & & & \cellcolor[gray]{0.8}F & $(L_L \, H^\dagger)$ & \hyperref[tab:classification:hybrid]{H2} & \\
	\cline{1-1} \cline{4-8}
	SU7 & & & \multirow{2}{*}{$(1,3,-2)^{N\geq2}$} & B & $(H^\dagger \, H^\dagger), (L_L \, L_L)$ & & $\checkmark (\alpha=\pm 1)$ \\
	\cline{1-1} \cline{5-8}
	\cellcolor[gray]{0.8}SU8 & & & &  \cellcolor[gray]{0.8}F & $(L_L \, H^\dagger)$ & \hyperref[tab:classification:hybrid]{H2} & \\
	\cline{1-1} \cline{3-8}
	SU9 & & $-4$ & $(1,1,-4)$ & B & $(\ell_R \, \ell_R)$ & &$\checkmark (\alpha=\pm 2)$ \\
	
	\noalign{\hrule height 1pt}
	
	SU10 & \multirow{4}{*}{$(1,N\pm1,\alpha)$} & \multirow{2}{*}{$-1$} & \multirow{2}{*}{$(1,2,-1)$} & B & $(d_R \, \overline{Q_L}), (\overline{u_R} \, Q_L), (\overline{L_L} \, \ell_R)$ & \hyperref[tab:classification:hybrid]{H3} &  \\
	\cline{1-1} \cline{5-8}
	\cellcolor[gray]{0.8}SU11 & & & & \cellcolor[gray]{0.8}F & $(\ell_R \, H)$ & \hyperref[tab:classification:hybrid]{H4} & \\
	\cline{1-1} \cline{3-8}
	SU12 & & \multirow{2}{*}{$-3$} & \multirow{2}{*}{$(1,2,-3)$} & B & $(L_L \, \ell_R)$ &  & \\
	\cline{1-1} \cline{5-8}
	\cellcolor[gray]{0.8}SU13 & & & & \cellcolor[gray]{0.8}F & $(\ell_R \, H^\dagger)$ & & \\
	
	\noalign{\hrule height 1pt}

	SU14 & \multirow{4}{*}{$(1,N\pm2,\alpha)$} & \multirow{2}{*}{$0$} & \multirow{2}{*}{$(1,3,0)$} & B & $(Q_L \, \overline{Q_L}), (L_L \, \overline{L_L}), (H \, H^\dagger)$ &   & $\checkmark (\alpha=0)$ \\
	\cline{1-1} \cline{5-8}
	\cellcolor[gray]{0.8}SU15 & & & & \cellcolor[gray]{0.8}F & $(L_L \, H)$ &  & \\
	\cline{1-1} \cline{3-8}
	SU16 & & \multirow{2}{*}{$-2$} & \multirow{2}{*}{$(1,3,-2)$} 
	& B & $(H^\dagger \, H^\dagger), (L_L \, L_L)$ & & $\checkmark (\alpha=\pm 1)$ \\
	\cline{1-1} \cline{5-8}
	\cellcolor[gray]{0.8}SU17 & & & & \cellcolor[gray]{0.8}F & $(L_L \, H^\dagger) $ & &  \\
	
	\noalign{\hrule height 1pt}
\end{tabular}

%% file: tables/classification_schannel_3_v4.tex
\begin{tabular}{ !{\vrule width 1pt} c | c | c | c | c !{\vrule width 1pt} c | c  | c !{\vrule width 1pt}} 
	\noalign{\hrule height 1pt}

	ID & X & $\alpha+\beta$ & M$_s$ & Spin & (SM$_1$ SM$_2$) &  X--DM--SM$_3$ & M$_s$--X--X \\

	\noalign{\hrule height 1pt}

	ST1 & \multirow{10}{*}{$(3,N,\alpha)$} & $\frac{10}{3}$ & $(3,1,\frac{10}{3})$ & B & $(u_R \, \overline{l_R})$ &  & $\checkmark (\alpha=-\tfrac{5}{3})$  \\
	\cline{1-1} \cline{3-8}
	ST2 & & \multirow{4}{*}{$\frac{4}{3}$} & \multirow{2}{*}{$(3,1,\frac{4}{3})$} & B & $(d_R \, \overline{\ell_R}), (Q_L \, \overline{L_L}), (\overline{d_R} \, \overline{d_R})$ & & $\checkmark (\alpha=-\tfrac{2}{3})$\\
	\cline{1-1} \cline{5-8}
	\cellcolor[gray]{0.8}ST3 & & & & \cellcolor[gray]{0.8}F & $(Q_L \, H)$ & \hyperref[tab:classification:hybrid]{H5} & \\
	\cline{1-1} \cline{4-8}
	ST4 & & & \multirow{2}{*}{$(3,3,\frac{4}{3})^{N\geq2}$} & B & $(Q_L \, \overline{L_L})$ &  & $\checkmark (\alpha=-\tfrac{2}{3})$ \\
	\cline{1-1} \cline{5-8}
	\cellcolor[gray]{0.8}ST5 & & & & \cellcolor[gray]{0.8}F & $(Q_L \, H)$ & \hyperref[tab:classification:hybrid]{H5} & \\
	\cline{1-1} \cline{3-8}
	ST6 & & \multirow{4}{*}{$-\frac{2}{3}$} & \multirow{2}{*}{$(3,1,-\frac{2}{3})$} & B & $(\overline{Q_L} \, \overline{Q_L}), (\overline{u_R} \, \overline{d_R)}, (u_R \, \ell_R), (Q_L \, L_L)$ & & $\checkmark (\alpha=\tfrac{1}{3})$ \\
	\cline{1-1} \cline{5-8}
	\cellcolor[gray]{0.8}ST7 & & & & \cellcolor[gray]{0.8}F & $(Q_L \, H^\dagger)$ & \hyperref[tab:classification:hybrid]{H6} & \\
	\cline{1-1} \cline{4-8}
	ST8 & & & \multirow{2}{*}{$(3,3,-\frac{2}{3})^{N\geq2}$} & B & $(\overline{Q_L} \, \overline{Q_L}), (Q_L \, L_L)$ &  & $\checkmark (\alpha=\tfrac{1}{3})$ \\
	\cline{1-1} \cline{5-8}
	\cellcolor[gray]{0.8}ST9 & & & & \cellcolor[gray]{0.8}F & $(Q_L \, H^\dagger) $ & \hyperref[tab:classification:hybrid]{H6} & \\
	\cline{1-1} \cline{3-8}
	ST10 & & $-\tfrac{8}{3}$ & $(3,1,-\tfrac{8}{3})$ & B & $(\overline{u_R} \, \overline{u_R}), (d_R \, \ell_R)$ &  & $\checkmark (\alpha=\tfrac{4}{3})$ \\

	\noalign{\hrule height 1pt}

	ST11 & \multirow{6}{*}{$(3,N\pm1,\alpha)$} & \multirow{2}{*}{$\frac{7}{3}$} & \multirow{2}{*}{$(3,2,\frac{7}{3})$} & B & $(Q_L \, \overline{\ell_R}), (u_R \, \overline{L_L})$ & & \\
	\cline{1-1} \cline{5-8}
	\cellcolor[gray]{0.8}ST12 & & & & \cellcolor[gray]{0.8}F & $(u_R \, H)$ & & \\
	\cline{1-1} \cline{3-8}
	ST13 & & \multirow{2}{*}{$\frac{1}{3}$} & \multirow{2}{*}{$(3,2,\frac{1}{3})$} & B & $(d_R \, \overline{L_L}), (\overline{Q_L} \, \overline{d_R}), (u_R \, L_L)$ & & \\
	\cline{1-1} \cline{5-8}
	\cellcolor[gray]{0.8}ST14 & & & & \cellcolor[gray]{0.8}F & $(u_R \, H^\dagger), (d_R \, H)$ & \hyperref[tab:classification:hybrid]{H7} & \\
	\cline{1-1} \cline{3-8}
	ST15 & & \multirow{2}{*}{$-\frac{5}{3}$} & \multirow{2}{*}{$(3,2,-\frac{5}{3})$} 	& B & $(\overline{Q_L} \, \overline{u_R}), (Q_L \, \ell_R), (d_R \, L_L)$ & & \\
	\cline{1-1} \cline{5-8}
	\cellcolor[gray]{0.8}ST16 & & & & \cellcolor[gray]{0.8}F & $(d_R \, H^\dagger)$ & & \\

	\noalign{\hrule height 1pt}

	ST17 & \multirow{4}{*}{$(3,N\pm2,\alpha)$} & \multirow{2}{*}{$\frac{4}{3}$} & \multirow{2}{*}{$(3,3,\frac{4}{3})$} & B & $(Q_L \, \overline{L_L})$ & &  $\checkmark (\alpha=-\tfrac{2}{3})$  \\
	\cline{1-1} \cline{5-8}
	\cellcolor[gray]{0.8}ST18 & & & & \cellcolor[gray]{0.8}F & $(Q_L \, H) $ & & \\
	\cline{1-1} \cline{3-8}
	ST19 & & \multirow{2}{*}{$-\frac{2}{3}$} & \multirow{2}{*}{$(3,3,-\frac{2}{3})$} & B & $(\overline{Q_L} \, \overline{Q_L}), (Q_L \, L_L)$ & & $\checkmark (\alpha=\tfrac{1}{3})$ \\
	\cline{1-1} \cline{5-8}
	\cellcolor[gray]{0.8}ST20 & & & & \cellcolor[gray]{0.8}F & $(Q_L \, H^\dagger)$ & & \\
	
	\noalign{\hrule height 1pt}
\end{tabular}

%% file: tables/classification_schannel_ex_v4.tex
\begin{tabular}{ !{\vrule width 1pt} c | c | c | c | c !{\vrule width 1pt} c | c  | c !{\vrule width 1pt}} 
	\noalign{\hrule height 1pt}
	
	ID & X & $\alpha+\beta$ & M$_s$ & Spin & (SM$_1$ SM$_2$) &  X--DM--SM$_3$ & M$_s$--X--X \\

	\noalign{\hrule height 1pt}

	SO1 & \multirow{3}{*}{$(8,N,\alpha)$} & \multirow{2}{*}{$0$} & $(8,1,0)$ & B & $(d_R \, \overline{d_R}), (u_R \, \overline{u_R}), (Q_L \, \overline{Q_L})$ & & $\checkmark (\alpha=0)$ \\
	\cline{1-1} \cline{4-8}
	SO2 & & & $(8,3,0)^{N\geq2}$ & B & $(Q_L \, \overline{Q_L})$ & & $\checkmark (\alpha=0)$ \\
	\cline{1-1} \cline{3-8}
	SO3 & & $-2$ & $(8,1,-2)$ & B & $(d_R \, \overline{u_R})$ & & $\checkmark (\alpha=\pm 1)$ \\
	
	\noalign{\hrule height 1pt}
	
	SO4 & $(8,N\pm1,\alpha)$ & $-1$ & $(8,2,-1)$ & B & $(d_R \, \overline{Q_L}), (Q_L \, \overline{u_R})$ & & \\
	
	\noalign{\hrule height 1pt}
	
	SO5 & $(8,N\pm2,\alpha)$ & $0$ & $(8,3,0)$ & B & $(Q_L \, \overline{Q_L})$ & & $\checkmark (\alpha=0)$ \\
	
	\noalign{\hrule height 1pt}
	
	SE1 & \multirow{4}{*}{$(6,N,\alpha)$} & $\tfrac{8}{3}$ & $(6,1,\tfrac{8}{3})$ & B & $(u_R \, u_R)$ & & $\checkmark (\alpha=-\tfrac{4}{3})$ \\
	\cline{1-1} \cline{3-8}
	SE2 & & \multirow{2}{*}{$\tfrac{2}{3}$} & $(6,1,\tfrac{2}{3})$ & B & $(Q_L \, Q_L), (u_R \, d_R)$ & & $\checkmark (\alpha=-\tfrac{1}{3})$ \\
	\cline{1-1} \cline{4-8}
	SE3 & & & $(6,3,\tfrac{2}{3})^{N\geq2}$ & B & $(Q_L \, Q_L)$ & & $\checkmark (\alpha=-\tfrac{1}{3})$ \\
	\cline{1-1} \cline{3-8}
	SE4 & & $-\tfrac{4}{3}$ & $(6,1,-\tfrac{4}{3})$ & B & $(d_R \, d_R)$ & & $\checkmark (\alpha=\tfrac{2}{3})$ \\

	\noalign{\hrule height 1pt}

	SE5 & \multirow{2}{*}{$(6,N\pm1,\alpha)$} & $\tfrac{5}{3}$ & $(6,2,\tfrac{5}{3})$ & B & $(Q_L \, u_R)$ & & \\
	\cline{1-1} \cline{3-8}
	SE6 & & $-\tfrac{1}{3}$ & $(6,2,-\tfrac{1}{3})$ & B & $(Q_L \, d_R)$ & & \\

	\noalign{\hrule height 1pt}

	SE7 & $(6,N\pm2,\alpha)$ & $\tfrac{2}{3}$ & $(6,3,\tfrac{2}{3})$ & B & $(Q_L \, Q_L)$ & & $\checkmark (\alpha=-\tfrac{1}{3})$ \\

	\noalign{\hrule height 1pt}
\end{tabular}

%% file: diagrams/spinassignment_tchannel_v2.tex
\raisebox{6mm}{\makebox[6mm]{
\begin{tikzpicture}[line width=1.4pt, scale=1]
	\node at (0,0) {I};
\end{tikzpicture}}}
\hspace{-2mm}
\resizebox{!}{16mm}{\begin{tikzpicture}[line width=1.4pt, scale=1]
	\draw[scalarna] (0.8,0.8)--(0,0.4);
	\draw[scalarna] (0.8,-0.8)--(0,-0.4);
	\draw[scalarna] (-0.8,0.8)--(0,0.4);
	\draw[scalarna] (-0.8,-0.8)--(0,-0.4);
	\draw[scalarna] (0,-0.4)--(0,0.4);
	\node at (-1.0,0.8) {$\mathrm{X}$};
	\node at (-1.2,-0.8) {$\mathrm{DM}$};
	\node at (1.25,0.8) {$\mathrm{SM_2}$};
	\node at (1.25,-0.8) {$\mathrm{SM_1}$};
	\node at (-0.45,0) {$\mathrm{M}_{t}$};
\end{tikzpicture}}
\resizebox{!}{16mm}{\begin{tikzpicture}[line width=1.4pt, scale=1]
	\draw[scalarna] (0.8,0.8)--(0,0.4);
	\draw[scalarna] (0.8,-0.8)--(0,-0.4);
	\draw[fermionna] (-0.8,0.8)--(0,0.4);
	\draw[fermionna] (-0.8,-0.8)--(0,-0.4);
	\draw[fermionna] (0,-0.4)--(0,0.4);
	\node at (-1.0,0.8) {$\mathrm{X}$};
	\node at (-1.2,-0.8) {$\mathrm{DM}$};
	\node at (1.25,0.8) {$\mathrm{SM_2}$};
	\node at (1.25,-0.8) {$\mathrm{SM_1}$};
	\node at (-0.45,0) {$\mathrm{M}_{t}$};
\end{tikzpicture}} \vspace{2mm}
\hspace{6mm}
\raisebox{6mm}{\makebox[6mm]{
\begin{tikzpicture}[line width=1.4pt, scale=1]
	\node at (0,0) {II};
\end{tikzpicture}}}
\hspace{-1mm}
\resizebox{!}{16mm}{\begin{tikzpicture}[line width=1.4pt, scale=1]
	\draw[scalarna] (0.8,0.8)--(0,0.4);
	\draw[fermionna] (0.8,-0.8)--(0,-0.4);
	\draw[scalarna] (-0.8,0.8)--(0,0.4);
	\draw[fermionna] (-0.8,-0.8)--(0,-0.4);
	\draw[scalarna] (0,-0.4)--(0,0.4);
	\node at (-1.0,0.8) {$\mathrm{X}$};
	\node at (-1.2,-0.8) {$\mathrm{DM}$};
	\node at (1.25,0.8) {$\mathrm{SM_2}$};
	\node at (1.25,-0.8) {$\mathrm{SM_1}$};
	\node at (-0.45,0) {$\mathrm{M}_{t}$};
\end{tikzpicture}}
\resizebox{!}{16mm}{\begin{tikzpicture}[line width=1.4pt, scale=1]
	\draw[scalarna] (0.8,0.8)--(0,0.4);
	\draw[fermionna] (0.8,-0.8)--(0,-0.4);
	\draw[fermionna] (-0.8,0.8)--(0,0.4);
	\draw[scalarna] (-0.8,-0.8)--(0,-0.4);
	\draw[fermionna] (0,-0.4)--(0,0.4);
	\node at (-1.0,0.8) {$\mathrm{X}$};
	\node at (-1.2,-0.8) {$\mathrm{DM}$};
	\node at (1.25,0.8) {$\mathrm{SM_2}$};
	\node at (1.25,-0.8) {$\mathrm{SM_1}$};
	\node at (-0.45,0) {$\mathrm{M}_{t}$};
\end{tikzpicture}} \vspace{2mm} \newline

\raisebox{6mm}{\makebox[6mm]{
\begin{tikzpicture}[line width=1.4pt, scale=1]
	\node at (0,0) {III};
\end{tikzpicture}}}
\hspace{-2mm}
\resizebox{!}{16mm}{\begin{tikzpicture}[line width=1.4pt, scale=1]
	\draw[fermionna] (0.8,0.8)--(0,0.4);
	\draw[scalarna] (0.8,-0.8)--(0,-0.4);
	\draw[fermionna] (-0.8,0.8)--(0,0.4);
	\draw[scalarna] (-0.8,-0.8)--(0,-0.4);
	\draw[scalarna] (0,-0.4)--(0,0.4);
	\node at (-1.0,0.8) {$\mathrm{X}$};
	\node at (-1.2,-0.8) {$\mathrm{DM}$};
	\node at (1.25,0.8) {$\mathrm{SM_2}$};
	\node at (1.25,-0.8) {$\mathrm{SM_1}$};
	\node at (-0.45,0) {$\mathrm{M}_{t}$};
\end{tikzpicture}}
\resizebox{!}{16mm}{\begin{tikzpicture}[line width=1.4pt, scale=1]
	\draw[fermionna] (0.8,0.8)--(0,0.4);
	\draw[scalarna] (0.8,-0.8)--(0,-0.4);
	\draw[scalarna] (-0.8,0.8)--(0,0.4);
	\draw[fermionna] (-0.8,-0.8)--(0,-0.4);
	\draw[fermionna] (0,-0.4)--(0,0.4);
	\node at (-1.0,0.8) {$\mathrm{X}$};
	\node at (-1.2,-0.8) {$\mathrm{DM}$};
	\node at (1.25,0.8) {$\mathrm{SM_2}$};
	\node at (1.25,-0.8) {$\mathrm{SM_1}$};
	\node at (-0.45,0) {$\mathrm{M}_{t}$};
\end{tikzpicture}} \vspace{2mm}
\hspace{6mm}
\raisebox{6mm}{\makebox[6mm]{
\begin{tikzpicture}[line width=1.4pt, scale=1]
	\node at (0,0) {IV};
\end{tikzpicture}}}
\hspace{-1mm}
\resizebox{!}{16mm}{\begin{tikzpicture}[line width=1.4pt, scale=1]
	\draw[fermionna] (0.8,0.8)--(0,0.4);
	\draw[fermionna] (0.8,-0.8)--(0,-0.4);
	\draw[fermionna] (-0.8,0.8)--(0,0.4);
	\draw[fermionna] (-0.8,-0.8)--(0,-0.4);
	\draw[scalarna] (0,-0.4)--(0,0.4);
	\node at (-1.0,0.8) {$\mathrm{X}$};
	\node at (-1.2,-0.8) {$\mathrm{DM}$};
	\node at (1.25,0.8) {$\mathrm{SM_2}$};
	\node at (1.25,-0.8) {$\mathrm{SM_1}$};
	\node at (-0.45,0) {$\mathrm{M}_{t}$};
\end{tikzpicture}}
\resizebox{!}{16mm}{\begin{tikzpicture}[line width=1.4pt, scale=1]
	\draw[fermionna] (0.8,0.8)--(0,0.4);
	\draw[fermionna] (0.8,-0.8)--(0,-0.4);
	\draw[scalarna] (-0.8,0.8)--(0,0.4);
	\draw[scalarna] (-0.8,-0.8)--(0,-0.4);
	\draw[fermionna] (0,-0.4)--(0,0.4);
	\node at (-1.0,0.8) {$\mathrm{X}$};
	\node at (-1.2,-0.8) {$\mathrm{DM}$};
	\node at (1.25,0.8) {$\mathrm{SM_2}$};
	\node at (1.25,-0.8) {$\mathrm{SM_1}$};
	\node at (-0.45,0) {$\mathrm{M}_{t}$};
\end{tikzpicture}}
\resizebox{!}{16mm}{\begin{tikzpicture}[line width=1.4pt, scale=1]
	\draw[fermionna] (0.8,0.8)--(0,0.4);
	\draw[fermionna] (0.8,-0.8)--(0,-0.4);
	\draw[fermionna] (-0.8,0.8)--(0,0.4);
	\draw[fermionna] (-0.8,-0.8)--(0,-0.4);
	\draw[scalarna] (0,-0.4)--(0,0.4);
	\node at (-1.0,0.8) {$\mathrm{X}$};
	\node at (-1.2,-0.8) {$\mathrm{DM}$};
	\node at (1.25,0.8) {$\mathrm{SM_2}$};
	\node at (1.25,-0.8) {$\mathrm{SM_1}$};
	\node at (-0.45,0) {$\mathrm{M}_{t}$};
\end{tikzpicture}}

%% file: tables/classification_tchannel_1_v2.tex
\begin{tabular}{!{\vrule width 1pt} c | c | c | c | c !{\vrule width 1pt} c | c !{\vrule width 1pt}}
	\noalign{\hrule height 1pt}

	ID & X & $\alpha+\beta$ & M$_t$ & Spin & (SM$_1$ SM$_2$) & X--DM--SM$_3$ \\

	\noalign{\hrule height 1pt}

	\cellcolor[gray]{0.8}TU1 & \multirow{15}{*}{$(1,N,\alpha)$} & \multirow{8}{*}{$0$} & $(1,N\pm1,\beta-1)$ & \cellcolor[gray]{0.8}I & $(H \, H^\dagger)$ &  \hyperref[tab:classification:hybrid]{H1} \\
	\cline{1-1} \cline{4-7}
	\cellcolor[gray]{0.95}TU2 & & & $(1,N\pm1,\beta+1)$ & \cellcolor[gray]{0.95}II & $(L_L \, H)$ & \\
	\cline{1-1} \cline{4-7}
	\cellcolor[gray]{0.6}TU3 & & & $(1,N\pm1,\beta-1)$ & \cellcolor[gray]{0.6}III & $(H \, L_L)$ & \\
	\cline{1-1} \cline{4-7}
	TU4 & & & $(\bar{3},N\pm1,\beta-\tfrac{1}{3})$ & IV & $(Q_L \, \overline{Q_L})$ & \hyperref[tab:classification:hybrid]{H1} \\
	\cline{1-1} \cline{4-7}
	TU5 & & & $(\bar{3},N,\beta-\tfrac{4}{3})$ & IV & $(u_R \, \overline{u_R})$ & \hyperref[tab:classification:hybrid]{H1} \\
	\cline{1-1} \cline{4-7}
	TU6 & & & $(\bar{3},N,\beta+\tfrac{2}{3})$ & IV & $(d_R \, \overline{d_R})$ & \hyperref[tab:classification:hybrid]{H1} \\
	\cline{1-1} \cline{4-7}
	TU7 & & & $(1,N\pm1,\beta+1)$ & IV & $(L_L \, \overline{L_L})$ & \hyperref[tab:classification:hybrid]{H1} \\
	\cline{1-1} \cline{4-7}
	TU8 & & & $(1,N,\beta+2)$ & IV & $(\ell_R \, \overline{\ell_R})$ &\hyperref[tab:classification:hybrid]{H1} \\
	\cline{1-1} \cline{3-7}
	\cellcolor[gray]{0.8}TU9 & & \multirow{6}{*}{$-2$} & $(1,N\pm1,\beta+1)$ & \cellcolor[gray]{0.8}I & $(H^\dagger \, H^\dagger)$ & \\
	\cline{1-1} \cline{4-7}
	\cellcolor[gray]{0.95}TU10 & & & $(1,N\pm1,\beta+1)$ & \cellcolor[gray]{0.95}II & $(L_L \, H^\dagger)$ &\hyperref[tab:classification:hybrid]{H2} \\
	\cline{1-1} \cline{4-7}
	\cellcolor[gray]{0.6}TU11 & & & $(1,N\pm1,\beta+1)$ & \cellcolor[gray]{0.6}III & $(H^\dagger \, L_L)$ & \hyperref[tab:classification:hybrid]{H2} \\
	\cline{1-1} \cline{4-7}
	TU12 & & & $(1,N\pm1,\beta+1)$ & IV & $(L_L \, L_L)$ & \\
	\cline{1-1} \cline{4-7}
	TU13 & & & $(3,N,\beta+\tfrac{4}{3})$ & IV & $(\overline{u_R} \, d_R)$ & \\
	\cline{1-1} \cline{4-7}
	TU14 & & & $(\bar{3},N,\beta+\tfrac{2}{3})$ & IV & $(d_R \, \overline{u_R})$ & \\
	\cline{1-1} \cline{3-7}
	TU15 & & $-4$ & $(1,N,\beta+2)$ & IV & $(\ell_R \, \ell_R)$ & \\

	\noalign{\hrule height 1pt}

	\cellcolor[gray]{0.8}TU16 & \multirow{10}{*}{$(1,N\pm1,\alpha)$} & \multirow{8}{*}{$-1$} & $(1,N,\beta+2)$ &\cellcolor[gray]{0.95}II & $(\ell_R \, H)$ & \hyperref[tab:classification:hybrid]{H4} \\
	\cline{1-1} \cline{4-7}
	\cellcolor[gray]{0.6}TU17 & & & $(1,N\pm1,\beta-1)$ &\cellcolor[gray]{0.6}III & $(H \, \ell_R)$ & \hyperref[tab:classification:hybrid]{H4} \\
	\cline{1-1} \cline{4-7}
	TU18 & & & $(1,N,\beta+2)$ & IV & $(\ell_R \, \overline{L_L})$ & \hyperref[tab:classification:hybrid]{H3} \\
	\cline{1-1} \cline{4-7}
	TU19 & & & $(1,N\pm1,\beta-1)$ & IV & $(\overline{L_L} \, \ell_R)$ & \hyperref[tab:classification:hybrid]{H3} \\
	\cline{1-1} \cline{4-7}
	TU20 & & & $(\bar{3},N,\beta+\tfrac{2}{3})$ & IV & $(d_R \, \overline{Q_L})$ & \hyperref[tab:classification:hybrid]{H3} \\
	\cline{1-1} \cline{4-7}
	TU21 & & & $(3,N\pm1,\beta+\tfrac{1}{3})$ & IV & $(\overline{Q_L} \, d_R)$ & \hyperref[tab:classification:hybrid]{H3} \\
	\cline{1-1} \cline{4-7}
	TU22 & & & $(\bar{3},N\pm1,\beta-\tfrac{1}{3})$ & IV & $(Q_L \, \overline{u_R})$ & \hyperref[tab:classification:hybrid]{H3} \\
	\cline{1-1} \cline{4-7}
	TU23 & & & $(3,N,\beta+\tfrac{4}{3})$ & IV & $(\overline{u_R} \, Q_L)$ & \hyperref[tab:classification:hybrid]{H3} \\
	\cline{1-1} \cline{3-7}
	TU24 & & \multirow{2}{*}{$-3$} & $(1,N\pm1,\beta+1)$ & IV & $(L_L \, \ell_R)$ & \\
	\cline{1-1} \cline{4-7}
	TU25 & & & $(1,N,\beta+2)$ & IV & $(\ell_R \, L_L)$ & \\

	\noalign{\hrule height 1pt}

	\cellcolor[gray]{0.8}TU26 & \multirow{8}{*}{$(1,N\pm2,\alpha)$} & \multirow{5}{*}{$0$} & $(1,N\pm1,\beta-1)$ & \cellcolor[gray]{0.8}I & $(H \, H^\dagger)$ &  \\
	\cline{1-1} \cline{4-7}
	\cellcolor[gray]{0.95}TU27 & & & $(1,N\pm1,\beta+1)$ &\cellcolor[gray]{0.95}II & $(L_L \, H)$ & \\
	\cline{1-1} \cline{4-7}
	\cellcolor[gray]{0.6}TU28 & & & $(1,N\pm1,\beta-1)$ &\cellcolor[gray]{0.6}III & $(H \, L_L)$ & \\
	\cline{1-1} \cline{4-7}
	TU29 & & & $(\bar{3},N\pm1,\beta-\tfrac{1}{3})$ & IV & $(Q_L \, \overline{Q_L})$ &   \\
	\cline{1-1} \cline{4-7}
	TU30 & & & $(1,N\pm1,\beta+1)$ & IV & $(L_L \, \overline{L_L})$ &  \\
	\cline{1-1} \cline{3-7}
	\cellcolor[gray]{0.8}TU31 & & \multirow{3}{*}{$-2$} & $(1,N\pm1,\beta+1)$ & \cellcolor[gray]{0.8}I & $(H^\dagger \, H^\dagger)$ & \\
	\cline{1-1} \cline{4-7}
	\cellcolor[gray]{0.95}TU32 & & & $(1,N\pm1,\beta+1)$ & \cellcolor[gray]{0.95}II & $(L_L \, H^\dagger)$ & \\
	\cline{1-1} \cline{4-7}
	\cellcolor[gray]{0.6}TU33 & & & $(1,N\pm1,\beta+1)$ & \cellcolor[gray]{0.6}III & $(H^\dagger \, L_L)$ & \\

	\noalign{\hrule height 1pt}
\end{tabular}

%% file: tables/classification_tchannel_3_v2.tex
\begin{tabular}{!{\vrule width 1pt} c | c | c | c | c !{\vrule width 1pt} c | c !{\vrule width 1pt}}
	\noalign{\hrule height 1pt}

	ID & X & $\alpha+\beta$ & M$_t$ & Spin & (SM$_1$ SM$_2$) &  X--DM--SM$_3$ \\

	\noalign{\hrule height 1pt}

	TT1 & \multirow{23}{*}{$(3,N,\alpha)$} & \multirow{2}{*}{$\tfrac{10}{3}$} & $(\bar{3},N,\beta-\tfrac{4}{3})$ & IV & $(u_R \, \overline{\ell_R})$ & \\
	\cline{1-1} \cline{4-7}
	TT2 & & & $(1,N,\beta-2)$ & IV & $(\overline{\ell_R} \, u_R)$ & \\
	\cline{1-1} \cline{3-7}
	\cellcolor[gray]{0.95}TT3 & & \multirow{7}{*}{$\tfrac{4}{3}$} & $(\bar{3},N\pm1,\beta-\tfrac{1}{3})$ & \cellcolor[gray]{0.95}II & $(Q_L \, H)$ & \hyperref[tab:classification:hybrid]{H5} \\
	\cline{1-1} \cline{4-7}
	\cellcolor[gray]{0.6}TT4 & & & $(1,N\pm1,\beta-1)$ & \cellcolor[gray]{0.6}III & $(H \, Q_L)$ & \hyperref[tab:classification:hybrid]{H5} \\
	\cline{1-1} \cline{4-7}
	TT5 & & & $(1,N,\beta-2)$ & IV & $(\overline{\ell_R} \, d_R)$ & \\
	\cline{1-1} \cline{4-7}
	TT6 & & & $(\bar{3},N\pm1,\beta-\tfrac{1}{3})$ & IV & $(Q_L \, \overline{L_L})$ & \\
	\cline{1-1} \cline{4-7}
	TT7 & & & $(1,N\pm1,\beta-1)$ & IV & $(\overline{L_L} \, Q_L)$ & \\
	\cline{1-1} \cline{4-7}
	TT8 & & & $(\bar{3},N,\beta+\tfrac{2}{3})$ & IV & $(d_R \, \overline{\ell_R})$ & \\
	\cline{1-1} \cline{4-7}
	TT9 & & & $(3,N,\beta-\tfrac{2}{3})$ & IV & $(\overline{d_R} \, \overline{d_R})$ & \\
	\cline{1-1} \cline{3-7}
	\cellcolor[gray]{0.95}TT10 & & \multirow{11}{*}{$-\tfrac{2}{3}$} & $(\bar{3},N\pm1,\beta-\tfrac{1}{3})$ & \cellcolor[gray]{0.95}II & $(Q_L \, H^\dagger)$ & \hyperref[tab:classification:hybrid]{H6} \\
	\cline{1-1} \cline{4-7}
	\cellcolor[gray]{0.6}TT11 & & & $(1,N\pm1,\beta+1)$ & \cellcolor[gray]{0.6}III & $(H^\dagger \, Q_L)$ & \hyperref[tab:classification:hybrid]{H6} \\
	\cline{1-1} \cline{4-7}
	TT12 & & & $(3,N,\beta+\tfrac{4}{3})$ & IV & $(\overline{u_R} \, \overline{d_R})$ & \\
	\cline{1-1} \cline{4-7}
	TT13 & & & $(3,N\pm1,\beta+\tfrac{1}{3})$ & IV & $(\overline{Q_L} \, \overline{Q_L})$ & \\
	\cline{1-1} \cline{4-7}
	TT14 & & & $(\bar{3},N,\beta-\tfrac{4}{3})$ & IV & $(u_R \, \ell_R)$ & \\
	\cline{1-1} \cline{4-7}
	TT15 & & & $(1,N,\beta+2)$ & IV & $(\ell_R \, u_R)$ & \\
	\cline{1-1} \cline{4-7}
	TT16 & & & $(\bar{3},N\pm1,\beta-\tfrac{1}{3})$ & IV & $(Q_L \, L_L)$ & \\
	\cline{1-1} \cline{4-7}
	TT17 & & & $(1,N\pm1,\beta+1)$ & IV & $(L_L \, Q_L)$ & \\
	\cline{1-1} \cline{4-7}
	TT18 & & & $(3,N,\beta-\tfrac{2}{3})$ & IV & $(\overline{d_R} \, \overline{u_R})$ & \\
	\cline{1-1} \cline{3-7}
	TT19 & & \multirow{3}{*}{$-\tfrac{8}{3}$} & $(3,N,\beta+\tfrac{4}{3})$ & IV & $(\overline{u_R} \, \overline{u_R})$ & \\
	\cline{1-1} \cline{4-7}
	TT20 & & & $(\bar{3},N,\beta+\tfrac{2}{3})$ & IV & $(d_R \, \ell_R)$ & \\
	\cline{1-1} \cline{4-7}
	TT21 & & & $(1,N,\beta+2)$ & IV & $(\ell_R \, d_R)$ & \\
	\noalign{\hrule height 1pt}
	\cellcolor[gray]{0.95}TT22 & \multirow{22}{*}{$(3,N\pm1,\alpha)$} & \multirow{6}{*}{$\tfrac{7}{3}$} & $(\bar{3},N,\beta-\tfrac{4}{3})$ & \cellcolor[gray]{0.95}II& $(u_R \, H)$ & \\
	\cline{1-1} \cline{4-7}
	\cellcolor[gray]{0.6}TT23 & & & $(1,N\pm1,\beta-1)$ & \cellcolor[gray]{0.6}III& $(H \, u_R)$ & \\
	\cline{1-1} \cline{4-7}
	TT24 & & & $(\bar{3},N,\beta-\tfrac{4}{3})$ & IV & $(u_R \, \overline{L_L})$ & \\
	\cline{1-1} \cline{4-7}
	TT25 & & & $(1,N\pm1,\beta-1)$ & IV & $(\overline{L_L} \, u_R)$ & \\
	\cline{1-1} \cline{4-7}
	TT26 & & & $(\bar{3},N\pm1,\beta-\tfrac{1}{3})$ & IV & $(Q_L \, \overline{\ell_R})$ & \\
	\cline{1-1} \cline{4-7}
	TT27 & & & $(1,N,\beta-2)$ & IV & $(\overline{\ell_R} \, Q_L)$ & \\
	\cline{1-1} \cline{3-7}
	\cellcolor[gray]{0.95}TT28 & & \multirow{8}{*}{$\tfrac{1}{3}$} & $(\bar{3},N,\beta-\tfrac{4}{3})$ & \cellcolor[gray]{0.95}II& $(u_R \, H^\dagger)$ & \hyperref[tab:classification:hybrid]{H7} \\
	\cline{1-1} \cline{4-7}
	\cellcolor[gray]{0.95}TT29 & & & $(\bar{3},N,\beta+\tfrac{2}{3})$ & \cellcolor[gray]{0.95}II& $(d_R \, H)$ & \hyperref[tab:classification:hybrid]{H7} \\
	\cline{1-1} \cline{4-7}
	\cellcolor[gray]{0.6}TT30 & & & $(1,N\pm1,\beta+1)$ & \cellcolor[gray]{0.6}III& $(H^\dagger \, u_R)$ & \hyperref[tab:classification:hybrid]{H7} \\
	\cline{1-1} \cline{4-7}
	\cellcolor[gray]{0.6}TT31 & & & $(1,N\pm1,\beta-1)$ & \cellcolor[gray]{0.6}III& $(H \, d_R)$ & \hyperref[tab:classification:hybrid]{H7} \\
	\cline{1-1} \cline{4-7}
	TT32 & & & $(\bar{3},N,\beta-\tfrac{4}{3})$ & IV & $(u_R \, L_L)$ & \\
	\cline{1-1} \cline{4-7}
	TT33 & & & $(1,N\pm1,\beta+1)$ & IV & $(L_L \, u_R)$ & \\
	\cline{1-1} \cline{4-7}
	TT34 & & & $3,N,\beta-\tfrac{2}{3})$ & IV & $(\overline{d_R} \, \overline{Q_L})$ & \\
	\cline{1-1} \cline{4-7}
	TT35 & & & $(3,N\pm1,\beta+\tfrac{1}{3})$ & IV & $(\overline{Q_L} \, \overline{d_R})$ & \\
	\cline{1-1} \cline{3-7}
	\cellcolor[gray]{0.95}TT36 & & \multirow{8}{*}{$-\tfrac{5}{3}$} & $(\bar{3},N,\beta+\tfrac{2}{3})$ & \cellcolor[gray]{0.95}II & $(d_R \, H^\dagger)$ & \\
	\cline{1-1} \cline{4-7}
	\cellcolor[gray]{0.6}TT37 & & & $(1,N\pm1,\beta+1)$ & \cellcolor[gray]{0.6}III & $(H^\dagger \, d_R)$ & \\
	\cline{1-1} \cline{4-7}
	TT38 & & & $(\bar{3},N,\beta+\tfrac{2}{3})$ & IV & $(d_R \, L_L)$ & \\
	\cline{1-1} \cline{4-7}
	TT39 & & & $(1,N\pm1,\beta+1)$ & IV & $(L_L \, d_R)$ & \\
	\cline{1-1} \cline{4-7}
	TT40 & & & $(\bar{3},N\pm1,\beta-\tfrac{1}{3})$ & IV & $(Q_L \, \ell_R)$ & \\
	\cline{1-1} \cline{4-7}
	TT41 & & & $(1,N,\beta+2)$ & IV & $(\ell_R \, Q_L)$ & \\
	\cline{1-1} \cline{4-7}
	TT42 & & & $(3,N,\beta+\tfrac{4}{3})$ & IV & $(\overline{u_R} \, \overline{Q_L})$ & \\
	\cline{1-1} \cline{4-7}
	TT43 & & & $(3,N\pm1,\beta+\tfrac{1}{3})$ & IV & $(\overline{Q_L} \, \overline{u_R})$ & \\
	
	\noalign{\hrule height 1pt}
	
	\cellcolor[gray]{0.95}TT44 & \multirow{9}{*}{$(3,N\pm2,\alpha)$} & \multirow{4}{*}{$\tfrac{4}{3}$} & $(\bar{3},N\pm1,\beta-\tfrac{1}{3})$ & \cellcolor[gray]{0.95}II & $(Q_L \, H)$ & \\
	\cline{1-1} \cline{4-7}
	\cellcolor[gray]{0.6}TT45 & & & $(1,N\pm1,\beta-1)$ & \cellcolor[gray]{0.6}III & $(H \, Q_L)$ & \\
	\cline{1-1} \cline{4-7}
	TT46 & & & $(\bar{3},N\pm1,\beta-\tfrac{1}{3})$ & IV & $(Q_L \, \overline{L_L})$ & \\
	\cline{1-1} \cline{4-7}
	TT47  & & \multirow{5}{*}{$-\tfrac{2}{3}$} & $(1,N\pm1,\beta-1)$ & IV & $(\overline{L_L} \, Q_L)$ & \\
	\cline{1-1} \cline{3-7}
	\cellcolor[gray]{0.95}TT48 & & & $(\bar{3},N\pm1,\beta-\tfrac{1}{3})$ & \cellcolor[gray]{0.95}II & $(Q_L \, H^\dagger)$ & \\
	\cline{1-1} \cline{4-7}
	\cellcolor[gray]{0.6}TT49 & & & $(1,N\pm1,\beta+1)$ & \cellcolor[gray]{0.6}III & $(H^\dagger \, Q_L)$ & \\
	\cline{1-1} \cline{4-7}
	TT50 & & & $(\bar{3},N\pm1,\beta-\tfrac{1}{3})$ & IV & $(Q_L \, L_L)$ & \\
	\cline{1-1} \cline{4-7}
	TT51 & & & $(1,N\pm1,\beta+1)$ & IV & $(L_L \, Q_L)$ & \\
	\cline{1-1} \cline{4-7}
	TT52 & & & $(3,N\pm1,\beta+\tfrac{1}{3})$ & IV & $(\overline{Q_L} \, \overline{Q_L})$ & \\
	
	\noalign{\hrule height 1pt}
\end{tabular}

%% file: tables/classification_tchannel_ex_v2.tex
\begin{tabular}{!{\vrule width 1pt} c | c | c | c | c !{\vrule width 1pt} c | c !{\vrule width 1pt}}
	\noalign{\hrule height 1pt}

	ID & X & $\alpha+\beta$ & M$_t$ & Spin & (SM$_1$ SM$_2$) &  X--DM--SM$_3$ \\

	\noalign{\hrule height 1pt}

	TO1 & \multirow{5}{*}{$(8,N,\alpha)$} & \multirow{3}{*}{$0$} & $(\bar{3},N\pm1,\beta-\tfrac{1}{3})$ & IV & $(Q_L \, \overline{Q_L})$ & \\
	\cline{1-1} \cline{4-7}
	TO2 & & & $(\bar{3},N,\beta-\tfrac{4}{3})$ & IV & $(u_R \, \overline{u_R})$ & \\
	\cline{1-1} \cline{4-7}
	TO3 & & & $(\bar{3},N,\beta+\tfrac{2}{3})$ & IV & $(d_R \, \overline{d_R})$ & \\
	\cline{1-1} \cline{3-7}
	TO4 & & \multirow{2}{*}{$-2$} & $(\bar{3},N,\beta+\tfrac{2}{3})$ & IV & $(d_R \, \overline{u_R})$ & \\
	\cline{1-1} \cline{4-7}
	TO5 & & & $(3,N,\beta+\tfrac{4}{3})$ & IV & $(\overline{u_R} \, d_R)$ & \\

	\noalign{\hrule height 1pt}

	TO6 & \multirow{4}{*}{$(8,N\pm1,\alpha)$} & \multirow{4}{*}{$-1$} & $(\bar{3},N,\beta+\tfrac{2}{3})$ & IV & $(d_R \, \overline{Q_L})$ & \\
	\cline{1-1} \cline{4-7}
	TO7 & & & $(3,N\pm1,\beta+\tfrac{1}{3})$ & IV & $(\overline{Q_L} \, d_R)$ & \\
	\cline{1-1} \cline{4-7}
	TO8 & & & $(\bar{3},N\pm1,\beta-\tfrac{1}{3})$ & IV & $(Q_L \, \overline{u_R})$ & \\
	\cline{1-1} \cline{4-7}
	TO9 & & & $(3,N,\beta+\tfrac{4}{3})$ & IV & $(\overline{u_R} \, Q_L)$ & \\

	\noalign{\hrule height 1pt}

	TO10 & $(8,N\pm2,\alpha)$ & $0$ & $(\bar{3},N\pm1,\beta-\tfrac{1}{3})$ & IV & $(Q_L \, \overline{Q_L})$ & \\

	\noalign{\hrule height 1pt}
	
	TE1 & \multirow{5}{*}{$(6,N,\alpha)$} & $\tfrac{8}{3}$ & $(\bar{3},N,\beta-\tfrac{4}{3})$ & IV & $(u_R \, u_R)$ & \\
	\cline{1-1} \cline{3-7}
	TE2 & & \multirow{3}{*}{$\tfrac{2}{3}$} & $(\bar{3},N\pm1,\beta-\tfrac{1}{3})$ & IV & $(Q_L \, Q_L)$ & \\
	\cline{1-1} \cline{4-7}
	TE3 & & & $(\bar{3},N,\beta-\tfrac{4}{3})$ & IV & $(u_R \, d_R)$ & \\
	\cline{1-1} \cline{4-7}
	TE4 & & & $(\bar{3},N,\beta+\tfrac{2}{3})$ & IV & $(d_R \, u_R)$ & \\
	\cline{1-1} \cline{3-7}
	TE5 & & $-\tfrac{4}{3}$ & $(\bar{3},N,\beta+\tfrac{2}{3})$ & IV & $(d_R \, d_R)$ & \\

	\noalign{\hrule height 1pt}

	TE6 & \multirow{4}{*}{$(6,N\pm1,\alpha)$} & \multirow{2}{*}{$\tfrac{5}{3}$} & $(\bar{3},N,\beta-\tfrac{4}{3})$ & IV & $(u_R \, Q_L)$ & \\
	\cline{1-1} \cline{4-7}
	TE7 & & & $(\bar{3},N\pm1,\beta-\tfrac{1}{3})$ & IV & $(Q_L \, u_R)$ & \\
	\cline{1-1} \cline{3-7}
	TE8 & & \multirow{2}{*}{-$\tfrac{1}{3}$} & $(\bar{3},N,\beta+\tfrac{2}{3})$ & IV & $(d_R \, Q_L)$ & \\
	\cline{1-1} \cline{4-7}
	TE9 & & & $(\bar{3},N\pm1,\beta-\tfrac{1}{3})$ & IV & $(Q_L \, d_R)$ & \\

	\noalign{\hrule height 1pt}

	TE10 & $(6,N\pm2,\alpha)$ & $\tfrac{2}{3}$ & $(\bar{3},N\pm1,\beta-\tfrac{1}{3})$ & IV & $(Q_L \, Q_L)$ & \\

	\noalign{\hrule height 1pt}
\end{tabular}

%% file: diagrams/externalvevs_v3.tex
\resizebox{!}{45mm}{
\begin{tikzpicture}[line width=1.4pt, scale=1]
	\draw[fermionna] (2.8,0.8)--(2,0);
	\draw[fermionna] (2.8,-0.8)--(2,0);
	\draw[fermionna] (2,0)--(0,0);
	\draw[fermionna] (-0.8,0.8)--(0,0);
	\draw[fermionna] (-0.8,-0.8)--(0,0);
	
	\draw[scalarna] (0.4,1.0583)--(0,0);
	\draw[scalarna] (-0.4,1.0583)--(0,0);
	
	\draw[scalarna] (2.4,1.0583)--(2,0);
	\draw[scalarna] (1.6,1.0583)--(2,0);
	
	\draw[fill=black] (0,0) circle (3.0mm);
	\draw[fill=white] (0,0) circle (2.9mm);
	\begin{scope}
		\clip (0,0) circle (3.0mm);
		\foreach \x in {-.9,-.75,...,.3}
		\draw[line width=1 pt] (\x,-.3) -- (\x+.6,.3);
	\end{scope}
	
	\draw[fill=black] (2,0) circle (3.0mm);
	\draw[fill=white] (2,0) circle (2.9mm);
	\begin{scope}
		\clip (2,0) circle (3.0mm);
		\foreach \x in {-.9,-.75,...,.3}
		\draw[line width=1 pt] (\x+2,-.3) -- (\x+2.6,.3);
	\end{scope}
	
	\node at (-1.0,0.8) {\scriptsize X};
	\node at (-1.15,-0.8) {\scriptsize DM};
	\node at (3.2,0.8) {\scriptsize SM$_2$};
	\node at (3.2,-0.8) {\scriptsize SM$_1$};
	\node at (1,0.2) {\scriptsize M$_s$};
	
	\node[vtxcircle, minimum size=2pt] at (0,0.915) {};
	\node[vtxcircle, minimum size=2pt] at (0.12,0.9) {};
	\node[vtxcircle, minimum size=2pt] at (-0.12,0.9) {};
	\node at (0.0,1.3) {\scriptsize $H, H^\dagger$};
	\node at (0.0,1.6) {\scriptsize $\overbrace{\hspace{8mm}}$};
	\node at (0.0,1.9) {\scriptsize $p$};
	
	\node[vtxcircle, minimum size=2pt] at (2,0.915) {};
	\node[vtxcircle, minimum size=2pt] at (2.12,0.9) {};
	\node[vtxcircle, minimum size=2pt] at (1.88,0.9) {};
	\node at (2.0,1.3) {\scriptsize $H, H^\dagger$};
	\node at (2.0,1.6) {\scriptsize $\overbrace{\hspace{8mm}}$};
	\node at (2.0,1.9) {\scriptsize $q$};
	
	\node at (1,-2.3) {\textbf{(a)}};
\end{tikzpicture}} 
\resizebox{!}{45mm}{\begin{tikzpicture}[line width=1.4pt, scale=1]
	\draw[fermionna] (0.8,-0.8)--(0,0);
	\draw[fermionna] (0.8,2.8)--(0,2);
	\draw[fermionna] (0,2)--(0,0);
	\draw[fermionna] (-0.8,-0.8)--(0,0);
	\draw[fermionna] (-0.8,2.8)--(0,2);
	
	\draw[scalarna] (1.0583,0.4)--(0,0);
	\draw[scalarna] (1.0583,-0.4)--(0,0);
	
	\draw[scalarna] (1.0583,2.4)--(0,2);
	\draw[scalarna] (1.0583,1.6)--(0,2);
	
	\draw[fill=black] (0,0) circle (3.0mm);
	\draw[fill=white] (0,0) circle (2.9mm);
	\begin{scope}
		\clip (0,0) circle (3.0mm);
		\foreach \x in {-.9,-.75,...,.3}
		\draw[line width=1 pt] (\x,-.3) -- (\x+.6,.3);
	\end{scope}
	
	\draw[fill=black] (0,2) circle (3.0mm);
	\draw[fill=white] (0,2) circle (2.9mm);
	\begin{scope}
		\clip (0,2) circle (3.0mm);
		\foreach \x in {-.9,-.75,...,.3}
		\draw[line width=1 pt] (\x,1.7) -- (\x+.6,2.3);
	\end{scope}
	
	\node at (-0.9,3) {\scriptsize X};
	\node at (-1.05,-1) {\scriptsize DM};
	\node at (1.1,3) {\scriptsize SM$_2$};
	\node at (1.1,-1) {\scriptsize SM$_1$};
	\node at (0.3,1) {\scriptsize M$_t$};
	
	\node[vtxcircle, minimum size=2pt] at (0.915,0) {};
	\node[vtxcircle, minimum size=2pt] at (0.9,0.12) {};
	\node[vtxcircle, minimum size=2pt] at (0.9,-0.12) {};
	\node at (1.8,0.0) {\scriptsize $H, H^\dagger$};
	\node at (1.8,-0.3) {\scriptsize $\underbrace{\hspace{8mm}}$};
	\node at (1.8,-0.6) {\scriptsize $p$};
	
	\node[vtxcircle, minimum size=2pt] at (0.915,2) {};
	\node[vtxcircle, minimum size=2pt] at (0.9,2.12) {};
	\node[vtxcircle, minimum size=2pt] at (0.9,1.88) {};
	\node at (1.8,2.0) {\scriptsize $H, H^\dagger$};
	\node at (1.8,1.7) {\scriptsize $\underbrace{\hspace{8mm}}$};
	\node at (1.8,1.4) {\scriptsize $q$};
	
	\node at (0,-1.3) {\textbf{(b)}};
\end{tikzpicture}} 
\resizebox{!}{45mm}{\begin{tikzpicture}[line width=1.4pt, scale=1]
	\draw[fermionna] (0.8,0)--(0,0);
	\draw[fermionna] (-0.6,0.6)--(0,0);
	\draw[fermionna] (-0.6,-0.6)--(0,0);
	
	\draw[scalarna] (0.4,1.0583)--(0,0);
	\draw[scalarna] (-0.4,1.0583)--(0,0);
	
	\draw[fill=black] (0,0) circle (3.0mm);
	\draw[fill=white] (0,0) circle (2.9mm);
	\begin{scope}
		\clip (0,0) circle (3.0mm);
		\foreach \x in {-.9,-.75,...,.3}
		\draw[line width=1 pt] (\x,-.3) -- (\x+.6,.3);
	\end{scope}
	
	\node at (-0.8,0.6) {\scriptsize X};
	\node at (-0.95,-0.6) {\scriptsize DM};
	\node at (1.15,0) {\scriptsize SM$_3$};
	
	\node[vtxcircle, minimum size=2pt] at (0,0.915) {};
	\node[vtxcircle, minimum size=2pt] at (0.12,0.9) {};
	\node[vtxcircle, minimum size=2pt] at (-0.12,0.9) {};
	\node at (0.0,1.3) {\scriptsize $H, H^\dagger$};
	\node at (0.0,1.6) {\scriptsize $\overbrace{\hspace{8mm}}$};
	\node at (0.0,1.9) {\scriptsize $p$};
	
	\node at (0,-2.3) {\textbf{(c)}};
\end{tikzpicture}} 
\resizebox{!}{45mm}{\begin{tikzpicture}[line width=1.4pt, scale=1]
	\draw[fermionna] (0.8,0.8)--(0,0);
	\draw[fermionna] (0.8,-0.8)--(0,0);
	\draw[fermionna] (-0.8,0.8)--(0,0);
	\draw[fermionna] (-0.8,-0.8)--(0,0);
	
	\draw[scalarna] (0.4,1.0583)--(0,0);
	\draw[scalarna] (-0.4,1.0583)--(0,0);
	
	\draw[fill=black] (0,0) circle (3.0mm);
	\draw[fill=white] (0,0) circle (2.9mm);
	\begin{scope}
		\clip (0,0) circle (3.0mm);
		\foreach \x in {-.9,-.75,...,.3}
		\draw[line width=1 pt] (\x,-.3) -- (\x+.6,.3);
	\end{scope}
	
	\node at (-1.0,0.8) {\scriptsize X};
	\node at (-1.15,-0.8) {\scriptsize DM};
	\node at (1.2,0.8) {\scriptsize SM$_2$};
	\node at (1.2,-0.8) {\scriptsize SM$_1$};
	
	\node[vtxcircle, minimum size=2pt] at (0,0.915) {};
	\node[vtxcircle, minimum size=2pt] at (0.12,0.9) {};
	\node[vtxcircle, minimum size=2pt] at (-0.12,0.9) {};
	\node at (0.0,1.3) {\scriptsize $H, H^\dagger$};
	\node at (0.0,1.6) {\scriptsize $\overbrace{\hspace{8mm}}$};
	\node at (0.0,1.9) {\scriptsize $p+q$};
	
	\node at (0,-2.3) {\textbf{(d)}};
\end{tikzpicture}}

%% file: diagrams/higgs_s_leftvertex_v2.tex
\begin{tikzpicture}[line width=1.4pt, scale=1]
	\draw[fermionna] (1.2,0)--(0,0);
	\draw[fermionna] (-0.9,0.9)--(0,0);
	\draw[fermionna] (-0.9,-0.9)--(0,0);
	
	\draw[scalarna] (0.4,1.0583)--(0,0);
	\draw[scalarna] (-0.4,1.0583)--(0,0);
	
	\draw[fill=black] (0,0) circle (3.0mm);
	\draw[fill=white] (0,0) circle (2.9mm);
	\begin{scope}
		\clip (0,0) circle (3.0mm);
		\foreach \x in {-.9,-.75,...,.3}
		\draw[line width=1 pt] (\x,-.3) -- (\x+.6,.3);
	\end{scope}
	
	\node at (-1.1,0.9) {\scriptsize X};
	\node at (-1.25,-0.9) {\scriptsize DM};
	\node at (1.45,0) {\scriptsize M$_s$};
	
	\node[vtxcircle, minimum size=2pt] at (0,0.915) {};
	\node[vtxcircle, minimum size=2pt] at (0.12,0.9) {};
	\node[vtxcircle, minimum size=2pt] at (-0.12,0.9) {};
	\node at (0.0,1.3) {\scriptsize $H, H^\dagger$};
	\node at (0.0,1.6) {\scriptsize $\overbrace{\hspace{8mm}}$};
	\node at (0.0,1.9) {\scriptsize $p$};
	
	\node at (0,-1.4) {\textbf{(a)}};
\end{tikzpicture} \hspace{-1mm}
\raisebox{14.3mm}{\begin{tikzpicture}[line width=1.4pt, scale=1]
	\node at (1.8,0.0) {\huge $\Rightarrow$};
\end{tikzpicture}} \hspace{-3mm}
\raisebox{-4.0mm}{\begin{tikzpicture}[line width=1.4pt, scale=1]
	\draw[fermionna] (1.4,0)--(0,0);
	\draw[fermionna] (-1,1)--(0,0);
	\draw[fermionna] (-1,-1)--(0,0);
	\node[vtxcircle] at (0,0) {};
	
	\draw[scalarna] (1,0.7)--(0.7,0);
	\draw[scalarna] (1.5,0.7)--(1.2,0);
	\node[vtxcircle, minimum size=1.6pt] at (1,0.35) {};
	\node[vtxcircle, minimum size=1.6pt] at (1.1,0.35) {};
	\node[vtxcircle, minimum size=1.6pt] at (1.2,0.35) {};
	
	\draw[scalarna] (-0.5,-0.5)--(-0.2,-1.2);
	\draw[scalarna] (-0.8,-0.8)--(-0.5,-1.5);
	\node[vtxcircle, minimum size=1.6pt] at (-0.425,-0.925) {};
	\node[vtxcircle, minimum size=1.6pt] at (-0.5,-1) {};
	\node[vtxcircle, minimum size=1.6pt] at (-0.575,-1.075) {};
	
	\draw[scalarna] (-0.5,0.5)--(-0.2,1.2);
	\draw[scalarna] (-0.8,0.8)--(-0.5,1.5);
	\node[vtxcircle, minimum size=1.6pt] at (-0.425,0.925) {};
	\node[vtxcircle, minimum size=1.6pt] at (-0.5,1) {};
	\node[vtxcircle, minimum size=1.6pt] at (-0.575,1.075) {};
	
	\node at (-1.2,1) {\scriptsize X};
	\node at (-1.35,-1) {\scriptsize DM};
	\node at (1.65,0) {\scriptsize M$_s$};

	\node at (-0.5,0.2) {\scriptsize X$'$};
	\node at (-0.58,-0.2) {\scriptsize DM$'$};
	\node at (0.4,-0.2) {\scriptsize M$_s'$};
	
	\node at (0,-1.8) {\textbf{(b)}};
\end{tikzpicture}} \hspace{3mm}
\raisebox{-4.0mm}{\begin{tikzpicture}[line width=1.4pt, scale=1]
	\draw[fermionna] (1.4,0)--(0,0);
	\draw[fermionna] (-1,1)--(0,0);
	\draw[fermionna] (-1,-1)--(0,0);
	\node[vtxcircle] at (0,0) {};
	
	\draw[scalarna] (0.6,1.4)--(0,0);
	\node at (0.7,1.6) {\scriptsize $H, H^\dagger$};
	
	\draw[scalarna] (1,0.7)--(0.7,0);
	\draw[scalarna] (1.5,0.7)--(1.2,0);
	\node[vtxcircle, minimum size=1.6pt] at (1,0.35) {};
	\node[vtxcircle, minimum size=1.6pt] at (1.1,0.35) {};
	\node[vtxcircle, minimum size=1.6pt] at (1.2,0.35) {};
	
	\draw[scalarna] (-0.5,-0.5)--(-0.2,-1.2);
	\draw[scalarna] (-0.8,-0.8)--(-0.5,-1.5);
	\node[vtxcircle, minimum size=1.6pt] at (-0.425,-0.925) {};
	\node[vtxcircle, minimum size=1.6pt] at (-0.5,-1) {};
	\node[vtxcircle, minimum size=1.6pt] at (-0.575,-1.075) {};
	
	\draw[scalarna] (-0.5,0.5)--(-0.2,1.2);
	\draw[scalarna] (-0.8,0.8)--(-0.5,1.5);
	\node[vtxcircle, minimum size=1.6pt] at (-0.425,0.925) {};
	\node[vtxcircle, minimum size=1.6pt] at (-0.5,1) {};
	\node[vtxcircle, minimum size=1.6pt] at (-0.575,1.075) {};
	
	\node at (-1.2,1) {\scriptsize X};
	\node at (-1.35,-1) {\scriptsize DM};
	\node at (1.65,0) {\scriptsize M$_s$};

	\node at (-0.5,0.2) {\scriptsize X$'$};
	\node at (-0.58,-0.2) {\scriptsize DM$'$};
	\node at (0.4,-0.2) {\scriptsize M$_s'$};
	
	\node at (0,-1.8) {\textbf{(c)}};
\end{tikzpicture}}

%% file: diagrams/higgs_s_rightvertex_v2.tex
\begin{tikzpicture}[line width=1.4pt, scale=1]
	\draw[fermionna] (-1.2,0)--(0,0);
	\draw[fermionna] (0.9,0.9)--(0,0);
	\draw[fermionna] (0.9,-0.9)--(0,0);
	
	\draw[scalarna] (0.4,1.0583)--(0,0);
	\draw[scalarna] (-0.4,1.0583)--(0,0);
	
	\draw[fill=black] (0,0) circle (3.0mm);
	\draw[fill=white] (0,0) circle (2.9mm);
	\begin{scope}
		\clip (0,0) circle (3.0mm);
		\foreach \x in {-.9,-.75,...,.3}
		\draw[line width=1 pt] (\x,-.3) -- (\x+.6,.3);
	\end{scope}
	
	\node at (1.25,0.9) {\scriptsize SM$_2$};
	\node at (1.25,-0.9) {\scriptsize SM$_1$};
	\node at (-1.45,0) {\scriptsize M$_s$};

	\node[vtxcircle, minimum size=2pt] at (0,0.915) {};
	\node[vtxcircle, minimum size=2pt] at (0.12,0.9) {};
	\node[vtxcircle, minimum size=2pt] at (-0.12,0.9) {};
	\node at (0.0,1.3) {\scriptsize $H, H^\dagger$};
	\node at (0.0,1.6) {\scriptsize $\overbrace{\hspace{8mm}}$};
	\node at (0.0,1.9) {\scriptsize $q$};
	
	\node at (0,-1.4) {\textbf{(a)}};
\end{tikzpicture} \hspace{-3mm}
\raisebox{14.3mm}{\begin{tikzpicture}[line width=1.4pt, scale=1]
	\node at (1.8,0.0) {\huge $\Rightarrow$};
\end{tikzpicture}} \hspace{1mm}
\raisebox{-4.0mm}{\begin{tikzpicture}[line width=1.4pt, scale=1]
	\draw[fermionna] (-1.4,0)--(0,0);
	\draw[fermionna] (1,1)--(0,0);
	\draw[fermionna] (1,-1)--(0,0);
	\node[vtxcircle] at (0,0) {};
	
	\draw[scalarna] (-1,0.7)--(-0.7,0);
	\draw[scalarna] (-1.5,0.7)--(-1.2,0);
	\node[vtxcircle, minimum size=1.6pt] at (-1,0.35) {};
	\node[vtxcircle, minimum size=1.6pt] at (-1.1,0.35) {};
	\node[vtxcircle, minimum size=1.6pt] at (-1.2,0.35) {};
	
	\draw[scalarna] (0.5,-0.5)--(0.2,-1.2);
	\draw[scalarna] (0.8,-0.8)--(0.5,-1.5);
	\node[vtxcircle, minimum size=1.6pt] at (0.425,-0.925) {};
	\node[vtxcircle, minimum size=1.6pt] at (0.5,-1) {};
	\node[vtxcircle, minimum size=1.6pt] at (0.575,-1.075) {};
	
	\draw[scalarna] (0.5,0.5)--(0.2,1.2);
	\draw[scalarna] (0.8,0.8)--(0.5,1.5);
	\node[vtxcircle, minimum size=1.6pt] at (0.425,0.925) {};
	\node[vtxcircle, minimum size=1.6pt] at (0.5,1) {};
	\node[vtxcircle, minimum size=1.6pt] at (0.575,1.075) {};	
	
	\node at (1.35,1) {\scriptsize SM$_2$};
	\node at (1.35,-1) {\scriptsize SM$_1$};
	\node at (-1.65,0) {\scriptsize M$_s$};

	\node at (0.65,0.2) {\scriptsize SM$_2'$};
	\node at (0.65,-0.2) {\scriptsize SM$_1'$};
	\node at (-0.4,-0.2) {\scriptsize M$_s''$};
	
	\node at (0,-1.8) {\textbf{(b)}};
\end{tikzpicture}} \hspace{3mm}
\raisebox{-4.0mm}{\begin{tikzpicture}[line width=1.4pt, scale=1]
	\draw[fermionna] (-1.4,0)--(0,0);
	\draw[fermionna] (1,1)--(0,0);
	\draw[fermionna] (1,-1)--(0,0);
	\node[vtxcircle] at (0,0) {};
	
	\draw[scalarna] (-0.6,1.4)--(0,0);
	\node at (-0.7,1.6) {\scriptsize $H, H^\dagger$};
	
	\draw[scalarna] (-1,0.7)--(-0.7,0);
	\draw[scalarna] (-1.5,0.7)--(-1.2,0);
	\node[vtxcircle, minimum size=1.6pt] at (-1,0.35) {};
	\node[vtxcircle, minimum size=1.6pt] at (-1.1,0.35) {};
	\node[vtxcircle, minimum size=1.6pt] at (-1.2,0.35) {};
	
	\draw[scalarna] (0.5,-0.5)--(0.2,-1.2);
	\draw[scalarna] (0.8,-0.8)--(0.5,-1.5);
	\node[vtxcircle, minimum size=1.6pt] at (0.425,-0.925) {};
	\node[vtxcircle, minimum size=1.6pt] at (0.5,-1) {};
	\node[vtxcircle, minimum size=1.6pt] at (0.575,-1.075) {};
	
	\draw[scalarna] (0.5,0.5)--(0.2,1.2);
	\draw[scalarna] (0.8,0.8)--(0.5,1.5);
	\node[vtxcircle, minimum size=1.6pt] at (0.425,0.925) {};
	\node[vtxcircle, minimum size=1.6pt] at (0.5,1) {};
	\node[vtxcircle, minimum size=1.6pt] at (0.575,1.075) {};
	
	\node at (1.35,1) {\scriptsize SM$_2$};
	\node at (1.35,-1) {\scriptsize SM$_1$};
	\node at (-1.65,0) {\scriptsize M$_s$};

	\node at (0.65,0.2) {\scriptsize SM$_2'$};
	\node at (0.65,-0.2) {\scriptsize SM$_1'$};
	\node at (-0.4,-0.2) {\scriptsize M$_s''$};
	
	\node at (0,-1.8) {\textbf{(c)}};
\end{tikzpicture}}

%% file: diagrams/higgs_t_uppervertex.tex
\begin{tikzpicture}[line width=1.4pt, scale=1]
	\draw[fermionna] (0,-1.2)--(0,0);
	\draw[fermionna] (-0.9,0.9)--(0,0);
	\draw[fermionna] (0.9,0.9)--(0,0);
	
	\draw[scalarna] (1.0583,0.4)--(0,0);
	\draw[scalarna] (1.0583,-0.4)--(0,0);
	
	\draw[fill=black] (0,0) circle (3.0mm);
	\draw[fill=white] (0,0) circle (2.9mm);
	\begin{scope}
		\clip (0,0) circle (3.0mm);
		\foreach \x in {-.9,-.75,...,.3}
		\draw[line width=1 pt] (\x,-.3) -- (\x+.6,.3);
	\end{scope}
	
	\node at (-1,1.1) {\scriptsize X};
	\node at (1.2,1.1) {\scriptsize SM$_2$};
	\node at (0,-1.4) {\scriptsize M$_t$};

	\node[vtxcircle, minimum size=2pt] at (0.915,0) {};
	\node[vtxcircle, minimum size=2pt] at (0.9,0.12) {};
	\node[vtxcircle, minimum size=2pt] at (0.9,-0.12) {};
	\node at (1.8,0) {\scriptsize $H, H^\dagger$};
	\node at (1.8,-0.3) {\scriptsize $\underbrace{\hspace{8mm}}$};
	\node at (1.8,-0.6) {\scriptsize $q$};
	
	\node at (0,-1.9) {\textbf{(a)}};
\end{tikzpicture} \hspace{1mm}
\raisebox{18.9mm}{\begin{tikzpicture}[line width=1.4pt, scale=1]
	\node at (1.8,0.0) {\huge $\Rightarrow$};
\end{tikzpicture}} \hspace{1mm}
\raisebox{-2.0mm}{\begin{tikzpicture}[line width=1.4pt, scale=1]
	\draw[fermionna] (0,-1.4)--(0,0);
	\draw[fermionna] (1,1)--(0,0);
	\draw[fermionna] (-1,1)--(0,0);
	\node[vtxcircle] at (0,0) {};
	
	\draw[scalarna] (0.7,-1)--(0,-0.7);
	\draw[scalarna] (0.7,-1.5)--(0,-1.2);
	\node[vtxcircle, minimum size=1.6pt] at (0.35,-1) {};
	\node[vtxcircle, minimum size=1.6pt] at (0.35,-1.1) {};
	\node[vtxcircle, minimum size=1.6pt] at (0.35,-1.2) {};
	
	\draw[scalarna] (0.5,0.5)--(1.2,0.2);
	\draw[scalarna] (0.8,0.8)--(1.5,0.5);
	\node[vtxcircle, minimum size=1.6pt] at (0.925,0.425) {};
	\node[vtxcircle, minimum size=1.6pt] at (1,0.5) {};
	\node[vtxcircle, minimum size=1.6pt] at (1.075,0.575) {};
	
	\draw[scalarna] (-0.5,0.5)--(-1.2,0.2);
	\draw[scalarna] (-0.8,0.8)--(-1.5,0.5);
	\node[vtxcircle, minimum size=1.6pt] at (-0.925,0.425) {};
	\node[vtxcircle, minimum size=1.6pt] at (-1,0.5) {};
	\node[vtxcircle, minimum size=1.6pt] at (-1.075,0.575) {};
	
	\node at (-1.1,1.2) {\scriptsize X};
	\node at (1.3,1.2) {\scriptsize SM$_2$};
	\node at (0,-1.6) {\scriptsize M$_t$};

	\node at (0.65,0.2) {\scriptsize SM$_2'$};
	\node at (-0.5,0.2) {\scriptsize X$'$};
	\node at (-0.25,-0.4) {\scriptsize M$_t'$};
	
	\node at (0,-2.1) {\textbf{(b)}};
\end{tikzpicture}} \hspace{3mm}
\raisebox{-2.0mm}{\begin{tikzpicture}[line width=1.4pt, scale=1]
	\draw[fermionna] (0,-1.4)--(0,0);
	\draw[fermionna] (1,1)--(0,0);
	\draw[fermionna] (-1,1)--(0,0);
	\node[vtxcircle] at (0,0) {};
	
	\draw[scalarna] (1.4,-0.6)--(0,0);
	\node at (1.5,-0.8) {\scriptsize $H, H^\dagger$};
	
	\draw[scalarna] (0.7,-1)--(0,-0.7);
	\draw[scalarna] (0.7,-1.5)--(0,-1.2);
	\node[vtxcircle, minimum size=1.6pt] at (0.35,-1) {};
	\node[vtxcircle, minimum size=1.6pt] at (0.35,-1.1) {};
	\node[vtxcircle, minimum size=1.6pt] at (0.35,-1.2) {};
	
	\draw[scalarna] (0.5,0.5)--(1.2,0.2);
	\draw[scalarna] (0.8,0.8)--(1.5,0.5);
	\node[vtxcircle, minimum size=1.6pt] at (0.925,0.425) {};
	\node[vtxcircle, minimum size=1.6pt] at (1,0.5) {};
	\node[vtxcircle, minimum size=1.6pt] at (1.075,0.575) {};
	
	\draw[scalarna] (-0.5,0.5)--(-1.2,0.2);
	\draw[scalarna] (-0.8,0.8)--(-1.5,0.5);
	\node[vtxcircle, minimum size=1.6pt] at (-0.925,0.425) {};
	\node[vtxcircle, minimum size=1.6pt] at (-1,0.5) {};
	\node[vtxcircle, minimum size=1.6pt] at (-1.075,0.575) {};
	
	\node at (-1.1,1.2) {\scriptsize X};
	\node at (1.3,1.2) {\scriptsize SM$_2$};
	\node at (0,-1.6) {\scriptsize M$_t$};

	\node at (0.65,0.2) {\scriptsize SM$_2'$};
	\node at (-0.5,0.2) {\scriptsize X$'$};
	\node at (-0.25,-0.4) {\scriptsize M$_t'$};
	
	\node at (0,-2.1) {\textbf{(c)}};
\end{tikzpicture}}

%% file: diagrams/higgs_t_lowervertex_v2.tex
\begin{tikzpicture}[line width=1.4pt, scale=1]
	\draw[fermionna] (0,1.2)--(0,0);
	\draw[fermionna] (-0.9,-0.9)--(0,0);
	\draw[fermionna] (0.9,-0.9)--(0,0);
	
	\draw[scalarna] (1.0583,0.4)--(0,0);
	\draw[scalarna] (1.0583,-0.4)--(0,0);
	
	\draw[fill=black] (0,0) circle (3.0mm);
	\draw[fill=white] (0,0) circle (2.9mm);
	\begin{scope}
		\clip (0,0) circle (3.0mm);
		\foreach \x in {-.9,-.75,...,.3}
		\draw[line width=1 pt] (\x,-.3) -- (\x+.6,.3);
	\end{scope}
	
	\node at (-1.05,-1.1) {\scriptsize DM};
	\node at (1.2,-1.1) {\scriptsize SM$_1$};
	\node at (0,1.4) {\scriptsize M$_t$};

	\node[vtxcircle, minimum size=2pt] at (0.915,0) {};
	\node[vtxcircle, minimum size=2pt] at (0.9,0.12) {};
	\node[vtxcircle, minimum size=2pt] at (0.9,-0.12) {};
	\node at (1.8,0) {\scriptsize $H, H^\dagger$};
	\node at (1.8,-0.3) {\scriptsize $\underbrace{\hspace{8mm}}$};
	\node at (1.8,-0.6) {\scriptsize $p$};
	
	\node at (0,-1.6) {\textbf{(a)}};
\end{tikzpicture} \hspace{1mm}
\raisebox{15.9mm}{\begin{tikzpicture}[line width=1.4pt, scale=1]
	\node at (1.8,0.0) {\huge $\Rightarrow$};
\end{tikzpicture}} \hspace{1mm}
\raisebox{-1.2mm}{\begin{tikzpicture}[line width=1.4pt, scale=1]
	\draw[fermionna] (0,1.4)--(0,0);
	\draw[fermionna] (1,-1)--(0,0);
	\draw[fermionna] (-1,-1)--(0,0);
	\node[vtxcircle] at (0,0) {};
	
	\draw[scalarna] (0.7,1)--(0,0.7);
	\draw[scalarna] (0.7,1.5)--(0,1.2);
	\node[vtxcircle, minimum size=1.6pt] at (0.35,1) {};
	\node[vtxcircle, minimum size=1.6pt] at (0.35,1.1) {};
	\node[vtxcircle, minimum size=1.6pt] at (0.35,1.2) {};
	
	\draw[scalarna] (0.5,-0.5)--(1.2,-0.2);
	\draw[scalarna] (0.8,-0.8)--(1.5,-0.5);
	\node[vtxcircle, minimum size=1.6pt] at (0.925,-0.425) {};
	\node[vtxcircle, minimum size=1.6pt] at (1,-0.5) {};
	\node[vtxcircle, minimum size=1.6pt] at (1.075,-0.575) {};
	
	\draw[scalarna] (-0.5,-0.5)--(-1.2,-0.2);
	\draw[scalarna] (-0.8,-0.8)--(-1.5,-0.5);
	\node[vtxcircle, minimum size=1.6pt] at (-0.925,-0.425) {};
	\node[vtxcircle, minimum size=1.6pt] at (-1,-0.5) {};
	\node[vtxcircle, minimum size=1.6pt] at (-1.075,-0.575) {};
	
	\node at (-1.1,-1.2) {\scriptsize DM};
	\node at (1.3,-1.2) {\scriptsize SM$_1$};
	\node at (0,1.6) {\scriptsize M$_t$};

	\node at (0.6,-0.2) {\scriptsize SM$_1'$};
	\node at (-0.55,-0.2) {\scriptsize DM$'$};
	\node at (-0.25,0.4) {\scriptsize M$_t''$};
	
	\node at (0,-1.7) {\textbf{(b)}};
\end{tikzpicture}} \hspace{3mm}
\raisebox{-1.2mm}{\begin{tikzpicture}[line width=1.4pt, scale=1]
	\draw[fermionna] (0,1.4)--(0,0);
	\draw[fermionna] (1,-1)--(0,0);
	\draw[fermionna] (-1,-1)--(0,0);
	\node[vtxcircle] at (0,0) {};
	
	\draw[scalarna] (1.4,0.6)--(0,0);
	\node at (1.5,0.8) {\scriptsize $H, H^\dagger$};
	
	\draw[scalarna] (0.7,1)--(0,0.7);
	\draw[scalarna] (0.7,1.5)--(0,1.2);
	\node[vtxcircle, minimum size=1.6pt] at (0.35,1) {};
	\node[vtxcircle, minimum size=1.6pt] at (0.35,1.1) {};
	\node[vtxcircle, minimum size=1.6pt] at (0.35,1.2) {};
	
	\draw[scalarna] (0.5,-0.5)--(1.2,-0.2);
	\draw[scalarna] (0.8,-0.8)--(1.5,-0.5);
	\node[vtxcircle, minimum size=1.6pt] at (0.925,-0.425) {};
	\node[vtxcircle, minimum size=1.6pt] at (1,-0.5) {};
	\node[vtxcircle, minimum size=1.6pt] at (1.075,-0.575) {};
	
	\draw[scalarna] (-0.5,-0.5)--(-1.2,-0.2);
	\draw[scalarna] (-0.8,-0.8)--(-1.5,-0.5);
	\node[vtxcircle, minimum size=1.6pt] at (-0.925,-0.425) {};
	\node[vtxcircle, minimum size=1.6pt] at (-1,-0.5) {};
	\node[vtxcircle, minimum size=1.6pt] at (-1.075,-0.575) {};
	
	\node at (-1.1,-1.2) {\scriptsize DM};
	\node at (1.3,-1.2) {\scriptsize SM$_1$};
	\node at (0,1.6) {\scriptsize M$_t$};

	\node at (0.6,-0.2) {\scriptsize SM$_1'$};
	\node at (-0.55,-0.2) {\scriptsize DM$'$};
	\node at (-0.25,0.4) {\scriptsize M$_t''$};
	
	\node at (0,-1.7) {\textbf{(c)}};
\end{tikzpicture}}

%% file: diagrams/higgs_hybrid.tex
\begin{tikzpicture}[line width=1.4pt, scale=1]
	\draw[fermionna] (1.2,0)--(0,0);
	\draw[fermionna] (-0.9,0.9)--(0,0);
	\draw[fermionna] (-0.9,-0.9)--(0,0);
	
	\draw[scalarna] (0.4,1.0583)--(0,0);
	\draw[scalarna] (-0.4,1.0583)--(0,0);
	
	\draw[fill=black] (0,0) circle (3.0mm);
	\draw[fill=white] (0,0) circle (2.9mm);
	\begin{scope}
		\clip (0,0) circle (3.0mm);
		\foreach \x in {-.9,-.75,...,.3}
		\draw[line width=1 pt] (\x,-.3) -- (\x+.6,.3);
	\end{scope}
	
	\node at (-1.1,0.9) {\scriptsize X};
	\node at (-1.25,-0.9) {\scriptsize DM};
	\node at (1.55,0) {\scriptsize SM$_3$};
	
	\node[vtxcircle, minimum size=2pt] at (0,0.915) {};
	\node[vtxcircle, minimum size=2pt] at (0.12,0.9) {};
	\node[vtxcircle, minimum size=2pt] at (-0.12,0.9) {};
	\node at (0.0,1.3) {\scriptsize $H, H^\dagger$};
	\node at (0.0,1.6) {\scriptsize $\overbrace{\hspace{8mm}}$};
	\node at (0.0,1.9) {\scriptsize $p$};
	
	\node at (0,-1.4) {\textbf{(a)}};
\end{tikzpicture} \hspace{-1mm}
\raisebox{14.3mm}{\begin{tikzpicture}[line width=1.4pt, scale=1]
	\node at (1.8,0.0) {\huge $\Rightarrow$};
\end{tikzpicture}} \hspace{-3mm}
\raisebox{-4.0mm}{\begin{tikzpicture}[line width=1.4pt, scale=1]
	\draw[fermionna] (1.4,0)--(0,0);
	\draw[fermionna] (-1,1)--(0,0);
	\draw[fermionna] (-1,-1)--(0,0);
	\node[vtxcircle] at (0,0) {};
	
	\draw[scalarna] (1,0.7)--(0.7,0);
	\draw[scalarna] (1.5,0.7)--(1.2,0);
	\node[vtxcircle, minimum size=1.6pt] at (1,0.35) {};
	\node[vtxcircle, minimum size=1.6pt] at (1.1,0.35) {};
	\node[vtxcircle, minimum size=1.6pt] at (1.2,0.35) {};
	
	\draw[scalarna] (-0.5,-0.5)--(-0.2,-1.2);
	\draw[scalarna] (-0.8,-0.8)--(-0.5,-1.5);
	\node[vtxcircle, minimum size=1.6pt] at (-0.425,-0.925) {};
	\node[vtxcircle, minimum size=1.6pt] at (-0.5,-1) {};
	\node[vtxcircle, minimum size=1.6pt] at (-0.575,-1.075) {};
	
	\draw[scalarna] (-0.5,0.5)--(-0.2,1.2);
	\draw[scalarna] (-0.8,0.8)--(-0.5,1.5);
	\node[vtxcircle, minimum size=1.6pt] at (-0.425,0.925) {};
	\node[vtxcircle, minimum size=1.6pt] at (-0.5,1) {};
	\node[vtxcircle, minimum size=1.6pt] at (-0.575,1.075) {};
	
	\node at (-1.2,1) {\scriptsize X};
	\node at (-1.35,-1) {\scriptsize DM};
	\node at (1.75,0) {\scriptsize SM$_3$};

	\node at (-0.5,0.2) {\scriptsize X$'$};
	\node at (-0.58,-0.2) {\scriptsize DM$'$};
	\node at (0.4,-0.2) {\scriptsize SM$_3'$};
	
	\node at (0,-1.8) {\textbf{(b)}};
\end{tikzpicture}} \hspace{3mm}
\raisebox{-4.0mm}{\begin{tikzpicture}[line width=1.4pt, scale=1]
	\draw[fermionna] (1.4,0)--(0,0);
	\draw[fermionna] (-1,1)--(0,0);
	\draw[fermionna] (-1,-1)--(0,0);
	\node[vtxcircle] at (0,0) {};
	
	\draw[scalarna] (0.6,1.4)--(0,0);
	\node at (0.7,1.6) {\scriptsize $H, H^\dagger$};
	
	\draw[scalarna] (1,0.7)--(0.7,0);
	\draw[scalarna] (1.5,0.7)--(1.2,0);
	\node[vtxcircle, minimum size=1.6pt] at (1,0.35) {};
	\node[vtxcircle, minimum size=1.6pt] at (1.1,0.35) {};
	\node[vtxcircle, minimum size=1.6pt] at (1.2,0.35) {};
	
	\draw[scalarna] (-0.5,-0.5)--(-0.2,-1.2);
	\draw[scalarna] (-0.8,-0.8)--(-0.5,-1.5);
	\node[vtxcircle, minimum size=1.6pt] at (-0.425,-0.925) {};
	\node[vtxcircle, minimum size=1.6pt] at (-0.5,-1) {};
	\node[vtxcircle, minimum size=1.6pt] at (-0.575,-1.075) {};
	
	\draw[scalarna] (-0.5,0.5)--(-0.2,1.2);
	\draw[scalarna] (-0.8,0.8)--(-0.5,1.5);
	\node[vtxcircle, minimum size=1.6pt] at (-0.425,0.925) {};
	\node[vtxcircle, minimum size=1.6pt] at (-0.5,1) {};
	\node[vtxcircle, minimum size=1.6pt] at (-0.575,1.075) {};
	
	\node at (-1.2,1) {\scriptsize X};
	\node at (-1.35,-1) {\scriptsize DM};
	\node at (1.75,0) {\scriptsize SM$_3$};

	\node at (-0.5,0.2) {\scriptsize X$'$};
	\node at (-0.58,-0.2) {\scriptsize DM$'$};
	\node at (0.4,-0.2) {\scriptsize SM$_3'$};
	
	\node at (0,-1.8) {\textbf{(c)}};
\end{tikzpicture}}

%% file: diagrams/higgs_exampleB.tex
	\begin{tikzpicture}[line width=1.4pt, scale=1]
		\draw[-] (1.6,0.8)--(1.2,0);
		\draw[-] (1.6,-0.8)--(1.2,0);
		\draw[-] (-1.1,0.8)--(-0.7,0);
		\draw[-] (-1.1,-0.8)--(-0.7,0);
		\draw[-] (-0.7,0)--(1.2,0);
		\draw[scalarna] (-0.3,0.8)--(-0.7,0);
		\draw[scalarna] (0.8,0.8)--(1.2,0);
		\draw[fill=black] (1.2,0) circle (3.0mm);
		\draw[fill=white] (1.2,0) circle (2.9mm);
		\begin{scope}
			\clip (1.2,0) circle (3.0mm);
			\foreach \x in {-1.9,-1.75,...,2}
			\draw[line width=1 pt] (\x,-.3) -- (\x+.6,.3);
		\end{scope}
		\draw[fill=black] (-0.7,0) circle (3.0mm);
		\draw[fill=white] (-0.7,0) circle (2.9mm);
		\begin{scope}
			\clip (-0.7,0) circle (3.0mm);
			\foreach \x in {-1.9,-1.75,...,1}
			\draw[line width=1 pt] (\x,-.3) -- (\x+.6,.3);
		\end{scope}
		\node at (-1.1,0.95) {\scriptsize X$^{(N,\alpha)}$};
		\node at (-1.15,-1) {\scriptsize DM$^{(N,\beta)}$};
		\node at (2.15,0.8) {\scriptsize $\ell_R^{(1,-2)}$};
		\node at (2.15,-0.8) {\scriptsize $\ell_R^{(1,-2)}$};
		\node at (0.3,-0.2) {\tiny M$_s^{(2,-3)}$};
		\node at (-0.2,1.02) {\scriptsize $H^\dagger$};
		\node at (0.8,1) {\scriptsize $H$};
         \node at (0,-1.5) {\textbf{(a)}};
	\end{tikzpicture}
	\hspace{0 mm}	
	\begin{tikzpicture}[line width=1.4pt, scale=1]
		\draw[-] (1.9,0.9)--(1.5,0);
		\draw[-] (1.9,-0.9)--(1.5,0);
		\draw[-] (-1.2,0.8)--(-0.8,0);
		\draw[-] (-1.2,-0.8)--(-0.8,0);
		\draw[-] (-0.8,0)--(1.5,0);
		\draw[scalarna] (0,0.8)--(0,0);
		\draw[scalarna] (0.7,0.8)--(0.7,0);
		\node[vtxcircle ] at (-0.8,0) {};
		\node[vtxcircle ] at (0,0) {};
		\node[vtxcircle ] at (0.7,0) {};
		\node[vtxcircle ] at (1.5,0) {};
		\node at (-1.1,0.95) {\scriptsize X$^{(N,\alpha)}$};
		\node at (-1.15,-1) {\scriptsize DM$^{(N,\beta)}$};
		\node at (1.8,0.95) {\scriptsize $\ell_R^{(1,-2)}$};
		\node at (1.8,-1.15) {\scriptsize $\ell_R^{(1,-2)}$};
		\node at (-0.425,0.3) {\tiny M$_s'^{(\!1\!,\!-\!4\!)}$};
		\node at (0.525,-0.3) {\tiny M$_s^{(2,-3)}$};
		\node at (1.2,0.3) {\tiny M$_s''^{(\!1\!,\!-\!4\!)}$};
		\node at (0, 1.02) {\scriptsize $H^\dagger$};
		\node at (0.7, 1) {\scriptsize $H$};
         \node at (0.35,-1.5) {\textbf{(b)}};
	\end{tikzpicture}
	\hspace{0 mm}	
	\begin{tikzpicture}[line width=1.4pt, scale=1]
		\draw[-] (2,0.8)--(1.6,0);
		\draw[-] (2,-0.8)--(1.6,0);
		\draw[-] (-0.7,0.6)--(-0.7,0);
		\draw[-] (-1.1,-0.8)--(-0.7,0);
		\draw[-] (-0.7,0)--(1.6,0);
		\draw[-] (-0.7,0.6)--(-1.1,1.2);
		\draw[scalarna] (-0.7,0.6)--(-0.3,1.2);
		\draw[scalarna] (0.7,0.8)--(0.7,0);
		\node[vtxcircle ] at (-0.7,0) {};
		\node[vtxcircle ] at (-0.7,0.6) {};
		\node[vtxcircle ] at (0.7,0) {};
		\node[vtxcircle ] at (1.6,0) {};
		\node at (-1.1,1.4) {\scriptsize X$^{(N,\alpha)}$};
		\node at (-1.15,-1) {\scriptsize DM$^{(N,\beta)}$};
		\node at (1.9,0.95) {\scriptsize $\ell_R^{(1,-2)}$};
		\node at (2,-1) {\scriptsize $\ell_R^{(1,-2)}$};
		\node at (-0.1,-0.3) {\tiny M$_s^{(2,-3)}$};
		\node at (1.1,-0.3) {\tiny M$_s''^{(1,-4)}$};
		\node at (0,0.35) {\tiny X$'^{(\!N\!\pm\!1,\alpha\!+\!1\!)}$};
		\node at (-0.15, 1.4) {\scriptsize $H^\dagger$};
		\node at (0.7, 1) {\scriptsize $H$};
         \node at (0.35,-1.5) {\textbf{(c)}};
	\end{tikzpicture}

%% file: diagrams/higgs_example_hybrid_v2.tex
\begin{tikzpicture}[line width=1.4pt, scale=1]
	\draw[fermionna] (-0.8,1.6)--(0,0);
	\draw[fermionna] (-0.4,-0.8)--(0,0);
	\draw[fermionna] (0,0)--(1.2,0);
	\node[vtxcircle] at (0,0) {};

	\draw[scalarna] (-0.6,1.2)--(0.1,1.55);
	\draw[scalarna] (-0.2,0.4)--(0.5,0.75);
	\node[vtxcircle] at (-0.6,1.2) {};
	\node[vtxcircle] at (-0.2,0.4) {};

	\node[vtxcircle, minimum size=1.6pt] at (-0.15,1.05) {};
	\node[vtxcircle, minimum size=1.6pt] at (-0.1,0.95) {};
	\node[vtxcircle, minimum size=1.6pt] at (-0.05,0.85) {};

	\node at (-1.1,1.8) {\scriptsize X$^{(3,\pmb{n}_{\mathrm{X}},-\beta+\frac{1}{3})}$};
	\node at (-0.6,-1) {\scriptsize DM$^{(1,N,\beta)}$};
	\node at (1.05,0.3) {\scriptsize X$'^{(3,N\pm1,-\beta+\frac{1}{3})}$};
	\node at (1.8,-0.15) {\scriptsize $Q_L^{(3,2,\frac{1}{3})}$};

	\node at (0.5,1.6) {\scriptsize ($H H^\dagger$)};
	\node at (0.9,0.8) {\scriptsize ($H H^\dagger$)};
	\node at (1.15,1.4) {\rotatebox{-60}{\scriptsize $\overbrace{\hspace{14mm}}$}};
	\node at (1.4,1.5) {\scriptsize $p$};
	
	\node at (0,-1.6) {\textbf{(a)}};
\end{tikzpicture} 
\hspace{-1mm}
\begin{tikzpicture}[line width=1.4pt, scale=1]
	\draw[fermionna] (3,0)--(0,0);
	\draw[fermionna] (-0.4,0.8)--(0,0);
	\draw[fermionna] (-0.4,-0.8)--(0,0);
	\node[vtxcircle] at (0,0) {};
	
	\draw[scalarna] (1.6,0.7)--(1.6,0);
	\draw[scalarna] (2.6,0.7)--(2.6,0);
	\node[vtxcircle] at (1.6,0) {};
	\node[vtxcircle] at (2.6,0) {};
	\node[vtxcircle, minimum size=1.6pt] at (2,0.35) {};
	\node[vtxcircle, minimum size=1.6pt] at (2.1,0.35) {};
	\node[vtxcircle, minimum size=1.6pt] at (2.2,0.35) {};
	
	\node at (-0.5,1) {\scriptsize X$^{(3,\pmb{n}_{\mathrm{X}},-\beta+\frac{1}{3})}$};
	\node at (-0.5,-1) {\scriptsize DM$^{(1,N,\beta)}$};
	\node at (0.8,-0.3) {\scriptsize SM$_3'^{(\!3,2p+2,\frac{1}{3}\!)}$};
	\node at (1.65,0.9) {\scriptsize ($H H^\dagger$)};
	\node at (2.65,0.9) {\scriptsize ($H H^\dagger$)};
	\node at (2.1,1.2) {\scriptsize $\overbrace{\hspace{16mm}}$};
	\node at (2.1,1.5) {\scriptsize $p$};
	\node at (3.6,0) {\scriptsize $Q_L^{(3,2,\frac{1}{3})}$};
	
	\node at (1,-1.8) {\textbf{(b)}};
\end{tikzpicture}

%% file: sections/phenomenology_v25.tex
\section{Phenomenology}
\label{sec:phenomenology}

In this section, we give an overview of the characteristics and phenomenological signatures of our simplified models. The plethora of possible models that were constructed following the assumptions and constraints detailed in section~\ref{sec:classification} comprises a rich set of new signals and discovery prospects.  We emphasize that our very general and conservative assumptions about possible annihilation mechanisms for dark matter, if true, underpin a framework that is guaranteed to include the true dark matter model of Nature.  The resulting set of simplified models, summarized in table~\ref{tab:classification:hybrid} for hybrid models, tables~\ref{tab:classification:s-channel:1}, \ref{tab:classification:s-channel:3}, \ref{tab:classification:s-channel:8ex} for $s$-channel mediators and tables~\ref{tab:classification:t-channel:1}, \ref{tab:classification:t-channel:3}, \ref{tab:classification:t-channel:8ex} for $t$-channel mediators, demonstrate that the full breadth of simplified dark matter models has yet to be systematically explored.

We remind the reader of the benefits, from a phenomenologist's point of view, of simplified models compared to frameworks augmenting the Standard Model by only a DM candidate and its effective interaction vertices. Since dark matter interactions to SM particles are constrained to be extremely weak, they are typically not mediated by SM gauge bosons but involve new vertices with a priori unknown coupling constants, e.g.,~Higgs portals.  Our ignorance about the coupling constants limits the predictability of models that feature only a DM candidate and its interactions.  Simplified models, on the other hand, typically allow for more robust phenomenological predictions. They contain, besides the DM particle, a minimal set of necessary additional ingredients.  Moreover, the additional particles typically carry Standard Model gauge charges, and thus they have guaranteed interactions with known coupling strengths and vertex structures.  This significantly enhances the prospects for discovering or excluding a simplified model compared to attempts to probe the DM particle in isolation. In addition, omitting the mediator of interactions between DM and SM particles and instead relying on effective operators severely limits the applicability of such effective models in a collider setting~\cite{Fox:2011pm, Busoni:2013lha, Buchmueller:2013dya, Busoni:2014haa, Busoni:2014sya, Racco:2015dxa}.

In contrast to a top-down reductionist approach (best exemplified by SUSY), we have pursued a bottom-up procedure.  We acknowledge that each model merits an independent study on its own, as an exemplary part of a more complicated dark matter sector involving additional fields and new interactions.  These could have an impact on the phenomenology, for example, by modifying DM interaction rates, opening up new DM production channels at the LHC, or adding totally new, subdominant components to the dark matter in the Universe.  Nevertheless, a large number of phenomenological traits can be characterized at the simplified model level, and this is the goal of this section.

We will first comment on the cosmology of our simplified models, the various direct detection and indirect detection signals and constraints, and then focus on the new bevy of LHC collider signatures.

\subsection{Coannihilation conditions}
\label{subsec:conditions}

One of the tenets of the present work is the assumption that coannihilation contributes significantly to the thermal freeze-out of dark matter, reducing its relic density to the measured value $\Omega h^2 = 0.1198 \pm 0.0026$~\cite{Ade:2015xua,Agashe:2014kda}.  As for models that only feature DM pair annihilation, the weakly interacting massive particle (WIMP) miracle~\cite{Kolb:1990vq} hypothesis can be invoked to argue that the DM, its coannihilation partner X, and the mediator M$_s$ or M$_t$ should have weak-scale masses and weak-scale couplings.  In this case, the relevant annihilation and coannihilation cross sections are parametrically within the correct range to yield the measured relic density.  Of course, the WIMP miracle can be avoided in many ways, for instance in scenarios with a ``WIMP-less miracle'' (see~\cite{Feng:2008ya}), allowing dark matter masses and couplings to span many decades.

In order for the coannihilation partner X to have a significant effect on the dark matter relic density, its number density should remain close to the dark matter number density during dark matter freeze-out.  Note that X should eventually decay, and due to the dark sector $\mathbb{Z}_2$ parity, the decay is into DM plus some SM particles. Since we are assuming standard thermal abundances for the dark sector particles in the early Universe, the number density of the coannihilating particle X tracks the DM number density during freeze-out, as long as the relative mass splitting
\begin{equation}\label{eq:defDelta}
 \Delta \equiv \frac{m_{\text{X}} - m_{\text{DM}}}{m_{\text{DM}}} \,
\end{equation}
is small.\footnote{More precisely, $m_{\text{DM}} \Delta$ should be small compared to the freeze-out temperature.}  The interactions that keep DM and X in equilibrium are of the form DM~DM $\leftrightarrow$ X~X, DM~SM $\leftrightarrow$ X~SM, and X $\leftrightarrow$ DM~SM~SM~\cite{Griest:1990kh}.

We now estimate the mass splitting $\Delta$ necessary for coannihilation to occur. The discussion is based on the seminal paper~\cite{Griest:1990kh} (for a recent treatment, see~\cite{Bell:2013wua}).  The mass splitting has direct implications on the collider signatures that we will discuss in section~\ref{subsec:LHCsigs}.

A naive estimate of the required $\Delta$ can be obtained as follows. The relic density, obtained by solving the Boltzmann equation, depends on the dimensionless variable $x = m_{ \rm DM} / T$. For cold, non-relativistic dark matter, freeze-out occurs at $x_F \equiv m_{\rm DM} / T_F \sim 20 - 30$, where $T_F$ and $x_F$ are the values at freeze-out. Coannihilation is expected to be important for $m_{\rm X} - m_{ \rm DM} \sim T_F$.  These two conditions imply that $\Delta \sim x_F^{-1} \sim$ 0.03$-$0.05.

This estimate can be further refined, since the relative strength of the coannihilation and annihilation processes also plays an important role. Assuming that, due to DM--X chemical equilibrium, the ratio of densities for dark matter and its coannihilation partner approximately maintains its equilibrium value, the problem can be reduced to solving a single Boltzmann equation for X and DM, with an effective cross section given by~\cite{Griest:1990kh}
\begin{align} \label{eq:sigmaeff}
  \sigma_{\text{eff}} = \frac{g_{\rm DM}^2}{g_{\text{eff}}^2} \bigg\{ & \sigma_{\rm DM \, DM} + 2 \sigma_{\rm DM \, X} \frac{g_{\rm X}}{g_{\rm DM}} (1 + \Delta)^{3/2} \exp(-x \Delta) \nonumber \\
  & + \sigma_{\rm X \, X} \frac{g_{\rm X}^2}{g_{\rm DM}^2} (1 + \Delta)^3 \exp(-2 x \Delta) \bigg\} \, .
\end{align}
Here $\sigma_{a \, b} = \sigma (a \, b \to \rm SM \, SM)$ and $g_{ \rm DM},\; g_{ \rm X}$ are the numbers of degrees of freedom of DM and X respectively, and
\begin{equation}
g_{\text{eff}} = g_{\rm DM} + g_{\rm X} (1+ \Delta)^{3/2} \exp ( -x \Delta) \, .
\end{equation}

At this point there is still a sizeable model dependence.  As our objective is to obtain an estimate of the mass splitting, we adopt a prototypical model.  Following~\cite{Griest:1990kh}, we consider a colored coannihilation partner $X$ and a weakly coupled DM, and neglect the temperature dependence of the relevant cross sections, allowing us to write $\sigma_{\rm X \, X} = A \sigma_{\rm DM \, X} = A^2 \sigma_{\rm DM \, DM}$, where $A \approx \alpha_s / \alpha \approx 20$ is the approximate relative weighting of cross sections for the various (co)annihilation scatterings. We obtain
\begin{equation}
\label{eq:relicratio}
  \frac{\sigma_{\text{eff}}}{\sigma_{\rm DM \, DM}} = \left( \frac{1 + A w}{1 + w} \right)^2 \ ,
\end{equation}
where $w = (1 + \Delta)^{3/2} \exp(-x \Delta) g_{\rm X} / g_{\rm DM} = (g_{\text{eff}} - g_{\rm DM}) / g_{\rm DM}$. We find numerically that for triplets, sextets and octets of $SU(3)_C$, $\Delta$ should be below 0.134, 0.163 and 0.175, respectively. We take, as a crude estimate, $\Delta \leq 0.2$.

Given that we made rough assumptions here (i.e., ratio of cross sections, velocity and temperature dependence, etc.) we will use this estimate as a rule of thumb when considering LHC signatures. Note that this is of capital importance when considering soft particles in the final state, in view of the experimental thresholds  for the different kinds of physical objects (electrons, muons, jets, \ldots). For the phenomenological case study discussed in section~\ref{sec:leptoquark}, the relic density will be computed numerically and used as a constraint in scanning the parameter space. 

\subsection{Direct detection}
\label{subsec:direct_detection}

Direct dark matter detection experiments have placed stringent constraints on the DM--nucleon scattering cross section. For spin-independent scattering, these experiments~\cite{Aprile:2013doa,Akerib:2013tjd,Agnese:2013rvf} have reached cross section limits as low as $7.6 \times 10^{-46}$~cm$^2$ for WIMP masses of 33 GeV~\cite{Akerib:2013tjd}. This rules out scattering through $Z$ exchange by orders of magnitude for DM particles with weak scale masses. In the MSSM, for instance, a typical cross section is $10^{-44}$~cm$^2$.

Since most of our simplified models feature dark matter candidates in non-trivial representations of $SU(2)_L$, the constraint on $Z$ exchange from direct detection alone would rule out many of them for DM masses not too far from the electroweak scale.  Therefore, one natural way to avoid direct detection constraints is to postulate a dark matter mass below the experimental detection threshold of typically a few GeV.  Note, however, that in this case it becomes more difficult to achieve the correct relic density since the naively expected magnitude of the annihilation cross section no longer matches the requirements for thermal freeze-out.  Even for heavier dark matter masses, $Z$ exchange does not play a role if the dark matter particle is the neutral component of an $SU(2)_L$ multiplet with hypercharge $Y=0$, which does not couple to the $Z$.  Another scenario in which direct detection bounds are significantly weaker is when dark matter is a Majorana fermion. Its vector couplings then vanish, so that $Z$ exchange only contributes an axial vector current to the dark matter--nucleus scattering cross section.  This current couples to the spin of the target nucleus rather than its mass or charges, and direct detection constraints on spin-dependent couplings are significantly weaker.  Finally, direct detection constraints can be relaxed by introducing non-minimal modifications. Many of these do not alter the associated collider phenomenology. For instance, an additional new particle with quantum numbers $(1, N, -\beta)$ could mix with the $(1, N, \beta)$ DM particle from our tables to yield a new dark matter candidate without a tree-level coupling to the $Z$ boson.
 
Beyond $Z$ exchange, bounds provided by direct detection experiments are highly model dependent.  The current WIMP--nucleon scattering cross sections probed by experiments lie in a region of parameter space that can be induced by Higgs exchange~\cite{Djouadi:2011aa,LopezHonorez:2012kv}.  Moreover, X, M$_s$, and M$_t$ could mediate additional scattering diagrams, and bounds on spin-dependent dark matter--nucleon scattering can also give important constraints.  Because of the freedom of our couplings, in particular in relation to the Higgs, the impact of these bounds should be assessed on an individual basis.

\subsection{Indirect detection}
In our simplified models, dark matter can be completely neutral under the Standard Model gauge symmetries or can carry weak and hypercharge quantum numbers.  In the first case, the possible indirect detection signals from pair annihilation of dark matter particles are highly model-specific, depending on unknown parameters or on additional interactions not included in the simplified model, such as the strength of a Higgs portal interaction or kinetic mixing between a hypothetical dark sector gauge boson and the SM hypercharge boson.  In the second case, the dark matter pairs may annihilate to SM particles via the weak interaction.  These possibilities have already been well-studied in the literature~\cite{Bertone:2004pz}, and we have no novel signatures to add in this regard.

One class of signatures that may become very interesting in scenarios with large $SU(2)_L$ representations are mono-energetic photons from dark matter annihilation to $\gamma\gamma$, $Z\gamma$ or $h\gamma$ final states.  These annihilation processes, which are necessarily loop-suppressed for electrically neutral DM, depend strongly on the electric charges of the particles propagating in the loop.  If they are doubly or even triply charged, the loop suppression is partially overcome, possibly raising the annihilation cross section to a level amenable to searches for gamma ray lines~\cite{Kopp:2013mi}.

Note that the additional particles in our simplified models, namely the coannihilation partner X, the mediator M$_s$ or M$_t$, and the charged components of the DM multiplet, in general have no indirect signatures since their number densities today vanish. We also remark that if the relic density were determined by dark matter annihilation alone instead of depending on coannihilation, we could establish a more direct connection between the observed relic density and a possible indirect detection signal.

\subsection{Production and decay at hadron colliders}
\label{subsec:collider}

The immediate significance of our complete classification is the plethora of new collider signatures from these simplified model constructions and the correlations between dark matter probes afforded by our framework.  In particular, our approach illustrates that many new dark matter model constructions are possible compared to what has been considered previously, and a systematic exploration of these signatures at the LHC is certainly warranted.

Given that our motivation is to test the dark matter (co)annihilation mechanism at the LHC, we discuss the prospects for finding the dark matter, its coannihilating partner, and the $s$-channel or $t$-channel mediator prescribed by the simplified model.  Depending on the Standard Model gauge quantum numbers of the new fields, strong or electroweak pair production of the dark particles can be allowed through gauge interactions. On the other hand, the vertices present in the coannihilation diagram serve as the minimally allowed decay channels for the coannihilating partner and the mediator. Relic density constraints determine the preferred regions of parameter space for the masses and couplings of the dark matter, its coannihilating partner, and the mediator. These guaranteed production modes, guaranteed decay channels, and characteristic kinematics combine to form a robust set of novel collider signatures that are well-motivated discovery possibilities for dark matter.

\subsubsection{Production processes}
\label{subsubsec:production}
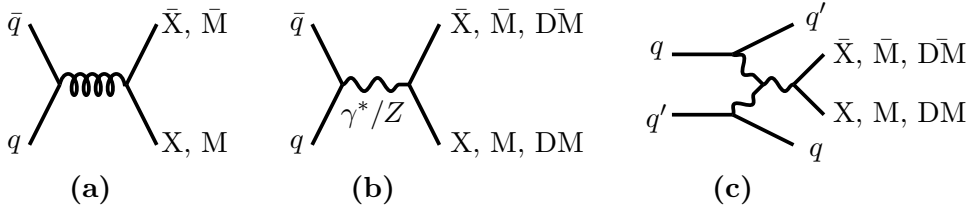
\begin{figure}[t]
	\centering
	\input{diagrams/doubleproduction_v5.tex}
        \caption{Representative diagrams for the pair production of new particles, relevant to both $s$-channel and $t$-channel simplified models.  Note that the final state jets in the vector boson fusion diagram {\bf (c)} are typically emitted in the forward direction.}
	\label{fig:doubleproduction}
\end{figure}

We begin with the production modes of the new particles in our simplified models. If charged under the Standard Model gauge group, DM, X, M$_s$ and M$_t$ can be pair produced via the kinetic terms. Thus, strong pair production of mediators M$_s$ or M$_t$ and coannihilating partners X is possible if they are colored, which obviously entails large rates at the LHC and important constraints from existing searches (e.g., figure~\subref{fig:doubleproduction}{(a)}). For mediators and coannihilation partners carrying only electroweak charges, production via $s$-channel electroweak gauge bosons, figure~\subref{fig:doubleproduction}{(b)}, as well as vector boson fusion processes, figure~\subref{fig:doubleproduction}{(c)}, are significant, especially if the masses of the new particles are lighter than or around the TeV scale.  If the dark matter particle is a component of a non-trivial $SU(2)_L$ multiplet, direct production of the other components of this multiplet is also possible. Note that bosonic DM, X and M also allow for four-point interactions with two gauge bosons stemming from the kinetic terms. These production modes only rely on Standard Model gauge charges and are common to all of our simplified models, irrespective of the given annihilation topology and the connection to dark matter physics. The universality of such production modes implies these production processes will be the main drivers for testing each individual simplified model at the LHC.

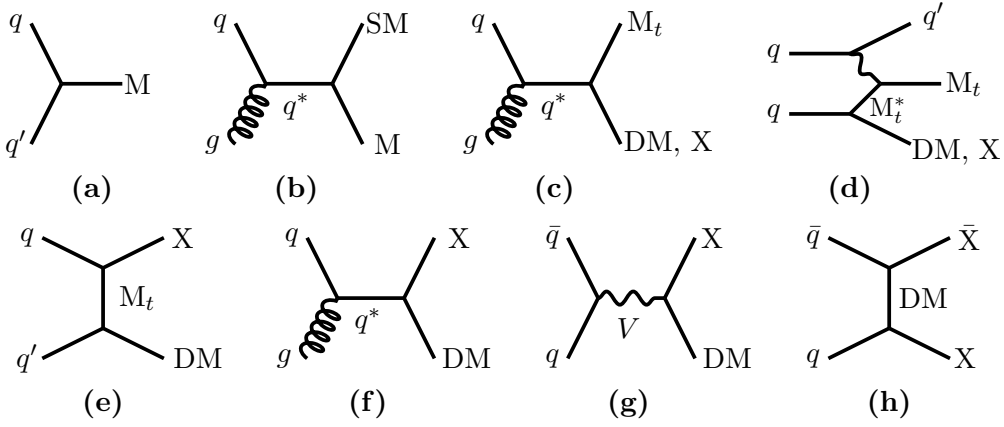
\begin{figure}[t]
	\centering
	\input{diagrams/doubleproduction_co_vertices.tex}
        \caption{Example production modes of new particles in $s$-channel, {\bf (a)} and {\bf (b)}, $t$-channel, {\bf (c)}--{\bf (e)}, and hybrid models, {\bf (f)}--{\bf (h)} (see section~\ref{subsubsec:hybrid}).  We use $V$ to denote an electroweak gauge boson.  Note the final state jet in diagram {\bf (d)} is typically emitted in the forward direction.}
	\label{fig:doubleproduction_co_vertices}
\end{figure}

Additionally, in $s$-channel models, crossing symmetry of the coannihilation diagram can lead to resonant mediator production if SM$_1$ and SM$_2$ are contained in the proton, as shown in figure~\subref{fig:doubleproduction_co_vertices}{(a)}.  Unlike in the pair production modes, here the production rate depends on the values of a priori unspecified coupling constants. If the coannihilation diagram has a final state quark or gauge boson, the mediator can also be produced in association with the other SM particle, as in figure~\subref{fig:doubleproduction_co_vertices}{(b)}. If DM and X are gauge singlets, mediator decay via the inverted coannihilation diagram is their only production mode.

In $t$-channel models, along with production via figures~\subref{fig:doubleproduction}{(a)--(c)}, X, DM and M$_t$ can be produced via combinations of gauge boson interactions and coannihilation diagram vertices.  The parity of the dark sector fields forbids single mediator production and so pairs of X, DM and M$_t$ must be produced.  If SM$_1$ or SM$_2$ is a quark, we can produce M$_t$ and DM or M$_t$ and X, respectively, for example via figure~\subref{fig:doubleproduction_co_vertices}{(c)}.  These could also be produced in association with a forward jet, which would afford an easier discrimination from backgrounds, but at the cost of a weak coupling factor, see figure~\subref{fig:doubleproduction_co_vertices}{(d)}.  Crossing symmetry of the coannihilation diagram can of course lead to an X DM final state if both SM$_1$ and SM$_2$ are quarks, figure~\subref{fig:doubleproduction_co_vertices}{(e)}. If both SM$_1$ and SM$_2$ are not part of the proton (quarks, gluons, or gauge bosons), then new production modes for DM, X, and M, aside from Standard Model kinetic terms, are highly model dependent.  Note that we have presented this discussion assuming the X, DM, and mediator fields to be unique.  If these fields coincide, then the mixed production modes become redundant. 

In models with a four-point vertex, X DM can again be produced by crossing symmetry of the coannihilation diagram if SM$_1$ and SM$_2$ are quarks. If only SM$_1$ (SM$_2$) is a parton, a three-body final state including X, DM and SM$_2$ (SM$_1$) would be possible. Again, if both SM$_1$ and SM$_2$ are not partons of the proton, new production modes for X and DM are model dependent.  Finally, for hybrid models, where a direct coupling between X, DM, and a SM particle exists, production of an DM--X final state is possible when SM$_3$ is part of the proton, as shown in figure~\subref{fig:doubleproduction_co_vertices}{(f)}, or an electroweak gauge boson, see figure~\subref{fig:doubleproduction_co_vertices}{(g)}. Also X pair production is possible, as depicted in figure~\subref{fig:doubleproduction_co_vertices}{(h)}.

Note that the production modes via Standard Model kinetic terms for DM, X, or the mediator only apply when these fields are not SM gauge singlets.  If any of these fields are SM gauge singlets, their only production modes require recycling of a coannihilation vertex or additional vertices which have not been specified.  Also note that any of these production modes can have additional initial state radiation which can be used to tag or trigger events.

\subsubsection{Decay modes}

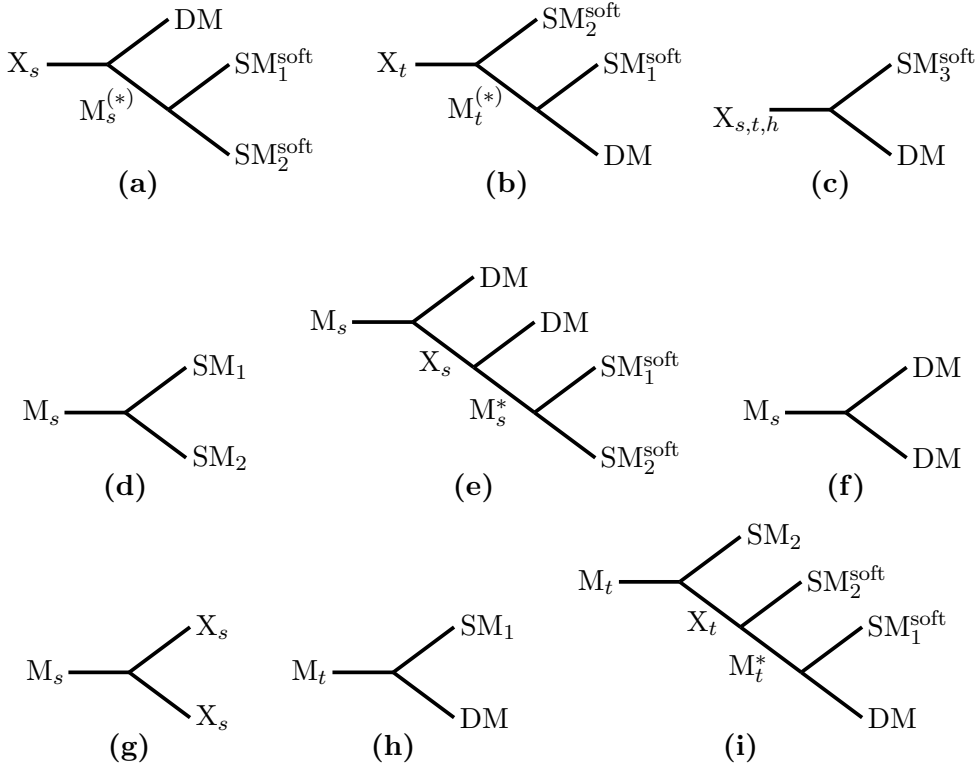
\begin{figure}[!tb]
   \centering
   \input{diagrams/decay_modes.tex}
   \caption{Decay modes of the mediator and the coannihilation partner in the different coannihilation models. Some final state SM particles are soft since the fractional mass splitting $\Delta$ is generally small. Diagrams \textbf{(a)}--\textbf{(c)} represent the standard decays of the coannihilation partner for the $s$-channel, $t$-channel and hybrid models. Diagrams \textbf{(d)} and \textbf{(e)} represent decays for the $s$-channel mediator, diagrams \textbf{(f)} and \textbf{(g)} show mediator decays arising from possible additional vertices.  Finally, diagrams \textbf{(h)} and \textbf{(i)} represent decays for the $t$-channel mediator.}
   \label{fig:decaymodes}
\end{figure}

For each of our simplified models, the minimal decay modes for each new particle are dictated by the coannihilation diagram.  We have ensured that dark matter decay is prevented by a technically natural $\mathbb{Z}_2$ parity, see section~\ref{subsubsec:decay}, and thus we only need to study the decays of the coannihilation partner X and the mediators M$_s$, M$_t$. All possible decay channels are summarized in figure~\ref{fig:decaymodes}. We note that, due to the coannihilation conditions discussed in section~\ref{subsec:conditions}, X and DM will tend to have a small mass splitting. This feature will show up in various decay channels, leading to soft SM particles in the final state. 

The coannihilation partner X has two possible ways of decaying: either via the crossed coannihilation diagram itself, or via an additional vertex DM--X--SM$_3$. In the first case we obtain DM accompanied by the two products of coannihilation, SM$_1$ and SM$_2$, as shown in figure~\subref{fig:decaymodes}{(a)--(b)}. In the second option, we have X $\to$ DM + SM$_3$, as depicted in figure~\subref{fig:decaymodes}{(c)}. Note that this decay mode is the only possible one in hybrid models. Furthermore, models with four-point interactions are similar but with a three body final state, X $\to$ DM + SM$_1$ + SM$_2$.

In $s$-channel models, the mediator minimally decays to SM$_1$ SM$_2$, as shown in figure~\subref{fig:decaymodes}{(d)} or to X DM, see figure~\subref{fig:decaymodes}{(e)}. In the latter case X has the two possible decay modes described in the previous paragraph. Additional vertices M$_s$--DM--DM and M$_s$--X--X provide new decay channels, figure~\subref{fig:decaymodes}{(f)--(g)}, thus making the branching ratios to the visible and semi-visible final states highly model-dependent.  Since the mediator can be pair-produced via Standard Model gauge interactions (except for the SM singlet cases \hyperref[tab:classification:s-channel:1]{SU1}, \hyperref[tab:classification:s-channel:1]{SU2}), this motivates a thorough exploration of two-body resonances in both fully visible and semi-visible decays.

For $t$-channel models, the main difference lies in M$_t$ now being odd under the $\mathbb{Z}_2$ parity. This inevitably forbids any fully visible signature, since M$_t$ is forced to decay into a final state containing the lightest $\mathbb{Z}_2$-odd particle DM. The final state can have one or three SM particles in addition, M$_t \to$ DM + SM$_1$, as shown in figure~\subref{fig:decaymodes}{(h)},  or M$_t \to$ X + SM$_2$, X $\to$ DM + SM$_1$ + SM$_2$, as depicted in figure~\subref{fig:decaymodes}{(i)}. Here it is important to stress that the mediator mass is not related to the coannihilation scale given by $m_{\text{X}}$ or $m_{\text{DM}}$. Hence, each decay mode of M$_t$ provides a particle that is potentially hard, depending on the actual mass of the mediator and on the SM particle. This leads to interesting cascade decay signatures at the LHC, which will be detailed in section~\ref{subsec:LHCsigs}. Again, the branching fraction for either decay channel is a free parameter, therefore the combination of different decay chain topologies is necessary in order to cover the full signature space.  In particular, the semi-visible X decay depends sensitively on the kinematics and particle type of SM$_1$ and SM$_2$. The $\mathbb{Z}_2$ parity explicitly forbids M$_t$ mediator couplings to X--X, and the only new possible vertex is a coupling between X, DM, and a SM field, as shown in figure~\subref{fig:decaymodes}{(c)}.

\subsection{Classes of LHC signatures}
\label{subsec:LHCsigs}
After having identified the main production and decay modes of the new particles in our simplified models, we now turn to the concrete LHC signatures of these new processes.  Our goal in this section is to illustrate the breadth of dark matter phenomenology in our simplified models and to highlight how the synergy among distinct experimental searches helps to draw a complete picture of the dark sector.

We will concentrate on the most generic signatures that are present in most of our simplified models.  We will not comment in detail on signatures that are relevant only in a particular corner of parameter space such as invisible Higgs decays, which give strong bounds on new DM states (and also X states since the decays of X are typically very soft) coupled to the Higgs~\cite{CMS:2015naa, Aad:2015pla}. However, these decays require new particles with masses $\lesssim 60$~GeV.  We will also not comment in detail on indirect constraints, in particular from electroweak precision data~\cite{Peskin:1991sw,Barbieri:2004qk,Cirelli:2005uq}, from loop-induced modifications to Drell-Yan processes~\cite{Altmannshofer:2014cla}, or from modifications to loop-mediated Higgs decay rates, in particular $h \to \gamma\gamma$ and $h \to \gamma Z$.  The EW precision constraints are particularly relevant for models with large electroweak multiplets that contain multiply charged components.  Modifications to Higgs branching ratios of course require new particles that couple to the Higgs, for instance through a scalar quartic coupling.  Note, however, that in such a case the value of the corresponding coupling constant is not necessarily related to the dark matter phenomenology. 

Another class of signatures we will not address in detail are those arising only after electroweak symmetry breaking. They include the cascade decays of the heavier components of an extended $SU(2)_L$ multiplet to the lighter ones via emission of an off-shell $W$ boson as well as the signatures arising from the Higgs-induced mixing of new particles with SM particles or among each other (see section~\ref{subsec:HiggsBreaking}).  Finally, we will not address in detail flavor constraints because our simplified models are constructed in such a way that they always admit a trivial flavor structure.

A general overview of signature classes for $s$-channel, $t$-channel, and hybrid models is given in table~\ref{tab:unifiedsignatures}. For each combination of production and decay processes of new particles, the table summarizes the resulting experimental signatures. The first part of the table contains processes that are common to $s$-channel, $t$-channel, and hybrid models, while the remainder of the table list processes that exist only in a specific class of models, as indicated in the first column. In the second column, we specify in each row a unique physical process, where the label ``soft'' indicates that a particle is highly produced close to threshold in the rest frame of its parent particle. (Note that, with this nomenclature, a soft particle can still achieve a large transverse momentum if its parent particle is highly boosted.) The label ``res'' indicates that two particles originate from the resonant decay of an on-shell particle, so that a bump is observable in their invariant mass distribution. The third column ``Prod.~via'' of table~\ref{tab:unifiedsignatures} gives the conditions under which the production processes in the second column exists.  Here, ``gauge int.'' means that the primary interaction products (before any decays) must carry Standard Model gauge quantum numbers. The fourth column exposes the relevant experimental signatures, and the fifth column references ATLAS and CMS searches for these final states where available. Note that a number of signatures are \emph{not} yet searched for by the experimental collaborations, and even for the signatures where many searches are listed, the different multiplicities of varied SM objects are not covered completely.  We also emphasize again that each simplified model leads to several new signatures, so that a potential discovery in one channel can be cross-checked in another channel.  Moreover, the combination of different search channels can be exploited to discriminate between different models.

\begin{table}
	\centering
	\ssmall
	\input{tables/unified_signatures_v9}

        \caption{Classification of LHC signatures for $s$-channel, $t$-channel, and hybrid simplified models. For each hard process relevant to these models (second column), the table lists the conditions under which it exists (third column) and the associated experimental signatures (fourth column). Where available, we also include references to existing experimental searches that are sensitive to or provide partial coverage of these signatures. In the list of processes the superscript ``soft'' indicates that a particle is produced close to threshold in the rest frame of its parent particle.  This is relevant in particular for the decay products of X, thanks to the small mass splitting between X and DM.  The superscript ``res'' indicates that a pair of particles is produced from the decay of a resonance, leading to a narrow invariant mass peak. Note that possible extra vertices appearing in specific models, such as M--X--X or M--DM--DM interactions, are not included here.}
	\label{tab:unifiedsignatures}
\vspace{-5mm}
\end{table}

\boldmath
\subsubsection{Signature class 1: mono-Y analyses and X--X, DM--X and DM--DM production}
\label{subsubsec:Xsoftdecay}
\unboldmath

The first class of experimental signatures we will discuss are those arising from the decay of the coannihilation partner via X $\to$ DM SM$_1$ SM$_2$ in processes of the form $p\,p \to$ X X or $p\,p \to$ X DM.  We will also briefly comment on the process $p\,p \to$ DM DM. These processes are ubiquitous in our models, and since X decay also appears as a sub-diagram in M$_s$ or M$_t$ decays, we discuss it first. As alluded to before, a particularly interesting aspect about $\text{X} \to \text{DM}\; \text{SM}_1\; \text{SM}_2$ is that the SM particles are typically very soft, thanks to the small fractional mass splitting $\Delta$ between X and DM (see section~\ref{subsec:conditions}).  For instance, for a typical mass splitting $\Delta \lesssim 0.2$ and typical DM masses of around 100 GeV, SM$_1$ and SM$_2$ particles would receive momenta of $\mathcal{O}(10\ \text{GeV})$.  The range of $\Delta$ preferred in coannihilation scenarios, from $\mathcal{O}(0.2)$ to the sub-per cent level, straddles the kinematic boundary between hard, resolvable objects and soft, marginal energy deposits in the event.

If the soft particles can be resolved, the expected signature is missing energy plus one SM$_1$ SM$_2$ pair slightly above the detection threshold for $p\,p \to$ X DM, and missing energy plus two SM$_1$ SM$_2$ pairs slightly above the detection threshold for $p\,p \to$ X X.

If the soft particles cannot be resolved, DM--X and X--X production are chracterized by pure missing energy and are thus indistinguishable from DM--DM pair production.  The traditional way to tag these events is to require initial state radiation (ISR), which leads to a mono-Y signature (Y=jet, photon, $Z$, $W$).  Here, monojet events typically offer a larger production cross section, while mono-photon, mono-$Z$ and mono-$W$ events lead to cleaner signatures. Note that in models where X is colored, the mono-jet signal is enhanced because not only ISR, but also final state radiation (FSR) contributes.

The extra ISR (or FSR) provides a signal to trigger on, but if it is hard enough, it will also endow the decay products of X with a boost that makes them more likely to pass the detection threshold.  This leads to signatures characterized by a hard ISR object, one or two relatively soft, but resolvable $\text{SM}_1 \, \text{SM}_2$ pairs, and large missing energy. Such a signature offers many handles to discriminate signal from background. We note that due to the large multiplicity of hadrons in a typical LHC event, leptons are easier to detect than jets; nevertheless, a multi-jet analysis could still be a viable option for models with large color representations, such as sextets and octets. The combination of ISR or FSR with the many possibilities for the soft SM$_{1,2}$ particles gives rise to a large and varied signature space and offers many handles to identify and discriminate between models.  

Note that in hybrid models (bottom part of table~\ref{tab:unifiedsignatures}), the final states of events involving DM--X (X--X) production need not involve two (four) SM particles. Since X can decay to SM$_3$ DM in this case, final states with only one (two) SM$_3$ particles are also possible.

We remark that the ISR jet + $\slashed{E}_T$ + soft lepton analysis has also been motivated from the MSSM context~\cite{Giudice:2010wb, Rolbiecki:2012gn, Schwaller:2013baa, Han:2014kaa, Low:2014cba, Bramante:2014dza, Bramante:2014tba}, which is included in our mono-Y + $\slashed{E}_T$ + $\leq 4$ SM and mono-Y + $\slashed{E}_T$ + $\leq 2$ SM signatures of table~\ref{tab:unifiedsignatures}.  One initial effort in this direction by CMS targets soft muons from pair-produced supersymmetric tops that undergo four-body decays~\cite{CMS:2015eoa}, which covers the single SM multiplicity in our signature class.  We advocate, however, that such searches be extended to include more multiplicities and different particle species, as motivated by the intrinsic kinematics of the coannihilation mechanism and the set of simplified models shown in tables~\ref{tab:classification:hybrid}--\ref{tab:classification:t-channel:8ex}.

\subsubsection{Signature class 2: resonances from $s$-channel models}
\label{subsubsec:resonances}

The second class of LHC signatures that we will discuss is related to the single or double production of the mediator M$_s$ in $s$-channel models. We will restrict our discussion to models where M$_s$ is a new particle. As mentioned before, the mediator can be pair-produced via strong or electroweak interactions unless it is a pure SM gauge singlet. Additionally, single or associated production of the mediator can be present if one of the coannihilation products is a SM quark.  Single mediator production via vector bosons is also possible in models with with extended gauge sectors after EWSB.

When these production modes are stitched together with the decays $\text{M}_s \to \text{SM}_1 \,\text{SM}_2$ and $\text{M}_s \to \text{X} \, \text{DM}$ followed by $\text{X} \to \text{SM}_1 \,\text{SM}_2 \, \text{DM}$, we find three different collider signatures for pair-produced mediators and four distinct single production signatures.

For pair-produced M$_s$, the signature that is easiest to reconstruct is the one where both M$_s$ particles decay fully visibly, $p p \to \text{M}_s \, \text{M}_s \to 2 (\text{SM}_1 \, \text{SM}_2)^\text{res}$, leading to paired resonances in a four-particle final state.  This is exemplified by strong pair production of leptoquarks (see our case study in section~\ref{sec:leptoquark}), and by dijet resonances (see models in the SO and SE categories).  The fully visible final state will dominate if Br$(\text{M}_s \to \text{SM}_1 \, \text{SM}_2)$ is much larger than Br$(\text{M}_s \to \text{X} ~ \text{DM})$.

In the opposite case, both mediators will mostly decay semi-visibly, $p p \to \text{M}_s \, \text{M}_s \to 2 \, \text{SM}_1^{\text{soft}} + 2 \, \text{SM}_2^{\text{soft}} + 4 \, \text{DM}$, and the resulting final state will be very similar to the mono-Y + soft particles final state discussed in the previous section. We note, however, that in the decay of M$_s$, X could receive a large boost if M$_s$ is much heavier than X and DM.  In this case, the SM$_1$ and SM$_2$ particles from X decay would no longer be soft.  Their detectability would be greatly enhanced, and moreover their boost would allow us to distinguish M$_s$ pair production from X--X or DM--X production.

The third signature of pair-produced $s$-channel mediators M$_s$ is the mixed decay, $p p \to \text{M}_s \, \text{M}_s \to (\text{SM}_1 \, \text{SM}_2)^\text{res} \, \text{SM}_1^\text{soft} \,\text{SM}_2^\text{soft} + 2 \, \text{DM}$.  This signature is especially interesting since the hard resonance recoils against the semi-visible decay, offering more kinematic handles than the usual resonance bump hunt.  For our case study in section~\ref{sec:leptoquark}, we will perform a detailed analysis of this novel signature.  Initial studies of the dilepton resonance+$\slashed{E}_T$, dijet resonance+$\slashed{E}_T$, and charged tracks+$\slashed{E}_T$ signatures have also been presented in the literature~\cite{Gupta:2015lfa, Autran:2015mfa, Bai:2015nfa}.

A secondary effect of the two possible decay modes of the $s$-channel mediator is the renewed impetus for the LHC to test a wide range of mediator masses with high luminosity data: even though the pair production rate of the mediator is only a function of its gauge charge and mass, the relative production rates of the fully visible, semi-visible, and mixed final states depend on additional free parameters. Thus, searches for relatively light mediators in the fully visible final state remain relevant with more luminosity, as the increased data affords tests of smaller branching fractions into that final state.  In the fortunate case of a discovery, measuring the decay rates will be critical for extracting the couplings of the mediator to SM particles and to the dark sector.

A singly produced $s$-channel mediator can appear alone in a Drell-Yan process if both SM$_1$ and SM$_2$ are constituents of the proton, or with an associated hard SM$_1$ (SM$_2$) particle if SM$_2$ (SM$_1$) is a constituent of the proton. Here, constituents of the proton are of course quarks and gluons, but also electroweak gauge bosons which could be radiated from a quark line in a vector boson fusion topology.  Vector boson fusion topologies are also interesting as they provide forward jets that help improve discrimination prospects between signal and background.  In contrast to the M$_s$ pair-production modes which proceed via Standard Model gauge couplings, the single production modes depend on the coannihilation diagram couplings. For resonant M$_s$ production in a Drell-Yan event, we expect dijet resonance searches to cover the space of colored mediators~\cite{Khachatryan:2015sja,Aad:2014aqa,Han:2010rf,Dobrescu:2013coa}. Resonances can be also produced in association with a hard SM particle. This hard SM particle can play an important role in tagging the event and reducing backgrounds, especially when the mediator decays semi-visibly to $\text{DM} + \text{X} \, (\to \text{SM}_1^{\text{soft}} \, \text{SM}_2^{\text{soft}} \, \text{DM})$, but also when it decays to SM$_1$ SM$_2$. As an example for this type of signatures, consider leptoquark mediated models, where the mediator can be singly produced in association with a lepton if the leptoquark coupling is large~\cite{Khachatryan:2015qda}.  We will discuss this particular example further in our case study in section~\ref{sec:leptoquark}. 

\subsubsection{Signature class 3: cascade decays from $t$-channel mediators}
\label{subsubsec:cascade}
Let us now turn our attention to signatures specific to models with a $t$-channel mediator M$_t$.  This mediator can again be pair-produced, or it can be produced in association with a DM or X particle. Its decay modes are $\text{M}_t \to \text{SM}_1 \, \text{DM}$ and $\text{M}_t \to \text{SM}_2 \, \text{X}$ followed by $\text{X} \to \text{SM}_1 \,\text{SM}_2 \, \text{DM}$. Since the branching fractions for M$_t$ to either two-body final state are not fixed a priori, a systematic search strategy should focus on both single step as well as multistep decay chains.  Depending on the mass splitting between M$_t$ and DM, which is a free parameter in our simplified models, the SM$_1$ or SM$_2$ particle produced in the decays of M$_t$ can be either soft or hard.  In the decay $\text{M}_t \to \text{SM}_2 \, \text{X}$, the possible boost of the X particle is also highly phenomenologically relevant. For boosted X, the SM particles produced in the subsequent decay of X will be boosted as well, making them easier to detect. For unboosted X, these SM particles are soft and may fall below the detection threshold.  Note that boosted X particles can also arise if a heavy M$_t$ particle is singly produced in association with an X. In this case, the X particle receives a boost from recoiling against M$_t$.

Signatures with several hard SM particles in the final state can be striking, depending on the identity of these SM particles. For example, in $t$-channel models with leptoquark mediators such as \hyperref[tab:classification:t-channel:3]{TT1} or \hyperref[tab:classification:t-channel:3]{TT8}, a promising final state involves a hard lepton, a hard jet, and missing energy from pair production of M$_t$, with one of the M$_t$ particles decaying to $q + \text{DM}$ and the other decaying to a lepton plus an unobserved X particle. The same models also give rise to a di-lepton plus $\slashed{E}_T$ final state if both M$_t$ particles decay via $\text{M}_t \to \text{X} \ell$. While this signature is characteristic also for slepton searches~\cite{Aad:2014vma, Khachatryan:1704963}, in the present case the M$_t$ production cross section is much larger thanks to the color charge of the mediator.

\subsubsection{Compelling examples}

By combining the general signature classes presented in table \ref{tab:unifiedsignatures} with specific information about DM, X and M$_s$ or M$_t$ fields and their interactions listed in tables~\ref{tab:classification:hybrid}--\ref{tab:classification:t-channel:8ex}, we can generate the complete landscape of LHC phenomenology stemming from the existence of the coannihilation diagram. Discussing every possible configuration of dark sector fields and $(\mathrm{SM}_1 \, \mathrm{SM}_2)$ pairs is beyond the scope of this work. We point out, however, several examples of models leading to potentially intriguing signatures at proton-proton colliders.

In the case of $s$-channel models, collider phenomenology is typically driven by the nature of the mediator. The decay products of M$_s$ dictate not only resonant signatures stemming from the couplings to the SM particles but also soft signatures characteristic for coannihilation models. They are related by the fact that the decay of X proceeds through an off-shell M$_s$. The possibilities for uncolored bosonic mediators include a SM singlet, a dark Higgs, a Higgs triplet and electroweak gauge boson partners. These are related to well-studied di-boson, di-lepton and di-jet resonances. However, double production of M$_s$ also motivates searches for \emph{mixed} di-particle signatures from this entire group of final states, leading to striking signatures such as a di-Higgs--di-lepton final state. Moreover, targeting the soft decay modes motivates mono-Y+$\slashed{E}_T$ searches that allow for additional reconstructed leptons or jets with moderate $p_T$. Uncolored bosonic mediators also include charge-two di-lepton resonances, namely a field that only couples to two leptons via a lepton number violating coupling. Such models can additionally lead to striking resonance signatures such as pairs of same-sign di-leptons. 

Uncolored M$_s$ fermionic mediators have a general resemblance to lepton partners, while colored fermionic M$_s$ exhibit phenomenology similar to that of quark partners. We stress that fermionic $s$-channel mediators are specific to coannihilating dark matter (X$\neq$DM) and absent for annihilation (X$\equiv$DM). While the LHC phenomenology of such resonances is well explored, the interpretation as mediators in coannihilating dark matter models is absent in the literature. In this framework, a complete phenomenological study of fermion partners combining direct LHC searches and indirect bounds would be appealing and is still lacking. Moreover, a lepton partner M$_s$ can be linked to existing multi-lepton+$\slashed{E}_T$ signatures, while a quark partner M$_s$ will give further motivation to quark partner searches.

The couplings of bosonic $s$-channel mediators to the visible sector can involve for instance two quarks, or a quark and a lepton. Both types of couplings should not appear simultaneously due to proton stability bounds. Models with colored mediators exist only in coannihilation models and necessarily require colored X particles. Hence both the mediator and the coannihilation partner will be strongly produced at the LHC, providing two qualitatively different complementary probes for the simplified model. Since the leptoquark model provides very non-standard signatures, which have not yet been interpreted in the context of dark matter coannihilation, we choose this model class for our case study in section~\ref{sec:leptoquark}.

In the case of the $t$-channel signatures, the mediator production will usually lead to signals with at least one resolvable SM particle and missing energy. These signatures are already covered by the SUSY simplified model searches, extensively studied at the LHC. Since the mediator has at least two different decay modes, however, its pair production followed by a mixed decay could lead to new unexplored signals, such as different flavor lepton pairs. If the coannihilating partner X is colored, some of the models in our classification also allow for strong production of hard and/or soft leptons in association with missing transverse energy. Such models can be considered as $t$-channel variants of the $s$-channel leptoquark model discussed above. For uncolored M$_t$, X pair production in association with one ISR jet would lead to a monojet signature with additional leptons and jets, which also arises for the $s$-channel leptoquark models. If both X and the mediator are colored, mediator pair production would lead to a spectacular MSSM slepton-like signature at colliders. For such models, mixed decays of the pair-produced mediators could also lead to signals with a hard lepton, a hard jet and missing energy. Although this signature is already studied at the LHC, a more targeted analysis of kinematic features (using for example $m_{T2}$~\cite{Lester:1999tx, Barr:2003rg, Lester:2014xy}) would allow direct access to the structure of the coannihilation diagram.

%% file: diagrams/doubleproduction_v5.tex
\begin{tikzpicture}[line width=1.4pt, scale=1]
	\draw[fermionna] (0.875,0.8)--(0.475,0);
	\draw[fermionna] (0.875,-0.8)--(0.475,0);
	\draw[fermionna] (-0.8,0.8)--(-0.4,0);
	\draw[fermionna] (-0.8,-0.8)--(-0.4,0);
	\draw[gluon] (-0.4,0)--(0.475,0);
	\node at (-1.0,0.8) {$\bar{q}$};
	\node at (-1.0,-0.8) {$q$};
	\node at (1.375,0.8) {$\bar{\mathrm{X}}$, $\bar{\mathrm{M}}$};
	\node at (1.375,-0.8) {X, M};
	\node at (0,-1.4) {\textbf{(a)}};
\end{tikzpicture} 
\hspace{3mm}
\begin{tikzpicture}[line width=1.4pt, scale=1]
	\draw[fermionna] (0.875,0.8)--(0.475,0);
	\draw[fermionna] (0.875,-0.8)--(0.475,0);
	\draw[fermionna] (-0.8,0.8)--(-0.4,0);
	\draw[fermionna] (-0.8,-0.8)--(-0.4,0);
	\draw[vector] (-0.4,0)--(0.475,0);
	\node at (-1.0,0.8) {$\bar{q}$};
	\node at (-1.0,-0.8) {$q$};
	\node at (0,-0.4) {$\gamma^* /Z$};
	\node at (1.9,0.8) {$\bar{\mathrm{X}}$, $\bar{\mathrm{M}}$, $\bar{\mathrm{DM}}$};
	\node at (1.9,-0.8) {X, M, DM};
	\node at (0,-1.4) {\textbf{(b)}};
\end{tikzpicture} 
\hspace{3mm}
\begin{tikzpicture}[line width=1.4pt, scale=1]
	\draw[fermionna] (0.8,0.8)--(0,0.4);
	\draw[fermionna] (0.8,-0.8)--(0,-0.4);
	\draw[fermionna] (-0.8,0.4)--(0,0.4);
	\draw[fermionna] (-0.8,-0.4)--(0,-0.4);
	\draw[vector] (0,0.4)--(0.4,0);
	\draw[vector] (0,-0.4)--(0.4,0);
	\draw[vector] (0.4,0)--(0.8,0);
	\draw[fermionna] (0.8,0)--(1.2,0.4);
	\draw[fermionna] (0.8,0)--(1.2,-0.4);
	\node at (-1.,0.45) {$q$};
	\node at (-1.,-0.45) {$q'$};
	\node at (1.1,0.9) {$q'$};
	\node at (1.1,-0.9) {$q$};
	\node at (2.2,0.4) {$\bar{\mathrm{X}}$, $\bar{\mathrm{M}}$, $\bar{\mathrm{DM}}$};
	\node at (2.2,-0.4) {X, M, DM};
	\node at (0,-1.4) {\textbf{(c)}};
\end{tikzpicture} 

%% file: diagrams/doubleproduction_co_vertices.tex
\begin{tikzpicture}[line width=1.4pt, scale=1]
	\draw[fermionna] (-0.8,0.8)--(-0.4,0);
	\draw[fermionna] (-0.8,-0.8)--(-0.4,0);
	\draw[fermionna] (-0.4,0)--(0.4,0);
	\node at (-1,0.8) {$q$};
	\node at (-1,-0.8) {$q'$};
	\node at (0.6,0) {$\mathrm{M}$};
	\node at (0,-1.4) {\textbf{(a)}};
\end{tikzpicture} 
\hspace{2mm}
\begin{tikzpicture}[line width=1.4pt, scale=1]
	\draw[fermionna] (0.875,0.8)--(0.475,0);
	\draw[fermionna] (0.875,-0.8)--(0.475,0);
	\draw[fermionna] (-0.8,0.8)--(-0.4,0);
	\draw[gluon] (-0.8,-0.8)--(-0.4,0);
	\draw[fermionna] (-0.4,0)--(0.475,0);
	\node at (-1.0,0.8) {$q$};
	\node at (-1.1,-0.8) {$g$};
	\node at (0,-0.3) {$q^*$};
	\node at (1.2,0.8) {SM};
	\node at (1.2,-0.8) {M};
	\node at (0,-1.4) {\textbf{(b)}};
\end{tikzpicture} 
\hspace{2mm}
\begin{tikzpicture}[line width=1.4pt, scale=1]
	\draw[fermionna] (0.875,0.8)--(0.475,0);
	\draw[fermionna] (0.875,-0.8)--(0.475,0);
	\draw[fermionna] (-0.8,0.8)--(-0.4,0);
	\draw[gluon] (-0.8,-0.8)--(-0.4,0);
	\draw[fermionna] (-0.4,0)--(0.475,0);
	\node at (-1.0,0.8) {$q$};
	\node at (-1.1,-0.8) {$g$};
	\node at (0,-0.3) {$q^*$};
	\node at (1.2,0.8) {M$_t$};
	\node at (1.5,-0.8) {DM, X};
	\node at (0,-1.4) {\textbf{(c)}};
\end{tikzpicture} 
\hspace{2mm}
\begin{tikzpicture}[line width=1.4pt, scale=1]
	\draw[fermionna] (0.8,0.8)--(0,0.4);
	\draw[fermionna] (0.8,-0.8)--(0,-0.4);
	\draw[fermionna] (-0.8,0.4)--(0,0.4);
	\draw[fermionna] (-0.8,-0.4)--(0,-0.4);
	\draw[vector] (0,0.4)--(0.4,0);
	\draw[fermionna] (0,-0.4)--(0.4,0);
	\draw[fermionna] (0.4,0)--(1.2,0);
	\node at (-1,0.4) {$q$};
	\node at (-1,-0.4) {$q$};
	\node at (1.1,0.9) {$q'$};
	\node at (1.4,-0.9) {DM, X};
	\node at (0.5,-0.35) {\small M$_t^*$};
	\node at (1.5,0) {M$_t$};
	\node at (0,-1.4) {\textbf{(d)}};
\end{tikzpicture} 
\hspace{2mm}
\begin{tikzpicture}[line width=1.4pt, scale=1]
	\draw[fermionna] (0.8,0.8)--(0,0.4);
	\draw[fermionna] (0.8,-0.8)--(0,-0.4);
	\draw[fermionna] (-0.8,0.8)--(0,0.4);
	\draw[fermionna] (-0.8,-0.8)--(0,-0.4);
	\draw[fermionna] (0,-0.4)--(0,0.4);
	\node at (-1,0.8) {$q$};
	\node at (-1,-0.8) {$q'$};
	\node at (1.05,0.8) {X};
	\node at (1.25,-0.8) {DM};
	\node at (0.45,0) {M$_t$};
	\node at (0,-1.4) {\textbf{(e)}};
\end{tikzpicture}
\hspace{2mm}
\begin{tikzpicture}[line width=1.4pt, scale=1]
	\draw[fermionna] (0.875,0.8)--(0.475,0);
	\draw[fermionna] (0.875,-0.8)--(0.475,0);
	\draw[fermionna] (-0.8,0.8)--(-0.4,0);
	\draw[gluon] (-0.8,-0.8)--(-0.4,0);
	\draw[fermionna] (-0.4,0)--(0.475,0);
	\node at (-1.0,0.8) {$q$};
	\node at (-1.1,-0.8) {$g$};
	\node at (0,-0.3) {$q^*$};
	\node at (1.2,0.8) {X};
	\node at (1.3,-0.8) {DM};
	\node at (0,-1.4) {\textbf{(f)}};
\end{tikzpicture} 
\hspace{2mm}
\begin{tikzpicture}[line width=1.4pt, scale=1]
	\draw[fermionna] (0.875,0.8)--(0.475,0);
	\draw[fermionna] (0.875,-0.8)--(0.475,0);
	\draw[fermionna] (-0.8,0.8)--(-0.4,0);
	\draw[fermionna] (-0.8,-0.8)--(-0.4,0);
	\draw[vector] (-0.4,0)--(0.475,0);
	\node at (-1.0,0.8) {$\bar{q}$};
	\node at (-1.0,-0.8) {$q$};
	\node at (0,-0.4) {$V$};
	\node at (1.1,0.8) {X};
	\node at (1.3,-0.8) {DM};
	\node at (0,-1.4) {\textbf{(g)}};
\end{tikzpicture} 
\hspace{2mm}
\begin{tikzpicture}[line width=1.4pt, scale=1]
	\draw[fermionna] (0.8,0.8)--(0,0.4);
	\draw[fermionna] (0.8,-0.8)--(0,-0.4);
	\draw[fermionna] (-0.8,0.8)--(0,0.4);
	\draw[fermionna] (-0.8,-0.8)--(0,-0.4);
	\draw[fermionna] (0,-0.4)--(0,0.4);
	\node at (-1,0.8) {$\bar{q}$};
	\node at (-1,-0.8) {$q$};
	\node at (1.05,0.8) {$\bar{\mathrm{X}}$};
	\node at (1,-0.8) {X};
	\node at (0.45,0) {DM};
	\node at (0,-1.4) {\textbf{(h)}};
\end{tikzpicture}

%% file: diagrams/decay_modes.tex
\begin{tikzpicture}[line width=1.4pt, scale=1]
	\draw[fermionna] (-0.8,0)--(0,0);
	\draw[fermionna] (0,0)--(0.8,0.6);
	\draw[fermionna] (0,0)--(0.8,-0.6);
	\draw[fermionna] (0.8,-0.6)--(1.6,0);
	\draw[fermionna] (0.8,-0.6)--(1.6,-1.2);
	\node at (-1.1,0) {X$_s$};
	\node at (1.2,0.6) {DM};
	\node at (0,-0.55) {M$_s^{(*)}$};
	\node at (2.2,0) {SM$_1^\textrm{soft}$};
	\node at (2.2,-1.2) {SM$_2^\textrm{soft}$};
	\node at (0.4,-1.6) {\textbf{(a)}};
\end{tikzpicture} 
\hspace{3mm}
\begin{tikzpicture}[line width=1.4pt, scale=1]
	\draw[fermionna] (-0.8,0)--(0,0);
	\draw[fermionna] (0,0)--(0.8,0.6);
	\draw[fermionna] (0,0)--(0.8,-0.6);
	\draw[fermionna] (0.8,-0.6)--(1.6,0);
	\draw[fermionna] (0.8,-0.6)--(1.6,-1.2);
	\node at (-1.1,0) {X$_t$};
	\node at (1.4,0.6) {SM$_2^\textrm{soft}$};
	\node at (0,-0.55) {M$_t^{(*)}$};
	\node at (2.2,0) {SM$_1^\textrm{soft}$};
	\node at (2.0,-1.2) {DM};
	\node at (0.4,-1.6) {\textbf{(b)}};
\end{tikzpicture}
\vspace{3mm}
\begin{tikzpicture}[line width=1.4pt, scale=1]
	\draw[fermionna] (-0.8,0)--(0,0);
	\draw[fermionna] (0,0)--(0.8,0.6);
	\draw[fermionna] (0,0)--(0.8,-0.6);
	\node at (-1.1,-0.15) {X$_{s,t,h}$};
	\node at (1.4,0.6) {SM$_3^\textrm{soft}$};
	\node at (1.2,-0.6) {DM};
	\node at (0,-1.0) {\textbf{(c)}};
\end{tikzpicture} \\
\vspace{3mm}
\begin{tikzpicture}[line width=1.4pt, scale=1]
	\draw[fermionna] (-0.8,0)--(0,0);
	\draw[fermionna] (0,0)--(0.8,0.6);
	\draw[fermionna] (0,0)--(0.8,-0.6);
	\node at (-1.1,0) {M$_s$};
	\node at (1.25,0.6) {SM$_1$};
	\node at (1.25,-0.6) {SM$_2$};
	\node at (0,-1.0) {\textbf{(d)}};
\end{tikzpicture} \hspace{3mm}
\begin{tikzpicture}[line width=1.4pt, scale=1]
	\draw[fermionna] (-0.8,0)--(0,0);
	\draw[fermionna] (0,0)--(0.8,0.6);
	\draw[fermionna] (0,0)--(0.8,-0.6);
	\draw[fermionna] (0.8,-0.6)--(1.6,0);
	\draw[fermionna] (0.8,-0.6)--(1.6,-1.2);
	\draw[fermionna] (1.6,-1.2)--(2.4,-0.6);
	\draw[fermionna] (1.6,-1.2)--(2.4,-1.8);
	\node at (-1.1,0) {M$_s$};
	\node at (1.2,0.6) {DM};
	\node at (0.3,-0.55) {X$_s$};
	\node at (2.0,0) {DM};
	\node at (1,-1.2) {M$_s^*$};
	\node at (3,-0.6) {SM$_1^\textrm{soft}$};
	\node at (3,-1.8) {SM$_2^\textrm{soft}$};
	\node at (0.8,-2.2) {\textbf{(e)}};
\end{tikzpicture} 
\hspace{3mm}
\begin{tikzpicture}[line width=1.4pt, scale=1]
	\draw[fermionna] (-0.8,0)--(0,0);
	\draw[fermionna] (0,0)--(0.8,0.6);
	\draw[fermionna] (0,0)--(0.8,-0.6);
	\node at (-1.1,0) {M$_s$};
	\node at (1.2,0.6) {DM};
	\node at (1.2,-0.6) {DM};
	\node at (0,-1.0) {\textbf{(f)}};
\end{tikzpicture} 
\hspace{3mm}
\begin{tikzpicture}[line width=1.4pt, scale=1]
	\draw[fermionna] (-0.8,0)--(0,0);
	\draw[fermionna] (0,0)--(0.8,0.6);
	\draw[fermionna] (0,0)--(0.8,-0.6);
	\node at (-1.1,0) {M$_s$};
	\node at (1.1,0.6) {X$_s$};
	\node at (1.1,-0.6) {X$_s$};
	\node at (0,-1.0) {\textbf{(g)}};
\end{tikzpicture}
\hspace{3mm}
\begin{tikzpicture}[line width=1.4pt, scale=1]
	\draw[fermionna] (-0.8,0)--(0,0);
	\draw[fermionna] (0,0)--(0.8,0.6);
	\draw[fermionna] (0,0)--(0.8,-0.6);
	\node at (-1.1,0) {M$_t$};
	\node at (1.25,0.6) {SM$_1$};
	\node at (1.2,-0.6) {DM};
	\node at (0,-1.0) {\textbf{(h)}};
\end{tikzpicture} 
\hspace{3mm}
\begin{tikzpicture}[line width=1.4pt, scale=1]
	\draw[fermionna] (-0.8,0)--(0,0);
	\draw[fermionna] (0,0)--(0.8,0.6);
	\draw[fermionna] (0,0)--(0.8,-0.6);
	\draw[fermionna] (0.8,-0.6)--(1.6,0);
	\draw[fermionna] (0.8,-0.6)--(1.6,-1.2);
	\draw[fermionna] (1.6,-1.2)--(2.4,-0.6);
	\draw[fermionna] (1.6,-1.2)--(2.4,-1.8);
	\node at (-1.1,0) {M$_t$};
	\node at (1.25,0.6) {SM$_2$};
	\node at (0.3,-0.55) {X$_t$};
	\node at (2.2,0) {SM$_2^\textrm{soft}$};
	\node at (0.9,-1.15) {M$_t^*$};
	\node at (3,-0.6) {SM$_1^\textrm{soft}$};
	\node at (2.8,-1.8) {DM};
	\node at (0.8,-2.2) {\textbf{(i)}};
\end{tikzpicture}

%% file: tables/unified_signatures_v9.tex
\setcitestyle{citesep={,\!\!}}
\hspace*{-0.2cm}
\begin{tabular}{!{\vrule width 1pt} c | l | c | l | c !{\vrule width 1pt}}
  \noalign{\hrule height 1pt}
  
  & $p p \to \ldots$ & Prod.~via & Signatures & Search \\

\noalign{\hrule height 1pt}
\multirow{8}{*}{\rotatebox[origin=c]{90}{common}}
  & \multirow{3}{*}{DM + DM + ISR}
    & gauge int.
    & \multirow{3}{*}{mono-Y + $\slashed{E}_T$}
    & \multirow{3}{*}{\cite{Aad:2015zva,Aad:2014tda,Aad:2014vka,Khachatryan:2014rwa,Khachatryan:2014rra}} \\
  & & or SM$_1 \in p$ & & \\
 & & for $t$-channel & & \\

  \cline{2-5}
  
  & X ($\to$ SM$_1^{\text{soft}}$ SM$_2^{\text{soft}}$ DM)
    & gauge int.
    & mono-Y + $\slashed{E}_T$
    & \cite{Aad:2015zva,Aad:2014tda,Aad:2014vka,Khachatryan:2014rwa,Khachatryan:2014rra}\\
  & \quad + X ($\to$ SM$_1^{\text{soft}}$ SM$_2^{\text{soft}}$ DM) + ISR
    & or SM$_2 \in p$
    & mono-Y + $\slashed{E}_T + \leq$ 4 SM
    & Partial coverage~\cite{CMS:2015eoa} \\
   & & for $t$-channel & & \\

  \cline{2-5}
  
  & \multirow{2}{*}{DM + X ($\to$ SM$_1^{\text{soft}}$ SM$_2^{\text{soft}}$ DM) + ISR}
    & \multirow{2}{*}{(SM$_1$ SM$_2$) $\in p$}
    & mono-Y + $\slashed{E}_T$
    & \cite{Aad:2015zva, Aad:2014tda, Aad:2014vka, Khachatryan:2014rwa, Khachatryan:2014rra} \\
  & 
    &                                   
    & mono-Y + $\slashed{E}_T + \leq$ 2 SM
    & Partial coverage~\cite{CMS:2015eoa} \\
  
\noalign{\hrule height 1pt}
\multirow{12}{*}{\rotatebox[origin=c]{90}{$s$-channel}}
  & M$_s$ ($\to$ [SM$_1$ SM$_2$]$^{\text{res}}$)
    & \multirow{6}{*}{gauge int.}
    & \multirow{2}{*}{2 resonances}
    &  \multirow{2}{*}{\cite{Aad:2015tba, Aad:2015kqa, Aad:2015asa, Aad:2014efa, Khachatryan:2014lpa, Aad:2015caa, Khachatryan:2015vaa}} \\
  & \quad + M$_s$ ($\to$ [SM$_1$ SM$_2$]$^\text{res}$)
    &
    &
    & \\
  
  \cline{2-2} \cline{4-5}
  
  & M$_s$ ($\to$ [SM$_1$ SM$_2$]$^{\text{res}}$)
    &
    & resonance + $\slashed{E}_T$
    & No search\\
  & \quad + M$_s$ ($\to$ DM + X ($\to$ SM$_1^{\text{soft}}$ SM$_2^{\text{soft}}$ DM))
    &
    & resonance + $\slashed{E}_T + \leq$ 2 SM
    & No search\\
  
  \cline{2-2} \cline{4-5}
  
  & M$_s$ ($\to$ DM + X ($\to$ SM$_1^{\text{soft}}$ SM$_2^{\text{soft}}$ DM))
    &
    & \multirow{2}{*}{$\slashed{E}_T + \leq 4$ SM}
    & \multirow{2}{*}{\cite{Aad:2014kra, Aad:2015wqa, Aad:2014iza, Aad:2014mha, Aad:2014qaa, Khachatryan:2033174, Aad:2015tin, Khachatryan:1989788, Chatrchyan:1631468, Chatrchyan:1527115, Aad:2015wqa, Chatrchyan:1662652, Aad:2014wea}}\\
  & \quad + M$_s$ ($\to$ DM + X ($\to$ SM$_1^{\text{soft}}$ SM$_2^{\text{soft}}$ DM))
    &
    &
    & \\
  
  \cline{2-5}

  & M$_s$ ($\to$ [SM$_1$ SM$_2$]$^{\text{res}}$)
    & \multirow{3}{*}{(SM$_1$ SM$_2$) $\in p$}
    & 1 resonance
    & \cite{Aad:2013gma, ATLAScollaboration:2014eb, CMS-PAS-EXO-12-061, ATLAS:2014wra, Khachatryan:2014tva, ATLAS-CONF-2013-066, ATLAScollaboration:2014uc, Khachatryan:2014xja, CMSCollaboration:2014df, CMSCollaboration:2014ke, Aad:2015owa, CMS-PAS-EXO-12-021, ATLAScollaboration:2014tj, ATLAScollaboration:2014ur, Aad:2014aqa, CMS-PAS-EXO-12-023, ATLAS-CONF-2013-052, CMS-PAS-B2G-12-005, ATLAScollaboration:2014va, ATLAScollaboration:2014ul, CMS-PAS-B2G-12-010, CMSCollaboration:2014wm} \\

  \cline{2-2} \cline{4-5}

  & \multirow{2}{*}{M$_s$ ($\to$ DM + X ($\to$ SM$_1^{\text{soft}}$ SM$_2^{\text{soft}}$ DM))}
    &
    & \multirow{2}{*}{$\slashed{E}_T + \leq$ 2 SM}
    & \cite{Khachatryan:1989788, Chatrchyan:1631468, Chatrchyan:1527115, Aad:2014wea} \\
&&&& \cite{Aad:2015hea, Aad:2013oja, Aad:2014vka, Aad:2015yga, Aad:2013ija, Aad:2014vma, Aad:2014pda, Aad:2014yka}\\

  \cline{2-5}

  & SM$_{1,2}$ + M$_s$ ($\to$ [SM$_1$ SM$_2$]$^{\text{res}}$)
    & \multirow{3}{*}{$\text{SM}_{2,1}\in p$}
    & 1 resonance + 1 SM
    & Partial coverage \cite{Khachatryan:2015qda, TheATLAScollaboration:2013gia} \\
  
  \cline{2-2} \cline{4-5}

  & SM$_{1,2}$
    & 
    & \multirow{2}{*}{$\slashed{E}_T$ + 1 $\leq$ 3 SM}
    & \cite{Chatrchyan:1631468, Khachatryan:1989788, Aad:2014wea, Chatrchyan:1662652, Chatrchyan:1527115, Aad:2015wqa}\\
  & \quad + M$_s$ ($\to$ DM + X ($\to$ SM$_1^{\text{soft}}$ SM$_2^{\text{soft}}$ DM)) 
    &
    &
    & \cite{Aad:2013oja, Aad:2015yga, Aad:2014vma, Aad:2013ija, Aad:2015hea, Aad:2014yka, Aad:2014pda, Aad:2014nua, Aad:2015jqa, Khachatryan:1704963} \\
  
\noalign{\hrule height 1pt}
\multirow{14}{*}{\rotatebox[origin=c]{90}{$t$-channel}} 

  & M$_t$ ($\to$ SM$_1$ DM)
    & \multirow{7}{*}{gauge int.}
    & \multirow{2}{*}{$\slashed{E}_T + \leq$ 2 SM}
       & \cite{Khachatryan:1989788, Chatrchyan:1631468, Chatrchyan:1527115, Aad:2014wea} \\
& \quad + M$_t$ ($\to$ SM$_1$ DM) &&& \cite{Aad:2015hea, Aad:2013oja, Aad:2014vka, Aad:2015yga, Aad:2013ija, Aad:2014vma, Aad:2014pda, Aad:2014yka}\\
  
  \cline{2-2} \cline{4-5}
  
  & M$_t$ ($\to$ SM$_1$ DM)
    &
    & \multirow{2}{*}{$\slashed{E}_T + \leq$ 4 SM}
    & \cite{Aad:2015tba, Aad:2015kqa, Aad:2015asa, Aad:2014efa, Khachatryan:2014lpa, Aad:2015caa, Khachatryan:2015vaa} \\
  & \quad + M$_t$ ($\to$ SM$_2$ + X ($\to$ SM$_1^{\text{soft}}$ SM$_2^{\text{soft}}$ DM))
    &
    &
    & \cite{Aad:2015tin, Khachatryan:1989788, Chatrchyan:1631468, Chatrchyan:1527115, Aad:2015wqa, Chatrchyan:1662652, Aad:2014wea}\\
  
  \cline{2-2} \cline{4-5}
  
  & M$_t$ ($\to$ SM$_2$ + X ($\to$ SM$_1^{\text{soft}}$ SM$_2^{\text{soft}}$ DM))
    &
    & \multirow{2}{*}{$\slashed{E}_T + \leq$ 6 SM}
    & \cite{Aad:2014kra, Khachatryan:1989788, Chatrchyan:1631468, Chatrchyan:1527115, Aad:2015wqa, Chatrchyan:1662652, Aad:2014wea}\\
  & \quad + M$_t$ ($\to$ SM$_2$ + X ($\to$ SM$_1^{\text{soft}}$ SM$_2^{\text{soft}}$ DM))
    &
    &
    & \cite{Aad:2014mha, Aad:2014qaa, Khachatryan:2033174, Khachatryan:1984165, Chatrchyan:1630049, Aad:2015mia, Khachatryan:2004846, Aad:2013wta} \\

  \cline{2-5}

  & \multirow{2}{*}{DM + M$_t$ ($\to$ SM$_1$ DM)}
    & \multirow{4}{*}{SM$_1 \in p$}
    & \multirow{2}{*}{$\slashed{E}_T + \leq$ 1 SM}
    & \cite{Aad:2015zva, Aad:2014tda, Khachatryan:2014rwa, Khachatryan:2014rra} \\
   &&&& \cite{Aad:2015yga, Aad:2014vka} \\
  
  \cline{2-2} \cline{4-5}
  
  & DM 
    &
    & \multirow{2}{*}{$\slashed{E}_T + \leq$ 3 SM}
    & \cite{Chatrchyan:1631468, Khachatryan:1989788, Aad:2014wea, Chatrchyan:1662652, Chatrchyan:1527115, Aad:2015wqa}\\  
  & \quad + M$_t$ ($\to$ SM$_2$ + X ($\to$ SM$_1^{\text{soft}}$ SM$_2^{\text{soft}}$ DM))
    &
    & 
    & \cite{Aad:2014yka, Aad:2014pda, Aad:2015jqa, Aad:2014nua, Khachatryan:1704963} \\

  \cline{2-5}

  & M$_t$ ($\to$ SM$_1$ DM)
    & \multirow{4}{*}{SM$_2 \in p$}
    & \multirow{2}{*}{$\slashed{E}_T + \leq$ 3 SM}
    & \cite{Aad:2014wea, Chatrchyan:1662652, Khachatryan:1989788, Chatrchyan:1631468, Aad:2015wqa, Chatrchyan:1527115} \\
  & \quad + X ($\to$ SM$_1^{\text{soft}}$ SM$_2^{\text{soft}}$ DM)
    &
    &
    & \cite{Aad:2014nua, Aad:2014pda, Aad:2015jqa, Aad:2014yka, Khachatryan:1704963}\\
  
  \cline{2-2} \cline{4-5}
  
  & M$_t$ ($\to$ SM$_2$ + X ($\to$ SM$_1^{\text{soft}}$ SM$_2^{\text{soft}}$ DM))
    &
    & \multirow{2}{*}{$\slashed{E}_T + \leq$ 5 SM}
    & \cite{Aad:2015tin, Aad:2014kra, Khachatryan:1989788, Chatrchyan:1631468, Chatrchyan:1527115, Aad:2015wqa, Chatrchyan:1662652, Aad:2014wea, Aad:2014mha, Aad:2014qaa, Khachatryan:2033174} \\
  & \quad + X ($\to$ SM$_1^{\text{soft}}$ SM$_2^{\text{soft}}$ DM)
    &
    &
    & \cite{Chatrchyan:1630049, Chatrchyan:1567175, Aad:2015mia, Khachatryan:1984165}\\
  
\noalign{\hrule height 1pt}
\multirow{4}{*}{\rotatebox[origin=c]{90}{hybrid}}
  & X ($\to$ DM + SM$_3^{\text{soft}}$)
    & gauge int.
    & \multirow{2}{*}{$\slashed{E}_T + \leq$ 2 SM}
      & \cite{Khachatryan:1989788, Chatrchyan:1631468, Chatrchyan:1527115, Aad:2014wea} \\
  &\quad + X ($\to$ DM + SM$_3^{\text{soft}}$) 
    & or SM$_3 \in p$ 
    & 
    &  \cite{Aad:2015hea, Aad:2013oja, Aad:2014vka, Aad:2015yga, Aad:2013ija, Aad:2014vma, Aad:2014pda, Aad:2014yka}\\

  \cline{2-5}

  & \multirow{2}{*}{DM + X ($\to$ DM + SM$_3^{\text{soft}}$)}
    & \multirow{2}{*}{SM$_3 \in p$}
    & \multirow{2}{*}{$\slashed{E}_T + \leq$ 1 SM}
    & \cite{ATLAS:2014wra, Khachatryan:2014tva, Aad:2015yga}\\
   &&&& \cite{Aad:2015zva, Aad:2014tda, Aad:2014vka, Khachatryan:2014rwa, Khachatryan:2014rra}\\

\noalign{\hrule height 1pt}
\end{tabular}
\setcitestyle{citesep={,}}

%% file: sections/casestudy_v35.tex
\section{Case study: leptoquark model}
\label{sec:leptoquark}

In this section, we will adopt the $s$-channel model \hyperref[tab:classification:s-channel:3]{ST11}, which features a leptoquark mediator, and we study its phenomenology in detail.  This will exemplify many of the general comments we have made in section~\ref{sec:phenomenology} regarding the phenomenology of our simplified models.  After introducing the Lagrangian and field content of model \hyperref[tab:classification:s-channel:3]{ST11} in section~\ref{subsec:Lagrangian}, we thoroughly discuss the physics of coannihilation and dark matter freeze-out in section~\ref{subsec:casestudy_cosmology}. There, we also briefly comment on direct and indirect detection prospects.  Finally, in section~\ref{subsec:casestudy:LHC}, we expound on the rich LHC phenomenology of the model. We use existing leptoquark and mono-jet searches to constrain the parameter space of the model, and we develop new search strategies for the mono-jet + leptons as well as the visible and invisible mixed leptoquark pair signatures.

\subsection{From the simplified model to the Lagrangian}
\label{subsec:Lagrangian}
The explicit field content of model \hyperref[tab:classification:s-channel:3]{ST11} is shown in table~\ref{tab:ST11fields}. As mentioned in section~\ref{sec:classification}, this is the minimal model content needed to realize the $s$-channel coannihilation diagram X DM $\rightarrow$ SM$_1$ SM$_2$.

\begin{table}[!tb]
	\centering
	\small
	\include{tables/casestudy_field_content}
	\caption{Field content, Standard Model gauge quantum numbers, and spin assignments for the case study \hyperref[tab:classification:s-channel:3]{ST11}.}
\label{tab:ST11fields}
\end{table}

For this model, we choose the DM to be a Majorana fermion, X a Dirac fermion, and M$_s$ a complex scalar.  The general dimension-four Lagrangian is then
\begin{align} \label{eq:SF11Lag}
  \mathcal{L} &= \frac{i}{2} \overline{\text{DM}} \slashed{\partial} \text{DM} + i \overline{\text{X}} \slashed{D} \text{X} + |D_{\mu} \text{M}_s|^2 - \dfrac{m_{\text{DM}}}{2} \overline{\text{DM}}\,\text{DM} - m_{\text{X}} \overline{\text{X}}\,\text{X} - V(\text{M}_s, H) \nonumber \\[0.2cm]
	&- \left( y_D \overline{\text{X}}\,\text{M}_s\,\text{DM} + y_{Q\ell}\,\overline{Q_L} \text{M}_s\, \ell_R + y_{Lu} \overline{L_L} \text{M}^c_s u_R + \text{h.c.} \right) ,
\end{align}
where
\begin{equation}
	V(\text{M}_s, H) = V(H) + m_{\text{M}_s}^2 \text{M}_s^\dagger \, \text{M} + \dfrac{1}{4} \lambda_{\text{M}_s} \big( \text{M}_s^\dagger \, \text{M}_s \big)^2 + \epsilon_{\text{M}_s} \text{M}_s^\dagger \, \text{M}_s \left( H^\dag H - \dfrac{v^2}{2} \right) ,
\end{equation}
where $\text{M}_s^c \equiv i \sigma^2 \text{M}_s^*$ is the charge-conjugate field and $V(H)$ is the SM Higgs potential.  Note that M$_s$ and $H$ do not mix as M$_s$ is colored.  We also note that the dark sector Yukawa coupling $y_D$ and the leptoquark Yukawa coupling matrices $y_{Q\ell}$ and $y_{Lu}$ are arbitrary in our construction.  This is no different than in traditional leptoquark models~\cite{Buchmuller:1986zs,Queiroz:2014pra}. The flavor structure of these matrices directly affects the prospects for LHC searches but is very well constrained by low energy probes.  To ensure compatibility with these probes, we adopt very minimal, flavor-safe configurations (see appendix~\ref{app:flavor} for a comprehensive discussion of the flavor structure of the model).  In particular, we take $y_{Lu} = 0$ and (a) $y_{Q\ell}^{11} \neq 0$ or (b) $y_{Q\ell}^{22} \neq 0$, with all other $y_{Q\ell}^{ij} = 0$. In case (a), the visible sector decay mode of the leptoquark is to a jet and an electron, whereas in case (b) it is to a jet and a muon.  We note that the Lagrangian in equation~\eqref{eq:SF11Lag} preserves the accidental global baryon and lepton number symmetries of the Standard Model if both M$_s$ and X carry baryon number $1/3$ and lepton number $-1$.  We will assume that these symmetries are also preserved in possible ultraviolet completions of our simplified model, ensuring that dangerous dimension-five operators inducing proton decay, such as $\frac{1}{\Lambda} d_R d_R \left( H^\dagger \, \text{M}_s \right)$, are absent.  Moreover, while our DM particle has the same gauge quantum numbers as a right-handed Majorana neutrino, the dark sector $\mathbb{Z}_2$ parity as well as the lepton number conservation forbid a Dirac neutrino mass term of the form $y_{\nu} \tilde{H} \, \overline{L_L} \, \text{DM}$.  Hence, we do not have any mixing between DM and SM neutrinos.

With the Lagrangian at hand, we now proceed to analyze the cosmology, direct detection, and indirect detection prospects for this model. In the rest of this section we denote the mediator M$_s$ as LQ, to indicate its leptoquark nature.

\subsection{Cosmology, direct detection, and indirect detection}
\label{subsec:casestudy_cosmology}
By construction, the coannihilation partner X and the $s$-channel mediator are vital ingredients to ensure the SM gauge singlet DM can attain the correct dark matter relic density, $\Omega h^2 = 0.1198 \pm 0.0026$~\cite{Agashe:2014kda,Ade:2015xua}.  In particular, DM and X readily stay in thermal contact with SM particles until freeze-out.  As X is colored, it efficiently annihilates, leading to the most important DM annihilation channels being X--X pair annihilation and DM--X coannihilation. As usual, the process that freezes out last determines the relic density. There are many processes, DM~DM $ \leftrightarrow$ X~X, DM~SM $ \leftrightarrow$ X~SM, and X $ \leftrightarrow$ DM~SM~SM, that keep the dark matter in chemical equilibrium with X~\cite{Griest:1990kh}. Thus, the DM relic density can be calculated by the effective annihilation cross-section given by equation~\eqref{eq:sigmaeff}.

\begin{figure}[tb!]
	\centering
	\includegraphics[width=0.7\textwidth]{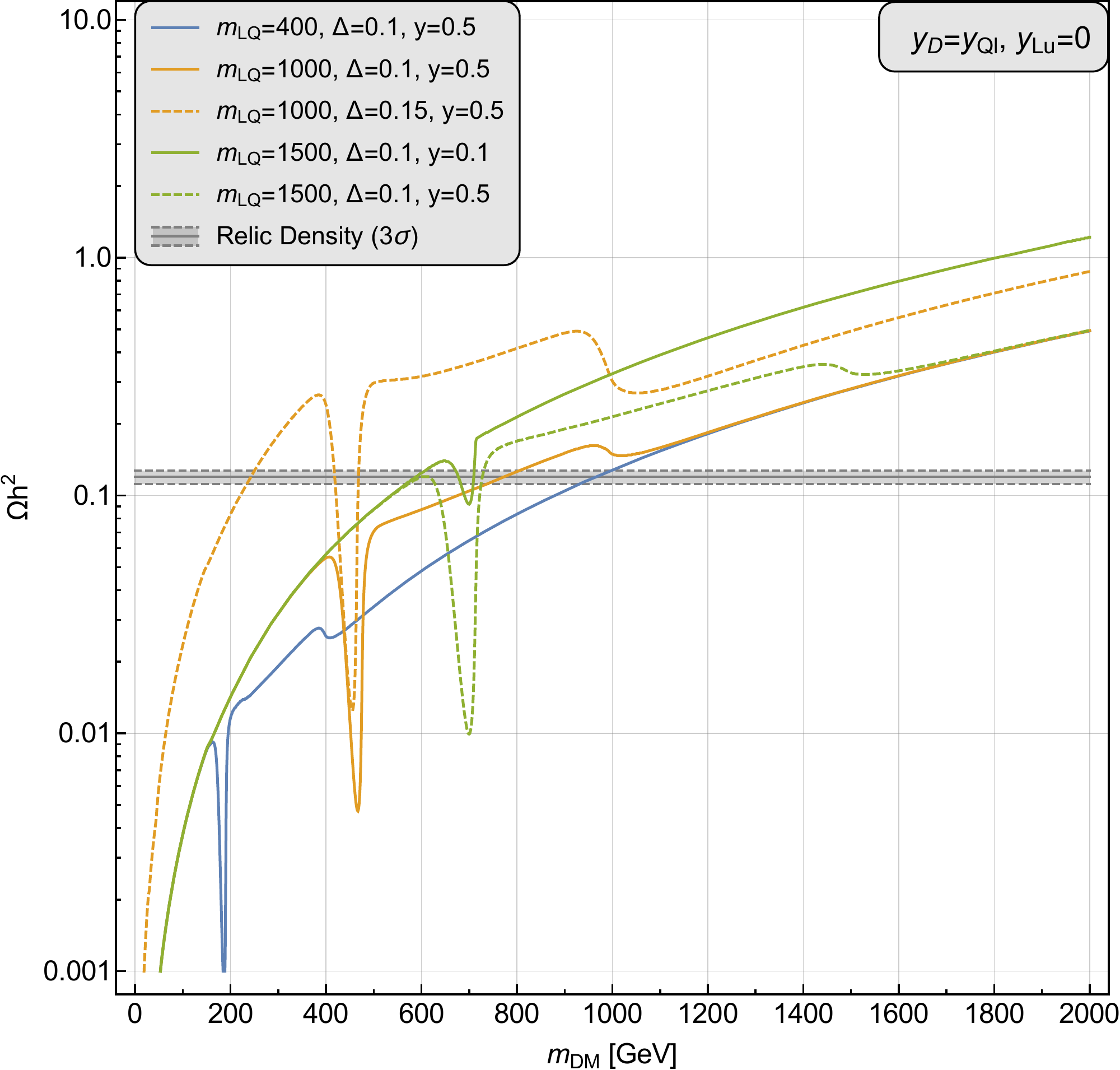}
	\caption{Relic density $\Omega h^2$ in model \hyperref[tab:classification:s-channel:3]{ST11} as a function of the dark matter mass $m_\text{DM}$, the leptoquark mediator mass $m_\text{LQ}$, the fractional mass splitting $\Delta$ (defined in equation~\eqref{eq:defDelta}), and the leptoquark Yukawa coupling $y \equiv y_D = y_{Q \ell}$. We always take $y_{Lu} = 0$. The horizontal band shows the measured DM relic density with its $3\sigma$ error band. The resonant coannihilation via $\text{DM}\,\text{X} \to q \ell$ is clearly visible as a dip in each curve.}
	\label{fig:omega:mdm}
\end{figure}

We analyze the coannihilation mechanism by writing the Lagrangian in equation~\eqref{eq:SF11Lag} as a model in \texttt{FeynRules v2.3}~\cite{Alloul:2013bka}, then using the model output in \texttt{CalcHEP v3.3.6}~\cite{Belyaev:2012qa} format for calculations in \texttt{micrOMEGAs v4.1.8}~\cite{Belanger:2014vza}.  We first show the DM relic density as a function of the DM mass for given choices of $m_\mathrm{LQ}$, $y \equiv y_D = y_{Q \ell}$, and the fractional mass difference $\Delta$ (defined in equation~\eqref{eq:defDelta}) in figure~\ref{fig:omega:mdm}. Note that, in this context, $y_{Q \ell}$ corresponds to either $y_{Q \ell}^{11}$ or $y_{Q \ell}^{22}$---the relic density is the same in both cases. In general, larger DM masses correspond to larger relic densities and vice versa, which simply follows from the fact that heavy particles freeze out earlier, when their comoving energy density is still higher. Each curve also exhibits resonant coannihilation, where the effective cross section in equation~\eqref{eq:sigmaeff} increases significantly for $m_{\text{DM}} \sim m_\mathrm{LQ} / 2$. We also show the dependence of the relic density on $y \equiv y_{Q\ell} = y_D$, observing that larger $y$ increases the coannihilation contribution to the effective cross section, thereby reducing the relic density. Finally, we exemplify the dependence on $\Delta$, which shifts the overall scale of the relic density via the exponential factors in equation~\eqref{eq:sigmaeff}. The small dip in the relic density around $m_{\text{DM}} \sim m_\mathrm{LQ}$, when visible, is because of the opening of the DM~DM $\to$ LQ~LQ and X~X $\to$ LQ~LQ annihilation channels. At large $m_\text{DM}$, and thus large freeze-out temperature, however, the relic density becomes insensitive to $m_\text{LQ}$.

\begin{figure}[tb!]
	\centering
	\includegraphics[width=0.7\textwidth]{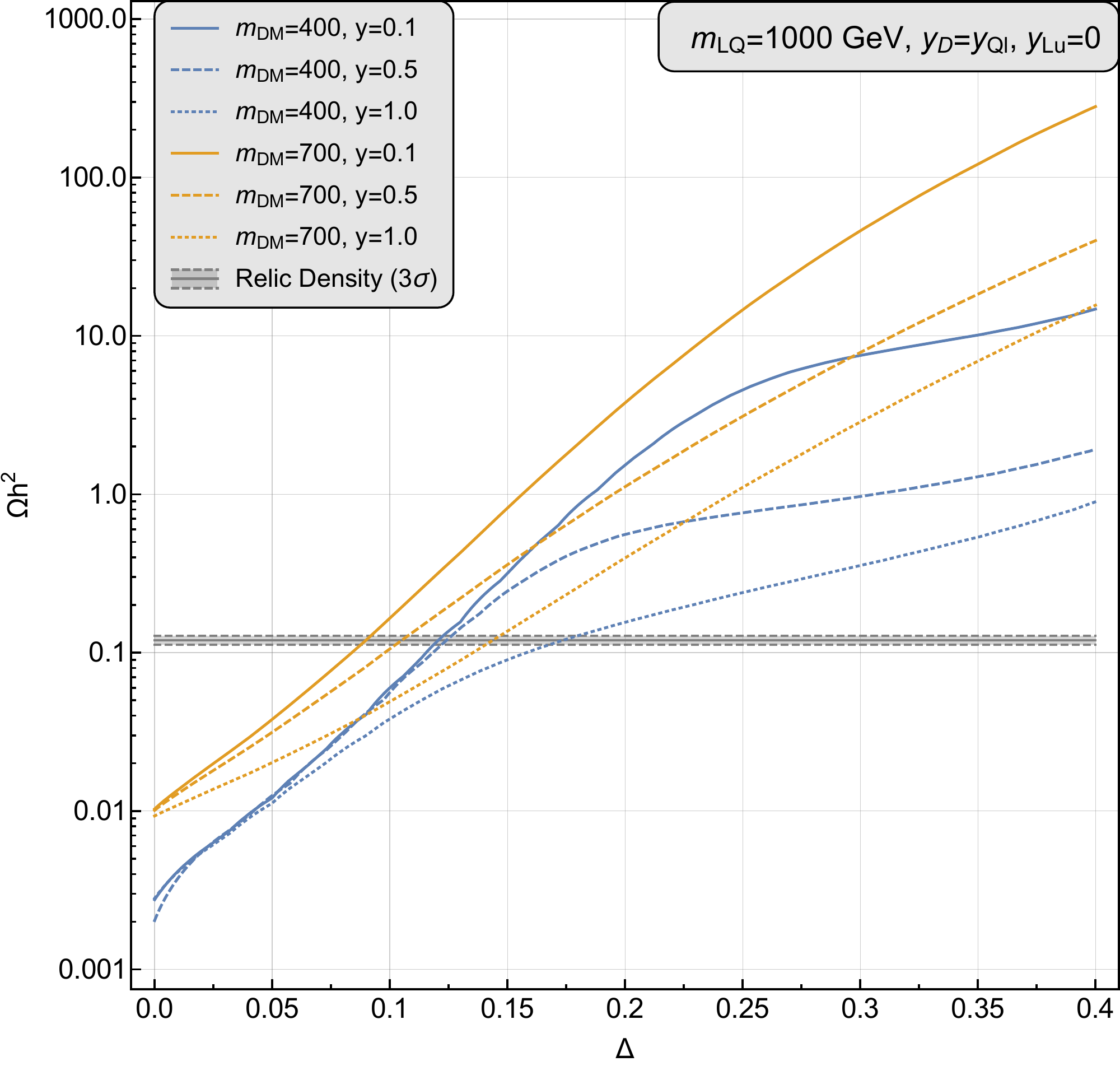}
	\caption{Relic density $\Omega h^2$ in model \hyperref[tab:classification:s-channel:3]{ST11} as a function of the fractional mass splitting $\Delta \equiv (m_{\text{X}} - m_{\text{DM}}) / m_{\text{DM}}$, using the indicated values for the leptoquark mediator mass $m_\text{LQ}$, the dark matter mass $m_\text{DM}$, and the leptoquark Yukawa coupling $y \equiv y_D = y_{Q \ell}$.  We always take $y_{Lu} = 0$. The horizontal band shows the measured DM relic density with its $3\sigma$ error band.  At large $\Delta$, strengthening the coannihilation process by increasing $y$ leads to a marked decrease in $\Omega h^2$.}
	\label{fig:omega:delta}
\end{figure}

Since $\Delta$ plays a critical role in the coannihilation calculation, we show the relic density for concrete choices of $m_\mathrm{LQ}$, $m_{\text{DM}}$, and $y_D = y_{Q \ell}$ as a function of $\Delta$ in figure~\ref{fig:omega:delta}.  For large $\Delta \gtrsim 0.4$, the coannihilation partner X plays only a small role in depleting the number density of DM, as the exponential factors in equation~\eqref{eq:sigmaeff} suppress the contributions of X--X pair annihilation and DM--X coannihilation to $\sigma_{\text{eff}}$ significantly.  In the leptoquark model, DM--DM pair annihilation is generally insufficient to avoid overclosure of the Universe, and therefore large $\Delta$ is generally ruled out.  At small $y$, DM pair annihilation to SM particles proceeds either via a triangle or box diagram with intermediate leptoquark mediator and X particles, or via $\text{DM} \, \text{DM} \to \text{LQ}^* \, \text{LQ}^* \to 2 \, \text{SM}_1 + 2 \, \text{SM}_2$.\footnote{The coannihilation process is suppressed by $e^{-\Delta x_F} \sim 0.05$ for $\Delta = 0.1$ and $x_F = 25$, which is larger than the characteristic loop suppression of $\alpha_s \alpha / (16 \pi^2)  \sim 5 \times 10^{-5}$ for $\text{DM} \, \text{DM} \to 2 \, \text{SM}$.  Moreover, the $\text{DM} \, \text{DM} \to \text{LQ}^* \, \text{LQ}^* \to 2 \, \text{SM}_1 + 2 \, \text{SM}_2$ process is suppressed by the phase space factor $\left( 1 / (4\pi)^2 \right)^2 \left( m_{\text{DM}} / (2 m_{\text{LQ}}) \right)^8  \sim 6 \times 10^{-4} \left( m_{\text{DM}}/ (2 m_{\text{LQ}}) \right)^8$, which is also smaller than the coannihilation process.  These two DM pair annihilation processes only begin to compete with the coannihilation process when $\Delta \gtrsim 0.4$ such that $e^{-\Delta x_F} \sim 5 \times 10^{-5}$ is of the same order as the loop suppression or the four-body phase space factor.}

As we can see in figure~\ref{fig:omega:delta}, increasing $y = y_D = y_{Q\ell}$ for a given $m_{\text{DM}}$ leads to smaller $\Omega h^2$, as the corresponding coannihilation cross section $\sigma_{\text{DM\,X}}$ in equation~\eqref{eq:sigmaeff} scales as $y^4$.  For small $\Delta \lesssim 0.05$, however, changing $y$ does not affect the relic density and the different curves in figure~\ref{fig:omega:delta} merge.  This is because the mass splitting is small enough that X--X pair annihilation, whose cross section scales as $\alpha_s^2 / m_{\text{X}}^2$, dominates over X--DM coannihilation.  The cross section for the latter process scales as $y^4 m_\text{DM}^2 / m_\text{LQ}^4$ for $m_{\text{DM}} \ll m_\text{LQ}$ and as $y^4 / m_{\text{DM}}^2$ for $m_{\text{DM}} \gg m_\text{LQ}$. For larger $m_{\text{DM}}$, we see the point of convergence of different $y$ lines shifts to smaller $\Delta$, which is a result of the DM mass scale dependence via $x$ in equation~\eqref{eq:sigmaeff}.

We note that for our case study, only a small $|y|$ is required to ensure that DM and X remain in chemical equilibrium. We can estimate the minimal $|y|$ for this to hold by requiring that the rate $n \, \langle \sigma v \rangle$ of $\text{X} \to \text{DM}$ conversion at the freeze-out temperature $T_F = m_\text{DM} / x_F$, where $n$ is the dark matter number density, is faster than the Hubble expansion rate:
\begin{equation}
	n \, \langle \sigma v \rangle \sim T_F^3 \frac{y^4 m_{\mathrm{DM}}^2}{m_\mathrm{LQ}^4} e^{- \Delta \, x_F} > H \sim \frac{T_F^2}{m_{\mathrm{Pl}}} \ .
\end{equation}
We thus find the following requirement
\begin{equation}
	y \gtrsim \left( \frac{m_\mathrm{LQ}^4 x_F}{m_\mathrm{Pl} m_{\mathrm{DM}}^3 } e^{\Delta \, x_F} \right)^\frac{1}{4} \, .
\end{equation}
We note that because of the $m_\mathrm{Pl}$ suppression in the denominator the minimal $y$ value is naively ${\cal O} (10^{-4})$ for $m_\text{DM} \sim m_\text{LQ} \sim 1$~TeV and small $\Delta$, but it can grow to ${\cal O} (10^{-2})$ for larger $\Delta$.

\begin{figure}[tb!]
	\centering
	\includegraphics[width=0.8\textwidth]{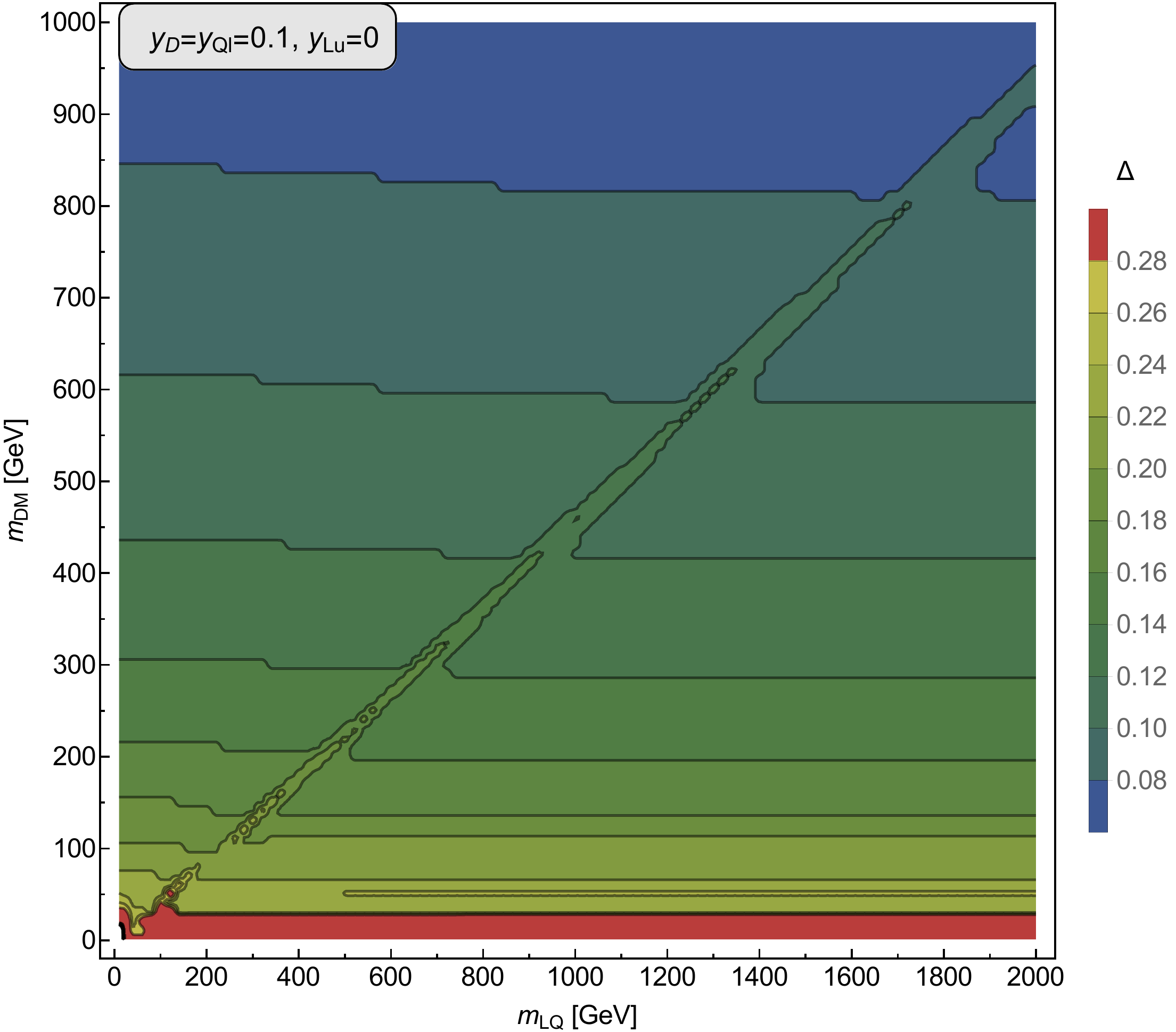}
	\caption{The value of $\Delta$ required to obtain the correct DM relic density as a function of the DM and leptoquark masses.  We fix $y_D = y_{Q \ell} = 0.1$.  The spikes along the diagonal correspond to the resonance coannihilation region.}
	\label{fig:omega:contours:delta}
\end{figure}

To see the interplay between coannihilation and X--X pair annihilation through the strong interaction, we scan the 2D parameter space spanned by $m_{\text{DM}}$ and $m_\mathrm{LQ}$, choosing $\Delta$ and $y$ such that the correct DM relic abundance is obtained.  In figure~\ref{fig:omega:contours:delta}, we fix $y = y_D = y_{Q\ell} = 0.1$, $y_{Lu} = 0$, and for each choice of $(m_{\text{DM}}, m_\mathrm{LQ})$, we show the required $\Delta$ to obtain the correct $\Omega h^2$. The relic density is rather insensitive to the leptoquark mass for given $\Delta$ since for the chosen value $y = 0.1$, X--X pair annihilation dominates over X--DM coannihilation, except for $m_\text{LQ} \sim m_\text{DM} +  m_\text{X}$. There the resonant mediator in the $s$-channel greatly enhances the coannihilation contribution to $\sigma_{\text{eff}}$, as shown by the upward spikes in $\Delta$. Although the leptoquark mediator can be lighter than the DM, the $\mathbb{Z}_2$ parity prevents DM from decaying.

\begin{figure}[tb!]
	\centering
	\includegraphics[width=0.8\textwidth]{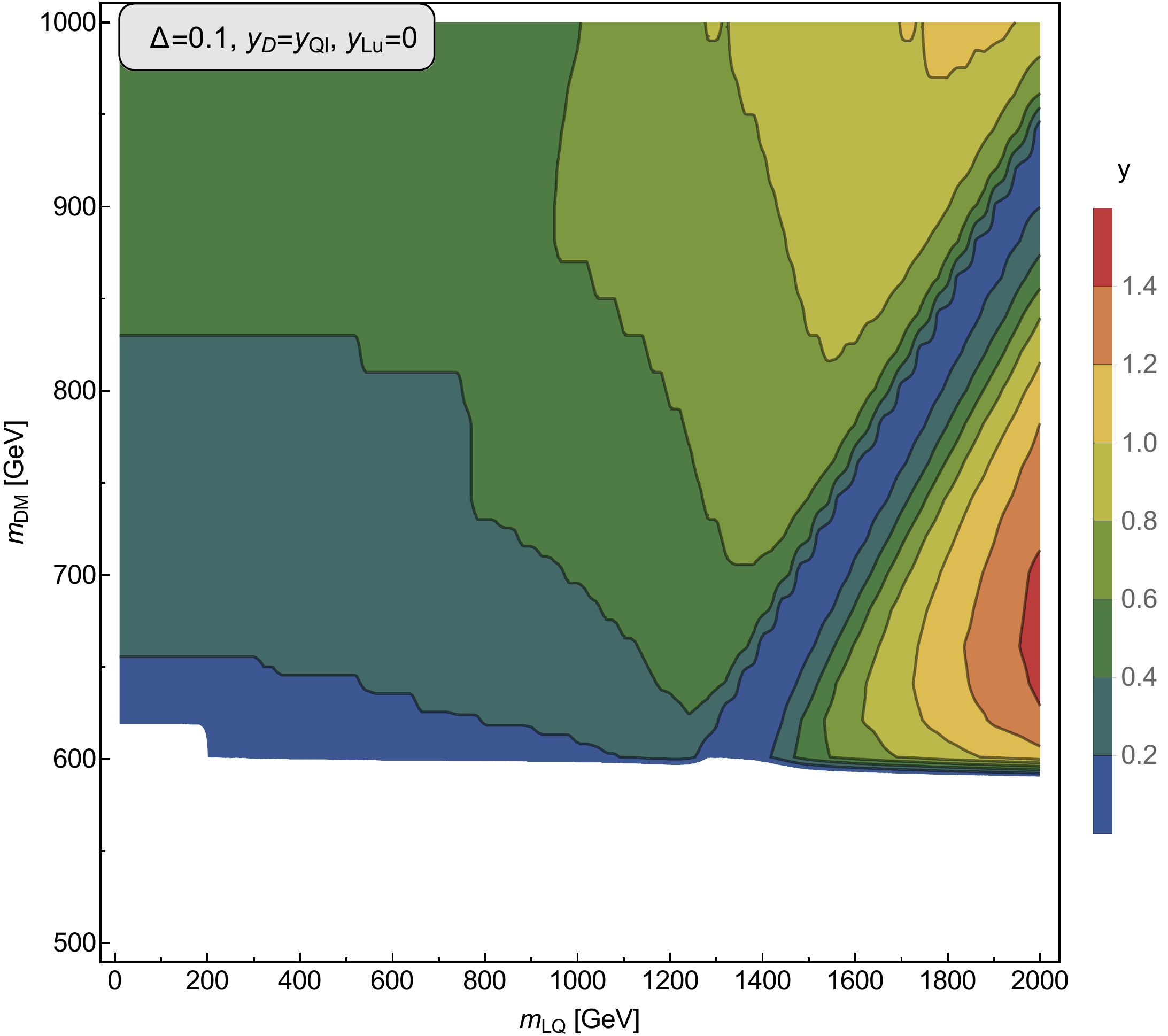}
	\caption{The value of $y = y_D = y_{Q \ell}$ required to obtain the correct DM relic density as a function of the DM and leptoquark masses. We fix $\Delta = 0.1$. The blue diagonal band corresponds to the resonant coannihilation region where $m_\mathrm{LQ} \sim m_\mathrm{DM} + m_\mathrm{X}$. In the white region at $m_\mathrm{DM} \lesssim 600$~GeV, no solution for $y$ exists as the calculated relic density is always smaller than the measured value.}
	\label{fig:omega:contours:y}
\end{figure}

In figure~\ref{fig:omega:contours:y}, we show again the 2D parameter space spanned by $m_{\text{DM}}$ and $m_\mathrm{LQ}$, but we now scan over $y$ and fix $\Delta = 0.1$.  As anticipated from figure~\ref{fig:omega:mdm}, the coannihilation resonance is evident in the blue diagonal region around $m_\mathrm{LQ} \sim m_\text{DM} + m_\text{X}$. This implies that only a small value of $y$ is needed to obtain the required relic density.  As alluded to in figure~\ref{fig:omega:delta}, for a fixed $m_\mathrm{LQ}$, the dependence on $y$ disappears for a small DM mass because $\sigma_{\text{eff}}$ is then dominated by X--X annihilation. Actually, for dark matter masses below 600~GeV, even with very small $y_D$ (but still large enough to keep X and DM in chemical equilibrium), the effective annihilation cross section is too large because of X--X pair annihilation. In this white region, no solution for $y_D$ can be found to reach the correct DM relic density. For a fixed DM mass, increasing the leptoquark mass requires a larger $y_D$ to attain the correct relic density, which is evident to the right of the black line in the figure.

\paragraph{Direct detection} As the typical momentum transfer in DM--nucleus scattering, $\mathcal{O}(100~\text{MeV})$, is very small compared to $m_\text{LQ}$ and $m_\text{X}$, we can match the full theory given by the Lagrangian~\eqref{eq:SF11Lag} onto an EFT where LQ and X are integrated out along with the high momentum modes of DM.  In this effective theory, the leading dimension-five operator connecting DM with the Standard Model is
\begin{align}
	Q_H = (\overline{\text{DM}} \, \text{DM}) \, (H^\dag H) \,.
\end{align}
This operator can contribute to DM--nucleon scattering at tree level after integrating out the Higgs. However, its Wilson coefficient is loop-suppressed and proportional to the Higgs portal coupling $\epsilon_{\text{M}_s}$, which is arbitrary and not constrained by the cosmological properties of dark matter.  Therefore, direct detection constraints can always be avoided by choosing $\epsilon_{\text{M}_s}$ small, and this is what we will assume in the following.  Note that dimension-six couplings between DM and SM fermions are absent. Vector and tensor currents of the DM field are forbidden by its Majorana nature, and a DM axial current is not generated due to the absence of chiral couplings in the dark sector. At dimension seven, the operator
\begin{align}
	Q_G = (\overline{\text{DM}} \, \text{DM}) \, G_{\mu\nu} G^{\mu\nu}
\end{align}
is generated. Since this operator is suppressed by a loop factor and by three powers of a heavy mass, its contribution to the direct detection cross section is small as well.  Hence, it is justified to disregard direct detection bounds in our study.

\paragraph{Indirect detection} If $m_{\text{DM}} > m_{\text{LQ}}$, then our dark matter particle can self-annihilate into LQ pairs via $t$-channel exchange of X.  This rate, proportional to $y_D^4$, leads to interesting kinematics in gamma ray spectra, positron spectra, and anti-proton spectra when the leptoquarks decay to SM particles~\cite{Cholis:2013psa, Martin:2014sxa, Elor:2015tva}.  If $m_{\text{DM}} < m_{\text{LQ}}$, on the other hand, the DM annihilates to SM pairs via loop-induced processes or to four SM particles via off-shell leptoquarks: the indirect detection signal is not promising in this case.

In summary, we see that electroweak scale DM, X, and LQ masses and perturbative $y_D$ and $y_{Q \ell}$ values in our simplified model case study can give the correct DM relic density and be consistent with direct and indirect bounds.  We are thus justified in motivating LHC physics from simplified models of dark matter coannihilation.

\subsection{LHC probes}
\label{subsec:casestudy:LHC}

We will now explore the LHC discovery prospects for our leptoquark-mediated dark matter coannihilation model.  The study of this particular model will exemplify how our framework can be used to test the coannihilation paradigm at the LHC.

As detailed in section~\ref{sec:phenomenology} on general grounds, X can be pair-produced in association with an ISR jet and studied using mono-jet techniques. However, compared to the traditional jet + $\slashed{E}_T$ signature, the decay products of X will also contribute soft leptons. Note that in our case, DM itself cannot be pair-produced at tree level because it is a total singlet under the Standard Model gauge group. We will therefore not consider direct DM production. Since X always decays to $\ell + j + \mathrm{DM}$, pair production of X will only lead to the signature
 \begin{itemize}
   \item[I.] $\text{A hard ISR jet}\,+\,2\,\ell\,+\,2\,j\,+\,\slashed{E}_T$: As we will show in section~\ref{sec:casestudy:future}, tagging the typically relatively soft leptons will lead to a significant improvement compared to the usual mono-jet searches~\cite{Aad:2015zva, Khachatryan:2014rra}.
 \end{itemize}
Our model also allows for the mediator to be either pair-produced through gluons or singly produced in association with a lepton. Since the mediator can either decay to a $(\ell j)$ pair or to X and DM, there are three possible signatures for pair-produced mediators:
\begin{itemize}
	\item[II.] Two $(\ell j)$ resonances, covered by the traditional LHC leptoquark searches~\cite{Aad:2015caa,Khachatryan:2015vaa}.
	\item[III.] $\text{A single $(\ell j)$ resonance}\,+\ell+\,j+\,\slashed{E}_T$. In section~\ref{sec:casestudy:future}, we will develop a search strategy for this mixed decay signature (introduced in section~\ref{subsec:LHCsigs}), which has not yet been studied by the LHC experiments.  Note that the non-resonant jet and lepton can be either soft or hard, depending on the mass of the leptoquark mediator.
	\item[IV.] $2\,\ell + 2\,j + \slashed{E}_T$: this final state has been searched for in \cite{Aad:2015wqa}. For $m_\text{LQ}$ not too much larger than $m_\text{DM}$, or for small $\Delta$, the leptons and jets are soft unless they recoil against a hard ISR jet. In the latter case, the final state becomes identical to signature I.
\end{itemize}
Associated single production of the leptoquark will lead to two possible signatures:
\begin{itemize}
	\item[V.] A single $(\ell j)$ resonance $+ \, \ell$: studied in~\cite{Mandal:2015vfa} and searched for by CMS in~\cite{Khachatryan:2015qda}.
	\item[VI.]$\ell + \ell + j + \slashed{E}_T$: one of the leptons is produced in association with the leptoquark and is hard. The other lepton and the jet originate from X decay and are hard only if $m_\text{X}$ and $\Delta$ are large. If this is the case, the search from \cite{Aad:2015wqa} is sensitive. If the second lepton and the jet are soft, the signature is equivalent to the mono-lepton final state discussed in~\cite{ATLAS:2014wra,Khachatryan:2014tva}.
\end{itemize}

In the following, we will discuss existing searches using signatures I, II and V in section~\ref{sec:casestudy:current}. We discuss signature IV and propose new searches probing signatures I and III in section~\ref{sec:casestudy:future}. The associated production rate of the leptoquark mediator is highly dependent on its couplings to the SM; thus, we focus on the existing single leptoquark searches and do not consider signature VI.

\subsubsection{Existing searches}
\label{sec:casestudy:current}

In this section we focus on current LHC searches testing our model: single and pair production of leptoquarks, as well as mono-jet searches. Moreover, we present projections of these limits for the $13$~TeV run.

\paragraph{Mediator production} We start with the phenomenology of the pair-produced leptoquark mediator $\mathrm{M}_s \equiv \mathrm{LQ}$.  As discussed in section~\ref{sec:phenomenology}, the $s$-channel mediator can generally be produced via strong or weak interactions and decay to the two-body final states X DM and SM$_1$ SM$_2$. Since the mediator mass is not tied to the DM mass, the visible SM$_1$ SM$_2$ decay products can be hard, leading to interesting two-body resonances.  In our case study, the process $p p \to \mathrm{LQ} \,\, \mathrm{LQ} \to \ell j \ell j$ leads to the well-known signature of pair-produced leptoquarks.  These have been searched for by ATLAS~\cite{Aad:2015caa} and CMS~\cite{Khachatryan:2015vaa} in the $2\,e + 2\,j$ and $2\,\mu + 2\,j$ final states.  In contrast to the models studied in these searches, however, the alternative decay LQ $\to$ DM + X in our model implies that we have to reweight the direct pair production bounds by the branching fractions $B \equiv \mathrm{Br} ( \mathrm{LQ} \,\to \ell j$) and $1 - B \equiv \mathrm{Br} ( \mathrm{LQ} \, \to$ DM + X). Singly produced leptoquarks have also been studied in \cite{Khachatryan:2015qda}; however, constraints obtained in such a search are more model dependent. The cross section for single leptoquark production is association with a lepton depends on the coupling between the mediator, the lepton and the quark. Therefore, constraints from single production can always be evaded by choosing a small value for this coupling. The constraints from both types of searches are presented below.

The branching fraction of the leptoquark to visible particles, $B$, can be expressed as a function of $y_{Q \ell}$, $y_D$, $m_\mathrm{LQ}$, $m_\mathrm{DM}$, and $\Delta$. Setting the lepton and quark masses to zero, we find
\begin{align}
  \begin{aligned}
	\Gamma \left( \mathrm{LQ} \to \ell \, j \right) & = \frac{y_{Q \ell}^2}{16 \pi} m_\mathrm{LQ} \, , \\
	\Gamma \left( \mathrm{LQ} \to \mathrm{DM} \,\, \mathrm{X} \right) & = \frac{y_D^2}{8 \pi}  m_\mathrm{LQ} \left(1- \Delta^2 \tau \right)^{1/2} \left[ 1 -  (2 + \Delta)^2 \tau   \right]^{3/2}  \equiv  \frac{y_D^2}{8 \pi}  m_\mathrm{\mathrm{LQ}} K( \Delta, \tau)  \, , \\
	B \equiv \mathrm{Br} \left( \text{LQ} \to \ell \, j \right) & = \frac{y_{Q \ell}^2}{y_{Q \ell}^2 + 2 y_D^2 K(\Delta, \tau)} = \frac{B_0}{B_0 + ( 1 - B_0 ) K(\Delta, \tau)} \, .
  \end{aligned}
  \label{eq:leptoquark:width:br}
\end{align}
Here, we have defined $\tau = m_{\mathrm{DM}}^2 / m_{\mathrm{LQ}}^2$ and $B_0 = \mathrm{Br}( \mathrm{LQ} \to \ell \, j)|_{m_\text{DM} = m_\text{X} = 0} = y_{Q \ell}^2 / (y_{Q \ell}^2 + 2 y_D^2)$.  Note that $B_0$ can be interpreted as the branching ratio of the decay $\mathrm{LQ} \to \ell \, j$ at zero DM and X mass.  For the particular choice $B_0 = 0.5$, which we will adopt in our plots, the mixed signature of LQ pair production, where one leptoquark decays to the SM and the other to the dark sector, has maximal rate. This signature is particularly interesting as it is not covered by existing LHC searches.

\begin{figure}[tb!]
	\centering
	\includegraphics[scale=0.6]{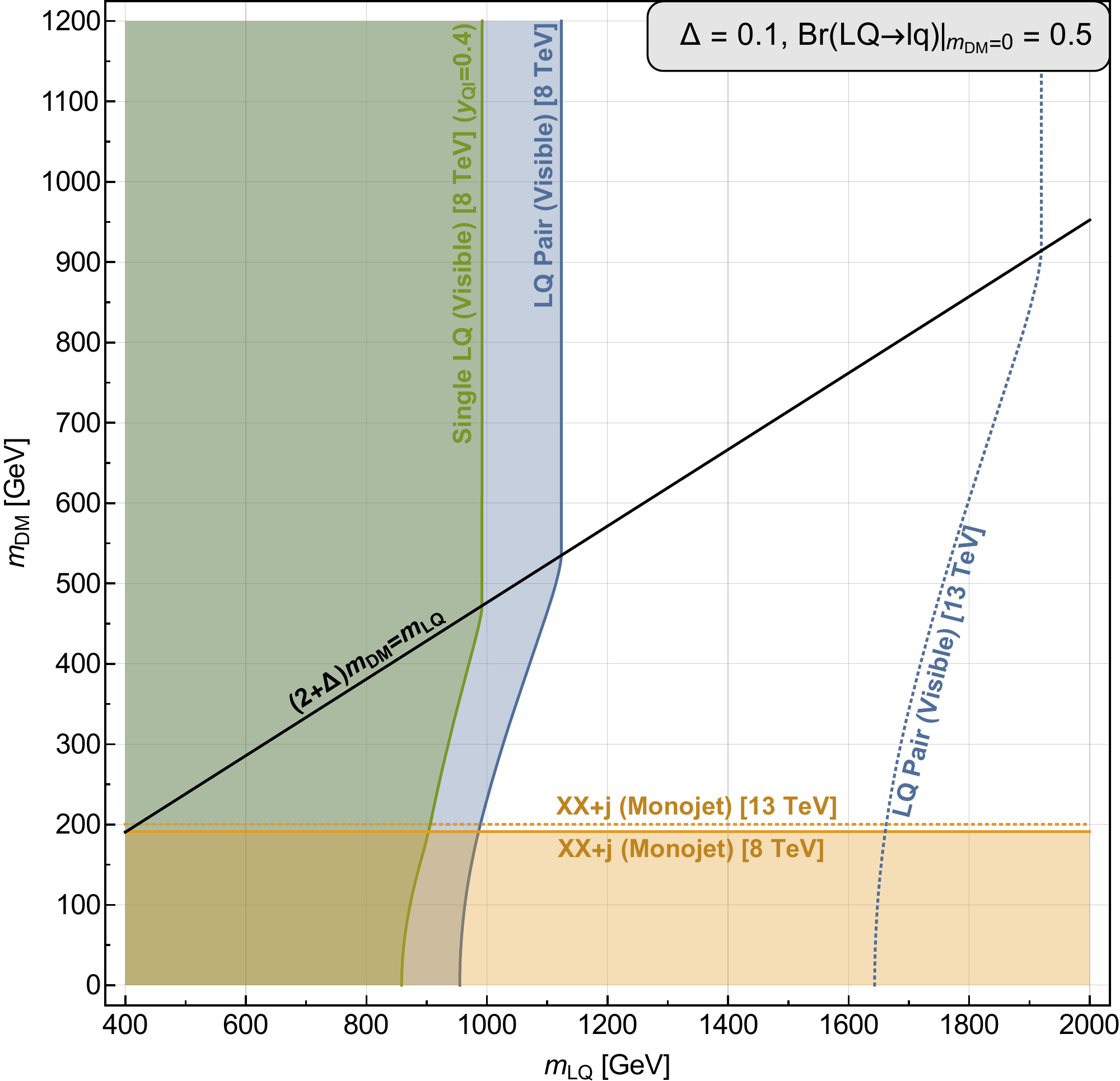}
	\caption{Leptoquark exclusion regions in the $m_\mathrm{DM}$ versus $m_\mathrm{LQ}$ plane for $\Delta = 0.1$ and $B_0 = 0.5$. The exclusion limit on single leptoquark production depends on an additional parameter, namely the coupling strength of the leptoquark to a lepton and a quark, which has been set to $y_{Q \ell} = 0.4$, to demonstrate the complementarity between the pair production and single production bounds. In the region above the black line the invisible decay of the leptoquark is kinematically forbidden. The dotted lines show projections for the $13$~TeV run, assuming $100$~fb$^{-1}$ of integrated luminosity.}
	\label{fig:leptoquark:exclusion:massplane}
\end{figure}

We first discuss pair production of leptoquarks. The kinematics of leptoquark production in our model is no different from that in the analyses~\cite{Khachatryan:2015vaa,Aad:2015caa}. However, we need to take into account the modified branching ratios (see above) and also the $SU(2)_L$ doublet nature of our leptoquark, which implies an enhancement of the production cross section by a factor of 2. Note that the components of the LQ doublet can be considered mass-degenerate for the purposes of an LHC search.  It is justified to assume the parameters $y_{Q \ell}$ and $y_D$ to be sufficiently small for production of LQ pairs to be dominated by strong interactions and for LQ to form a narrow resonance, $\left( \Gamma_{\text{LQ}} / m_\text{LQ} \right) < 0.2$.  In order to recast the experimental limits, which were presented in the $B$--$m_\mathrm{LQ}$ plane in~\cite{Khachatryan:2015vaa,Aad:2015caa}, we assume a fixed value for $B_0$ and then use equation~\eqref{eq:leptoquark:width:br} to rescale the exclusion bounds by the appropriate branching ratio for each combination of $m_\text{LQ}$, $m_\text{DM}$ and $\Delta$.  The resulting limits on our model, for $\Delta = 0.1$, are presented in figure~\ref{fig:leptoquark:exclusion:massplane} as the blue shaded region.  We remark that the dependence on $\Delta$ is very weak for this signature.  Moreover, the pair production limits asymptote once the LQ mediator decay to X DM is kinematically forbidden above the black diagonal line in figure~\ref{fig:leptoquark:exclusion:massplane}, since the LQ mediator then always decays to SM final states.

Single leptoquark production is strongly model dependent since the production cross section depends on $y_{Q \ell}$. The most recent constraints come from the analysis in~\cite{Khachatryan:2015qda}, which looks for single leptoquark production in association with a lepton in interactions of a quark and a gluon from the two colliding protons. The production cross section is proportional to $y_{Q \ell}^2$ for the dominant production channels considered in \cite{Khachatryan:2015qda}. We again need to ensure that $\left( \Gamma_\mathrm{LQ} / m_\text{LQ} \right) < 0.2$ and take into account the branching ratio of the leptoquark into a lepton and a quark. To recast the limits from \cite{Khachatryan:2015qda} into the $m_\mathrm{DM}$ versus $m_\mathrm{LQ}$ plane, we use the same procedure as for double leptoquark production.  The results are shown in figure~\ref{fig:leptoquark:exclusion:massplane}, for the specific choice $y_{Q\ell} = 0.4$. The constraints on single leptoquark production can become much stronger than those from leptoquark pair production for larger $y_{Q\ell}$. However, for the case of first generation leptoquarks, large values of $y_{Q\ell}$ are strongly excluded by atomic parity violation experiments (see appendix~\ref{app:flavor}). For our purposes, we show the single leptoquark bound to demonstrate the complementary reach between the two search strategies. Note that ref.~\cite{Khachatryan:2015qda} considers a leptoquark that couples only to the up-type quarks, while the leptoquark in our model is an $SU(2)_L$ doublet and therefore has a down component as well. In order to account for this, we rescale the production cross section quoted in~\cite{Khachatryan:2015qda} by the ratio of the down- and up-quark parton distribution functions (PDFs).\footnote{More precisely, the relation between the cross section $\sigma_\text{CMS}$ constrained in~\cite{Khachatryan:2015qda} and the leptoquark doublet production cross section $\sigma_\text{LQ}$ in our model is given by
\begin{align*}
	\sigma_\text{LQ} \approx \sigma_\text{CMS} \times \left[ 1 + \frac{\int_{x_\text{min}}^1 \! dx \, \Big( f_g(x,m_\text{LQ}) \, f_d(x_\text{min}/x, m_\text{LQ}) + f_g(x_\text{min}/x,m_\text{LQ}) \, f_d(x, m_\text{LQ}) \Big)}
		{\int_{x_\text{min}}^1 \! dx \, \Big( f_g(x,m_\text{LQ}) \, f_u(x_\text{min}/x, m_\text{LQ}) + f_g(x_\text{min}/x,m_\text{LQ}) \, f_u(x, m_\text{LQ}) \Big)} \right] \,,
\end{align*}
where $f_g$, $f_d$ and $f_u$ are the gluon, up-quark and down-quark PDFs from MSTW2008~\cite{Martin:2009iq} and $x_\text{min} \equiv (m_\text{LQ} / \sqrt{s})^2$.}

\paragraph{Mono-jet searches} The traditional mono-jet analysis searches for events with large missing transverse energy and an accompanying hard jet. It is designed to capture the process $p p \to \mathrm{DM} \,\, \mathrm{DM} \, j$, where the jet comes from initial state radiation. After hard cuts on the missing transverse energy and the jet transverse momentum, events with reconstructed leptons, and for CMS~\cite{Khachatryan:2014rra} also events with more than one additional hard jet, are vetoed. We note that direct $\mathrm{DM} \,\, \mathrm{DM} \, j$ production is not possible in our model at tree level. Mono-jet searches also impose bounds on the $\mathrm{X}\,\mathrm{X}\,j$ final state, however, provided that the additional jets and leptons from the decay $\mathrm{X} \to \mathrm{DM} \, \ell \, j$ are not observed or otherwise vetoed. We emphasize that these mono-jet and related searches have additional impetus in our coannihilation framework, as they probe directly the coannihilation partner X without relying on the mediator, giving complementary senstivity of the dark matter coannihilation mechanism compared to mediator searches. The reconstruction efficiency for the decay products of X strongly depend on $\Delta$ and the DM and X mass scale.

In order to estimate the reach of current mono-jet searches using the full $8$~TeV dataset, we recast the latest ATLAS~\cite{Aad:2015zva} search into a search for $\mathrm{X}\,\mathrm{X} + j$ production using \texttt{CheckMATE v1.2.1}~\cite{Drees:2013wra,deFavereau:2013fsa,Cacciari:2011ma,Cacciari:2005hq,Cacciari:2008gp,Read:2002hq}. This search leads to bounds on $m_{\text{X}}$ of $300$~GeV, $210$~GeV, and $180$~GeV, respectively, for $\Delta = 0.05$, $0.1$ and $0.2$, irrespective of the leptoquark mass.\footnote{In principle, this search would also constrain $\mathrm{LQ}\,\mathrm{LQ}\,j$ production, followed by the decay $\text{LQ} \to \text{X} \, \text{DM}$. However, in those regions of parameter space where this decay is kinematically allowed, the cross section for $\mathrm{LQ}\,\mathrm{LQ}\,j$ production is only a few percent of the $\mathrm{X}\,\mathrm{X}\,j$ production cross section.} Recasting the latest CMS search~\cite{Khachatryan:2014rra} leads to bounds of $230$~GeV, $190$~GeV and $140$~GeV for $\Delta = 0.05$, $0.1$ and $0.2$, respectively.

We display our results in figure~\ref{fig:leptoquark:exclusion:massplane} in the $m_{\mathrm{DM}}$ versus $m_\mathrm{LQ}$ plane. Recall that the bound from current mono-jet searches is independent of $m_\text{LQ}$ and $B_0$, therefore it simply becomes a horizontal line in the plot. For concreteness, we have set $\Delta = 0.1$ in this plot.

\paragraph{Projections} We estimate the reach of the ATLAS and CMS searches for pair-produced leptoquarks at $13$~TeV using \texttt{Collider Reach}~\cite{Salam:2014xy}, assuming a luminosity of $100$~fb$^{-1}$. The result is shown as a blue dashed line in figure~\ref{fig:leptoquark:exclusion:massplane}. The projected $13$~TeV exclusion bounds on $\text{X}\,\text{X}\,j$ production are computed by recasting the latest CMS mono-jet search~\cite{Khachatryan:2014rra} and are shown as an orange dashed line. This search requires a hard jet with $p_T \ge 110$~GeV and vetoes on events with leptons as well as events with more than two jets. In order to reduce the QCD background, the angular separation for the leading two jets (in events with two jets) is required to be $\Delta \phi_{j_1j_2} < 2.5$. The minimum $\slashed{E}_T$ requirement is then varied from $250$~GeV to $500$~GeV in steps of $50$~GeV to find the optimal expected limit for each $m_\text{X}$. Here, we use the same cuts as for LHC8, and we have simulated the $W$ + jets, $Z(\rightarrow\nu\nu, \ell^+\ell^-)$ + jets, single top and di-boson backgrounds at $13$~TeV (see section~\ref{sec:casestudy:future} for details on our simulation procedure). For $\Delta = 0.1$, we obtain a 95\% confidence level exclusion bound on $m_\text{X}$ of $220$~GeV and a $5\sigma$ discovery reach of about $120$~GeV. For $\Delta \sim 0.2$, the exclusion bound goes down to around $150$~GeV and a $5\sigma$ discovery becomes very challenging. For $\Delta\sim 0.05$, the exclusion bound is around $270$~GeV and the discovery reach is around $180$~GeV.

Since the cuts that we used for the mono-jet search were optimized for $8$~TeV, the bound we obtained by recasting the CMS mono-jet search can be considered pessimistic. These cuts, especially the veto on hard leptons, prevent the bounds from significantly improving at the $13$~TeV LHC. We also compute the bounds for the case where only leptons with $p_T > 25$~GeV (as opposed to the 10~GeV baseline cut implemented in the default description of the CMS detector in \texttt{Delphes v3.2.0}, not shown in figure~\ref{fig:leptoquark:exclusion:massplane}) are vetoed in the $13$~TeV scenario. With this relaxed lepton veto, the exclusion bound on $m_\text{X}$ for $\Delta = 0.1$ is $310$~GeV and the discovery reach is $150$~GeV. For $\Delta = 0.2$, the exclusion bound does not change significantly and is around $160$~GeV, since many signal events are still killed by the lepton veto. For $\Delta = 0.05$, the exclusion bound and the discovery reach are $390$~GeV and $180$~GeV, respectively.

\subsubsection{Future searches}
\label{sec:casestudy:future}
In this section we explore two new suggested searches: (1) the {\it mixed} signature for leptoquark pairs from $p\,p \to \text{LQ} (\to \ell \, j)\,+\, \text{LQ} (\to \text{DM} + \text{X} (\to \text{DM} \, \ell \, j))$, involving a $\ell j$ resonance plus $\slashed{E}_T$ in the final state. We consider in particular the small $\Delta$ regime where the lepton and the jet from X decay fall below the detection threshold; (2) the mono-jet topology with the addition of leptons. This final state will be sensitive to $\text{X} \, \text{X} + j$ production. We emphasize again that these final states are not included in the current portfolio of LHC searches (see table~\ref{tab:unifiedsignatures}), although a search for mono-jets plus soft muons has been presented in~\cite{CMS:2015eoa}. Therefore, we study the expected sensitivity using 100~fb$^{-1}$ of $13$~TeV LHC data. 

\paragraph{Event generation} Both signal and background events are simulated in \texttt{MadGraph5 v1.5.14}~\cite{Alwall:2014hca} using CTEQ6L1 parton distribution functions~\cite{Pumplin:2002vw}, interfaced with \texttt{Pythia v6.4}~\cite{Sjostrand:2006za} for parton showering and hadronization. For the signal, we implement the Lagrangian in equation~\eqref{eq:SF11Lag} in \texttt{FeynRules}~\cite{Alloul:2013bka} with \texttt{Universal FeynRules Output}~\cite{Degrande:2011ua}. Basic detector simulation is performed in \texttt{Delphes}~\cite{deFavereau:2013fsa}, with the default implementation of the CMS detector. In order to accommodate second generation leptoquarks, whose results are shown in appendix~\ref{app:muons}, we modify the default \texttt{Delphes} $\slashed{E}_T$ computation to include muons.  For first generation leptoquarks, our searches include a muon veto and therefore muon contributions to the $\slashed{E}_T$ will be negligible.

In computing the production cross section for $\mathrm{LQ} \,\, \mathrm{LQ}$ and $\text{X} \, \text{X} + j$, we apply $K$-factors of 1.5~\cite{Mandal:2015lca} to account for next-to-leading order (NLO) corrections.  To estimate the acceptance, we simulate events, with all decays implemented at the \texttt{MadGraph} level. Since the leptons and jets originating from X decay can be very soft, we loosen the jet and lepton $p_T$ cuts at generator level down to a few GeV. Thus, the generator level cross section for the signal can be approximated by the cross section of the hard process $p \, p \to \text{LQ} \, \text{LQ}$ or $p \, p \to \text{X} \, \text{X} + j$, multiplied by the appropriate branching ratio factors.

For LQ pair production with the mixed decay topology, there are eight main categories of backgrounds that could mimic our signature of a hard jet, hard lepton, and large $\slashed{E}_t$:
\begin{itemize}
	\item QCD: We simulate a matched sample of two- and three-jet events. The $\slashed{E}_T$ as well as the lepton arise from mismeasured or misreconstructed jets.
	\item $W^\pm + \text{jets}$, followed by $W^\pm \to \ell^\pm \nu$, where $\ell = e$ or $\mu$: these samples are matched up to two jets. The leptonic decay of the $W$ provides a real lepton and real $\slashed{E}_T$.
	\item $Z + $ jets, $Z \to \nu \nu$: these samples are matched up to two jets.  The $\slashed{E}_T$ comes from the invisible decay of the $Z$. The hard lepton in our analysis can arise when one of the additional jets fakes a lepton.
	\item $Z + $ 1 jet, $Z \to \tau^+ \tau^-$: We generate an unmatched sample for $Z \to \tau^+ \tau^-$, where the leptonic decay of the $\tau$ can provide a real source of $\slashed{E}_T$ and a hard lepton.
	\item Semileptonic $t \overline{t}$: We simulate an unmatched sample with no additional jet. Similar to $W + \text{jets}$, the semileptonic $t \overline{t}$ background provides a real lepton and real $\slashed{E}_T$.
	\item Semileptonic $W^+ W^-$: We generate an unmatched sample.  In the same way as semileptonic $t \overline{t}$ events, this background will give a hard lepton and real $\slashed{E}_T$.
	\item $W^\pm Z$, $W^\pm \to \ell^\pm \nu$, $Z \to j j$: Another diboson background that gives a hard lepton and real $\slashed{E}_T$. We do not use jet matching for this background.
	\item $W^\pm Z$ + 0 or 1 jet, $W^\pm \to \ell^\pm \nu$, $Z \to \nu \nu$: We generate a matched sample for this diboson background, which has a real lepton and real missing transverse energy.
\end{itemize}

For the mono-jet search, we use the $W^\pm + \text{jets}$ sample described above, as well as the following additional backgrounds:
\begin{itemize}
	\item $Z$ + 1 or 2 jets, $Z \to \ell^+ \ell^-$: This sample is matched up to two jets, where $\slashed{E}_T$ can arise from the mismeasurement of a jet. This background is largely subdominant in traditional mono-jet searches but can become more important if leptons are required.
	\item $t\bar t$: We generate unmatched events with no additional jets and decay the top quarks in \texttt{Pythia}.
	\item $W^+ W^-$, $W^\pm Z$, $ZZ$: We use unmatched events, where $\slashed{E}_T$ arises from invisible or leptonic decays of one of the vector bosons.
	\item $\text{single-$t$} + \text{1 or 2~jets}$: We generate a matched sample with up to two jets. The semileptonic top decay generates a real lepton and real $\slashed{E}_T$.
\end{itemize}

For the matched samples, \texttt{MadGraph} and \texttt{Pythia}~\cite{Sjostrand:2006za} are interfaced for MLM matching~\cite{Mangano:2002ea} with the $k_T$ shower scheme~\cite{Alwall:2008qv}.  The matching scale is chosen to be $40$~GeV.

In order to populate efficiently the tails of the background distributions, we split each background into bins of the variable $S_T^*$, which is defined as the scalar sum of the $p_T$ of all generator level particles. Following the procedure detailed in~\cite{Avetisyan:2013onh}, we modify \texttt{MadGraph} to implement a cut on $S_T^*$ at generator level and require each bin to satisfy
\begin{align}
	\sigma_i = \sigma(\mathtt{htmax}_i > S_T^* > \mathtt{htmin}_i) \gtrsim 0.9 \times \sigma(S_T^* > \mathtt{htmin}_i)
\end{align}
where $\mathtt{htmax}_i$, $\mathtt{htmin}_i$ are the edges of the $i$-th bin. The final overflow bin has to satisfy $N > 10 \times \sigma$, where $N$ is the total number of events to be generated in the bin and $\mathcal{L}$ is the luminosity. For each background category, we generated $5\times 10^5$ events in each $S_T^*$ bin. These events are then showered and passed through \texttt{Delphes} independently, before being weighted by the cross section in each bin and combined.  Each background category is also reweighted by an NLO $K$-factor.  The different categories of backgrounds, as well as their cross sections are shown in table~\ref{tab:xsections}, together with the $K$-factors. The latter are based on \cite{Giele:1994gf,Ellis:1992en} for the QCD multijet background and computed using \texttt{MCFM v6.8}~\cite{Campbell:2010ff} for all other backgrounds.  For the $W^\pm + \text{jets}$ background, an extra rescaling factor of 2 is applied to adjust our prediction to the events rates predicted in~\cite{Khachatryan:2014rra}, based on measurements in control regions.  The discrepancy with our results is likely due to the impact of hadronic $\tau$ decays on the lepton veto. We have also compared our predictions for the $Z_{\nu\nu} + \text{jets}$ background to the predictions from~\cite{Khachatryan:2014rra} and find agreement within $10$\%. For the QCD multijet background, our $13$~TeV Monte Carlo sample is validated against the $13$~TeV ATLAS dijet search~\cite{ATLAS-CONF-2015-042}, and the event rates and shapes have been found to agree within $20\%$ with the ones in this search. As an additional cross check of our simulation of the QCD background, we have also compared our predictions with those from \cite{Chatrchyan:2012me}, a $7$~TeV CMS mono-jet search. In contrast to later searches, ref.~\cite{Chatrchyan:2012me} offers a detailed cut flow table for the QCD background. At $7$~TeV, our estimates are consistently larger than the ones by CMS. Therefore, they can be considered conservative. Although QCD is a largely subdominant background for the mono-jet search, it will be dominant for our \emph{mixed} signature study. Since this signature involves an $\ell\,j$ resonance, though, the total background in an actual experimental study could be directly estimated from control regions. This is a second reason why the bounds that we provide for this search in our study can therefore be considered extremely conservative and are likely to become tighter when a data-driven analysis is performed.

\begin{table}[tb!]
	\centering
	\input{tables/casestudy_xsec_backgrounds_v2}
	\caption{Leading order cross sections and NLO $K$-factors at the $13$~TeV LHC, including the corresponding leptonic branching ratios, for the different backgrounds simulated for our $\text{LQ} \, \text{LQ}$ and $\text{X} \, \text{X} + j$ searches. The QCD $K$-factor is adopted from~\cite{Giele:1994gf,Ellis:1992en} while the $K$-factors for the other backgrounds have been computed using \texttt{MCFM}~\cite{Campbell:2010ff}.}
	\label{tab:xsections}
\end{table}

\boldmath \paragraph{LHC prospects for \texorpdfstring{LQ LQ $\to \ell \, j$ X DM}{LQ LQ to l j X DM}} \unboldmath The mixed decay signature $p\,p \to $ LQ LQ $\to \ell j$ X DM, with $\ell j$ being a leptoquark resonance and X $\to \ell j$ DM possibly giving an additional lepton and jet, explicitly establishes the connection between the dark sector and the SM. Moreover, this signature of a single leptoquark resonance with large missing transverse energy has not been searched for at the LHC. In particular, together with results from traditional leptoquark searches involving two $\ell j$ resonances, measurements in this channel can be used to extract the ratio of the Lagrangian couplings $y_{Q \ell}$ and $y_D$.

We now develop an LHC search strategy for this final state and estimate its sensitivity and discovery prospects in 100~fb$^{-1}$ of $13$~TeV data. The signal is characterized by a hard leptoquark resonance and missing transverse energy.  Depending on the mass splitting between X and DM, the lepton and quark from the three-body decay of X may also be visible in the detector.  We focus our cut-based analysis on identifying the leptoquark resonance over a smoothly falling background, amenable to data-driven background estimation techniques.  Our cuts are designed to identify one hard, isolated lepton, one hard jet, and large missing energy. In what follows, we will focus on first generation leptoquarks, and therefore consider only final states with electrons. Results for a second generation leptoquark are shown in appendix~\ref{app:muons}.

As mentioned before, the dominant backgrounds for this search are: QCD multijet production; leptonic $W^\pm + \text{jets}$; $Z + \text{jets}$ with $Z\rightarrow \nu\nu$; $t\overline{t}$; $Z + \text{jets}$ with $Z \to \tau^+ \tau^-$; semi-leptonic $W^+ W^-$; $W^\pm Z + \text{jets}$ with $W^\pm \to \ell^\pm \nu$, $Z \rightarrow jj$ and $W^\pm Z + \text{jets}$ with $W^\pm \to \ell^\pm \nu$, $Z \to \nu \nu$. None of these backgrounds produce resonances that mimic our leptoquark signal. In table~\ref{tab:cutflow}, we show the cut flow for the background and a benchmark signal using a $950$~GeV leptoquark, a $405$~GeV DM particle and a $445$~GeV coannihilation partner X.

The baseline transverse momentum and rapidity cuts for SM objects are those defined in the default \texttt{Delphes} card for the CMS detector. In particular, we require $p_T > 10$~GeV for leptons and $p_T > 20$~GeV for jets. We first impose a preselection cut of $p_T > 50$~GeV for the leading jet and require exactly one hard electron with $p_T > 30$~GeV in the event. The total number of leptons is required to be $\leq 2$ and the total number of jets is required to be $\leq 3$, with no $b$-jets or muons allowed. In order to reject events coming from the leptonic decay of a $Z$, we veto all events where the invariant mass of the two leptons is within $10$~GeV of the $Z$ mass. Next, we cut on $\slashed{E}_T$ and the transverse mass $m_T$ of the leading lepton and the missing energy.  Both cuts are optimized for each LQ mass point.  In addition, we require the invariant mass of the leading lepton and the leading jet to be in a $40$~GeV wide mass window centered on $m_\text{LQ}$. The probability for a jet to fake an electron is taken to be $0.0023$, following~\cite{ATLAS:2014wga}.

As mentioned before, our cuts leave the QCD multijet background as the dominant background for our search.  We remark that this background is mainly reduced by two drastic cuts: the single, hard, isolated electron requirement and the large $\slashed{E}_T$ cut.  We expect that refinements to both our treatment of jet faking electron rates, including momentum and rapidity dependence, as well as a more sophisticated treatment of jet energy scale uncertainties, will modify our projections for this mixed decay signature.  In the end, however, since we are performing a resonance search amenable to data-driven background estimates, we believe our projections will not be drastically affected by our simplified treatment of such detector effects.

\begin{table}[tb!]
	\centering
	\scriptsize
	\input{tables/casestudy_cutflow_mixed_v2}
	\caption{Cut flow for the background and for a $950$~GeV leptoquark with a $405$~GeV DM particle and a $445$~GeV $\mathrm{X}$. The subscript for the $Z$-boson indicates its decay channel, while ``signal'' refers to leptoquark pair production. $N^{\mathrm{h}}_{e}$ is the number of hard electrons, $p_T(e) > 30$~GeV. The mass window cut corresponds to $\left|m_{\ell j} - 950~\mathrm{GeV}\right| < 20$~GeV. The numbers of events quoted correspond to a center of mass energy of $13$~TeV and a luminosity of $100$~fb$^{-1}$.}
	\label{tab:cutflow}
\end{table}

We evaluate the background within the mass window by considering $20$~GeV sidebands on each side of the window. For our study, such small sidebands are sufficient, while in a real experimental analysis, the sidebands would typically be chosen larger, or more sophisticated analysis techniques would be used. We compute the total background in each of them and take the average.  We then compute the signal significance using Poisson statistics, adding the statistical and systematic uncertainties in quadrature. The significance, $r$, can then be written as
\begin{align}
	r = \frac{s}{\sqrt{(\delta b)^2  + (\mathrm{sys}\,b)^2}} \ ,
\end{align}
where $s$ is the number of signal events, $b$ is the number of background events, ``sys'' is the relative systematic uncertainty, which we take to be $20\%$, and $\delta b$ is the confidence interval. We estimate the achievable exclusion bound by choosing $\delta b$ such that the probability of observing more than $b + \delta b$ events in the background-only hypothesis is less than $5\%$. We use a $5\sigma$ threshold for the discovery reach.

\begin{figure}[tb!]
	\centering
	\includegraphics[width=0.48\linewidth]{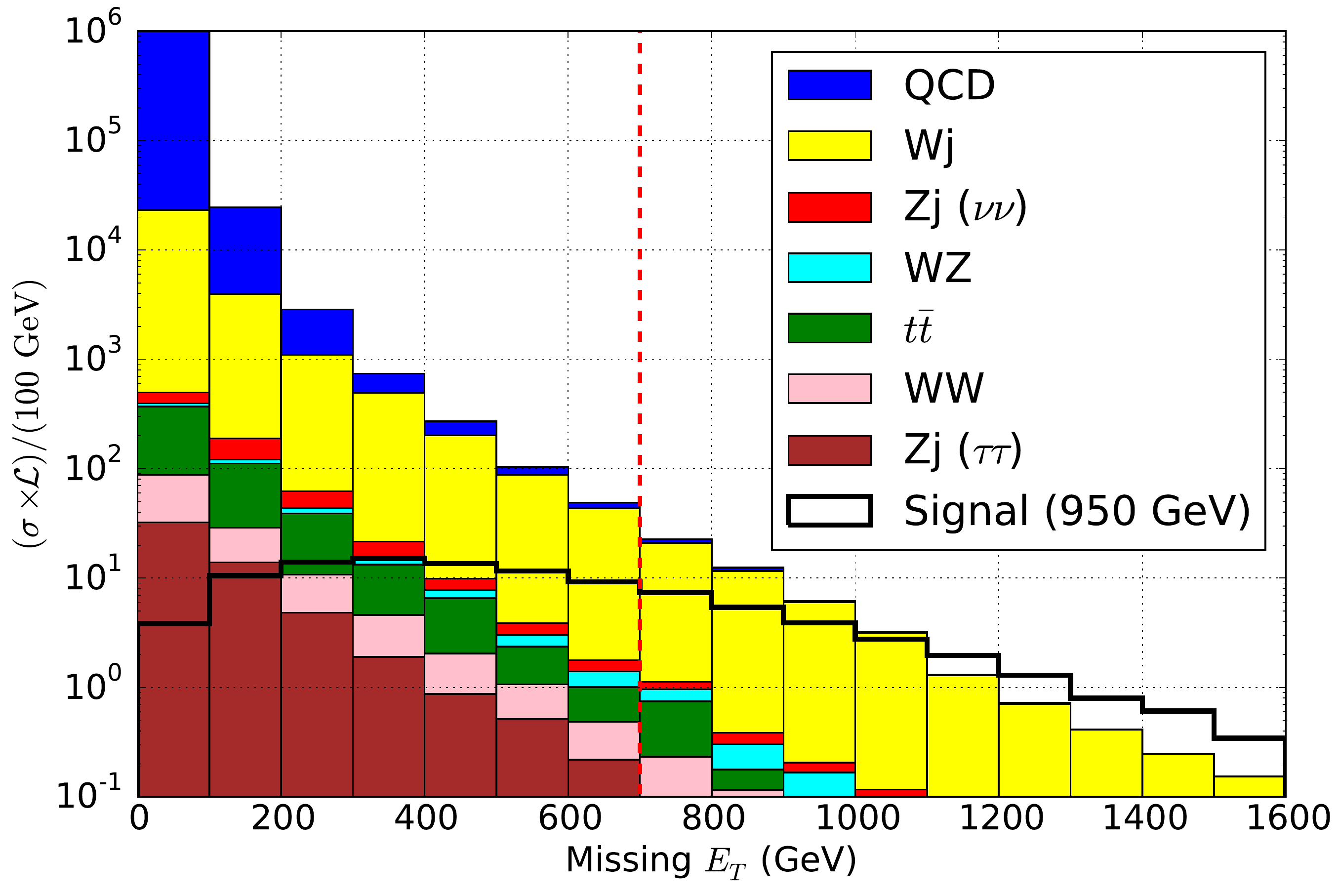}
	\includegraphics[width=0.48\linewidth]{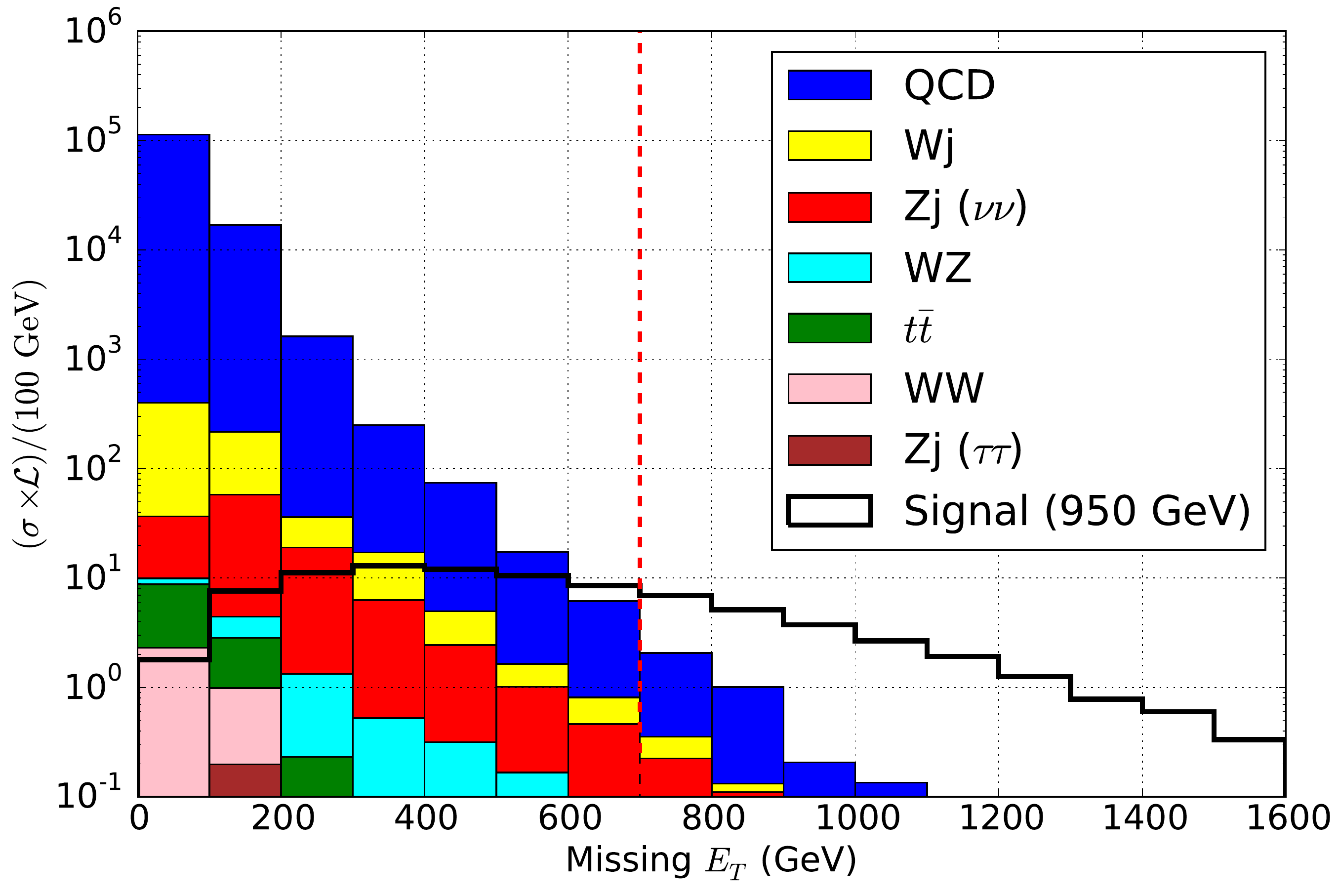}
	\caption{$\slashed{E}_T$ distributions for the background (stacked) and for a $950$~GeV leptoquark with a $405$~GeV DM particle and a $445$~GeV coannihilation partner X. The invariant mass of the leading lepton and the leading jet is required to lie within $20$~GeV of the leptoquark mass. On the left, no $m_T$ cut is applied, on the right, a $m_T$ cut of $150$~GeV is applied. The red line corresponds to the $\slashed{E}_T > 700$~GeV cut that is the optimal cut for the mass point shown here. }
	\label{fig:dismT0}
\end{figure}

\begin{figure}[tb!]
	\centering
	\includegraphics[width=0.6\linewidth]{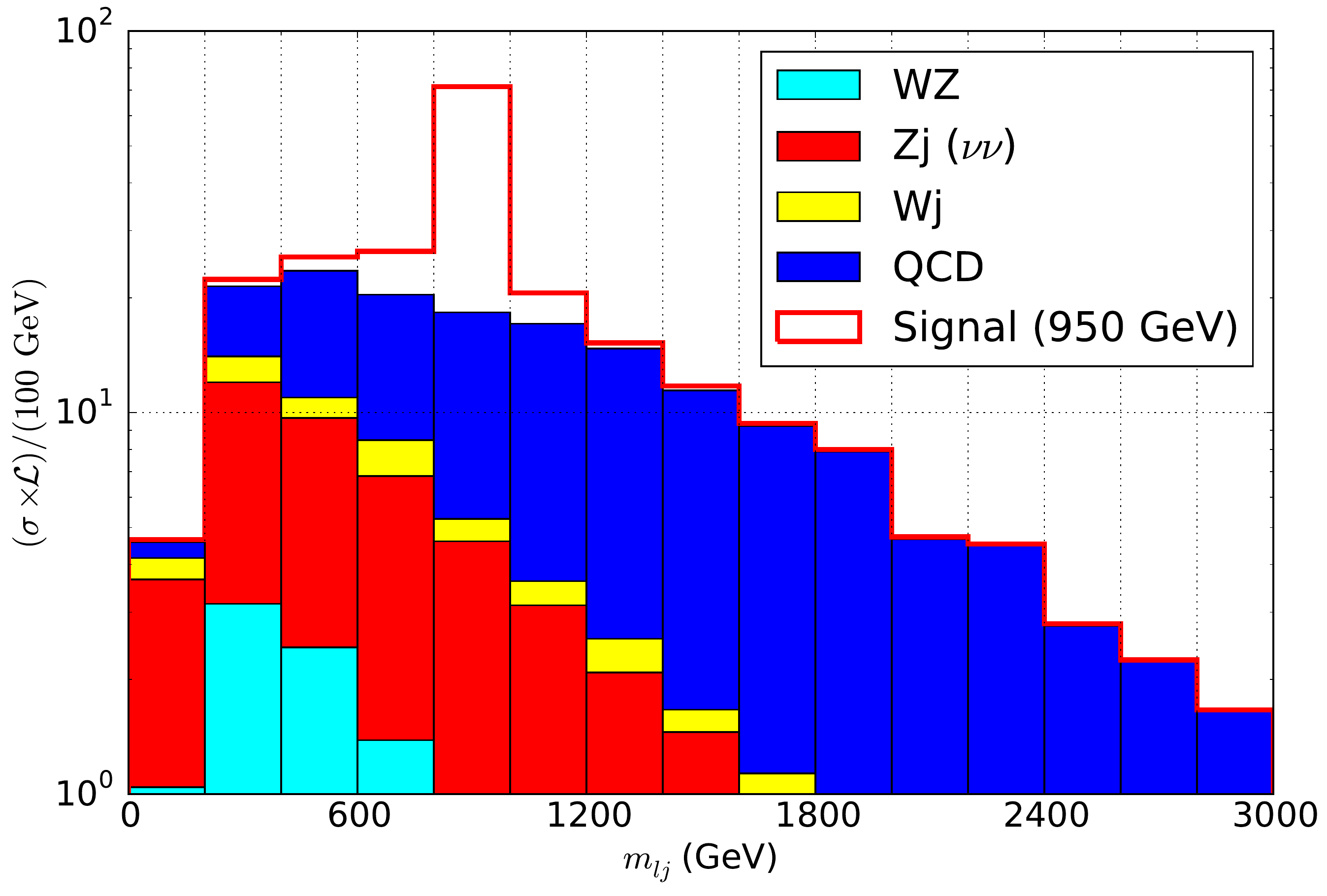}
	\caption{Distribution of the invariant mass of the leading lepton and the leading jet for a $950$~GeV leptoquark with a $405$~GeV DM particle and a $445$~GeV X and for the dominant backgrounds (stacked). We have imposed cuts on $\slashed{E}_T > 700$~GeV and $m_T > 150$~GeV.}
	\label{fig:invm}
\end{figure}

Figure~\ref{fig:dismT0} shows the signal and background $\slashed{E}_T$ distributions for the benchmark mass point presented in table~\ref{tab:cutflow}. The invariant mass of the leading lepton and the leading jet is required to be within the leptoquark mass window. This figure also shows the impact of the cut on the transverse mass $m_T$ of the leading lepton and the $\slashed{E}_T$, this cut being applied for the distribution shown in the right panel. Figure~\ref{fig:invm} shows the invariant mass distribution for the signal and the background after having applied the $\slashed{E}_T$ and $m_T$ cuts shown in table \ref{tab:cutflow}.  The leptoquark resonance is clearly visible above the $\ell j$ continuum background.

\boldmath \paragraph{LHC prospects for X~X + \texorpdfstring{$j$}{j}, \texorpdfstring{X $\to \ell \,  j $ DM}{X to l j DM}} \unboldmath We next study the pair production of X where X undergoes a three-body decay to a lepton, a jet and a DM particle.  As emphasized in section~\ref{sec:phenomenology}, the three-body decay of X is a generic signature of coannihilating dark matter models, since X typically decays through an $s$- or $t$-channel mediator that is generally off-shell. Since we are interested in coannihilating dark matter models, we will study regions of the parameter space where the splitting between X and DM is small. The SM decay products of X are then expected to be soft and are likely not reconstructed. We should therefore consider topologies that allow for X to be boosted.

In our model, X can be pair-produced through two different channels: pair production of $\text{LQ} \rightarrow $ X DM and direct pair production through strong interactions. The former production mode allows for X to be boosted if the leptoquark is significantly heavier than $m_{\text{X}} + m_{\mathrm{DM}}$. For such masses, however, the leptoquark pair production cross section will be orders of magnitude smaller than the X pair production cross section. This difference in the production rates is further exacerbated by the fact that, in our model, X is a fermion and LQ is a scalar. Pair production of X through leptoquarks will also be suppressed by the leptoquark invisible branching ratio. Therefore, in the rest of this study, we will focus on direct X pair production.

In order to boost the final states and obtain a reasonable amount of $\slashed{E}_T$, we consider X pair production in association with one additional jet from ISR.  If the leptons and the jets from X decay are very soft and escape detection, this signal can be probed using traditional mono-jet plus $\slashed{E}_T$ searches. For large DM and X masses, however, even low values of $\Delta$ allow for leptons hard enough to be detected. The corresponding signature will be a variant of the traditional mono-jet signature. This type of signal is not captured by the existing searches, which typically veto on hard objects other than the ISR jet. Requiring two leptons in addition to the hard jet and the $\slashed{E}_T$ cut then allows significant improvement over traditional mono-jet searches.

We consider a modified mono-jet plus $\slashed{E}_T$ search where the leading jet $p_T$ and $\slashed{E}_T$ cuts are supplemented by requiring two leptons with $p_T > 25$~GeV and $|\eta| < 2.5$. We use the same set of cuts as in~\cite{Khachatryan:2014rra} but optimize cuts on $\slashed{E}_T$ and $\rho$ for each parameter point, where $\rho$ is ratio of the leading jet $p_T$ over $\slashed{E}_T$, $\rho = p_{T\,j_1}/\slashed{E}_T$. We do not impose a veto on additional jets. The baseline cuts on the different objects are those defined in the default \texttt{Delphes} card for the CMS detector.  We first apply a mild selection cut, requiring the $p_T$ of the leading jet and the $\slashed{E}_T$ to be both larger than $50$~GeV.  In order to reduce the QCD background, we also require $\Delta\phi_{j_1j_2} < 2.5$. We veto on $b$-jets and muons and apply the same $Z$ veto as in the leptoquark mixed decay study. We then apply our new lepton requirement.

We generate $W^\pm + $ jets, $t\bar t$, diboson, $Z_{\ell^+\ell^-}$ + jets and single top + jets backgrounds. Because of the two lepton requirement, QCD and $Z_{\nu\nu}$ + jets backgrounds can be neglected. The $W^\pm$ + jets and single top backgrounds will need to have one jet misidentified as a lepton in order to pass the lepton requirements. As in the mixed decay study, we choose the electron fake rate to be $0.0023$ for jets with $p_T > 25$~GeV and $|\eta| < 2.5$. Table~\ref{tab:xx_cutflow} shows the cut flow for all these backgrounds, as well as for a benchmark signal with $m_{\text{X}} = 660$~GeV, $m_{\text{LQ}} = 1.7$~TeV and $m_{\text{DM}} = 600$~GeV ($\Delta \sim 0.1$).  The $\slashed{E}_T$ and $\rho$ cuts shown in the table have been optimized for discovery for this particular benchmark point. Prior to these cuts, we apply mild $\slashed{E}_T$ and $p_T^{\mathrm{lead}}$ cuts, the angular $\Delta\phi_{j_1j_2}$ cut for events with two jets or more, a $Z$ and muon veto, and require two leptons. The $95\%$ confidence limits on $m_{\text{X}}$ are $700$~GeV for $\Delta \sim 0.05$ and about $870$~GeV and $1$~TeV for $\Delta\sim $ $0.1$ and $0.2$ respectively. The discovery reach for $m_{\text{X}}$ are about $500$~GeV for $\Delta \sim 0.05$ and about $730$~GeV and $850$~GeV for $\Delta = 0.1$ and $0.2$ respectively.  As expected, the bounds become weaker when the mass splitting decreases.

Table~\ref{tab:cutlep} shows how the bounds on $m_{\text{X}}$ evolve for different values of the baseline $p_T$ requirement on the leptons, given $\Delta = 0.05, 0.1$ or $0.2$. As expected, tagging on softer leptons significantly increases the exclusion and discovery reach for small values of $\Delta$ but has little effect once $\Delta$ is large.

\begin{table}[tb!]
	\centering
	\scriptsize
	\input{tables/casestudy_cutflow_xxj_v2}
	\caption{Cut flow for the backgrounds and for a signal with $m_\mathrm{X} = 660$~GeV, $m_\mathrm{LQ} = 1.7$~TeV and $m_{DM} = 600$~GeV ($\Delta \sim 0.1$). These cuts correspond to a mono-jet  plus $\slashed{E}_T$ search where we require two leptons. The numbers of events quoted correspond to a center of mass energy of $13$~TeV and a luminosity of $100$~fb$^{-1}$.}
	\label{tab:xx_cutflow}
\end{table}

\begin{table}[tb!]
	\centering
	\input{tables/casestudy_exclusions_xxj_cutlep_v2}
	\caption{Exclusion bounds (discovery reach) on the $m_{\text{X}}$ in~GeV for $\Delta = 0.05, 0.1, 0.2$ and lepton $p_T$ requirements of $10,15,25$~GeV.}
	\label{tab:cutlep}
\end{table}

\subsubsection{Summary of the results}
\label{sec:casestudy:summary}

\begin{figure}[tb!]
	\centering
	\includegraphics[scale=0.6]{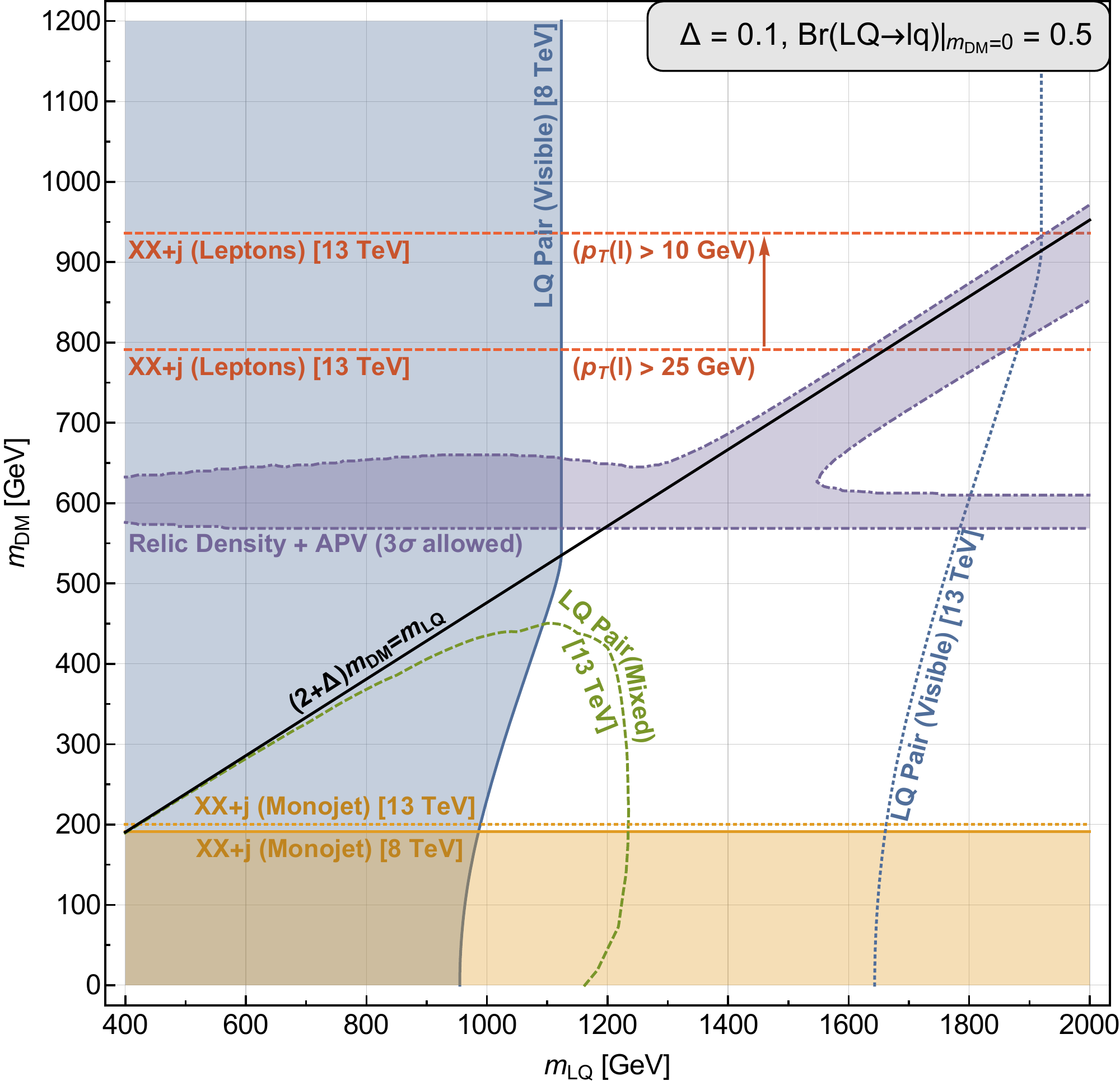}
	\caption{Leptoquark exclusion regions, the $2\sigma$ projection for the mixed decay topology, the $2\sigma$ projection for XX + $j$ production, and the parameter space region consistent with the DM relic abundance requirement within $3\sigma$, presented in the $m_\mathrm{DM}$ versus $m_\mathrm{LQ}$ plane.  We have chosen $\Delta = 0.1$ and $B_0 = 0.5$. In the region above the black line the invisible decay of the leptoquark is kinematically forbidden. The dashed lines corresponding to the solid regions show the projections for the same searches at $13$~TeV and $100$~fb$^{-1}$.}
	\label{fig:leptoquark:exclusion:projection}
\end{figure}

\begin{figure}[tb!]
	\centering
	\includegraphics[scale=0.6]{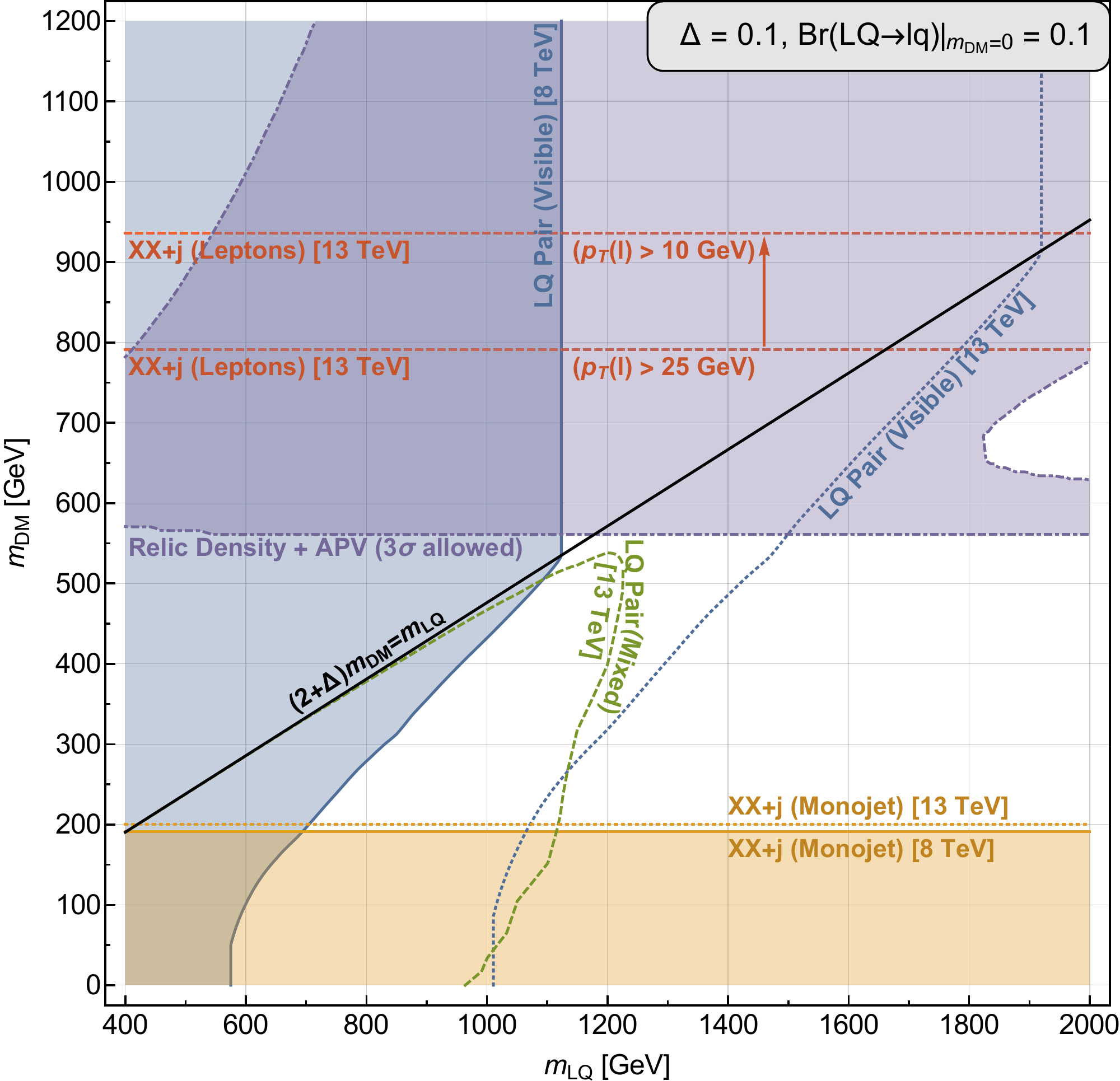}
	\caption{Leptoquark exclusion regions, the $2\sigma$ projection for the mixed decay topology, the $2\sigma$ projection for XX + $j$ production, and the parameter space region consistent with the DM relic abundance requirement within $3\sigma$, presented in the $m_\mathrm{DM}$ versus $m_\mathrm{LQ}$ plane.  We have chosen $\Delta = 0.1$ and $B_0 = 0.1$. In the region above the black line the invisible decay of the leptoquark is kinematically forbidden. The dashed lines corresponding to the solid regions show the projections for the same searches at $13$~TeV and $100$~fb$^{-1}$.}
	\label{fig:leptoquark:exclusion:projection:inverted}
\end{figure}

Figure~\ref{fig:leptoquark:exclusion:projection} shows the different exclusion bounds on our leptoquark model from double resonance searches at $8$ and $13$~TeV, the leptoquark mixed signature at $13$~TeV, the mono-jet plus lepton signature at $13$~TeV, and the traditional mono-jet searches at $8$ and $13$~TeV, as well as the preferred region where the DM relic density requirement can be satisfied (within $3\sigma$) simultaneously with the bound on $y_{Q\ell}^{11}$ from atomic parity violation (see appendix~\ref{app:flavor}). For $B_0 = 0.5$ (see equation~\eqref{eq:leptoquark:width:br}), the exclusion bounds for the double resonance signature are stronger than for the mixed one. The latter signature can however become the primary discovery channel for smaller leptoquark visible branching ratios, as is shown in figure~\ref{fig:leptoquark:exclusion:projection:inverted} (for $B_0 = 0.1$).

Adding a lepton requirement to the traditional mono-jet search leads to a huge improvement of the sensitivity. Bounds from this improved mono-jet search are also much more powerful than bounds coming from the leptoquark pair production searches. The leptoquark mixed decay, however, remains the best probe of the coannihilation mechanism, since it allows to directly establish the connection between the dark and the visible sector. In addition, the performance of the modified mono-jet search is highly dependent on the spin of X and the bounds on $m_{\mathrm{X}}$ would be significantly weakened in models where X is a scalar. Finally, as $\Delta$ becomes smaller, the $\mathrm{XX+j}$ signature becomes increasingly similar to a simple mono-jet plus $\slashed{E}_T$ signature, leading to increasingly weaker bounds. Conversely, bounds coming from pair-produced mediator searches have a much weaker dependence on $\Delta$, since the visible decay products of X are not tagged.

As evident from figures~\ref{fig:leptoquark:exclusion:projection} and~\ref{fig:leptoquark:exclusion:projection:inverted}, the combination of direct searches at the 13 TeV LHC covers a large region of the parameter space that gives the correct DM relic density.  The shapes of these favored regions of parameter space follow the behavior shown in figures~\ref{fig:omega:mdm}--\ref{fig:omega:contours:y}.  We note that our choices $B_0 = 0.5$ for figure~\ref{fig:leptoquark:exclusion:projection} and $B_0 = 0.1$ for figure~\ref{fig:leptoquark:exclusion:projection:inverted} only constrain the combination of dark and visible Yukawas $B_0 = y_{Q\ell}^2 / (y_{Q\ell}^2 + 2 y_D^2)$.  Respecting the atomic parity violation bound, $|y_{Q\ell}^{11}| < 0.4 \, ( m_{\text{LQ}} / 1 \, \text{TeV})$ (see appendix~\ref{app:flavor}), then fixes $y_D$ and also determines $\Omega h^2$.  Hence, while the shaded region corresponds to many possible choices of $y_{Q\ell}$ and $y_D$, we emphasize that the direct searches for pair-produced LQ and X particles, by nature, are largely insensitive to the magnitude of these couplings as long as the respective decays are prompt and the $B_0$ relations are satisfied.  This feature attractively illustrates the importance of the gauge interaction production modes presented in table~\ref{tab:unifiedsignatures} as drivers for the LHC signature classes discussed in section~\ref{subsec:LHCsigs}.  With this case study, we have demonstrated that the LHC can test the mechanism of dark matter coannihilation, affording us greater reach and flexibility in determining the particle nature of dark matter.

%% file: tables/casestudy_field_content.tex
\begin{tabular}{!{\vrule width 1pt} c | c | c !{\vrule width 1pt}}
	\noalign{\hrule height 1pt}
	Field & $(SU(3), SU(2), U(1))$ & Spin assignment \\
	\hline
	\text{DM}    & (1, 1, 0)   & \text{Majorana fermion} \\
	\text{X}     & (3, 2, 7/3) & \text{Dirac fermion} \\
	$\text{M}_s$ & (3, 2, 7/3) & \text{Scalar} \\
	\noalign{\hrule height 1pt}
\end{tabular}

%% file: tables/casestudy_xsec_backgrounds_v2.tex
\begin{tabular}{!{\vrule width 1pt} c!{\vrule width 1pt}c!{\vrule width 1pt}c!{\vrule width 1pt}}
	\noalign{\hrule height 1pt}
	Background & Cross section (pb) & NLO $K$-factor ($\times$ ``extra'' factor)\\
	\noalign{\hrule height 1pt}
	QCD, 2-3 jets & $2.1\times 10^7$ & 1.3 \cite{Giele:1994gf,Ellis:1992en} \\
	Leptonic $W^\pm + 1$, $2$ jets & $2222$ & 1.15 ($\times 2$) \\
	$Z (\rightarrow \nu\nu) + 1,2j$ & $736$ & 1.15 \\
	$t\overline t$ (all modes) & 465 & 1.67\\
	$Z (\rightarrow \ell^+ \ell^-) + 1,2j$ & $370$ & 1.15 \\
	$Z (\rightarrow \tau^+ \tau^-) + 1j$ & $163$ & 1.15 \\
	Semileptonic $t \overline{t}$ & $124$ & 1.67 \\
	$W^+ W^-, W^\pm Z,ZZ$ & 37 & 1.7\\
	$t + 1,2j$ & $16.9$ & 1.07\\
	Semileptonic $W^+ W^-$ & $9.8$ & 1.5 \\
	$W^\pm (\rightarrow \ell^\pm \nu) + Z (\rightarrow jj) + 0,1j$ & $2.2$ & 1.7 \\	
	$W^\pm (\rightarrow \ell^\pm \nu) + Z (\rightarrow \nu \nu) + j$ & $2.2$ & 1.7  \\
	\noalign{\hrule height 1pt}
\end{tabular}

%% file: tables/casestudy_cutflow_mixed_v2.tex
\begin{tabular}{!{\vrule width 1pt}c !{\vrule width 1pt} ccccccccc!{\vrule width 1pt}}
	\noalign{\hrule height 1pt}
	& QCD & $W + 1,2j$ & $t\bar t$ & $Z_{\nu\nu}  + j$ & $Z_{\tau\tau}  + j$ & $W^+W^-$ & $W Z_{\nu \nu} + j$ & $W Z_{jj}$ & signal \\
	\noalign{\hrule height 1pt}
	$p_T(j_1) > 50$ GeV & $2.1 \! \times \! 10^{12}$ & $4.4 \! \times \! 10^8$ & $1.3 \! \times \! 10^8$ & $7.0 \! \times \! 10^7$ & $1.3 \! \times \! 10^7$ & $1.2 \! \times \! 10^6$ &  $1.3 \! \times \! 10^5$ & $3.1 \! \times \! 10^5$ & $600$\\
	$N^{\mathrm{h}}_{e} = 1$, $N_{e} \le 2$ & $4.8 \! \times \! 10^9$ & $8.8 \! \times \! 10^7$ & $1.2 \! \times \! 10^7$ & $8.6 \! \times \! 10^4$ & $4.8 \! \times \! 10^5$ &  $2.4 \! \times \! 10^5$ & $1.9 \! \times \! 10^4$ & $6.1 \! \times \! 10^4$ & $415$\\
	$b$-jet veto & $4.0 \! \times \! 10^9$ & $8.2 \! \times \! 10^7$ & $5.0 \! \times \! 10^6$ & $8.2 \! \times \! 10^4$ & $4.6 \! \times \! 10^5$ & $2.2 \! \times \! 10^5$ & $1.9 \! \times \! 10^4$ & $5.4 \! \times \! 10^4$ & $395$\\
	$N_{\mathrm{hard\;jets}} \le 3$ & $3.9 \! \times \! 10^9$ & $8.2 \! \times \! 10^7$ & $4.3 \! \times \! 10^6$ & $8.2 \! \times \! 10^4$ & $4.6 \! \times \! 10^5$ & $2.2 \! \times \! 10^5$ & $1.9 \! \times \! 10^4$ & $5.4 \! \times \! 10^4$ & $335$\\
	$Z$ veto & $3.9 \! \times \! 10^9$ & $8.2 \! \times \! 10^7$ & $1.7 \! \times \! 10^6$ & $8.2 \! \times \! 10^4$ & $4.6 \! \times \! 10^5$ & $2.2 \! \times \! 10^5$ & $1.9 \! \times \! 10^4$ & $5.4 \! \times \! 10^4$ & $326$\\
	$\slashed{E}_T > 700$ GeV & $133$ & $1738$ & $15$ & $19$ & $9$ & $10$ & $27$ &$2$ & $75$\\
	$m_T > 150$ GeV & $132$ & $16$ & $10^{-3}$ & $18$ & $0.005$ & $0.01$ & $10$ & $0.001$ & $67$\\
	mass window & $3$ & $0.2$ & $< 10^{-5}$ & $0.3$ & $10^{-5}$ & $10^{-5}$ & $0.1$  &$10^{-5}$ & $24$\\
	\noalign{\hrule height 1pt}
\end{tabular}

%% file: tables/casestudy_cutflow_xxj_v2.tex
\begin{tabular}{!{\vrule width 1pt}c !{\vrule width 1pt} cccccc!{\vrule width 1pt}}
	\noalign{\hrule height 1pt}
	& $t\bar t$ & $Z_{\ell\ell} + j$ & Diboson & $W_{\ell\nu} + j$ & $t+j$ & Signal\\
	\noalign{\hrule height 1pt}
	$\slashed{E}_T > 50$ GeV & $1.9\times 10^7$ & $7.9\times 10^6$ & $1.1\times 10^6$ & $1.9\times 10^8$ & $5.6\times 10^5$ & $8.5\times 10^4$\\
	$p_T^{\mathrm{lead}} > 50$ GeV & $1.8\times 10^7$ & $6.1\times 10^6$ & $5.9\times 10^5$ & $1.5\times 10^8$ & $4.6\times 10^5$ & $7.1\times 10^4$\\
	$\Delta\phi_{j_1j_2} < 2.5$ & $1.2\times 10^7$ & $4.2\times 10^6$ & $5.0\times 10^5$ & $1.1\times 10^8$ & $2.9\times 10^5$ & $5.4\times 10^4$\\
	$Z$ and $\mu$ veto & $8.5\times 10^6$ & $2.7\times 10^6$& $4.0\times 10^5$ & $8.6\times 10^7$ & $1.9\times 10^5$ & $5.2\times 10^4$\\
	$b$ veto & $3.6\times 10^6$ & $2.6\times 10^6$ & $3.7\times 10^5$ & $8.2\times 10^7$ & $1.1\times 10^5$ & $2.0\times 10^4$\\
	$N_l \ge 2$ & $2.5\times 10^4$ & $4371$ & $1076$ & $9.8\times 10^4$ & $382$ & $1748$ \\
	$\slashed{E}_T > 400$ GeV & $12$ & $11$ & $0.07$ & $780$ & $2$ & $118$\\
	$\left|\dfrac{p_{T\, j_1}}{\slashed{E}_T} - 1\right| < 0.2$ & $1$ & $11$ & $0.07$ & $148$ & $0.2$ & $85$\\
	\noalign{\hrule height 1pt}
\end{tabular}

%% file: tables/casestudy_exclusions_xxj_cutlep_v2.tex
\begin{tabular}{!{\vrule width 1pt}c !{\vrule width 1pt} c c c !{\vrule width 1pt}}
	\noalign{\hrule height 1pt}
	& $p_T > 10$ GeV & $p_T > 15$ GeV & $p_T > 25$ GeV\\
	\noalign{\hrule height 1pt}
	$\Delta  = 0.05$ &  $1030$ ($860$) & $930$ ($790$) & $700$ ($500$)\\
	$\Delta = 0.1$ &  $1030$ ($860$) & $1000$ ($830$)& $870$ ($730$)\\
	$\Delta = 0.2$ &  $1030$ ($860$) & $1020$ ($870$)& $1000$ ($850$)\\
	\noalign{\hrule height 1pt}
\end{tabular}

%% file: sections/conclusions_v10.tex
\section{Conclusions}
\label{sec:conclusion}

We have classified and studied simplified dark matter models featuring coannihilation.  Each model is defined by the quantum numbers of the dark matter particle DM, its coannihilation partner X, a possible mediator particle, and the products SM$_1$, SM$_2$ of the coannihilation process $\text{DM} + \text{X} \to \text{SM}_1 + \text{SM}_2$.  We operated under two basic assumptions, namely: (a) that dark matter is a thermal relic, colorless and electrically neutral and (b) that the relevant interactions (three- and four-point vertices) arise from operators with a canonical dimension of four or less, which preserve Lorentz and gauge invariance. To the best of our knowledge, this is the first time that such a general and complete taxonomy of coannihilating dark matter models and a thorough survey of their collider signatures has been carried out. 

An additional ingredient in our framework was the existence of a $\mathbb{Z}_2$ symmetry that stabilizes the dark matter and is ubiquitous in dark matter model building, e.g.~$R$-parity in the MSSM or $T$-parity in Little Higgs models. This parity split our classification into $s$-channel and $t$-channel mediated models. If the $s$-channel mediator M$_s$ is a SM particle, only two new fields (DM, X) and one new coupling are necessary, leading to hybrid models, which feature simultaneously $s$- and $t$-channel mediated (co)annihilation.

Our classification was performed in the unbroken phase of electroweak symmetry, which led to compact results. We have extensively discussed the effects of moving to the broken phase, explaining how our simplified models transmute after electroweak symmetry breaking and which classes of ultraviolet completions are included in our classification. Except for limited cases, which we mention explicitly in section~\ref{subsec:HiggsBreaking}, our coannihilation classification is preserved even in models that require EWSB. Hence, with a few exceptions, the phenomenology discussed in the unbroken phase is directly applicable to models after electroweak symmetry breaking.

After building our catalog of simplified models, we have discussed the main phenomenological features of these models with regard to cosmology, direct and indirect detection, and especially collider phenomenology.  We acknowledge that there are limitations to any model-independent statements made at a generic level without exploring in detail the parameter space of each simplified model: low energy flavor probes, electroweak precision data, loop-induced processes, etc., need to be addressed on a case-by-case basis.  On the other hand, we stress that the collider phenomenology is strongly model-independent since the pair production rates of the new particles are typically dictated by gauge couplings.  We offer a complete compilation of those LHC signatures which directly stem from the existence of the coannihilation process $\text{DM} + \text{X} \to \text{SM}_1 + \text{SM}_2$ (see table~\ref{tab:unifiedsignatures}), and we emphasize that many of these signatures are not covered by present search strategies.

Among the signatures characteristic for coannihilation models is the decay $\text{X} \to \text{DM} + \text{SM}_1 + \text{SM}_2$, where SM$_1$ and SM$_2$ may be relatively soft because coannihilation requires a small mass splitting between $\text{X}$ and $\text{DM}$ (compressed spectrum).  Accessing these soft particles (for instance by looking for events in which the X particle recoils against initial or final state radiation) would provide additional handles to unravel the dynamics of coannihilating dark matter. In order to achieve this goal, new searches going beyond the traditional mono-jet plus $\slashed{E}_T$ paradigm, specifically targeting these soft decays, must be designed.  An initial effort in this direction by CMS in an MSSM context has been done for soft muons~\cite{CMS:2015eoa}.

A second generic feature of our coannihilation models is the possibility of single or pair production of the mediator particle, followed by its decay to two SM particles or two dark sector particles in the case of $s$-channel coannihilation, or to one SM particle and one dark sector particle in the case of the $t$-channel topology.  The phenomenology of these decays is very rich, leading to final states with various combinations of two-body resonances, soft and hard SM particles, and typically large missing energy.

Examples of final states that are not efficiently covered by the present portfolio of ATLAS and CMS analyses are (1) $\text{mono-jets} + \slashed{E}_T + \text{several soft particles}$, arising for instance from $\text{DM} + \text{X}$ production together with initial state radiation, followed by a relatively soft decay of the X particle; (2) $\text{a two-body resonance} + \slashed{E}_T$, generated for instance by pair production of the $s$-channel mediator particle M$_s$, with one of the M$_s$ particles decaying to the Standard Model sector, the other to the dark sector; (3) $\slashed{E}_T$ + \text{various combinations of non-resonant SM particles}, not all of which are fully covered by current searches. In particular, a dedicated kinematic analysis in some of these final states would, for example, improve the sensitivity to pair-produced $t$-channel mediators in their mixed decay signature.

In the last part of the paper, we have illustrated the usefulness of our classification by studying the phenomenology of one case: leptoquark mediated dark matter. We chose this particular model because, while well-motivated and viable, it is not prevalent in the theoretical literature and features experimental signatures that are not yet being searched for at the LHC. Our study of this model illustrates how to move from our general classification to actual phenomenological predictions and constraints, spanning both cosmological and collider implications. We have discussed the intricacies of coannihilation and have shown that over wide ranges of parameter space, the correct dark matter relic density is obtained. We have argued that direct detection is loop-suppressed and therefore does not impose relevant constraints, while indirect searches are only promising if the leptoquark mediator is lighter than the DM particles. On the collider side, we have demonstrated that existing constraints in the $\text{mono-jet} + \slashed{E}_T$ channel force the DM mass to be $> 200$~GeV, while searches for pair production of leptoquarks require their mass to be $\gtrsim 1$~TeV.  In 100~fb$^{-1}$ of 13~TeV data, conventional mono-jet searches will not improve prospects dramatically, but our proposed searches for a hard jet, accompanied by leptons can improve the constraints on the DM mass to at least 800~GeV.  The mass reach of leptoquark searches will be improved up to $\sim $1.6-1.9~TeV, depending on the DM mass. Complementary to these searches, we propose a novel analysis aimed at the mixed decay signature of pair-produced leptoquarks, with one of them decaying to the dark sector and the other decaying directly to SM particles. This search would explicitly probe dark matter properties via its coannihilation mechanism.

In conclusion, this paper offers to the phenomenologist and model builder a complete classification of simplified coannihilation models onto which more UV-complete scenarios can be easily mapped. Within the simplified model, the rich spectrum of cosmological, astrophysical and collider signals can then be studied in a straightforward way. In particular, LHC final states that directly test the ingredients of coannihilation can be read off from our table~\ref{tab:unifiedsignatures}.  To the experimentalist, our work provides a comprehensive summary of possible signatures of coannihilation models at the LHC, aiding in the selection of promising targets for future searches.

%% file: sections/acknowledgements_v2.tex
\acknowledgments
We would like to thank Martin Gorbahn, Roni Harnik, Gordan Krnjaic, Josh Ruderman, Matt Strassler, Sascha Turczyk, and Jure Zupan for valuable discussions.  This research is supported by the Cluster of Excellence Precision Physics, Fundamental Interactions and Structure of Matter (PRISMA-EXC 1098), by the ERC Advanced Grant EFT4LHC of the European Research Council, and by the Mainz Institute for Theoretical Physics.  The work of MB, JK and JL is supported by the German Research Foundation (DFG) in the framework of the Research Unit ``New Physics at the Large Hadron Collider'' (FOR~2239) and of Grant No.\ \mbox{KO~4820/1--1}.  This work was supported in part by the National Science Foundation under Grant No. PHYS-1066293 and the hospitality of the Aspen Center for Physics.  FY would like to acknowledge the hospitality of the Munich Institute for Astro- and Particle Physics (MIAPP) of the DFG cluster of excellence ``Origin and Structure of the Universe.''  SEH, AK, AT, FY, and JZ would also like to acknowledge the hospitality of the Gallileo Gallilei Institute for Theoretical Physics.  SEH would also like to acknowledge the hospitality of the Berkeley Center for Theoretical Physics and the Theoretical Physics Department at LBNL.

%% file: sections/appendix_flavor_v9.tex
\section{Flavor physics in the leptoquark case study}
\label{app:flavor}

In this appendix, we expand our discussion of non-minimal flavor structures for the Yukawa coupling matrices $y_{Q\ell}$ and $y_{Lu}$ in equation~\eqref{eq:SF11Lag}.

The full Lagrangian relevant for flavor constraints is given by
\begin{equation}
	\mathcal{L} = \mathcal{L}_\text{gauge}^\text{SM} + \mathcal{L}_Y^\text{SM} - \left( y_{Q\ell} \overline{Q_L} \text{M}_s~ \ell_R + y_{Lu} \overline{L_L} \text{M}_s^c u_R + \text{h.c.} \right) \ ,
\end{equation}
where
\begin{equation}
	\mathcal{L}_Y^\text{SM} = - \left( Y^u \overline{Q_L} \tilde{H} u_R + Y^d \overline{Q_L} H d_R + Y^e \overline{L_L} H \ell_R \right) + \text{h.c.}
\end{equation}
is the Standard Model Yukawa Lagrangian that, after electroweak symmetry breaking, gives rise to the fermion mass terms. We can choose to diagonalize $Y^u$ and $Y^e$ using the field rotations $L_L^\prime = S_e L_L$, $\ell_R^\prime = R_e \ell_R$, $Q_L^\prime = S_u Q_L$, $u_R^\prime = R_u u_R$, $d_R^\prime = R_d d_R$, with unitary rotation matrices $S_e$, $R_e$, $S_u$, $R_u$, $R_d$ (see, e.g.,~\cite{Anikeev:2001rk}). The down Yukawa can then be diagonalized as $Y^d_\text{diag} = S_d Y^d R_d^\dagger$, where $S_d$ is another unitary matrix. This introduces the Cabibbo-Kobayashi-Maskawa (CKM) matrix $V \equiv S_u S_d^\dagger$ that will appear in the weak charged current. In the mass basis, the leptoquark interactions are then given by
\begin{equation}\label{eq:rel}
	y_{Q\ell}^\prime = S_u y_{Q\ell} R_e^\dagger\,, \quad y_{Lu}^\prime = S_e y_{Lu} R_u^\dagger \,.
\end{equation}
In the following, we will assume that this rotation to the mass basis has been performed, and will drop all primes. It is important, however, to keep in mind that the relation between the parameters in the Lagrangian~\eqref{eq:SF11Lag} and the phenomenological parameters used in our collider and flavor studies are given by the nontrivial relations~\eqref{eq:rel}.

In our collider analyses, we choose the values $y_{Lu}=0$, and $y_{Q\ell}^{ii}$ diagonal and nonzero for exactly one generation (either $i=1$ or $i=2$). This corresponds to a cancellation of terms in the $y_{Q\ell}$ Yuakwa matrix in the original Lagrangian~\eqref{eq:SF11Lag}. In the remainder of this section, we will show that precision observables do not allow us to deviate appreciably from this idealistic case (apart from possibly including the third generation) if we want to keep one of the couplings of order unity.

Mixing of neutral $K$, $D$, $B$, and $B_s$ mesons leads to numerous constraints on various off-diagonal elements of the couplings $y_{Q\ell}$ and $y_{Lu}$. For instance, constraints from neutral kaon mixing lead to the bound $y_{Q\ell}^{11} \, y_{Q\ell}^{12} \lesssim 0.1$, for mediator masses of the order of 1~TeV~\cite{Bona:2007vi}.  For simplicity, we assume henceforth that $y_{Q\ell}^{ij} = 0$ for $i \neq j$.

To study bounds from low energy processes on the remaining couplings, it is easiest to integrate out the heavy leptoquark and to work in the effective field theory that contains only the light degrees of freedom.
The interactions of the mediator with SM particles, after electroweak symmetry breaking, read
\begin{align} \label{eq:Lfull}
	{\mathcal L} & \supset -\left[ y_{Q\ell}^{ij} \left( \overline{u_L}^i \phi_{5/3} \ell_R^j  + \overline{d_L}^i \phi_{2/3} \ell_R^j \right) + \text{h.c.} \right] \nonumber \\
	& \quad - \left[ y_{Lu}^{ij} \left( \overline{\nu_L}^i \phi_{2/3}^* u_R^j - \overline{\ell_L}^i \phi_{5/3}^* u_R^j \right) + \text{h.c.} \right] \ ,
\end{align}
where $i$, $j$ are flavor indices.  We have discarded the operators involving neutrino fields and assumed all couplings are real (no $CP$ violation).  Moreover, we wrote the $SU(2)_L$ components of the mediator explicitly, M$_s \equiv (\phi_{5/3}, \phi_{2/3})^T$, and assumed that they have approximately equal mass $m_\text{LQ}$.

The effective Lagrangian obtained from equation~\eqref{eq:Lfull} via tree-level matching reads
\begin{equation} \label{eq:Leff}
	{\mathcal L} = \sum_{a=1,2,3} C_a^{ijkl} Q_a^{ijkl} + \text{h.c.} + \cdots \,, 
\end{equation}
where ``+\,h.c.'' denotes the addition of the hermitian conjugate term
where appropriate, and the operators are defined as
\begin{equation}
	Q_1^{ijkl} = (\overline{u_L}^i \ell_R^j)(\overline{\ell_L}^k u_R^l) \,, \quad Q_2^{ijkl} = (\overline{u_L}^i \ell_R^j)(\overline{\ell_R}^k u_L^l) \,, \quad Q_3^{ijkl} = (\overline{d_L}^i \ell_R^j) (\overline{\ell_R}^k d_L^l) \,.  
\end{equation}
Their Wilson coefficients are given to leading order by
\begin{equation}
	C_1^{ijkl} = - \frac{y_{Q\ell}^{ij} \, y_{Lu}^{kl}}{m_\text{LQ}^2} \,, \quad C_2^{ijkl} = \frac{y_{Q\ell}^{ij} \, y_{Q\ell}^{kl}}{m_\text{LQ}^2} \,, \quad C_3^{ijkl} = \frac{y_{Q\ell}^{ij} \, y_{Q\ell}^{kl}}{m_\text{LQ}^2} \ .
\end{equation}

The Wilson coefficient $C_3^{1122}$ is strongly constrained by the smallness of the measured branching ratio for the decay $K_L \to \mu e$.  Neglecting the tiny amount of CP violation in the neutral kaon system, we write $| K_L \rangle = (| K^0 \rangle - | \overline{K^0} \rangle)/\sqrt{2}$.  Since the charge of the leptons tags the flavor of the kaons, we can write $\text{Br}(K_L \to \mu^+ e^-) = \text{Br}(K^0 \to \mu^+ e^-)/2$.  Neglecting the electron mass, equation~\eqref{eq:Leff} yields
\begin{equation}
	\text{Br}(K_L \to \mu^+ e^-) = \tau_{K_L} \frac{ \left( y_{Q\ell}^{11} \, y_{Q\ell}^{22} \right)^2}{512\pi} \frac{m_\mu^2 M_K f_K^2}{m_\text{LQ}^4} \left( 1 - \frac{m_\mu^2}{M_K^2} \right)^2 \ ,
\end{equation}
where $f_K = 156.1$ MeV is the kaon decay constant~\cite{Laiho:2009eu}, and $\tau_{K_L} = 5.116 \times 10^{-8}$ s is the $K_L$ lifetime, $M_K = 497.614$ MeV the neutral kaon mass, and $m_\mu = 105.658$ MeV the muon mass~\cite{Agashe:2014kda}.

Using the bound $\text{Br}(K_L \to \mu^\pm e^\mp) < 4.7 \times 10^{-12}$ at 90\% confidence level~\cite{Agashe:2014kda}, we find
\begin{equation}
y_{Q\ell}^{11} y_{Q\ell}^{22} \left( \frac{1 \, \text{TeV}}{m_\text{LQ}} \right)^2 \lesssim 2.74 \times 10^{-5} \ ,
\end{equation}
so at least one coupling should be very tiny or the mediator is very heavy.  This bound motivates our phenomenological studies for {\bf (a)} $y_{Q\ell}^{11} \neq 0$ and {\bf (b)} $y_{Q\ell}^{22} \neq 0$, as discussed in section~\ref{subsec:Lagrangian}.

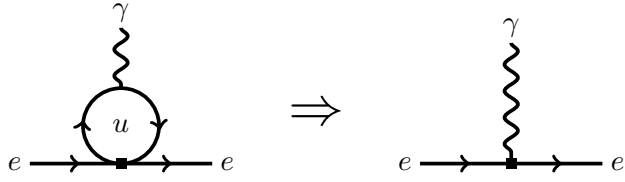
\begin{figure}[t]
	\centering
	\input{diagrams/mdm_mixing}
	\caption{Mixing of the four-fermion operator $Q_1^{1111}$ into the magnetic dipole operator $Q_\text{MDM}$.}
	\label{fig:mix}
\end{figure}

We will now show, using the measurement of the anomalous magnetic moment of the electron, that a similar restriction applies to the combination $y_{Q\ell}^{11} \, y_{Lu}^{11}$. We can obtain a rough estimate of the expected size of the effect by calculating the leading-logarithmic contribution. For this we need to calculate the mixing of the four-fermion operator $Q_1^{1111}$ into the dipole operator $Q_\text{MDM}$ (see figure~\ref{fig:mix}), defined by 
\begin{equation}
  Q_\text{MDM}  = \frac{m_u}{e} (\overline{e} \sigma_{\mu\nu} e) F^{\mu\nu} \,. 
\end{equation}
Adding the contribution of the Hermitian conjugated operator and using the well-known leading-order relation for the renormalization group evolution equations for the Wilson coefficients, $C_i(\mu) = C_i(\mu_0) + \tfrac{\alpha}{4\pi} Z_{ji} C_j(\mu_0) \log \tfrac{\mu_0^2}{\mu^2}$, where the anomalous dimension coefficient is $Z_{Q_1^{1111}, \ Q_\text{MDM}}^{(1)} = 1/4$, we find
\begin{equation}
  C_\text{MDM} = \frac{y_{Q\ell}^{11} y_{Lu}^{11}}{m_\text{LQ}^2} \frac{\alpha}{8\pi} \log \frac{\mu^2}{m_\text{LQ}^2} \ .
\end{equation}
It could be worthwhile to resum the large logarithm to all orders in the strong coupling constant. 

The Wilson coefficient $C_\text{MDM}$ is related to the shift $\Delta a_e$ in the anomalous magnetic dipole moment of the electron, defined by $a_e \equiv (g-2)_e/2$, via
\begin{equation}
	\Delta a_e = - \frac{m_e m_u}{\pi\alpha} C_\text{MDM} = - \frac{y_{Q\ell}^{11} \, y_{Lu}^{11}}{8\pi^2} \frac{m_e m_u}{m_\text{LQ}^2} \log \frac{\mu^2}{m_\text{LQ}^2} \,.
\end{equation}

To get a rough estimate, we take the scale $\mu$ to be of the order of $1\,$GeV where the transition to the meson picture takes place, and use $m_\text{LQ} = 1000\,$GeV, to find 
\begin{equation}
 |\Delta a_e| \simeq 1.6 \times 10^{-10} \,\, y_{Q\ell}^{11} \, y_{Lu}^{11}\,. 
\end{equation}

The anomaly in the gyromagnetic ratio of the electron, $a_e \equiv (g-2)_e/2$, is conventionally used to determine the fine-structure constant $\alpha$~\cite{Hanneke:2008tm,Aoyama:2014sxa}. However, as pointed out in ref.~\cite{Giudice:2012ms}, the recent precise independent measurements of the fine-structure constant in atomic physics experiments can be used to obtain a Standard Model prediction for $a_e$ with an uncertainty that is only a factor of few larger than error of the experimental measurement. Therefore, the anomalous magnetic moment of the electron can be used as a probe of new physics.

We employ the value $\alpha^{-1} = 137.035999037(91)$ from the most recent determination of the fine-structure constant using a measurement of the ratio between the Planck constant and the mass of the $^{87}$Rb atom~\cite{Bouchendira:2010es}.  Using the corresponding uncertainty induced on $a_e$ around the Standard Model value, we obtain the allowed range for the new physics contribution to $a_e$
\begin{equation}
 |\Delta a_e| < 8.1 \times 10^{-13} \,.
\end{equation}
This translates into the bound
\begin{equation}
 |y_{Q\ell}^{11} \, y_{Lu}^{11}| < 5.0 \times 10^{-3} \,.
\end{equation}
We see that by setting $y_{Lu}$ to zero and keeping only one of either $y_{Q\ell}^{11}$ or $y_{Q\ell}^{22}$ nonzero, we can satisfy the above precision constraints.

A further bound on $y_{Q\ell}^{11}$ alone is obtained from the measurement of atomic parity violation. To this end, we rewrite the operator $Q_2^{1111}$, using Fierz relations~\cite{Nishi:2004st}, as
\begin{equation} \label{eq:fierz:transformation}
	Q_2^{1111} = - \frac{1}{2} (\overline{u_L} \gamma_\mu u_L) (\overline{e_R} \gamma^\mu e_R) = - \frac{1}{8} \left[ (\overline{u} \gamma_\mu u) (\overline{e} \gamma^\mu \gamma_5 e) - (\overline{u} \gamma_\mu \gamma_5 u) (\overline{e} \gamma^\mu e) \right] + \ldots \ .
\end{equation}
The effective Lagrangian leading to atomic parity violation can be written as~\cite{Gresham:2012wc}
\begin{equation} \label{eq:LAPV}
	{\mathcal L}^\text{APV} = \frac{G_F}{\sqrt{2}} \sum_{q=u,d} \left[ C_{1q} (\overline{q} \gamma_\mu q) (\overline{e} \gamma^\mu \gamma_5 e) + C_{2q} (\overline{q} \gamma_\mu \gamma_5 q) (\overline{e} \gamma^\mu e) \right] \,.
\end{equation}
The precise measurement of atomic parity violation in cesium ($^{133}$Cs) atoms together with the precision prediction of the corresponding Standard Model contributions constrains new physics contributions to $C_{1u}$ to the $10^{-3}$ level. The electroweak physics is contained in the nuclear weak charge~\cite{Agashe:2014kda, Gresham:2012wc} 
\begin{equation}
	Q_W(Z,N) = -2 \left[ (2Z+N) C_{1u} + (2N+Z) C_{1d} \right] \,,
\end{equation}
where $Z$ and $N$ are the numbers of protons and neutrons in the nucleus, respectively. The most recent determination, including a recent update of the atomic structure calculation~\cite{Dzuba:2012kx}, yields $Q_W(\text{Cs}) = -72.62(43)$~\cite{Wood:1997zq, Guena:2004sq, Agashe:2014kda}, where the parentheses indicate $1\sigma$ experimental uncertainty. This agrees within $1.5 \sigma$ with the Standard Model prediction $Q_W^\text{SM}(\text{Cs}) \simeq -73.26$~\cite{Agashe:2014kda}. As advocated in~\cite{Barger:2000gv} this slight deviation can actually be alleviated by leptoquark models, which is also true in our leptoquark model. Therefore we require the Standard Model contribution plus the leptoquark contribution to $Q_W(\text{Cs})$ to lie within $3\sigma$ of the experimentally measured value. Comparing equation~\eqref{eq:Leff} and equation~\eqref{eq:LAPV} we then find
\begin{equation} \label{eq:apv:bound}
	|y_{Q\ell}^{11}| < 0.40 \, \left( \frac{m_\text{LQ}}{1 \, \text{TeV}} \right) \ ,
\end{equation}
at $3\sigma$ confidence level.

Besides atomic parity violation there are bounds directly on the operators $(\overline{u_L} \gamma_\mu u_L) (\overline{e_R} \gamma^\mu e_R)$ and $(\overline{d_L} \gamma_\mu d_L) (\overline{e_R} \gamma^\mu e_R)$ (see equation~\eqref{eq:fierz:transformation}) from experiments operating at different energies. A combined fit of several experiments~\cite{Agashe:2014kda,Cheung:2001wx}, significantly driven by Tevatron data~\cite{Abe:1997gt,Abbott:1998rr}, gives a limit similar to that in equation~\eqref{eq:apv:bound}. Moreover, there are ATLAS~\cite{Aad:2014wca} and CMS~\cite{CMS:2014aea} analyses constraining this contact interaction, giving a sligthly more constraining bound than equation~\eqref{eq:apv:bound}. However, the validity of the effective description in terms of four-fermion operators for the mass range of the leptoquark we are considering is limited~\cite{deVries:2014apa}. A rescaling of the bounds is possible, but would result in less constraining bounds than from atomic parity violation. We therefore discard the bounds from ATLAS and CMS as well as the bound from the combined fit.

%% file: diagrams/mdm_mixing.tex
\begin{tikzpicture}[line width=1.4pt, scale=1]
	\draw[fermionbar] (1.2,0)--(0,0);
	\draw[fermion] (-1.2,0)--(0,0);
	\draw[fermionbar] (0,0) arc (-90:90:0.5);
	\draw[fermion] (0,0) arc (270:90:0.5);
	\draw[vector] (0,1.8)--(0,1);
	
	\node[vtxsquare] at (0,0) {};
	
	\node at (1.4,0) {$e$};
	\node at (-1.4,0) {$e$};
	\node at (0,0.5) {$u$};
	\node at (0,2.0) {$\gamma$};
\end{tikzpicture} \hspace{2mm}
\raisebox{6mm}{\begin{tikzpicture}[line width=1.4pt, scale=1]
	\node at (1.8,0.0) {\huge $\Rightarrow$};
\end{tikzpicture}} \hspace{2mm}
\begin{tikzpicture}[line width=1.4pt, scale=1]
	\draw[fermionbar] (1.2,0)--(0,0);
	\draw[fermion] (-1.2,0)--(0,0);
	\draw[vector] (0,1.6)--(0,0);
	
	\node[vtxsquare] at (0,0) {};
	
	\node at (1.4,0) {$e$};
	\node at (-1.4,0) {$e$};
	\node at (0,1.8) {$\gamma$};
\end{tikzpicture}

%% file: sections/appendix_muons_v11.tex
\section{Mixed signature for second generation leptoquarks}
\label{app:muons}

\begin{figure}[tb!]
	\centering
	\includegraphics[width=0.48\linewidth]{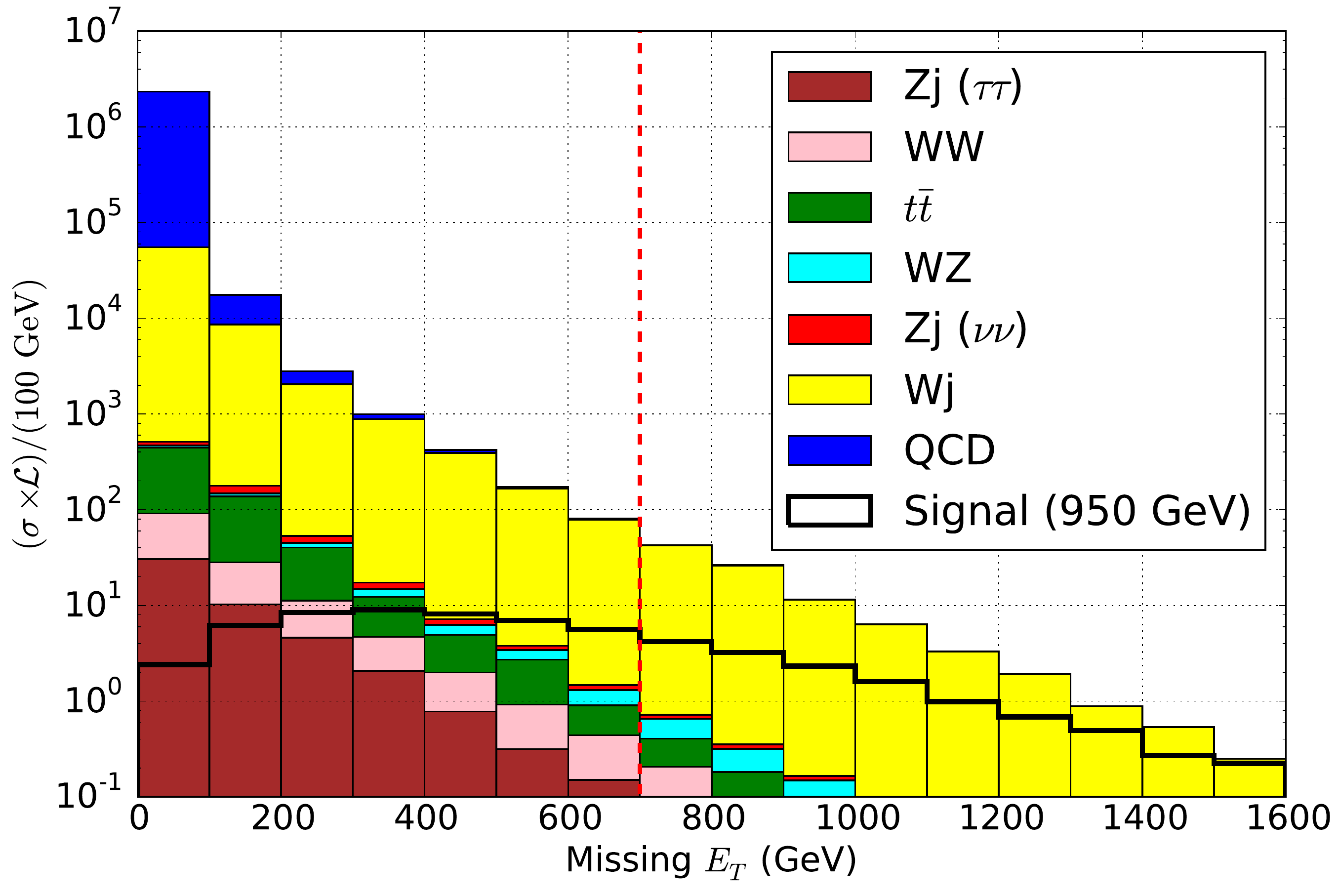}
	\includegraphics[width=0.48\linewidth]{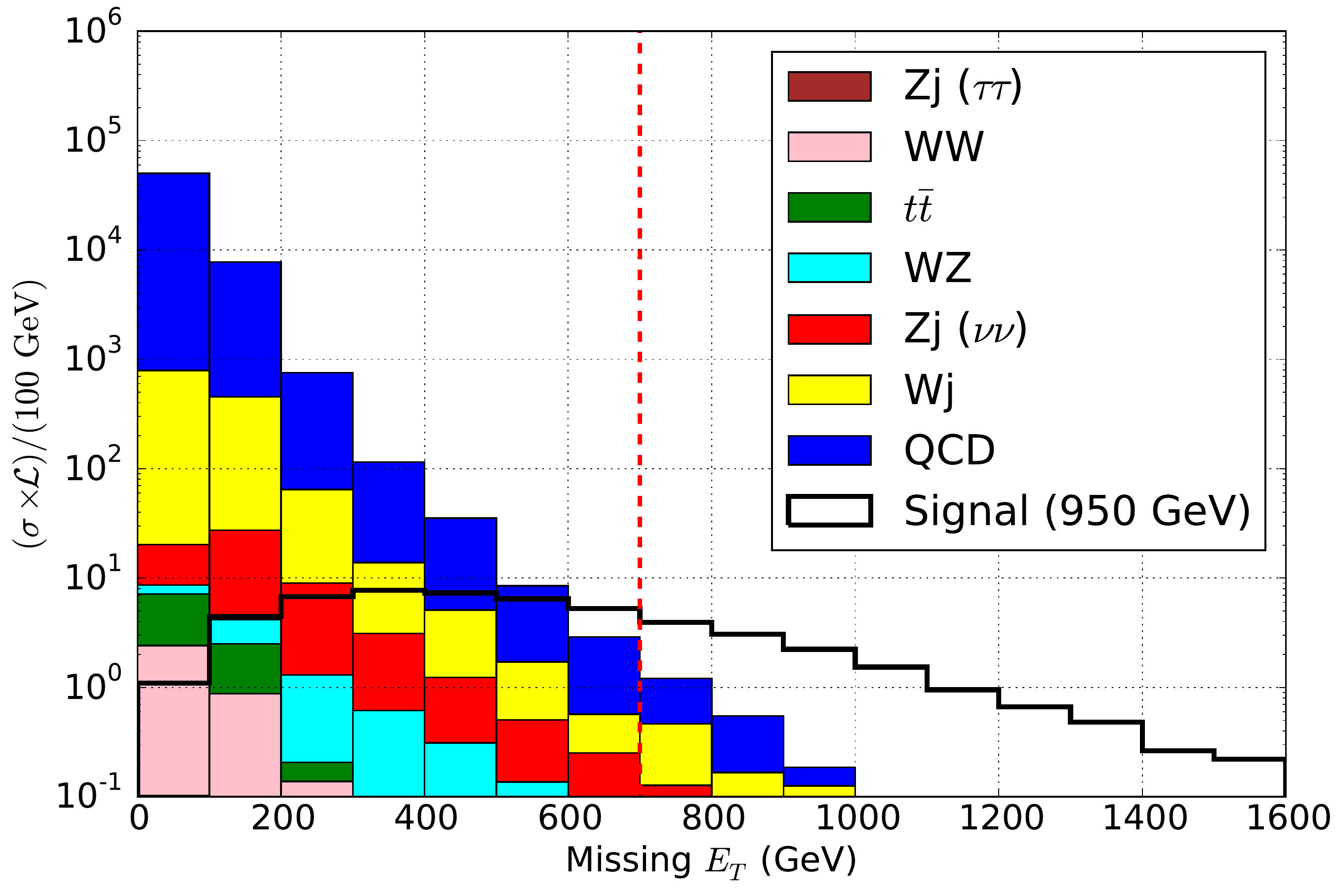}
	\caption{$\slashed{E}_T$ distributions for the background (stacked) and for a scenario with a $405$~GeV DM particle, a $445$~GeV coannihilation partner X, and a $950$~GeV leptoquark mediator. The invariant mass of the leading lepton and the leading jet is required to be within $30$~GeV of the leptoquark mass. On the left, no cut is applied on the transverse mass $m_T$ of the leading lepton and the missing transverse energy, on the right, a $m_T$ cut of $150$~GeV is applied. The red line corresponds to the optimal $\slashed{E}_T > 700$~GeV cut.}
	\label{fig:dismT0_mu}
\end{figure}

\begin{figure}[tb!]
	\centering
	\includegraphics[width=0.6\linewidth]{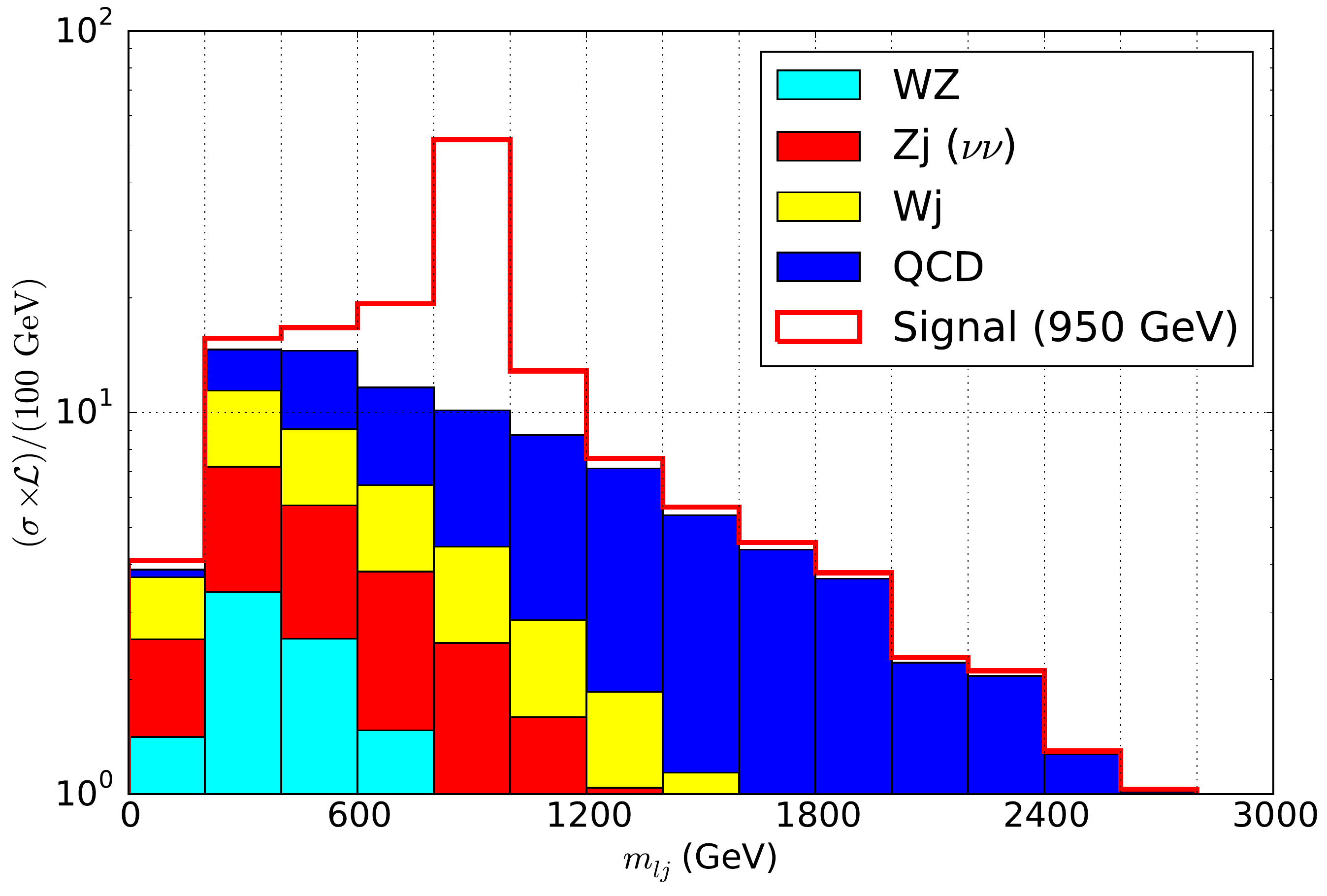}
	\caption{Stacked histogram showing the invariant mass distribution of the leading muon and the leading jet for the total background and a signal scenario with a $405$~GeV DM particle, a $445$~GeV coannihilation partner X, and a $950$~GeV second generation leptoquark.  We cut on $\slashed{E}_T > 700$~GeV and $m_T > 150$~GeV.}
	\label{fig:invm_mu}
\end{figure}

In this section, we study the process $p \, p \to$ LQ LQ $\rightarrow \left(\mu j\right)^{\mathrm{res}} + \mathrm{X \, DM}$. We use the same backgrounds and sets of cuts as in section~\ref{sec:leptoquark}, except we conservatively adopt a value of 0.001 for the probability of a jet to be misidentified as a muon.  We also enlarge the mass window for the muon--jet resonance to $60$ GeV since the muon--jet invariant mass resolution is worse than in the electron--jet case. Table~\ref{tab:cutflow_mu} shows the cut flow for the backgrounds and for a $950$ GeV second generation leptoquark, a 405~GeV DM particle, and $\Delta = 0.1$. Except for the wider leptoquark invariant mass window, the cuts used here are the same as for the first generation leptoquarks studied in section~\ref{sec:leptoquark}.  Figures~\ref{fig:dismT0_mu} and~\ref{fig:invm_mu} show the signal and background event distributions for the same signal as the one in table~\ref{tab:cutflow_mu}. 

\begin{table}[tb!]
	\centering
	\scriptsize
	\input{tables/casestudy_cutflow_mixed_muons_v2}
	\caption{Cut flow for the background and for a $950$~GeV leptoquark with a $405$~GeV DM particle and a $445$~GeV X.  The subscript for the $Z$-boson indicates its decay channel, and ``signal'' refers to the process $p \, p \to$ LQ LQ with a mixed decay topology.  $N^{\mathrm{h}}_{\mu}$ is the number of hard muons, $p_T(\mu) > 30$~GeV.  The mass window cut corresponds to $\left|m_{\ell j} - 950~\mathrm{GeV}\right| < 30$~GeV.  The numbers of events quoted correspond to a center of mass energy of $13$~TeV and a luminosity of $100$~fb$^{-1}$.}
	\label{tab:cutflow_mu}
\end{table}

\begin{figure}[tb!]
	\centering
	\includegraphics[scale=0.6]{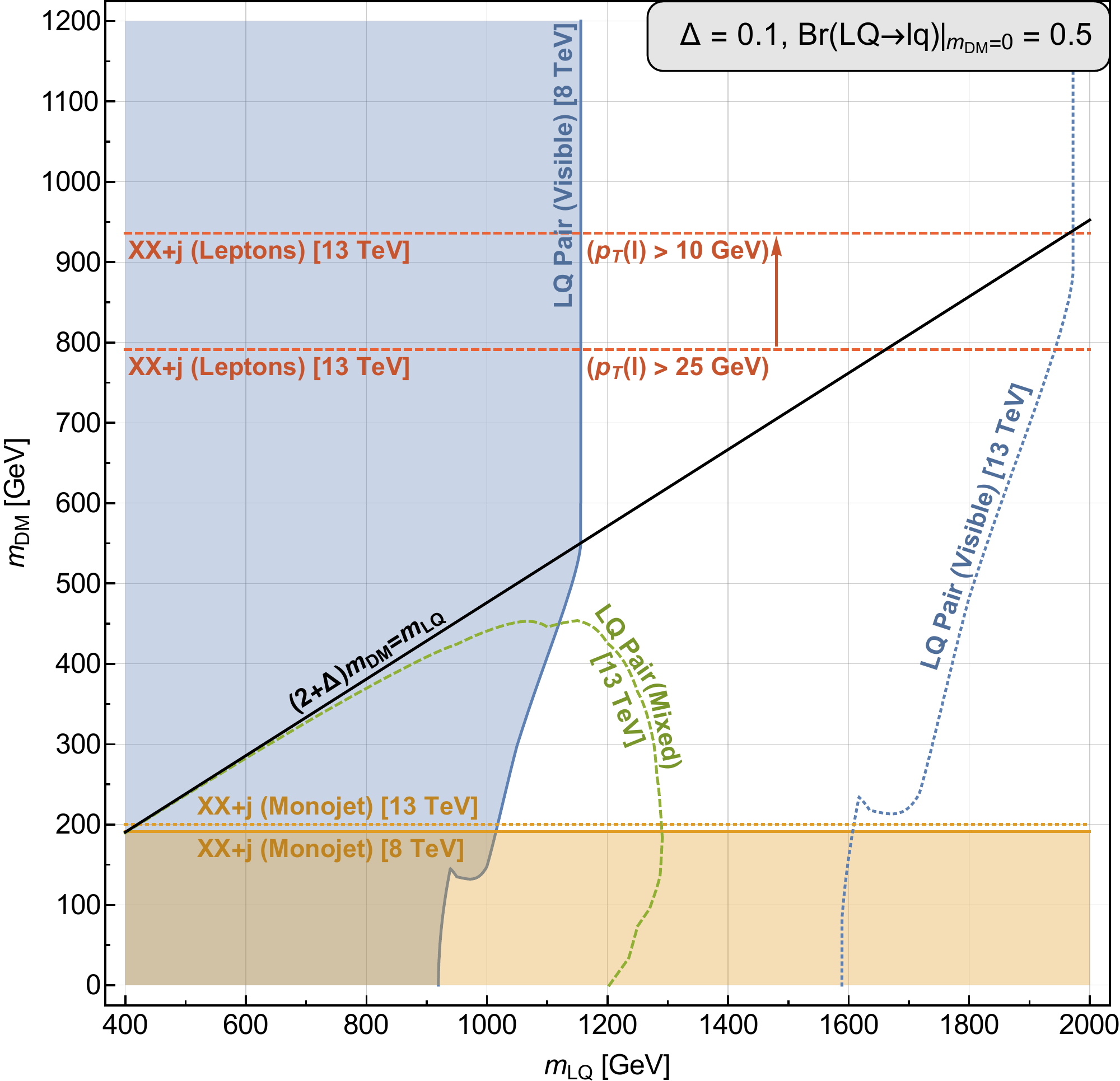}
	\caption{Second generation leptoquark exclusion regions combined with the $2\sigma$ projections for the mixed decay topology and XX + $j$ production in the $m_\mathrm{DM}$ versus $m_\mathrm{LQ}$ plane, with $\Delta = 0.1$ and $B_0 = 0.5$. In the region above the black line, the $\text{LQ} \to \text{X} \, \text{DM}$ decay of the leptoquark is kinematically forbidden. The dashed lines corresponding to the solid regions show the projections for the same searches at $13$ TeV and $100$ fb$^{-1}$.  This figure, which uses a second generation leptoquark coupling for the signal, is the counterpart of figure~\ref{fig:leptoquark:exclusion:projection}, which assumed a first generation leptoquark coupling.}
	\label{fig:leptoquark:exclusion:projection:muons}
\end{figure}

As for the first generation leptoquark, the exclusion bounds for this mixed signature are shown in the $m_\mathrm{DM}$ versus $m_\mathrm{LQ}$ plane in figure~\ref{fig:leptoquark:exclusion:projection:muons}. Although the fake muon rate is lower than the fake electron rate, the bounds for a second generation leptoquark are not significantly different from the bounds on a first generation leptoquark. This is due to the fact that the $\ell j$ invariant mass distribution is less peaked for muons than for electrons and therefore the invariant mass cut is not as efficient. For X~X~$+j$ signatures, the fake lepton rate plays a less important role than in this study and no sophisticated kinematic cut is applied to the leptons. The bounds for first and second generation leptoquarks are then expected to be nearly identical for this signature.

The remaining limits in figure~\ref{fig:leptoquark:exclusion:projection:muons} are similar to the ones for a first generation leptoquark. Pair production of second generation leptoquarks is constrained by analyses from ATLAS~\cite{Aad:2015caa} and CMS~\cite{CMS:zva} and results in similar constraints as obtained in section~\ref{sec:casestudy:current}. The X~X~ $+ j$ cross section is approximately independent of the $y_{Q\ell}$ coupling, since the production rate arises from strong interactions.  In addition, we expect the existing monojet searches to be insensitive to the $y_{Q\ell}$ coupling.  For the search proposed in section~\ref{sec:casestudy:future}, which targets a lepton accompanied by a hard jet and large missing transverse energy, we expect we can use the same $p_T$ thresholds for electrons and muons in order to extract the soft lepton signal.  Hence, the mono-jet + lepton + $\slashed{E}_T$ analysis is also expected to have similar sensitivity for first and second generation leptoquarks.

In contrast to the first generation leptoquark case, though, there is no corresponding atomic parity violation bound for second generation leptoquarks.  Thus, the entire parameter space for $m_{\text{DM}} > 570$~GeV can obtain the dark matter relic density, subject to suitable choices of $y_D$ and $y_{Q\ell}^{22}$.

We remark that an experimental analysis targeting soft muons in an MSSM context has already been performed by CMS~\cite{CMS:2015eoa}.  In this search, the four-body decay of pair-produced supersymmetric tops to soft bottom quarks, muons, neutrinos, and neutralinos has been probed.  This analysis is complementary to the soft lepton analysis we propose in section~\ref{sec:casestudy:future}, since an upper $p_T$ threshold of 30~GeV is used for the soft muons.  We do not perform a recasting of this search to show its sensitivity to our case study, but we expect extending this search will also probe the same parameter space as shown in figure~\ref{fig:leptoquark:exclusion:projection:muons}.

%% file: tables/casestudy_cutflow_mixed_muons_v2.tex
\begin{tabular}{!{\vrule width 1pt}c!{\vrule width 1pt}ccccccccc!{\vrule width 1pt}}
	\noalign{\hrule height 1pt}
	& QCD & $W + 1,2j$ & $t\bar t$ & $Z_{\nu\nu}  + j$ & $Z_{\tau\tau}  + j$ & $W^+W^-$ & $W Z_{\nu \nu} + j$ & $W Z_{jj}$ & signal \\
	\noalign{\hrule height 1pt}
	$p_T(j_1) > 50$ GeV & $2.1 \! \times \! 10^{12}$ & $4.4 \! \times \! 10^8$ & $1.3 \! \times \! 10^8$ & $7.0 \! \times \! 10^7$ & $1.3 \! \times \! 10^7$ & $1.2 \! \times \! 10^6$ &  $1.3 \! \times \! 10^5$ & $3.1 \! \times \! 10^5$ & $600$\\
	$N^{\mathrm{h}}_{\mu} = 1$, $N_{\mu} \le 2$ & $4.8 \! \times \! 10^9$ & $8.8 \! \times \! 10^7$ & $1.2 \! \times \! 10^7$ & $8.6 \! \times \! 10^4$ & $4.8 \! \times \! 10^5$ &  $2.4 \! \times \! 10^5$ & $1.9 \! \times \! 10^4$ & $6.1 \! \times \! 10^4$ & $502$\\
	$b$-jet veto & $4.0 \! \times \! 10^9$ & $8.2 \! \times \! 10^7$ & $5.0 \! \times \! 10^6$ & $8.2 \! \times \! 10^4$ & $4.6 \! \times \! 10^5$ & $2.2 \! \times \! 10^5$ & $1.9 \! \times \! 10^4$ & $5.4 \! \times \! 10^4$ & $360$\\
	$N_{\mathrm{hard\;jets}} \le 3$ & $3.9 \! \times \! 10^9$ & $8.2 \! \times \! 10^7$ & $4.3 \! \times \! 10^6$ & $8.2 \! \times \! 10^4$ & $4.6 \! \times \! 10^5$ & $2.2 \! \times \! 10^5$ & $1.9 \! \times \! 10^4$ & $5.4 \! \times \! 10^4$ & $306$\\
	$Z$ veto & $3.9 \! \times \! 10^9$ & $8.2 \! \times \! 10^7$ & $1.7 \! \times \! 10^6$ & $8.2 \! \times \! 10^4$ & $4.6 \! \times \! 10^5$ & $2.2 \! \times \! 10^5$ & $1.9 \! \times \! 10^4$ & $5.4 \! \times \! 10^4$ & $297$\\
	$\slashed{E}_T > 700$ GeV & $133$ & $1738$ & $15$ & $19$ & $9$ & $10$ & $27$ &$2$ & $62$\\
	$m_T > 150$ GeV & $132$ & $16$ & $10^{-3}$ & $18$ & $0.005$ & $0.01$ & $10$ & $0.001$ & $58$\\
	mass window & $3$ & $0.2$ & $< 10^{-5}$ & $0.3$ & $10^{-5}$ & $10^{-5}$ & $0.1$  &$10^{-5}$ & $13$\\
	\noalign{\hrule height 1pt}
\end{tabular}